\documentclass{article}

\usepackage{amsfonts}
\usepackage{bbm}
\usepackage{amsmath}
\usepackage{amssymb}
\usepackage{calc}
\usepackage{epsfig}
\usepackage{psfrag}
\usepackage{latexsym}

\usepackage{epubtk}
\usepackage{graphicx}

\showlistoftablesfalse

\newcommand{\inter}{{\lrcorner}}
\newcommand{\n}{\nonumber}

\newcommand{\be}{\nopagebreak[3]\begin{equation}}
\newcommand{\ee}{\end{equation}}
\newcommand{\ba}{\nopagebreak[3]\begin{eqnarray}}
\newcommand{\ea}{\end{eqnarray}}
\newcommand{\phys}{\ \ =_{ \left.\right. _{\left.\right._{\!\!\!\!\!\!\!\!\!\!\!\!\!phys}}} \ \ }
\newcommand{\C}{\mathbb{C}}

\newcommand{\N}{\mathbb{N}}
\newcommand{\va}{\scriptscriptstyle}
\newcommand{\vani}{\scriptstyle}
\newcommand{\R}{\mathbb{R}}

\newcommand{\Hhp}{{\cal H}_{phys}}

\newcommand{\Hk}{{\cal H}_{ kin}}
\newcommand{\PP}{{ P}}

\newcommand{\Pp}{{\mathbbm P}(\gamma;+)}
\newcommand{\Pm}{{\mathbbm P}(\gamma;-)}
\newcommand{\Pmd}{\dot{\mathbbm P}(\gamma;-)}															
\newcommand{\Ppm}{{\mathbbm P}(\gamma;\pm)}

\newcommand{\Hm}{{\mathbbm H}(\gamma;-)}
\newcommand{\Hpm}{{\mathbbm H}(\gamma;\pm)}

\newcommand{\Pmp}{{\mathbbm P}(\frac{1}{\gamma};-)}

\newcommand{\Hmp}{{\mathbbm H}(\frac{1}{\gamma};-)}

\newcommand{\op}{{{}^{\va +}{\mathbbm A}}}
\newcommand{\opd}{{{}^{\va +}\dot{\mathbbm A}}}
\newcommand{\opm}{{}^{\va \pm}{\mathbbm A}}
\newcommand{\om}{{}^{\va -}{\mathbbm A}}
\newcommand{\omd}{{}^{\va -}\!\!\dot{\mathbbm A}}

\DeclareFontFamily{U}{rsfs}{}         
\DeclareFontShape{U}{rsfs}{m}{n}{<5> rsfs5 <6><7> rsfs7          %
  <8><9><10><10.95><12><14.4><17.28><20.74><24.88> rsfs10}{}     %
\DeclareMathAlphabet{\mathfs}{U}{rsfs}{m}{n}                     %
\newcommand{\mfs}[1]{\mathfs {#1}}                               %
\newcommand{\van}{\scriptstyle}

\newcommand{\sK}{{\mfs K}}

\newcommand{\sY}{{\mfs Y}}

\newcommand{\sH}{{\mfs H}}
\newcommand{\sL}{{\mfs L}}

\newcommand{\sM}{{\mfs M}}

\newcommand{\sO}{{\mfs O}}

\def\pb#1{\rlap{\lower1.5ex\hbox{$\longleftarrow$}}{#1}}
\def\dpb#1{\rlap{\lower1.5ex\hbox{$\Longleftarrow$}}{#1}}
\def\spb#1{\rlap{\lower1.5ex\hbox{$\leftarrow$}}{#1}}
\def\sdpb#1{\rlap{\lower1.5ex\hbox{$\Leftarrow$}}{#1}}

\newcommand{\bn}{\mathbf{n}}
\newcommand{\bm}{\mathbf{m}}

\newcommand{\la}{\langle}
\newcommand{\ra}{\rangle}

\newcommand{\SU}{\mathrm{SU}}

\newcommand{\Spin}{\mathrm{Spin}}

\newcommand{\ftj}{\mbox{15}j}   

\newcommand{\nb}{\mathbf{n}} 

\newcommand{\rd}{\mathrm{d}}

\def\lsim{\
  \lower-2.0pt\vbox{\hbox{\rlap{$<$}\lower5.5pt\vbox{\hbox{$\sim$}}}}\ }
\def\gsim{\
  \lower-2.0pt\vbox{\hbox{\rlap{$>$}\lower5.5pt\vbox{\hbox{$\sim$}}}}\ }

\begin{document}

\title{The Spin Foam Approach to Quantum Gravity}

\author{
  \epubtkAuthorData{Alejandro Perez}{%
    Centre de Physique Th\'eorique\\
    Unit\'e Mixte de Recherche (UMR 6207) du CNRS et des\\
    Universit\'es Aix-Marseille I, Aix-Marseille II, et du Sud Toulon-Var;\\
    laboratoire afili\'e \`a la FRUMAM (FR 2291)}{%
    perez@cpt.univ-mrs.fr}{%
  }
}

\date{}
\maketitle

\begin{abstract}
This article reviews the present status of the \emph{spin foam
  approach} to the quantization of gravity. Special attention is payed
to the pedagogical presentation of the recently introduced new models
for four dimensional quantum gravity. The models are motivated by a
suitable implementation of the path integral quantization of the Plebanski
formulation of gravity on a simplicial regularization. The article
also includes a self-contained treatment of the 2+1 gravity. The
simple nature of the latter provides  the basis and a perspective for
the analysis of both conceptual and technical issues that remain open
in four dimensions.
\end{abstract}

\epubtkKeywords{quantum gravity}


\newpage
\part{Introduction}
\label{intro}
\clearpage

\section{Quantum Gravity: A Unifying Framework for Fundamental Physics}

The revolution brought by Einstein's theory of gravity lays more
in the discovery of the principle of \emph{general covariance} than
in the form of the dynamical equations of general relativity.
General covariance brings the relational character of nature into
our description of physics as an essential ingredient for the
understanding of the gravitational force. In general relativity
the gravitational field is encoded in the dynamical geometry of
space-time, implying a strong form of universality that precludes
the existence of any non-dynamical reference system -- or
non-dynamical background -- on top of which things occur. This
leaves no room for the old view where fields evolve on an rigid
preestablished space-time geometry (e.g., Minkowski space-time): to
understand gravity one must describe the dynamics of fields with
respect to one another, and independently of any background
structure.

General relativity realizes the requirements of general covariance
as a classical theory, i.e., for $\hbar=0$. Einstein's theory is,
in this sense, incomplete as a fundamental description of nature.
A clear indication of such incompleteness is the generic
prediction of space-time \emph{singularities} in the context of
gravitational collapse. Near space-time \emph{singularities} the
space-time curvature and energy density become so large that any
classical description turns inconsistent. This is reminiscent of
the foundational examples of quantum mechanics -- such as the UV
catastrophe of black body radiation or the instability of the
classical model of the hydrogen atom -- where similar singularities
appear if quantum effects are not appropriately taken into
account. General relativity must be replaced by a more fundamental
description that appropriately includes the quantum degrees of
freedom of gravity.

At first sight the candidate would be a suitable generalization of
the formalism of quantum field theory (QFT). However, the standard
QFT's used to describe other fundamental forces are not
appropriate to tackle the problem of quantum gravity. Firstly,
because standard QFT's are not generally covariant as they can
only be defined if a non-dynamical space-time geometry is
provided: the notion of particle, Fourier modes, vacuum, Poincare
invariance are essential tools that can only be constructed on a
given space-time geometry. This is a strong limitation when it
comes to quantum gravity since the very notion of space-time
geometry is most likely not defined in the deep quantum regime.
Secondly, quantum field theory is plagued by singularities too (UV
divergences) coming from the contribution of arbitrary high energy
quantum processes. This limitation of standard QFT's is expected
to disappear once the quantum fluctuations of the gravitational
field, involving the dynamical treatment of spacetime geometry,
are appropriately taken into account. But because of its
intrinsically background dependent definition, standard QFT cannot
be used to shed light on this issue. A general covariant approach
to the quantization of gravity is needed.

This is obviously not an easy challenge as in the construction of a
general covariant QFT one must abandon from the starting point most of
the concepts that are essential in the description of
`no-gravitational' physics. One has to learn to formulate a quantum
theory in the absence of preferred reference systems or pre-existent
notion of space and time. Loop quantum gravity (LQG) is a framework to
address this task. The degrees of freedom of gravity are quantized in
accordance to the principles of general covariance. At the present
stage there are some indications that  both the singularity problems
of classical general relativity as well as the UV problem of standard
QFT's might vanish in the framework.

However, these indications are not conclusive mainly because the
systematic description of dynamical processes remain an open problem
at the present stage. In this article we review the status of the
construction of a promising approach for handling the difficult
dynamical question in LQG: the \emph{spin foam formulation}.

\subsection{Why non-perturbative quantum gravity?}

Let us make some observations about the problems of standard
perturbative quantum gravity. In doing so we will revisit the
general discussion above, in a special situation. In standard
perturbative approaches to quantum gravity one attempts to
describe the gravitational interaction using the same techniques
applied to the definition of the standard model. As these
techniques require a notion of non dynamical background one
(arbitrarily) separates the degrees of freedom of the
gravitational field in terms of a background geometry $\eta_{ab}$
for $a,b=1\cdots 4$ -- fixed once and for all -- and dynamical
metric fluctuations $h_{ab}$. Explicitly, one writes the spacetime
metric as
\begin{equation}
\label{one}
g_{ab} = \eta_{ab}+h_{ab}.
\end{equation}
Notice that the previous separation of degrees of freedom has no
intrinsic meaning in general relativity. In other words, for a generic
spacetime metric $g_{ab}$ we can write
\begin{equation}
g_{ab}=\eta_{ab}+h_{ab} = \tilde \eta_{ab}+\tilde h_{ab},
\end{equation}
where $\eta_{ab}$ and $\tilde \eta_{ab}$ can lead to different
background light-cone structures of the underlying spacetime
$(M,g_{ab})$; equally natural choices of flat background metrics lead
to different Minkowski metrics in this sense. This is quite dangerous
if we want to give any physical meaning to the background, e.g., the
light cone structures of the two `natural' backgrounds will be
generally different providing different notions of causality!
Equation~(\ref{one}) is used in the classical theory in very special
situations when one considers perturbations of a given background
$\eta_{ab}$. In quantum gravity one has to deal with arbitrary
superpositions of spacetimes; the above splitting can at best be
meaningful for very special semi-classical states `peaked', so to say,
around the classical geometry $\eta_{ab}$ with small fluctuations,
i.e., it might only make sense only \emph{a posteriori} in the
framework of a quantum gravity theory defined independently of any
background geometry. Consequently, even when perturbations should be
remain useful to approximate special situations, the splitting in
Equation~(\ref{one}) is inconsistent when considering general states
with their arbitrary quantum excitations at all scales.

This is specially so in gravity due to the dual role of the
gravitational field that simultaneously describes the geometry and its
own dynamical degrees of freedom. More explicitly, in the standard
background dependent quantization the existence of a fixed background
geometry is fundamental in the definition of the theory. In the usual
treatment one imposes  fields at space-like separated points to
commute alluding to standard causality considerations. Even when this
is certainly justified for matter fields (at least in the range of
applicability of the standard model) it makes no sense in the case of
gravity: one would be using the causal structure provided by the
unphysical background $\eta_{ab}$.  This difficulty has been raised
several times (see, for instance, \cite{Wald:1984rg}).

Therefore, treating the gravitational field according to the splitting
given in Equation~(\ref{one}) is inconsistent with the fundamental
nature of the gravitational degrees of freedom. Treating $h_{ab}$ with
the standard perturbative techniques is equivalent to viewing $h_{ab}$
as another matter field with no special properties. As mentioned
above, gravitons would propagate respecting the causal structure of
the un physical background \emph{to all orders in perturbation
  theory!}\footnote{One can in principle avoid this by carefully devising a theory of perturbations for generally covariant systems. For some ideas on this see \cite{Dittrich:2006ee, Khavkine:2011kj}.} Radiative corrections do not affect causality. Even though
such thing is quite sensible when dealing with matter fields in the
regime where gravity is neglected it would be clearly wrong for the
gravitational interaction.

\epubtkImage{}{%
\begin{figure}
\centerline{\includegraphics[height=5cm]{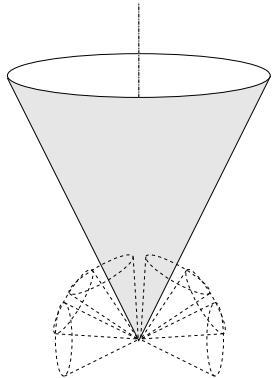}}
\caption{The larger cone represents the light-cone at a point
  according to the ad hoc background $\eta_{ab}$. The smaller cones
  are a cartoon representation of the fluctuations of the true
  gravitational field represented by $g_{ab}$. Gravitons respect the
  causal structure of the un physical metric to \emph{all orders} in
  perturbation theory.}
\label{spinox}
\end{figure}}

\subsection{Non-renormalizability of general relativity}
\label{nrr}

If we ignore all these issues and try to setup a naive
perturbative quantization of gravity we find that the theory is
non renormalizable. This can be expected from dimensional analysis
as the quantity playing the role of the coupling constant turns
out to be the Planck length $\ell_p$.  The non renormalizability
of perturbative gravity is often explained through an analogy with
the (non-renormalizable) Fermi's four fermion effective
description of the weak interaction. Fermi's four
fermions theory is known to be an effective description of the
(renormalizable) Weinberg--Salam theory.  The non renormalizable UV
behavior of Fermi's four fermion interaction is a consequence of
neglecting the degrees of freedom of the exchanged massive gauge
bosons which are otherwise disguised as the dimension-full
coupling $\Lambda_{\va \rm Fermi}\approx 1/m_W^2$ at momentum
transfer much lower than the mass of the $W$ particle
($q^2<<m_W^2$).  A similar view is applied to gravity to promote
the search of a more fundamental theory which is renormalizable or
finite (in the perturbative sense) and reduces to general
relativity at low energies. From this perspective it is argued
that the quantization of general relativity is a hopeless attempt
to quantizing a theory that does not contain the fundamental
degrees of freedom.

These arguments, based on background dependent concepts, seem at the
very least questionable in the case of gravity. Although one should
expect the notion of a background geometry to be useful in certain
semi-classical situations, the assumption that such structure exists
all the way down to the Planck scale is inconsistent with what we know
about gravity and quantum mechanics. General considerations indicate
that standard notions of space and time are expected to fail near the
Planck scale $\ell_p$\epubtkFootnote{For instance a typical example is
  to use a photon to measure distance. The energy of the photon in our
  lab frame is given by $E_{\gamma}=hc/\lambda$. We put the photon in
  a cavity and produce a standing wave measuring the dimensions of the
  cavity in units of the wavelength. The best possible precision is
  attained when the Schwarzschild radius corresponding to energy of
  the photon is of the order of its wavelength. Beyond that the photon
  can collapse to form a black hole around some of the maxima of the
  standing wave. This happens for a value $\lambda_c$ for which
  $\lambda_c\approx G E_{\gamma}/c^4=h G/(\lambda_c c^3)$. The
  solution is $\lambda_c\approx \sqrt{hG/c^3}$ which is Planck
  length.}.  In the field of loop quantum gravity people usually tend
to interpret  the severe divergences of perturbative quantum gravity
as an indication of the inconsistency of the separation of degrees of
freedom in Equation~(\ref{one}). According to concrete results in LQG
the nature of spacetime is very different from the classical notion in
quantum gravity. The treatment that uses Equation~(\ref{one}) as the
starting point is assuming a well defined notion of background
geometry at all scales which directly contradicts these results.

It is possible that new degrees of freedom would become important at
more fundamental scales. It is also possible that including these
degrees of freedom might be very important for the consistency of the
theory of quantum gravity. However, there is a constraint that seem
hardly avoidable: if we want to get a quantum theory that reproduces
gravity in the semi-classical limit we should have a background
independent formalism. In loop quantum gravity one stresses this
viewpoint. The hope is that learning how to define quantum field
theory in the absence of a background is a key ingredient in a recipe
for quantum gravity.

One of the achievements of the background independent formulation of
LQG is the discovered intrinsic discreteness of quantum
geometry. Geometric operators have discrete spectra yielding a
physical mechanism for the cut-off of UV degrees of freedom around the
Planck scale. In this way LQG is free of UV divergences. Is this a
complete answer to the non-renormalizability problem of gravity
mentioned above? Unfortunately, the answer to this question is  in the
negative at this stage. The reason is that the true answer to the
question of renormalizabity or not renormalizability is not in the
presence of divergences but in the degree of intrinsic ambiguity of
the quantization recipe applied to a field theory.

In standard background dependent quantum field theories, in order to
avoid UV divergences one has to provide a regularization prescription
(e.g., an UV cutoff, dimensional regularization, point splitting,
etc.). Removing the regulator is a subtle task involving the tuning of
certain terms in the Lagrangian (counter-terms) that ensure finite
results when the regulator is removed. In fact by taking special care
in the mathematical definition of the products of distributions at the
same point, in good cases, one can provide a definition of the quantum
theory which is completely free of UV
divergences~\cite{Epstein:1973gw} (see also~\cite{Scharf:1996zi,
  Hollands:2001fb, Hollands:2002ux}). However, any of these
regularization procedures is intrinsically ambiguous. The dimension of
the parameter space of ambiguities depends on the structure of the
theory. The right theory must be fixed by comparing predictions with
observations (by the so-called renormalization conditions). In loop
quantum gravity there is strong indications that  the mathematical
framework of the theory provides a regularization of divergences. It
remains to settle the crucial issue of how to fix the associated
ambiguities.
 
According to the previous discussion, ambiguities associated to the UV
regularization allow for the classification of theories as
renormalizable or non nonrenormalizable quantum field theories. In a
renormalizable theory such as QED there are finitely many ambiguities
which can be fixed by a finite number of renormalization conditions,
i.e., one selects the suitable theory by appropriate tuning of the
ambiguity parameters in order to match observations. In a
nonrenormalizable theory (e.g., perturbative quantum gravity) the
situation is similar except for the fact that there are infinitely
many parameters to be fixed by renormalization conditions. As the
latter must be specified by observations, a non-renormalizable theory
has very limited predictive power. In the case of gravity there is
evidence indicating that the theory might be non-perturbatively
renormalizable coming from the investigations of non trivial fix
points of the renormalization group flow in truncated models
(see~\cite{Reuter:2007rv} and references therin).

Removing UV divergences by a regularization procedure and ambiguities
are two sides of the same coin quantum field theory. Although this can
happen in different ways in particular formulations, the problem is
intrinsic to the formalism of quantum field theory. In this respect,
it is illustrative to analyze the non-perturbative treatment of gauge
theories in the context of lattice gauge theory (where the true theory
is studied by means of a regulated theory defined on a space-time
discretization or lattice). It is well known that here too the
regulating procedure leads to ambiguities; the relevance of the
example resides in the fact that these ambiguities resemble in nature
those appearing in loop quantum gravity. More precisely, let us take
for concreteness $SU(2)$ Yang--Mills theory which can be analyzed
non-perturbatively using the standard (lattice) Wilson action
\be
S_{\va \rm LYM}=\frac{1}{g^2}\sum_{p} \left(1-\frac{1}{4}{\rm
  Tr}[U_{p}+U_{p}^{\dagger}]\right).
\label{two}
\ee
In the previous
equation $U_{p}\in SU(2)$ is the holonomy around plaquettes $p$,
and the sum is over all plaquettes of a regulating (hyper-cubic)
lattice. It is easy to check that the previous action approximates
the Yang-Mills action when the lattice is shrunk to zero for a
fixed smooth field configuration.  This property is referred to as
the \emph{naive continuum limit}. Moreover, the quantum theory
associated to the previous action is free of any UV problem due to
the UV cut-off provided by the underlying lattice.

Is this procedure unique? As it is well known the answer is no: one can regulate
Yang--Mills theory equally well using the following action instead
of Equation~(\ref{two}):
\be S^{(m)}_{\va \rm LYM}\propto
\frac{1}{g^2}\sum_{p} \left(1-\frac{1}{2(2m+1)}{\rm Tr}^{\va
(m)}[\Pi^{\va (m)}(U_{p})+\Pi^{\va (m)}
(U_{p}^{\dagger})]\right),
\ee
where $\Pi^{\va (m)}(U_{p})$ denotes
to the $SU(2)$ unitary irreducible representation matrix (of spin
$m$) evaluated on the plaquette holonomy $U_{p}$. Or more
generally one can consider suitable linear combinations \be S_{\va
\rm LYM}=\sum_{m}\ a_m \ S^{(m)}_{\va \mathrm{LYM}}. \ee From the view
point of the classical continuum theory all these actions are
equally good as they all satisfy the \emph{naive continuum limit}.
Do these theories approximate in a suitable sense the continuum
quantum field theory as well? and are these ambiguities
un-important in describing the physics of quantum Yang--Mills
theory? The answer to both of these questions is yes and the
crucial property that leads to this is the renormalizability of
Yang-Mills theory. Different choices of actions lead indeed to
different discrete theories.  However, in the low energy effective
action the differences appear only in local operators of dimension
five or higher. This simple dimensional argument leads to the expectation that in the
continuum limit (i.e., when the regulating lattice dependence is
removed by shrinking it to zero) all the above theories should produce 
the same predictions in the sense that can safely ignore
non-renormalizable contributions. This is indeed the case 
in usual cases, and thus the ambiguities at
the {`microscopic level'} do not have any effect at low energies
where one recovers quantum Yang--Mills theory.

Similar ambiguities are present in LQG, they appear in the
regularization of the constraints in the canonical
formulation~\cite{Perez:2005fn}. These ambiguities are also present in
the spin foam formulation which is the central subject of this
review. The simple dimensional argument given for Yang--Mills theory
is no longer valid in the case of gravity, and a full dynamical
investigation of their effect in physical predictions must be
investigated. The spin foam approach is best suited for getting the
necessary insights into this important question.

\section{Why Spin Foams?}

We now provide the basic motivation for the study and definition of
spin foams. We present this motivation from the perspective of the
canonical quantization approach known as \emph{loop quantum gravity}
(see~\cite{Rovelli:2004tv, bookt, ash10}). In such context spin foams appear as
the natural tool for studying the dynamics of the canonically defined
quantum theory from a covariant perspective. In order to introduce the
spin foam approach it is therefore convenient to start with a short
introduction of the quantum canonical formulation.

\subsection{Loop quantum gravity}

Loop quantum gravity is an attempt to define a quantization of gravity
paying special attention to the conceptual lessons of general
relativity. The theory is explicitly formulated in a background
independent, and therefore, non perturbative fashion. The theory is
based on the Hamiltonian (or canonical) quantization of general
relativity in terms of variables that are different from the standard
metric variables. In terms of these variables general relativity is
cast into the form of a background independent $SU(2)$ gauge theory
whose phase space structure is similar in various ways to that of
$SU(2)$ Yang--Mills theory (the key difference being the absence of
any background metric structure).  The main prediction of loop quantum
gravity (LQG) is the discreteness~\cite{lee1} of the spectrum of
geometrical operators such as area and volume. The discreteness
becomes important at the Planck scale while the spectrum of geometric
operators crowds very rapidly at `low energy scales' (large
geometries). This property of the spectrum of geometric operators is
consistent with the smooth spacetime picture of classical general
relativity.

Thus, from the perspective of LQG, it is not surprising that
perturbative approaches would lead to inconsistencies. In splitting
the gravitational field degrees of freedom as in Equation~(\ref{one})
one is assuming the existence of a background geometry which is smooth
all the way down to the Planck scale. As we consider contributions
from `higher energies', this assumption is increasingly inconsistent
with the fundamental structure discovered in the non perturbative
treatment. However, despite of the many  achievements of LQG there
remain important issues to be addressed. At the heart of the
completion of the definition of the theory the clear-cut definition of
the  quantum dynamics remains open. The spin foam approach is one of
the main avenues to exploring this problem.

The background independence of general relativity implies that the
canonical formulation of the field theory is that of a gauge theory
with diffeomorphism as part of the gauge group. LQG is constructed by
quantizing a phase space formulation of general relativity in terms of
$SU(2)$ connection variables. This introduces an extra $SU(2)$ gauge
symmetry group. The presence of gauge symmetries implies the existence
of relations among phase space variables -- defined on a spacelike
initial value hypersurface -- known as constraints. These constraints
define the Poisson algebra of infinitesimal generators of gauge
transformations. There are three local constraints $G^i$ -- the Gauss
constraints -- generating $SU(2)$ gauge transformations, three local
constraints $V_a$ -- the vector constraints -- generating three
dimensional diffeomorphisms of the initial spacelike hypersurface, and
finally a scalar local constraint $S$ related to the remaining gauge
symmetry related to the four-diffeomorphism symmetry of the Lagrangian
formulation.
   
The canonical quantization of systems with gauge symmetries is often
called the Dirac program. The Dirac
program~\cite{dirac,Henneaux:1992ig} applied to the quantization of
general relativity in connection variables leads to the LQG approach.

The first step in the recipe  consists in finding a representation of
the phase space variables of the theory as operators in a kinematical
Hilbert space $\Hk$ satisfying the standard commutation relations,
i.e., $\{\ ,\ \}\rightarrow -i/\hbar [\ ,\ ] $. This step has been
succesfully completed in LQG. One chooses the polarization where the
$SU(2)$ connection is the configuration variable. Unconstrained phase
space variables are replaced by the so-called holonomy-flux algebra
which is represented by associated operators in a kinematical Hilbert
space of  suitable  functionals of the generalized-connection
$\psi[A]$ which are square integrable with respect to the so-called
Ashtekar--Lewandowski~\cite{ash3} (gauge invariant and diffeomorphism
invariant) measure $d\mu_{\va AL}[A]$.  A key input is the use of the holonomy flux algebra as a starting point for quantization. 
Many peculiar properties of LQG follow from this choice this choice is motivated by 
having an observable algebra leading to simple (diffeomorphism covariant) Poisson brackets \footnote{Interestingly, and as a side remark, the structure of the phase space of gravity with boundaries used in the description of concrete physical models seem to provide an extra justification of the use of the holonomy flux  variables (see section IV E in \cite{Engle:2010kt} for a more detailed discussion of this intriguing fact).}.

The kinematical inner product
is given by
\begin{equation}
\label{alm}
<\psi,\phi>=\mu_{AL}[\overline \psi \phi]=\int d\mu_{\va AL}[A]\
\overline \psi[A] \phi[A].
\end{equation}
The representation of the basic unconstrained phase space variables as
suitable operators in $\Hk$ used in LQG -- which has the additional
key property of containing a special state that is diffeomorphism
invariant -- has been shown to be unique~\cite{lost}.

The next step is to promote the constraints to (self-adjoint)
operators in $\Hk$ in such a way that the classical Poisson algebra is
respected by the appropriate quantum commutator algebra (if this last
part is achieved the quantization is said to be non-anomalous). In the
case of gravity one must quantize the seven constraints $G_i$, $V_a$,
and $S$. Both the Gauss constraint and (finite) diffeomorphism
transformations have a natural (unitary) action on states on
$\Hk$. For that reason the quantization (and subsequent solution) is
rather direct. The quantization of the scalar constraint presents
difficulties. Concrete quantizations producing well defined operators
are available: no UV divergences are encountered by these proposals;
the fundamental discreteness of quantum geometry in LQG plays the role
physical regulator at the Planck scale~\cite{Thiemann:1997rt}. Despite
of this partial success, problems partly related to special features
of the constraint algebra (field dependent structure constant) and
partly related to the non-existence of an infinitesimal generator of
diffeomorphism at the quantum  level make the issue of whether any of
the proposed quantizations of the constraints are anomaly-free a
difficult open question.

In the third step one needs to characterize the space of solutions of
the constraints and define the corresponding inner product that
defines a notion of physical probability. This defines the so-called
physical Hilbert space ${\cal H}_{\va ph}$.  In LQG  physical states
are those who satisfy the quantum constraints which, in this sense,
could be called {Quantum Einstein's equations}. More precisely $\Psi
\in {\cal H}_{\va ph}$ if
\begin{eqnarray}
\label{diracy}\nonumber
&& \widehat G_i(A,E)|\Psi> 
=0\\
&&\nonumber\widehat V_a(A,E)|\Psi> 
=0,\\ &&\widehat S(A,E)|\Psi> 
=0.
\label{QEE}
\end{eqnarray}
The space of solutions of the first six equations is well
understood. The space of solutions of quantum scalar constraint
remains open. For some mathematically consistent definitions of
$\widehat S$ the characterization of the solutions is well
understood~\cite{ash10}. The definition of the physical inner product
is still an open issue. The completion of this step is intimately
dependent on the previously mentioned consistency of the quantum
constraint algebra.

Finally,  physical interpretation necessitates the definition of a
(complete) set of gauge invariant observables, i.e., operators
commuting with the constraints. They represent the questions that are
well-posed in a generally covariant quantum theory. Already in
classical gravity the construction of gauge independent quantities is
a subtle issue. At the present stage of the approach physical
observables are explicitly  known only in some special
cases. Understanding the set of physical observables is however
intimately related with the problem of characterizing the solutions of
the scalar constraint described before.

The spin foam approach was constructed as a means to tackle the
difficult question of dynamics and the definition of observable
quantities in LQG. It is an attempt to address the difficulties
associated last three steps of the list given above  by rethinking the
problem of dynamics from a path integral covariant  perspective. So we
need first to briefly discuss the special features of the path
integral  formulation in the case of generally covariant systems.

\subsection{The path integral for generally covariant systems}
\label{valin}

\begin{figure}[h]\!\!\!\!\!\!
\centerline{\hspace{0.5cm} \(
\begin{array}{c}
\includegraphics[height=5cm]{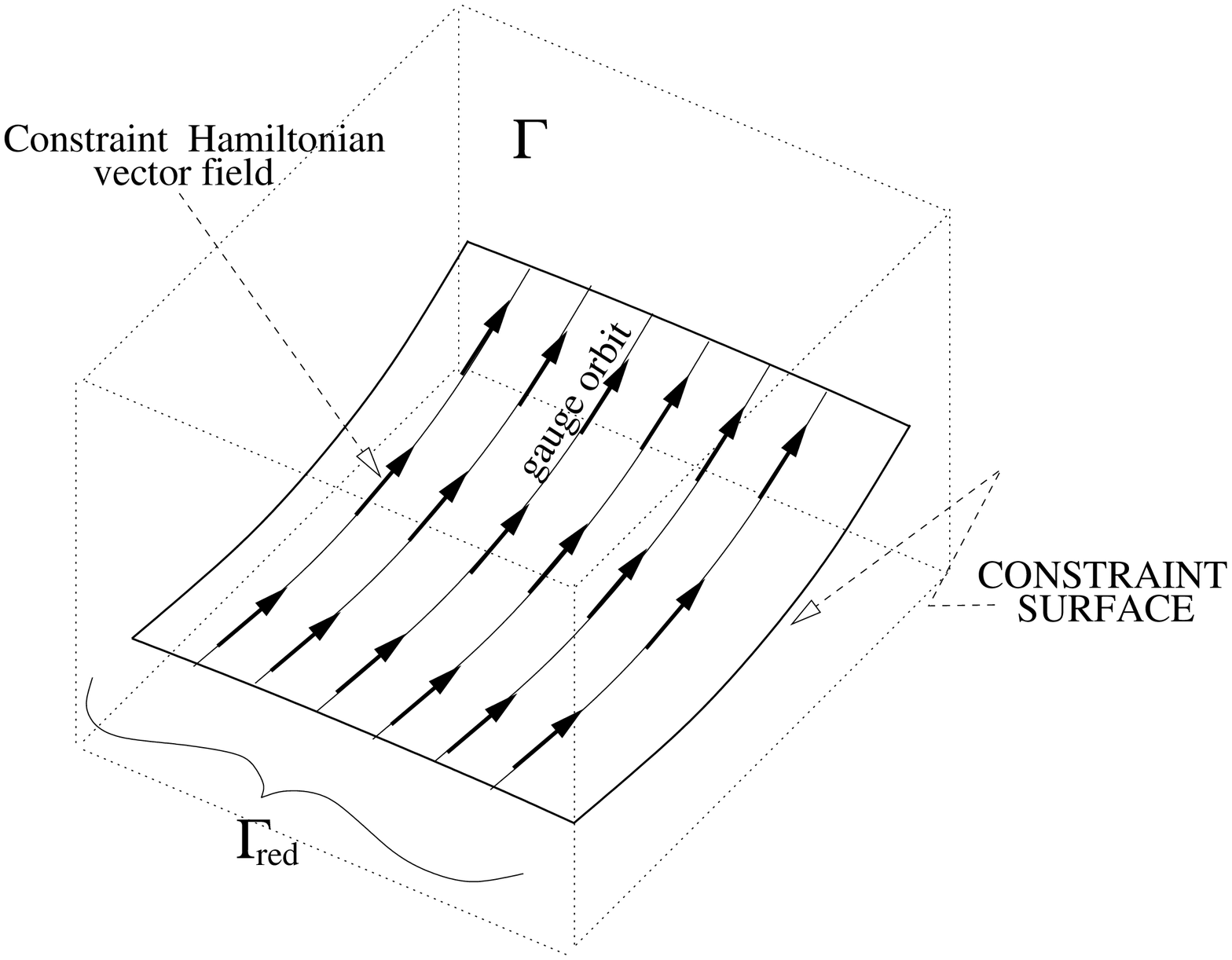}
\end{array}\ \ \ \begin{array}{c}
\includegraphics[height=3cm]{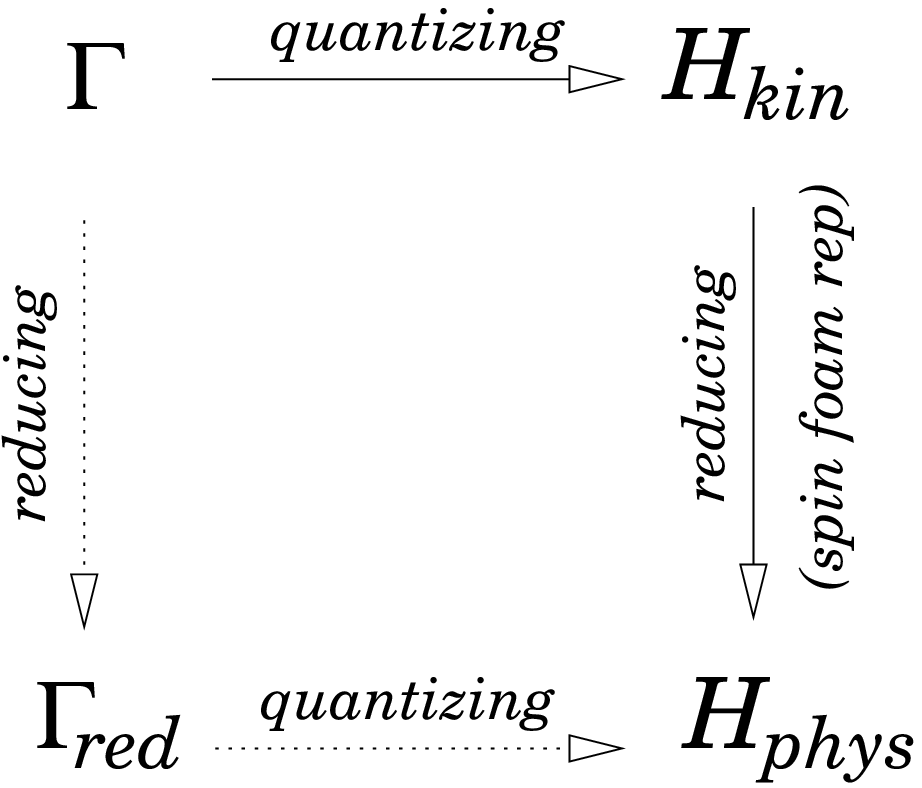}
\end{array}\) } \caption{On the left: the geometry of phase space in gauge
  theories. On the right: the quantization path of LQG (continuous arrows).}
\label{phase}
\end{figure}

LQG is based on the canonical (Hamiltonian) quantization of general
relativity whose gauge symmetry is diffeomorphism invariance. In the
Hamiltonian formulation the presence of gauge symmetries~\cite{dirac}
gives rise the constraints mentioned in the previous section. In the
present discussion we will schematically represent the latter by the
equation $C(p,q)=0$ for $(p,q)\in \Gamma$. The constraints restrict
the set of possible states of the theory by requiring them to lay on
the constraint hyper-surface. In addition, through the Poisson
bracket, the constraints generate motion associated to gauge
transformations on the constraint surface (see
Figure~(\ref{phase})). The set of physical states (the so called
reduced phase space $\Gamma_{red}$) is isomorphic to the space of
orbits, i.e., two points on the same gauge orbit represent the same
state in $\Gamma_{red}$ described in different gauges
(Figure~\ref{phase}).

In general relativity the absence of a preferred notion of time
implies that the Hamiltonian of gravity is a linear combination of
constraints. This means that Hamilton equations cannot be interpreted
as time evolution and rather correspond to motion along gauge orbits
of general relativity. In generally covariant systems conventional
time evolution is pure gauge: from an initial data satisfying the
constraints one recovers a spacetime by selecting a particular
one-parameter family of gauge-transformations (in the standard ADM
context this amounts to choosing a particular lapse field $N(x,t)$ and
shift $N^a(x,t)$).

From this perspective the notion of spacetime becomes secondary and
the dynamical interpretation of the the theory seems problematic (in
the quantum theory this is refered to as the \emph{``problem of
time''}).  A reason for this is the central
role played by the spacetime representation of classical gravity
solutions useful only if can be interpreted via the notion of \emph{test observers} (or more
generally \emph{test fields})\epubtkFootnote{Most (if not all) of the applications of
general relativity make use of this concept together with the
knowledge of certain exact solutions. In special situations there are
even preferred coordinate systems based on this notion which greatly simplify
interpretation (e.g., co-moving observers in cosmology, or
observers at infinity for isolated systems).}. Due to the fact that
this idealization is a good approximation to the (classical) process
of observation the notion of spacetime is a useful concept only in
classical gravity.

As emphasized by Einstein with his hole argument (see~\cite{Rovelli:2004tv} for
a modern explanation) only the information in relational statements
(independent of any spacetime representation) have physical
meaning. In classical gravity it remains useful to have a spacetime
representation when dealing with idealized test observers. For
instance to solve the geodesic equation and then ask
\emph{diff-invariant-questions} such as: {what is the proper time
  elapsed on particle 1 between two succesive crossings with particle
  2?}  However, already in the classical theory the advantage of the
spacetime picture becomes, by far, less clear if the test particles
are replaced by real objects coupling to the gravitational
field\epubtkFootnote{In this case one would need first to solve the
  constraints of general relativity in order to find the initial data
  representing the self-gravitating objects. Then one would have
  essentially two choices: 1) Fix a lapse $N(t)$ and a shift $N^a(t)$,
  evolve with the constraints, obtain a spacetime (out of the data) in
  a particular gauge, and finally ask the
  \emph{diff-invariant-question}; or 2) try to answer the question by
  simply studying the data itself (without t-evolution). It is far
  from obvious whether the first option (the conventional one) is any
  easier than the second.}.

The possibility  of using the notion of test observers and test fields
to construct our interpretative framework is no longer available in
quantum gravity. At the Planck scale ($\ell_p$) the quantum
fluctuations of the gravitational field become so important that there
is no way (not even in principle\epubtkFootnote{In order to make a
  Planck scale observation we need a Planck energy probe (think of a
  Planck energy photon). It would be absurd to suppose that one can
  disregard the interaction of such photon with the gravitational
  field treating it as test photon.}) to make observations without
affecting the gravitational field. The notion of test fields and test
observer is lost and with it the usual spacetime representation. In
this context there cannot be any, \emph{a priori}, notion of time and
hence no notion of spacetime is possible at the fundamental level. A
spacetime picture would only arise in the semi-classical regime with
the identification of some subsystems that approximate the notion of
test observers.

What is then the meaning of the path integral in such background
independent context? The previous discussion rules out the
conventional interpretation of the path integral. There is no
meaningful notion of transition amplitude between states at different
times $t_1>t_0$ or equivalently a notion of \emph{``unitary time
  evolution''} represented by an operator $U(t_1-t_0)$. Nevertheless,
a path integral representation of generally covariant systems exists
and arises as a tool for implementing the constraints in the quantum
theory as we argue below.

Due to the difficulty associated with the explicit description of the
reduced phase space $\Gamma_{\rm red}$, in LQG one follows Dirac's
prescription. One starts by quantizing unconstrained phase space
$\Gamma$, representing the canonical variables as self-adjoint
operators in a \emph{kinematical Hilbert space} $\Hk$. Poisson
brackets are replaced by commutators in the standard way, and the
constraints are promoted to self-adjoint operators (see
Figure~\ref{phase}). If there are no anomalies the Poisson algebra of
classical constraints is represented by the commutator algebra of the
associated quantum constraints. In this way the quantum constraints
become the infinitesimal generators of gauge transformations in
$\Hk$. The physical Hilbert space $\Hhp$ is defined as the kernel of
the constraints, and hence to \emph{gauge invariant} states. Assuming
for simplicity that there is only one constraint we have
\[\psi \in \Hhp\ \ \ {\rm iff}\ \ \
\exp[iN \hat C]|\psi\rangle=|\psi\rangle\ \ \
\forall \ \ \ N\in\R,
\]
where $U(N)=\exp[iN \hat C]$ is the unitary operator associated to the
gauge transformation generated by the constraint $C$ with parameter
$N$. One can characterize the set of gauge invariant states, and hence
construct $\Hhp$, by appropriately defining a notion of `averaging'
along the orbits generated by the constraints in $\Hk$. For instance
if one can make sense of the \emph{projector}
\be P:\Hk \rightarrow
\Hhp \ \ \ {\rm where } \ \ \ P:=\int dN\ U(N).
\label{pipi}
\ee
It is apparent from the definition that for any $\psi\in \Hk$ then
$P\psi\in\Hhp$. The path integral representation arises in the
representation of the unitary operator $U(N)$ as a sum over
\emph{gauge-histories} in a way which is technically analogous to
standard path integral in quantum mechanics. The physical
interpretation is however quite different. The \emph{spin foam
  representation} arises naturally as the path integral representation
of the field theoretical analog of $P$ in the context of LQG. Needles
is to say that many mathematical subtleties appear when one applies
the above formal construction to concrete examples
(see~\cite{Marolf:2000iq} and references therein). In the second part
of this review we will illustrate in  complete detail all the features
of the path integral approach for generally covariant systems,
formally described here, in the 2+1 pure gravity solvable model. In
four dimension the spinfoam approach is a program in progress where
various issues still remain open. We will describe the various models
in the first part of this article.

\subsection{A brief history of spin foams in four dimensions}
\label{sfm4d}

In this section, we briefly describe the various spin foam models for
quantum gravity in the literature (for previous reviews on spin foams
see~\cite{ori2, baez5, Perez:2004hj, Perez:2003vx, Livine:2010zx,
  Alexandrov:2010un}).

\subsubsection{The Reisenberger model}
\label{Reise}

According to Plebanski~\cite{pleb} the action of self dual Riemannian
gravity can be written as a constrained $SU(2)$ BF theory
\begin{equation}
\label{pleb00}
S(B,A)=\int {\rm Tr}\left[B\wedge F(A)\right]-\psi_{ij}\left[
B^i\wedge B^j-\frac{1}{3}\delta^{ij} B^{k}\wedge B_k\right],
\end{equation}
where variations with respect to the symmetric (Lagrange multiplier)
tensor $\psi_{ij}$ imposes the constraints
\begin{equation}
\label{forty}
\Omega^{ij}=B^i\wedge B^j-\frac{1}{3}\delta^{ij} B^{k}\wedge
B_k=0.
\end{equation}
When $B$ is non degenerate the constraints are satisfied if and
only if $B^i=\pm(e^0\wedge e^i+{\vani \frac{1}{2}}\epsilon^i_{\
jk}e^{j}\wedge e^k)$ which reduces the previous action to that of
self-dual general relativity. Reisenberger studied the simplicial
discretization of this action in~\cite{reis6} as a preliminary
step toward the definition of the corresponding spin foam model.
The consistency of the simplicial action is argued by showing that
the simplicial theory converges to the continuum formulation when
the triangulation is refined: both the action and its variations
(equations of motion) converge to those of the continuum theory.

In reference~\cite{reis4}, Reisenberger constructs a spin foam
model for this simplicial theory by imposing the constraints
$\Omega^{ij}$ directly on the $SU(2)$ BF amplitudes. The spin foam
path integral for BF theory is obtained as in Section~\ref{sfm3d}.
The constraints are imposed by promoting the $B^i$ to operators
${\chi}^i$ (combinations of left/right invariant vector fields)
acting on the discrete connection\epubtkFootnote{Notice that (for
example) the right invariant vector field ${\cal J}^{i}(U)=
{\sigma}^{i\ A }_{\ B}U^{B}_{\  C} {\partial}/{\partial U^{A}_{\
C}}$ has a well defined action at the level of
Equation~(\ref{coloring}) and acts as a B operator at the level
of~(\ref{Zdiscrete}) since
\begin{eqnarray}\label{RIV}
-i{\chi}^{i}(U)\left[ e^{i{\rm Tr}[BU]}\right]|_{U\sim 1} = {\rm
Tr}[\sigma^{i}UB] e^{i{\rm Tr}[BU]}|_{U\sim 1}\sim B^{i}e^{i{\rm
Tr}[BU]},
\end{eqnarray}
where $\sigma^{i}$ are Pauli matrices.}. The model is defined as
\begin{equation}
Z_{GR}={\underbrace{\int \ \ \prod_{e \in {\cal J}_{\va \Delta}}
dg_e}_{\int {\cal D}[A]} } \ \overbrace{\delta ( \hat \Omega^{ij} )}^{\int {\cal D}[\psi] \  e^{i \psi_{ij}\Omega^{ij} }}\!\!
\underbrace{\sum \limits_{{\cal C}:\{j\} \rightarrow \{ f\}}
\prod_{f \in {\cal J}_{\va \Delta}} \Delta_{j_f} \ {\rm
Tr}\left[j_f(g^1_e\dots g^{\va N}_e)\right]}_{\int {\cal D}[B] \
e^{i\int {\rm Tr}[B\wedge F(A)]}},
\end{equation}
where $\hat \Omega={\cal J}^i\wedge {\cal
J}^j-\frac{1}{3}\delta^{ij} {\cal J}^{k}\wedge {\cal J}_k $ and we
have indicated the correspondence of the different terms with the
continuum formulation. The preceding equation is rather formal; for
the rigorous implementation see~\cite{reis4}. Reisenberger uses
locality so that constraints are implemented on a single
4-simplex amplitude. There is however a difficulty with the
this procedure: the algebra of operators $\hat \Omega^{ij}$ do
not close so that imposing the constraints sharply becomes a too
strong condition on the BF configurations\epubtkFootnote{This
  difficulty also arises in the Barrett--Crane model as we shall see
  in Section~\ref{BCM}.}. In order to avoid this, Reisenberger defines
a one-parameter family of models by inserting the operator
\begin{equation}
\label{sw}
e^{-\frac{1}{2z^2}\hat \Omega^2}
\end{equation}
instead of the delta function above. In the limit $z\rightarrow
\infty$ the constraints are sharply imposed. This introduces an
extra parameter to the model. The properties of the kernel of
$\hat \Omega$ have not been studied in detail.

\subsubsection{The Freidel--Krasnov prescription}
\label{Freide}

Freidel and Krasnov~\cite{fre5} define a general framework to
construct spin foam models corresponding to theories whose action has
the general form
\begin{equation}
S(B,A)=\int {\rm Tr}\left[B\wedge F(A)\right] + \Phi(B),
\end{equation}
where the first term is the BF action while $\Phi(B)$ is a certain
polynomial function of the $B$ field. The formulation is
constructed for compact internal groups. The definition is based on
the formal equation
\begin{equation}
\int {\cal D}[{B}]{\cal D}[{ A}]\  e^{i\int {\rm Tr}\left[B\wedge F(A)\right] + \Phi(B)}:=
\left.e^{i\int \Phi(\frac {\delta}{\delta J})}   Z[J]\right|_{J=0},
\end{equation}
where the {\em generating functional} $Z[J]$ is defined as
\begin{equation}
Z[J]:=\int {\cal D}[{B}]{\cal D}[{A}] \ e^{i\int {\rm Tr}\left[B\wedge F(A)\right] + {\rm Tr}\left[B\wedge J\right]},
\end{equation}
where $J$ is an algebra valued 2-form field. They provide a
rigorous definition of the generating functional by introducing a
discretization of $\cal M$ in the same spirit of the other spin
foam models discussed here. Their formulation can be used to
describe various theories of interest such as BF theories with
cosmological terms, Yang--Mills theories (in 2 dimensions) and
Riemannian self-dual gravity. In the case of self dual gravity $B$
and $A$ are valued in $su(2)$, while
\begin{equation}
\Phi(B)= \int \psi_{ij}\left[ B^i \wedge B^j - \frac{1}{3} \delta^{ij} B^k \wedge B_k\right]
\end{equation}
according to Equation~(\ref{pleb00}). The model obtained in this way
is very similar to Reisenberger's one. There are however some
technical differences. One of the more obvious one is that the
non-commutative invariant vector fields ${\cal J}^i$ representing $B^i$ are replaced
here by the commutative functional derivatives $\delta/\delta
J^{i}$. 

The idea of using such generating functional techniques have regain
interest in the context of the so called McDowell--Mansouri
formulation of general relativity~\cite{MacDowell:1977jt}. The
interest of the latter formulation is that it provides an action of
gravity that is given by a BF term plus a genuine potential term
instead of constraints. The implementation of the spin foam
quantization of such formulation is investigated
in~\cite{Starodubtsev:2005mf, Smolin:2003qu, Freidel:2005ak}, see
also~\cite{Rovelli:2005qb} for an important remark on the
approach. More recently similar techniques have been used
in~\cite{Mikovic:2011fr}.

\subsubsection{The Iwasaki model}

Iwasaki defines a spin foam model of self dual Riemannian
gravity\epubtkFootnote{Iwasaki defines another model involving
  multiple cellular complexes to provide a simpler representation of
  wedge products in the continuum action. A more detail presentation
  of this model would require the introduction of various
  technicalities at this stage so we refer the reader to~\cite{iwa2}.}
by a direct lattice discretization of the continuous Ashtekar
formulation of general relativity. The action is
\begin{equation}
S(e,A)=\int dx^4\ \epsilon^{\mu \nu \lambda \sigma}\left[2\
e^{0}_{[\mu} e_{\nu ] i} + \epsilon^{0}_{\ ijk}
e^{j}_{\mu}e^{k}_{\nu} \right]\left[ 2\ \partial_{[\lambda}
A_{\sigma ]}^i + \epsilon^{0i}_{\ \ lm}
A^{l}_{\lambda}A^{m}_{\sigma}\right],
\end{equation}
where $A^i_a$ is an $SU(2)$ connection. The fundamental
observation of~\cite{iwa1} is that one can write the discrete
action in a very compact form if we encode part of the degrees of
freedom of the tetrad in an $SU(2)$ group element. More precisely,
if we take $g_{\mu\mu}=e^{i}_{\mu}e^{j}_{\mu}\delta_{ij}=1$ we can
define ${\mathbf e}_{\mu}:=e_{\mu}^0+i\sigma_i e^i_{\mu}\in SU(2)$
where $\sigma_i$ are the Pauli matrices. In this parameterization
of the `angular' components of the tetrad and using a hypercubic
lattice the discrete action becomes
\begin{equation}
S_{\Delta}=-\beta \sum_{v\in \Delta}
\epsilon^{\mu\nu\lambda\sigma} {r_{\mu} r_{\nu}} \ {\rm
Tr}\left[{\mathbf e}^{\dagger}_{\mu}{\mathbf e}_{\nu} U_{\lambda
\sigma} \right],
\end{equation}
where $r_{\mu}:=(\beta^{1/2} \ell_p)^{-1}\epsilon
\sqrt{g_{\mu\mu}}$, $U_{\mu \nu}$ is the holonomy around the
$\mu\nu$-plaquette, $\epsilon$ the lattice constant and $\beta$ is
a cutoff for $r_{\mu}$ used as a regulator ($r_{\mu}\le
\beta^{1/2}\ell_p \epsilon^{-1}$). The lattice path integral is
defined by using the Haar measure both for the connection and the
`spherical' part of the tetrad $\mathbf e$'s and the radial part
$dr_{\mu}:= dr_{\mu} r^{3}_{\mu}$. The key formula to obtain an
expression involving spin foams is
\begin{equation}
e^{i{x}{\rm Tr}[U]}=\sum_j (2j+1) \frac{J_{2j+1}(2x)}{x} \chi_j(U).
\end{equation}
Iwasaki writes down an expression for the spin foam amplitudes in
which the integration over the connection and the $\mathbf e$'s
can be computed explicitly. Unfortunately, the integration over
the radial variables $r$ involves products of Bessel functions and
its behavior is not analyzed in detail. In 3 dimensions the
radial integration can be done and the corresponding amplitudes
coincide with the results of Section~\ref{sec:qbf}.

\subsubsection{The Barrett--Crane model}

The appealing feature of the previous models is the clear connection
to loop quantum gravity, since they are defined directly using the
self dual formulation of gravity (boundary states are $SU(2)$-spin
networks). The drawback is the lack of closed simple expressions for
the amplitudes which complicates their analysis. There is however a
simple model that can be obtained as a systematic quantization of
simplicial $SO(4)$ Plebanski's action. This model was introduced by
Barrett and Crane in~\cite{BC2} and further motivated by Baez
in~\cite{baez7}. The basic idea behind the definition was that of the
\emph{quantum tetrahedron} introduced by Barbieri in~\cite{barb2} and
generalized to 4d in~\cite{baez6}. The beauty of the model resides in
its remarkable simplicity. This has stimulated a great deal of
explorations and produced many interesting results.

\subsubsection{Markopoulou--Smolin causal spin networks}
\label{fotin}

Using the kinematical setting of LQG with the assumption of the
existence of a micro-local (in the sense of Planck scale) causal
structure Markopoulou and Smolin define a general class of (causal)
spin foam models for gravity~\cite{fot1,fot2} (see
also~\cite{fot3}). The elementary transition amplitude
$A_{s_I\rightarrow s_{I+1}}$ from an initial spin network $s_{I}$ to
another spin network $s_{I+1}$ is defined by a set of simple
combinatorial rules based on a definition of causal propagation of the
information at nodes. The rules and amplitudes have to satisfy certain
causal restrictions (motivated by the standard concepts in classical
Lorentzian physics). These rules generate surface-like excitations of
the same kind we encounter in the more standard spin foam model but
endow the foam with a notion of causality. Spin foams ${\cal
  F}^{N}_{s_i\rightarrow s_{f}}$ are labeled by the number of times
these elementary transitions take place. Transition amplitudes are
defined as
\begin{equation}
\left<s_i,s_f\right>=\sum_{N} \prod \limits^{N-1}_{I=0} A_{s_I\rightarrow s_{I+1}}.
\end{equation}
The models are not related to any continuum action. The only
guiding principles are the restrictions imposed by causality,
simplicity and the requirement of the existence of a non-trivial
critical behavior that would reproduce general relativity at large
scales. Some indirect evidence of a possible non trivial continuum
limit has been obtained in some versions of the model in 1+1
dimensions.

\subsubsection{Gambini--Pullin model}

Gambini and Pullin~\cite{pul2} introduced a very
simple model obtained by modification of the $BF$ theory skein
relations. As we argued in Section~\ref{sfm3d} skein relations
defining the physical Hilbert space of BF theory are implied by
the spin foam transition amplitudes. These relations reduce the
large kinematical Hilbert space of $BF$ theory (analogous to that of
quantum gravity) to a physical Hilbert space corresponding to the
quantization of a finite number of degrees of freedom. Gambini and
Pullin define a model by modifying these amplitudes so that some
of the skein relations are now forbidden. This simple modification
frees local excitations of a field theory. A remarkable feature
is that the corresponding physical states are (in a certain sense)
solutions to various regularizations of the scalar constraint for
(Riemannian) LQG. The fact that physical states of BF theory solve
the scalar constraint is well known~\cite{th6}, since roughly
$F(A)=0$ implies $EEF(A)=0$. The situation here is of a similar
nature, and -- as the authors argue -- one should interpret this
result as an indication that some `degenerate' sector of quantum
gravity might be represented by this model. The definition of this
spin foam model is not explicit since the theory is directly
defined by the physical skein relations.

\subsubsection{Capovilla--Dell--Jacobson theory on the lattice}

The main technical difficulty that we gain in going from
3-dimensional general relativity to the 4-dimensional one is
that the integration over the $e$'s becomes intricate. In the
Capovilla--Dell--Jacobson~\cite{cap1,cap2} formulation of general
relativity this `integration' is partially performed at the
continuum level. The action is
\begin{equation}
S(\eta,A)=\int \eta {\rm Tr}\left[\epsilon \cdot F(A) \wedge F(A) \epsilon \cdot F(A)\wedge F(A) \right],
\end{equation}
where $\epsilon \cdot F \wedge F:=\epsilon^{abcd}F_{ab}
F_{cd}$. Integration over $\eta$ can be formally performed in the path
integral and we obtain
\begin{equation}
Z=\int \prod_x \delta\left({\rm Tr}\left[\epsilon \cdot F(A) \wedge F(A) \epsilon \cdot F(A)\wedge F(A) \right]\right),
\end{equation}
One serious problem of this formulation is that it corresponds to a
sector of gravity where the Weyl tensor satisfy certain algebraic
requirements. In particular flat geometries are not contained in this
sector.

\subsubsection{The Engle--Pereira--Rovelli--Livine (EPRL)}

A modification of the Barrett--Crane model was recently introduced
in~\cite{Engle:2007uq, Engle:2007qf}  and extended for arbitrary
Immirzi parameter in~\cite{Engle:2007wy}. The basic idea was to relax
the imposition of the Plebanski constraints that reduce BF theory to
general relativity  in the quantum theory.  The anomalous commutation
relations of the $B$\footnote{The six components $B^{IJ}$ are associated to invariant vector fields in the Lorentz group as a direct consequence of the discretization
procedure in spin foams. This makes them non commutative and render the simplicity constraints (functionals of the $B$ field alone) non commutative. The origin of the non commutativity of $B$ fields is similar to the non commutativity of fluxed in canonical LQG \cite{Ashtekar:1998ak}.} field in the quantum theory imply that the
commutation of the  Plebanski constraints does not define a closed
algebra. Imposing the constraints strongly as in the Barrett--Crane
model implies the imposition of additional conditions that are not
present in the classical theory. There is a natural way to relax the
constraints and this leads to a simple model that has more clear
relationship with the canonical picture provided by LQG. The detailed
description of this model will be the main subject of the following
discussion of spin foam models for general relativity in 4 dimensions.

\subsubsection{Freidel--Krasnov (FK)}

A very similar set of models were independently introduced by Freidel
and Krasnov in~\cite{Freidel:2007py}. Indeed these models are
arguably the same as the EPRL model for a suitable range of the
Immirzi parameter (the vertex amplitudes coincide for $\gamma<1$ as
we will discuss in the sequel). However, the logic  used in the
derivation is different. The idea is to express the BF path integral
in terms of the coherent state representation in order to impose the
Plebanski constraints semiclassically  in terms of expectation
values. The  coherent intertwiner basis  relevant for the definition
of  spin foams and the 4d quantum gravity amplitude was also derived
by Livine and Speziale in~\cite{Livine:2007vk}. Freidel and Krasnov
introduced the linear version of Plebanski constraints that where
later used in order to provide a simpler derivation and generalization
of the EPRL model to arbitrary Immirzi parameter.

The last two models are going to be the subject of most of our
discussions in what follows. They are the most developed candidates in
four dimensions and we will review their properties in detail in the
following part.

\clearpage 

\part{Preliminaries: LQG and the canonical quantization of 4d gravity}
\label{cano}

In this part we briefly review the basic results of the loop quantum
gravity approach to the canonical quantization of gravity. This part
is relevant for the interpretation of the new spin foams models
presented in the next part. The reader interested in the canonical
formulation of general relativity in terms of connection variables is
referred to the text books~\cite{Rovelli:2004tv, bookt, ash} and the review
article~\cite{ash10}. For a pedagogical introduction
see~\cite{Perez:2004hj}.

In the following section we review the canonical analysis of general
relativity formulated as a constrained BF theory, i.e., in  the
so-called Plebanski formulation~\cite{pleb}. The study is done in
terms of the variables that, on the one hand, allow for the closest
comparison with the discrete variables used in the spin foam approach,
while, on the other hand, allow  for the introduction the basic
elements of the canonical quantization program of loop quantum
gravity. The latter being, in my view, the clearest possible setting
in which spin foams are to be interpreted.

In Section~\ref{lqg} we present the basic elements of loop quantum
gravity (LQG). In Section~\ref{sec:intsf} we give a short description
of the spin foam representation of the path integral of
gravity. Concrete realizations of this will be reviewed in 4d in
Part~\ref{thenew} and in 3d in Part~\ref{part3d}.

\clearpage
\section{Classical General Relativity in Connection Variables}
\label{canocal}

The Hamiltonian analysis of general relativity is the basic starting point for canonical quantization.
Loop quantum gravity and spin foams are based on the possibility of formulating Hamiltonian general 
relativity in terms of Yang--Mills-like connection variables. The primitive ancestor of these formulations is
Ashtekar's self-dual \emph{complex} connection formulation of general relativity~\cite{ash1}. Modern, LQG and spin foams are 
based on a certain relative of these variables that is often referred to as \emph{ Ashtekar--Barbero} variables~\cite{barbero, immi}. 
These variables can be introduced directly by means of a canonical transformation of the phase space of the Palatini formulation
of general relativity~\cite{Thiemann:2007zz}, or can be derived directly from the canonical analysis of the so called Holst action for 
general relativity~\cite{Holst:1995pc}. More simply the new variables also follow from the addition of the Nieh--Yan topological invariant to the
Palatini action~\cite{Date:2008rb}. The Ashtekar--Barbero connection parametrization of the phase space of general relativity also arises 
naturally from the consideration of the most general diffeomorphism invariant action principle that can be written for the field content of 
the Palatini first order formulation~\cite{Rezende:2009sv} (i.e. a Lorentz connection $\omega^{IJ}_a=-\omega^{JI}_a$ and a co-tetrad $e^I_a$ with $a$ 
spacetime indeces and $I,J=0,1,2,3$ internal Lorentz indeces).
 
The new spin foam models are based on Plebanski's realization that one can obtain classical  general relativity by suitable constraining
the variations of the $B$ field in the simple kind of topological theory called BF theory~\cite{pleb}. For that reason, 
the best suited action principle of gravity for the study of the new spin foam models corresponds to one of the Plebanski type.
However, the details of such treatment are only partially presented in the literature (see for instance~\cite{Gielen:2010cu},~\cite{Wieland:2010ec}).
The material of the following subsection is meant to complete this void, the study is largely inspired in the perspective
exploited in references~\cite{monvel09,vel11}  for the canonical formulation of constrained BF theory.  As we will show below, the hamiltonian analysis of such formulation of gravity is at the heart of the replacement of the 
Plebanski quadratic constraints by linear ones that has been so useful in the definition of the new vertex amplitudes~\cite{Freidel:2007py}.

\subsection{Gravity as constrained BF theory}

In order to keep the presentation simple (and not to bother with the $\pm$ signs appearing in dealing with raising and lowering of Lorentz indices) we present here
the Hamiltonian formulation of Riemannian gravity. This is enough to illustrate the algebraic structures that are necessary to introduce 
in the present context\epubtkFootnote{I am grateful to M.~Montesinos and M.~Velazquez for discussions and contributions to the present section.}. 

Our starting point is Plebanski's Riemannian action for general relativity which  can be thought of as $Spin(4)$
BF theory plus constraints on the $B$-field.
It depends on an $so(4)$ connection
$\omega$, a Lie-algebra-valued 2-form $B$ and Lagrange multiplier
fields $\lambda$ and $\mu$. Writing explicitly the Lie-algebra
indexes, the action is given by
\begin{equation}\label{pleb}
S[B,\omega,\lambda]=\frac{1}{\kappa}\int_{\sM} \left[({}^{\star}B^{IJ}+\frac{1}{\gamma}   B^{IJ})\wedge F_{IJ}(\omega) + \lambda_{IJKL}
\ B^{IJ} \wedge B^{KL} \right],
\end{equation}
where $\mu$ is a 4-form and
$\lambda_{IJKL}=-\lambda_{JIKL}=-\lambda_{IJLK}=\lambda_{KLIJ}$ is
a tensor in the internal space satisfying  $\epsilon^{IJKL}\lambda_{IJKL}=0$, and $\gamma$ is the Immirzi parameter. The Lagrange multiplier tensor $\lambda_{IJKL}$
has then 20 independent components. The previous action is closely related to the one introduced in~\cite{merced1, vel11, monvel22} (there is a simple analog of the previous action in 3d~\cite{Bonzom:2008tq}).  Variation with respect to
$\lambda$ imposes 20 algebraic equations on the 36 components
of $B$. They are
\begin{equation}\label{ito}
 \epsilon^{\mu\nu\rho\sigma} B^{IJ}_{\mu\nu}B^{KL}_{\rho\sigma}=e \ \epsilon^{IJKL}
\end{equation}
where
$e=\frac{1}{4!}\epsilon_{OPQR}B^{OP}_{\mu\nu}B^{QR}_{\rho\sigma}\epsilon^{\mu\nu\rho\sigma}$
\cite{fre6}. The  solutions to these equations are
\begin{equation}\label{ambi}
B=\pm {}^*( e \wedge e), \ \ \ {\rm and}\ \ \ B=\pm e\wedge e,
\end{equation}
in terms of the $16$ remaining degrees of freedom of the tetrad
field $e^I_a$. If one substitutes the first solution into the
original action one obtains Holst's formulation of general
relativity~\cite{Holst:1995pc}
\begin{equation}\label{pala}
S[e,\omega]=\frac{1}{\kappa}\int_{\sM} {\rm Tr}\left[({}^{\star}e\wedge e +\frac{1}{\gamma} e\wedge e) \wedge F(A)\right].
\end{equation}
This property is the key to the definition of 
the spin foam model for gravity of the next section.

\subsection{Canonical analysis} \label{canolys}

Now one  performs the usual 3+1 decomposition by assuming the existence of a global time function $t$ 
whose level hypersurfaces $\Sigma_t$ define a foliation of the spacetime manifold $\sM$. The previous action takes 
 the following form
\ba S[\Pi,H,\omega,\lambda]=\int dt \int_{\Sigma_t} {\rm Tr}[\Pi\wedge  \dot \omega + \omega_{0}\  D_{\omega}\wedge \Pi + H\wedge  F + \lambda^{\prime}
\ (\gamma {}^{\star} H-H)\wedge(\gamma {}^{\star} \Pi-\Pi)],
\ea
where
\ba\n
\Pi_{ab}^{IJ}&=&{}^{\star} B_{ab}^{IJ}+\frac{1}{\gamma}   B_{ab}^{IJ} \\ 
H_{a}^{IJ}&=&{}^{\star} B_{0a}^{IJ}+\frac{1}{\gamma}   B_{0a}^{IJ},\label{defidifi}
\ea
 for $a,b=1,2,3$ are  $\Sigma_t$-cotangent-space abstract indices, $\lambda^{\prime}={\gamma^2}/{(1-\gamma^2)^2} \lambda$, and the trace is the obvious contraction of internal indices. 
By choosing an internal direction $n^I$ (and in the gauge where $n^I=(1,0,0,0)$) we can now introduce a convenient reparametrization of the unconstrained phase space by writing the 18 canonical pairs $(\Pi^{IJ}_{ab},\omega^{KL}_{c})$ by $(\Ppm^i_{ab},\opm^j_{c})$  
where
\ba \label{newmom}  {\Ppm}^i_{ab}&\equiv&
\frac{1}{4}\epsilon^{i}_{\ jk} \Pi_{ab}^{jk}\pm
\frac{1}{2\gamma}\Pi_{ab}^{0i}, \ \ \ \ \ \  \opm_c^{ j} =  \frac{1}{2} \epsilon^{j}_{\ m
    n} \omega_c^{m n}\pm {\gamma}\ \omega_c^{ j0} . \ea 
 For notational convenience we will also introduce
\ba \nonumber {\Hpm}^i_{a}&\equiv&
\frac{1}{4}\epsilon^{i}_{\ jk} H_{a}^{jk}\pm
\frac{1}{2\gamma}H_{a}^{0i}. \ea     
We can rewrite the previous action as
\ba && \n S[\Ppm,\opm, \lambda]=\int dt \int_{\Sigma} \Pp_i \wedge  \opd^i+ \Pm_i \wedge   \omd^i +N_i \  {\mathbbm G}^i+\eta_i \ {\mathbbm B}^i+\\ && + H\wedge  F + \lambda^{\prime}
\ (\gamma {}^{\star}H-H)\wedge(\gamma {}^{\star}\Pi-\Pi)],\label{axi}
\ea
where $N^i$, $\eta^i$, and $\lambda$ are Lagrange multipliers imposing the constraints:
\ba\n &&
{\mathbbm B}^i=(D_{\frac{(\op)+(\om)}{2}}\wedge \Pp^i )\approx 0\\
\n && {\mathbbm G}^i= \frac{1}{2\gamma}\epsilon^{ijk} (\op-\om)_j \wedge \Pp_k\approx 0,\\ 
&& \n \left. \begin{array} {lll}
 (\frac{\gamma}{2}\epsilon^{i}_{\ jk}H^{jk} -H^{0i})\wedge (\frac{\gamma}{2}\epsilon^{i}_{\ jk}\Pi^{jk}-\Pi^{0i})+(i \leftrightarrow j)\approx 0\\
 \frac{1}{4}\epsilon^{(p}_{\ ij} (\frac{\gamma}{2}\epsilon^{ij}_{\ k}H^{0k}-H^{ij})\wedge \epsilon^{q)}_{\ lm}(\frac{\gamma}{2}\epsilon^{lm}_{\ s}\Pi^{0s}-\Pi^{lm})\approx 0 \\
(\frac{\gamma}{2}\epsilon^{i}_{\ jk}H^{jk} -H^{0i})\wedge (\frac{\gamma}{2}\epsilon^{lm}_{\ s}\Pi^{0s}-\Pi^{lm})+\\ +(\frac{\gamma}{2}\epsilon^{lm}_{\ k}H^{0k}-H^{lm})\wedge (\frac{\gamma}{2}\epsilon^{i}_{\ jk}\Pi^{jk}-\Pi^{0i})-v \epsilon^{0ilm} \approx 0
\end{array}\right\}  \epsilon^{\mu\nu\rho\sigma} B^{IJ}_{\mu\nu}B^{KL}_{\rho\sigma}-e \ \epsilon^{IJKL}\approx 0. \ \ 
  \ea
 In the previous equation we explicitly recall that the last three equations are a rewriting of the simplicity constraints (\ref{ito}).
We can rewrite them in terms of $\Pm$ and $\Pmp$ as
\ba \n &&
{\mathbbm B}^i=D_{\frac{(\om)+(\op)}{2}}\wedge \Pp^i \approx 0,\\
\n  && {\mathbbm G}^i = \frac{1}{2\gamma}\epsilon^{ijk} (\op-\om)_j \wedge \Pp_k\approx 0,\\
\n&& {\mathbbm I}^{ij}=\Hm^{(j}\wedge \Pm^{i)}\approx 0,\\
\n  &&{\mathbbm {II}}^{ij}= \Hmp^{(i} \wedge \Pmp^{j)}\approx 0,\\
\n  &&{\mathbbm {III}}^{ij}=\Hm^i \wedge \Pmp^j+\Hmp^{i} \wedge \Pm^j\\ && \ \ \ \ \ \  -\frac{1}{3} [\Hm_k \wedge \Pmp^k+\Hmp_{k} \wedge \Pm^k] \delta^{ij} \approx 0.
\ea
The 20 constraints ${\mathbbm I}^{ij}$, ${\mathbbm {II}}^{ij}$, and ${\mathbbm {III}}^{ij}$ are referred to as the \emph{simplicity constraints}.
The general solution of the simplicity constraints (\ref{ambi}) requires, in the 3+1 setting, the introduction of four new fields: the lapse $N$ and the shift vectors $N^a$. The solution is 
\ba &&
\Pm^i_{ab}=0,\n \\ \n && \Pmp^i_{ab}=\frac{(1-\gamma^2)}{2} \Pp^i_{ab},\\ \n &&
\Hm^{i}_{a}=-N\ \frac{(1-\gamma^2)}{2\gamma^2} e_a^i(\Pp),\\ \n
&& \Hmp_a^i=\frac{(1-\gamma^2)}{2} N^c \Pp_{ca}^i,\\
&& {\rm with} \n \\ &&
e_a^i(\Pp)=\frac{\gamma^{1/2}}{8} \frac{\epsilon_{abc}\epsilon^{i}_{jk} \epsilon^{bde}\epsilon^{bfg} \Pp^j_{de}\Pp_{fg}^k }{\sqrt{\det(\Pp)}}. \label{patatra} \ea
This solution breaks the internal Lorentz gauge as the first condition  
above does not commute with the boost constraint ${\mathbbm B}^i$: as it will become clear below it  amounts to choosing the  so-called
\emph{time gauge}. The above equations parametrize the solutions of the Plebanski constraints in terms of the 9 components of $\Pp^i_{ab}$ plus 
4 extra parameters given by a scalar $N$ and a space tangent vector $N^a\in T(\Sigma)$ and the 3 parameters in the choice of an internal direction $n^I$ (here implicitly taken as $n^I=(1,0,0,0)$). These are exactly the 16 parameters in the co-tetrad $e_{\mu}^I$ in (\ref{ambi}). 

The components of $H$ that are necessary to write the term $H\wedge F$ in the action (\ref{axi}) are\ba\n &&
H^{0i}_{a}=\gamma \Pp^{i}_{ca}N^c+\frac{N}{\gamma}e_a^i\\ \n
&& \frac{1}{2}\epsilon^k_{\ ij} H^{ij}_{a}=\Pp^{k}_{ca}N^c+N e_a^i.
\ea
The conservation in time of the set of constraints above lead to 
6 secondary constraints. As shown in~\cite{Rezende:2009sv} these follow from 
\be\label{tency}
{\mathbbm C}^{ij}=e^{(j}_{}\wedge \Pmd^{i)}\propto N e^{(j}_{}\wedge \left[d e^{i)}+\frac{1}{2}\epsilon^{i)}_{\ lm} (\op+\om)^l\wedge e^{m}_{}\right]\approx 0
\ee
Notice now that the set of 9 constraints ${\mathbbm B}^i\approx 0$ and ${\mathbbm C}^{(ij)}\approx 0$ are indeed equivalent to
the following simple condition
\be\label{35} 
{\mathbbm {IV}}^i_a=\frac{1}{2}(\op+\om)_a^i-\Gamma^i_{a}(e)\approx 0,
\ee
where  $\Gamma^{i}_a$  is the torsion free spin connection  compatible with the triad $e^i_a$, i.e. the unique  solution of Cartan's structure equations  \begin{equation}\label{cartan}
\partial_{[a} e^i_{b]} + \epsilon^{i}_{\ jk}\Gamma^j_{[a}
e^k_{b]} = 0. \end{equation}
 \vskip.2cm
\noindent {\bf Remark:} \emph{The primary constraint $\Pm\approx 0$ (first line in Equation (\ref{patatra})) will have central 
importance in the definition of the new spin foam models. Explicitly, from (\ref{newmom}), one has
\be\label{piripipi}
{\Pm}^i_{ab} \equiv
\frac{1}{4}\epsilon^{i}_{\ jk} \Pi_{ab}^{jk}-
\frac{1}{2\gamma}\Pi_{ab}^{0i}\approx 0.
\ee This is exactly the linear simplicity constraints that we will rediscover in the simplicial setting of Sections \ref{eprl-r} and \ref{eprl-l}, and which will be imposed 
in a suitable sense at all tetrahedra of the cellular decomposition in order to satisfy the Plebanski constraints and produce a state sum model for gravity out of that of BF theory. The consistency requirement (\ref{tency}) are expected to hold in the discrete setting from the fact that one is imposing the linear simplicity constraint for all tetrahedra, thus $\Pm\approx 0$ is valid for all times.}
\vskip.2cm

Now from  the fact that \be \{{\mathbbm {IV}}^i_a(x),\Pm_{bc}^j(y)\}=\frac{1}{2} \epsilon_{abc}\delta^{ij}\delta(x,y)\ee we conclude that the given pair are second class. We can substitute their solution in the action. In particular, from eqs. (\ref{newmom}) and (\ref{35}),  $\op_a^i$ becomes the Ashtekar--Barbero connection
\be
\op_a^i=\Gamma^i_a+\gamma K_a^i
\ee
where we have defined $K_a^i=(\op^i_a-\om^i_a)/2$.
With all this
\ba && \n \!\!\!\!\!\!\!\!\!\!\!\!\!S[\Pp,\op,N,\vec N,N_i ]=\\ && \!\!\!\!\!\!\!\!\!\!\!\!\!=\frac{1}{\kappa}\int dt \int_{\Sigma} \Pp_i \wedge  \dot\op^i+\n \\  && + N_i \  D_{\op}\wedge \Pp^i+N\inter \Pp^{k}\wedge (\gamma  F_{0i}+F^{jk}\epsilon_{ijk})+N e^i \wedge (\frac{F_{0i}}{\gamma}+F^{jk}\epsilon_{ijk}).\label{axi}
\ea
The previous action corresponds to the standard Hamiltonian formulation of general relativity in terms of $SU(2)$ connection variables.
This becomes more transparent if we now change to a more standard notation by introducing the socalled \emph{densitized triad} \be
E^a_i=\gamma \epsilon^{abc} \Pp^{i}_{bc}
\ee in terms of which the last line of Equation (\ref{patatra}) becomes \begin{equation}\label{twenty} e_a^i=\frac{1}{2}\frac{\epsilon_{abc}
\epsilon^{ijk} E_j^b E_k^c}{\sqrt{|{\rm
    det}(E)|}} \ \ \ {\rm and} \ \ \ e^a_i=\frac{{\rm sgn}({\rm
    det}(E))\ E^a_i }{\sqrt{|{\rm
    det}(E)}|}.
\end{equation} 
From now on we denote the Ashtekar--Barbero\cite{Barbero:1995ud}
 connection
$A_a^i$ given by \begin{equation}\label{immi} A_a^i=\op_a^i=\Gamma_a^i +\gamma K_a^i.
\end{equation} The Poisson brackets of the new variables are \begin{equation}
\left\{E^{a}_j(x),A_{b}^i(y)\right\}=\kappa \, \gamma
\delta^a_{b}\delta^i_{j} \delta(x,y), \ \ \ \
\left\{E^{a}_j(x),E^{b}_i(y)\right\}=\left\{A_{a}^j(x),A_{b}^i(y)\right\}=0.
\end{equation} All the previous equations follow explicitly from (\ref{axi})
except for $\left\{A_{a}^j(x),A_{b}^i(y)\right\}=0$ which follows from the special property of
$\Gamma_a^i$ in three dimensions that
\begin{equation}\label{19} \Gamma^i_a=\frac{\delta W[E]}{\delta E^a_i}, \end{equation}
where $W[E]$ is the generating functional for the spin connection\epubtkFootnote{In
fact equation (\ref{cartan}) implies that $\Gamma^i_a[E]=\Gamma^i_a[\lambda
E]$ for a constant $\lambda$. This homogeneity property plus the existence of
$W[E]$ imply that $W[E]=\int E^a_i \Gamma^i_a$ as it can be verified by direct
calculation.}. 
The action (\ref{axi}) becomes \begin{eqnarray}\label{NEW}
&&\nonumber S[P,A,N,\vec N,N_i ]= \\ && \frac{1}{\gamma \kappa}
\int dt \int_{\Sigma} dx^3\left[  E^a_i \dot A^a_i - N^b V_b
(E^a_j,A_a^j)- N S(E^a_j,A_a^j)-N^i G_i(E^a_j,A_a^j)\right], \end{eqnarray}
where the constraints are explicitly given by: \begin{equation}\label{uno}
 V_b (E^a_j,A_a^j)=E_j^a F_{ab}-(1+\gamma^2) K_a^i G_i
\end{equation}
\begin{equation}
\label{dos}
S(E^a_j,A_a^j)= \frac{E^{a}_iE^b_j}{\sqrt{{\rm
    det}(E)}} \left(\epsilon^{ij}_{\ \ k} F^k_{ab}-2(1+\gamma^2)
K^i_{[a}K^j_{b]} \right)
\end{equation}
\begin{equation}\label{tres}
 G_i(E^a_j,A_a^j)=D_aE^a_i,
\end{equation} 
where $F_{ab}=\partial_a A_b^i-\partial_b A_a^i+\epsilon^i_{\
jk}A_a^jA^k_b$ is the curvature of the connection $A_a^i$ and
$D_aE^a_i=\partial_a E^a_i+\epsilon_{ij}^{\ \ k} A^j_aE^a_k$ is
the covariant divergence of the densitized triad. We have seven
(first class) constraints for the 18 phase space variables
$(A_a^i, E^b_j)$. In addition to imposing conditions among the
canonical variables, first class constraints are generating
functionals of (infinitesimal) gauge transformations. From the
18-dimensional phase space of general relativity we end up with 11
fields necessary to coordinatize the constraint surface on which
the above seven conditions hold. On that 11-dimensional constraint
surface, the above constraint generate a seven-parameter-family of
gauge transformations. The reduce phase space is four dimensional
and therefore the resulting number of physical degrees of freedom
is \emph{two}, as expected.

The constraint (\ref{tres}) coincides with the standard Gauss law
of Yang--Mills theory (e.g. $\vec{\nabla}\cdot \vec E=0$ in
electromagnetism). In fact if we ignore  (\ref{uno}) and
(\ref{dos}) the phase space variables $(A_a^i, E^b_j)$ together with the
Gauss law (\ref{tres}) characterize the physical phase space of an
$SU(2)$\epubtkFootnote{The constraint structure does not distinguish
$SO(3)$ from $SU(2)$ as both groups have the same Lie algebra.
From now on we choose to work with the more fundamental (universal
covering) group $SU(2)$. In fact this choice is physically
motivated as $SU(2)$ is the gauge group if we want to include
fermionic matter\cite{c10bis}.} Yang--Mills (YM) theory. The
gauge field is given by the connection $A_a^i$ and its conjugate
momentum is the electric field $E_j^b$. Yang--Mills theory is
a theory defined on a background spacetime geometry. Dynamics in
such a theory is described by a non vanishing Hamiltonian -- the
Hamiltonian density of YM theory being ${\cal
H}=E_a^iE^a_i+B_a^iB^a_i$. General relativity is a generally
covariant theory and coordinate time plays no physical role.  The
Hamiltonian is a linear combination of constraints.\epubtkFootnote{In the physics of the standard model we are used to identifying the coordinate $t$
with the physical time of a suitable family of observers. In the general covariant context of
gravitational physics the coordinate time $t$ plays the role of a label with no physical
relevance. One can arbitrarily change the way we coordinatize spacetime without affecting the physics.
This redundancy in the description of the physics (gauge symmetry) induces the appearance of constraints
in the canonical formulation. The constraints in turn are the generating functions of these
gauge symmetries. The Hamiltonian generates evolution in coordinate time $t$ but because redefinition of $t$ is pure gauge, the
Hamiltonian is a constraint itself, i.e. ${\cal H}=0$ on shell\cite{dirac,Henneaux:1992ig}.
More on this in the next section.} Dynamics is
encoded in the constraint equations (\ref{uno}),(\ref{dos}), and (\ref{tres}). In this sense we can
regard general relativity in the new variables as a background
independent relative of $SU(2)$ Yang--Mills theory. We will see in
the sequel that the close similarity between these theories will
allow for the implementation of techniques that are very natural
in the context of YM theory.

To conclude this section let us point out that the real connection formulation of general relativity 
presented here is a peculiar property of four dimensions due to the special property (\ref{19}). Nevertheless, there are means to obtaining
real connection formulations for gravity and supergravity in higher dimensions as recently 
shown by Thiemann and collaborators~\cite{Bodendorfer:2011nv, Bodendorfer:2011nw, Bodendorfer:2011nx, Bodendorfer:2011ny, Bodendorfer:2011pa, Bodendorfer:2011pb, Bodendorfer:2011pc, Bodendorfer:2011hs}. 

\subsubsection{Constraints algebra}\label{contalgebra}

Here we simply present the structure of the constraint algebra of
general relativity in the new variables. \begin{equation}
\left\{G(\alpha),G(\beta) \right\}=G([\alpha,\beta]), \end{equation} where
$\alpha=\alpha^{i}\tau_i\in su(2)$, $\beta=\beta^{i}\tau_i\in
su(2)$ and $[\alpha,\beta]$ is the commutator in $su(2)$. \begin{equation}
\left\{G(\alpha),V(N^a) \right\}= -G({\sL}_{N}\alpha). \end{equation} \begin{equation}
\left\{G(\alpha),S(N) \right\}= 0. \end{equation} \begin{equation}\left\{V(N^a),V(M^a)
\right\}= V([N,M]^a), \end{equation} where
$[N,M]^a=N^b\partial_bM^a-M^b\partial_bN^a$ is the vector field
commutator, and $\sL_{N}$ denotes the Lie derivative along the vector field $N^a$. The previous constraints define the subalgebra od spacial diffeomorphisms and $SU(2)$ internal gauge transformations. 
This property allows to implement them in the quantum theory separately from the scalar constraint. If we include the 
scalar constraint the remaining Poisson brackets are
 \begin{equation}\left\{S(N),V(N^a) \right\}= -S({\sL}_{N}N), 
\end{equation}  and finally \begin{equation}\label{prob} \left\{S(N),S(M) \right\}=V(S^a)+
{\rm terms \ proportional\ to \ the \ Gauss
  \ constraint} ,
\end{equation} where for simplicity we are ignoring the terms proportional to
the Gauss law (the complete expression can be found in
\cite{ash10}) and \begin{equation}S^{a}= \frac{E^a_iE^b_j \delta^{ij}}{|{\rm
det} E|}(N\partial_b M-M\partial_b N).\end{equation} Notice
that  instead
of structure constants, the r.h.s.\ of (\ref{prob}) is written in
terms of field dependent structure functions. For this reason it
is said that the constraint algebra does not close in the BRS
sense.

\subsection{Geometric interpretation of the new variables}\label{geo}

The geometric interpretation of the connection $A_a^i$, defined in
(\ref{immi}), is standard. The connection provides a definition of
\emph{parallel transport} of $SU(2)$ spinors on the space manifold
$\Sigma$. The natural object is the $SU(2)$ element defining
parallel transport along a path $e\subset \Sigma$  also called \emph{holonomy}
denoted  $h_{e}[A]$, or more explicitly  \begin{equation}\label{hol}h_{e}[A]=P
\exp-\int \limits_{e} A,\end{equation} where $P$ denotes a
path-order-exponential.

The densitized triad -- or electric field -- $E_i^a$ also has a
simple geometrical meaning.  As it follows from (\ref{twenty}),  $E_i^a$ encodes the full background
independent Riemannian geometry of $\Sigma$. Therefore, any geometrical quantity in space can
be written as a functional of $E^a_i$. One of the simplest is the
area $A_{S}[E^a_i]$ of a surface $S\subset \Sigma$ whose
expression we derive in what follows. Given a two dimensional
surface in $S\subset \Sigma$ -- with normal \begin{equation}n_a=\frac{\partial
x^b}{\partial \sigma^1}\frac{\partial x^c}{\partial \sigma^2}
\epsilon_{abc}\end{equation} where $\sigma^1$ and $\sigma^2$ are local
coordinates on $S$ -- its area is given by  \begin{equation}A_S[q^{ab}]=\int_S
\sqrt{h} \ d\sigma^1d\sigma^2, \end{equation} where $h={\rm det}(h_{ab})$
is the determinant of the metric $h_{ab}=q_{ab}-n^{-2}n_an_b$ induced on $S$ by
$q^{ab}$. From equation (\ref{twenty}) it follows that $q q^{ab}=E^a_iE^{bi}$ so that 
${\rm det}(q^{ab})={\rm det}(E^a_i)$.  Contracting the previous equality
with $n_an_b$, namely \begin{equation}\label{dete} q
q^{ab}n_an_b= E^a_iE^b_j\delta^{ij}n_an_b. \end{equation} Now observe that
$q^{nn}=q^{ab}n_an_b$ is the $nn$-matrix element of the inverse of
$q_{ab}$. Through the well known formula for components of the
inverse matrix we have that \begin{equation}
q^{nn}=\frac{{\det}(q_{ab}-n^{-2}n_an_b)}{{\det}(q_{ab})}=\frac{h}{q}.\end{equation}
But $q_{ab}-n^{-2} n_an_b$ is precisely the induced metric
$h_{ab}$. Replacing $q^{nn}$ back into (\ref{dete}) we conclude
that \begin{equation}h= E^a_iE^b_j\delta^{ij}n_an_b. \end{equation} Finally we can write
the area of $S$ as an explicit functional of $E^a_i$:
\begin{equation}\label{areac} A_S[E^a_i]=\int_S
\sqrt{E^a_iE^b_j\delta^{ij}n_an_b}\ d\sigma^1d\sigma^2. \end{equation} This
simple expression for the area of a surface will be very important
in the quantum theory.

\section{Loop Quantum Gravity and Quantum Geometry in a Nutshell}
\label{lqg}

Loop quantum gravity is a proposal for the implementation of the quantization
program established in the 1960s by Dirac, Wheeler, and DeWitt, among
others (for recent reviews see~\cite{ash10,Thiemann:2007zz,c9}). The technical
difficulties of Wheeler's `geometrodynamics' are circumvent by the
use of connection variables instead of metrics~\cite{ash,
ash1,barbero}. At the kinematical level, the formulation is
similar to that of standard gauge theories. The fundamental
difference is however the absence of any non-dynamical background
field in the theory.

The configuration variable is an $SU(2)$-connection $A_a^i$
on a 3-manifold $\Sigma$ representing space. The canonical momenta
are given by the densitized triad $E_i^a$. The latter encode the
(fully dynamical) Riemannian geometry of $\Sigma$ and are the
analog of the `electric fields' of Yang--Mills theory.

In addition to diffeomorphisms there is the local $SU(2)$ gauge freedom that
rotates the triad and transforms the connection
in the usual way. According to Dirac, gauge freedoms result in
constraints among the phase space variables which conversely are
the generating functionals of infinitesimal gauge transformations.
In terms of connection variables the constraints are
\begin{equation}\label{constro}
{\cal G}_{i}={\cal D}_a E^a_i=0,\ \ \ \ {\cal C}_a=E_k^b
F^k_{ba}=0, \ \ \ \ {\cal S}=\epsilon^{ijk}E^a_iE^b_j F_{ab\, k}+
\cdots=0,
\end{equation}
where ${\cal D}_a$ is the covariant derivative and $F_{ba}$ is the
curvature of $A_a^i$.  ${\cal G}_{i}$ is the familiar Gauss
constraint -- analogous to the Gauss law of electromagnetism -- generating 
infinitesimal $SU(2)$ gauge transformations, ${\cal
C}_{a}$ is the vector constraint generating space-diffeomorphism, 
and ${\cal S}$ is the scalar constraint generating
`time' reparameterization (there is an additional term that 
we have omitted here for simplicity -- see Equations~(\ref{uno}) to (\ref{tres}), and Section~\ref{contalgebra} for the precise form of the constraints and their relationship with the 
gauge symmetry groups).

Loop quantum gravity is defined using Dirac quantization. One
first represents (\ref{constro}) as operators in an auxiliary
Hilbert space $\cal H$  and then solves the constraint equations
\begin{equation}\label{constroq}
\hat {\cal G}_{i}\Psi=0,\ \ \ \ \hat {\cal C}_a \Psi=0, \ \ \ \
\hat {\cal S}\Psi=0.
\end{equation}
The Hilbert space of solutions is the so-called physical Hilbert
space ${\cal H}_{phys}$. In a generally covariant system quantum
dynamics is fully governed by constraint equations. In the case of
loop quantum gravity they represent \emph{quantum Einstein's
equations}.

States in the auxiliary Hilbert space are represented by wave
functionals of the connection $\Psi(A)$ which are square
integrable with respect to a natural diffeomorphism invariant
measure, the Ashtekar-Lewandowski measure~\cite{ash3} (we denote
it ${\sL}^2[{\cal A}]$ where ${\cal A}$ is the space of
(generalized) connections). Roughly speaking \footnote{The construction is rather a projective limit ~\cite{ash3}.  This leads to non-trivial requirements (cylindrical consistency) on the structure (operators) of the theory. This point is very important for spin foams specially in considering the refining limit of amplitudes. We will revisit this issue in Section \ref{anofree}.}, this space can be decomposed into a
direct sum of orthogonal subspaces ${\cal H}=\bigoplus_{\gamma}
{\cal H}_{\gamma}$ labeled by a graph $\gamma$ in $\Sigma$. The
fundamental excitations are given by the holonomy $h_{\ell}(A)\in
SU(2)$ along a path $\ell$ in $\Sigma$:
\begin{equation}
h_{\ell}(A)={\cal P}\ {\rm exp}\ \int_{\ell} A.
\end{equation}
Elements of ${\cal H}_{\gamma}$ are given by functions
\begin{equation}
\Psi_{f,\gamma}(A)=f(h_{\ell_{1}}(A),h_{\ell_{2}}(A),\dots,h_{\ell_{n}}(A)),
\end{equation}
where $h_{\ell}$ is the holonomy along the links $\ell\in \gamma$
and $f:SU(2)^n\rightarrow \C$ is (Haar measure) square integrable.
They are called \emph{cylindrical functions} and represent a dense
set in $\cal H$ denoted $Cyl$. The``momenta" conjugate to the holonomies are given by the so-called flux operators
\be E(S,\alpha)=\int_{S} \alpha^iE_i
\ee acros a 2-suface $s\subset \Sigma$ and labelled by a smearing field
$\alpha\in su(2)$. It has been shown that the associated (Poisson) holonomy-flux algebra
admits a unique quantization in a Hilbert space with a diffeomorphism invariant states~\cite{Fleischhack:2004jc, lost}.
The Hilbert space $\cal H$ mentioned above is precisely that unique representation. 

Gauge transformations generated by the Gauss constraint act
non-trivially at the endpoints of the holonomy, i.e., at nodes of
graphs. The Gauss constraint (in (\ref{constro})) is solved by
looking at $SU(2)$ gauge invariant functionals of the connection
(${\sL}^2[{\cal A}]/{\cal G}$). The fundamental gauge invariant
quantity is given by the holonomy around closed loops. An
orthonormal basis of the kernel of the Gauss constraint is defined
by the so called spin network states 
$\Psi_{\va \gamma, \{j_{\ell}\},\{\iota_{n}\}}(A)$~\cite{reis8, c4, baez10}. Spin-networks\epubtkFootnote{Spin-networks were introduced by 
Penrose~\cite{pen1, pen2, pen3, pen4} in a attempt to define 3-dimensional
Euclidean quantum geometry from the combinatorics of angular momentum in QM. 
Independently they have been used in lattice gauge theory as a natural basis
for gauge invariant functions on the lattice. 
For an account of their applications in various contexts see~\cite{Smolin:1997aa}.}
are defined by a graph $\gamma$ in $\Sigma$, a collection of spins
$\{j_{\ell}\}$ -- unitary irreducible representations of $SU(2)$ -- associated 
with links $\ell\in \gamma$ and a collection of $SU(2)$
intertwiners $\{\iota_{n}\}$ associated to nodes $n \in \gamma$
(see Figure~\ref{spinn}). The spin-network gauge invariant wave
functional $\Psi_{\va \gamma, \{j_{\ell}\},\{\iota_{n}\}}(A)$ is
constructed by first associating an $SU(2)$ matrix in the
$j_{\ell}$-representation to the holonomies $h_{\ell}(A)$
corresponding to the link $\ell$, and then contracting the
representation matrices at nodes with the corresponding
intertwiners $\iota_n$, namely
\begin{equation}\label{cyly}
\Psi_{\va \gamma, \{j_{\ell}\},\{\iota_{n}\}}(A)=\prod_{n \in
\gamma} \ \iota_n\ \prod_{\ell \in \gamma}\  j_{\ell}[h_{\va
\ell}(A)],
\end{equation}
where $ j_{\ell}[h_{\va
\ell}(A)]$ denotes the corresponding $j_{\ell}$-representation matrix
evaluated at corresponding link holonomy
and the matrix index contraction is left implicit.

\epubtkImage{}{%
  \begin{figure}[htbp]
    \centerline{\includegraphics[width=10cm]{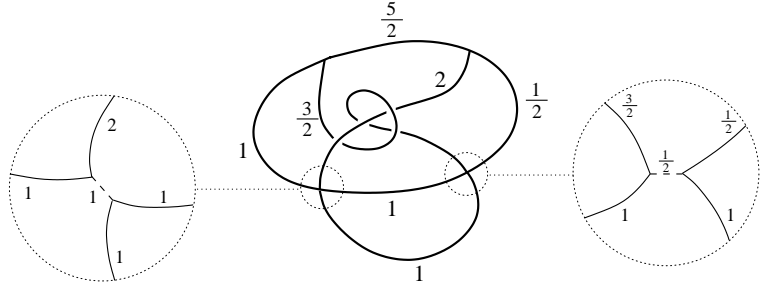}}
    \caption{Spin-network state: At 3-valent nodes the intertwiner is
      uniquely specified by the corresponding spins. At 4 or higher
      valent nodes an intertwiner has to be specified. Choosing an
      intertwiner corresponds to decompose the $n$-valent node in
      terms of 3-valent ones adding new virtual links (dashed lines)
      and their corresponding spins. This is illustrated explicitly in
      the figure for the two 4-valent nodes.}
    \label{spinn}
\end{figure}}

The solution of the vector constraint is more subtle~\cite{ash3}.
One uses group averaging techniques together with the 
diffeomorphism invariance of the kinematical inner product in $\cal H$. The diffeomorphism constraint does not exist in the quantum theory. Only finite diffeomorphisms can be defined. As a result
solutions (diffeomorphism invariant states) correspond to generalized states. These are not in
$\cal H$ but are elements of the topological dual
$Cyl^*$\epubtkFootnote{According to the triple $Cyl\subset {\cal H}
\subset Cyl^*$.}. However, the intuitive idea is quite simple: 
solutions to the vector constraint are given by equivalence classes
of spin-network states up to diffeomorphism. Two spin-network states
are considered equivalent if their underlying graphs can be deformed into each 
other by the action of a diffeomorphism. 

This can be regarded as an indication that the smooth spin-network category
could be replaced by something which is more combinatorial in nature so that 
diffeomorphism invariance becomes a derived property of the classical limit. 
LQG has been modified along these lines by replacing the smooth 
manifold structure of the standard theory by
the weaker concept of piecewise linear manifold~\cite{za2}. In this context,
graphs defining spin-network states can be completely characterized using 
the combinatorics of cellular decompositions of space. Only a discrete 
analog of the diffeomorphism symmetry survives which can be 
dealt with in a fully combinatorial manner.
We will take this point of view when we introduce the notion 
of spin foam in the following section.

\subsection{Quantum geometry}

The generalized states described above solve all of the constraints
(\ref{constro}) but the scalar constraint. They are regarded as
quantum states of the Riemannian geometry on $\Sigma$. They define
the kinematical sector of the theory known as \emph{quantum
geometry}.

Geometric operators acting on spin network states  can
be defined in terms of the fundamental triad 
operators $\hat E^a_i$. The simplest of such operators is the area of a surface
$S$ classically given by
\begin{equation}
A_{S}(E)=\int_S dx^2 \sqrt{{\rm Tr}[n_an_bE^aE^b]}
\end{equation}
where $n$ is a co normal. The geometric operator $\hat A_{S}(E)$
can be rigorously defined by its action on spin network states~\cite{lee1,c3,ash2}. 
The area operator gives a clear geometrical
interpretation to spin-network states: the fundamental
1-dimensional excitations defining a spin-network state can be
thought of as quantized `flux lines' of area. More precisely, if
the surface $S\subset \Sigma$ is punctured by a spin-network link
carrying a spin $j$, this state is an eigenstate of $\hat A_S(E)$
with eigenvalue proportional to $\ell^2_P \sqrt{j(j+1)} $. In
the generic sector -- where no node lies on the surface -- the
spectrum takes the simple form
\begin{equation}\label{aarreeaa}
a_S(\{j\})=8 \pi \iota \ell^2_P \sum_i \sqrt{j_i(j_i+1)},
\end{equation}
where $i$ labels punctures and $\iota$ is the Imirzi parameter~\cite{immi}\epubtkFootnote{The
Imirzi parameter $\iota$ is a free parameter in the theory. This ambiguity is purely quantum
mechanical (it disappears in the classical limit). It has to be
fixed in terms of physical predictions. The computation of black hole
entropy in LQG fixes the value of $\iota$ (see~\cite{Ashtekar:2000eq}).}. $a_S(\{j\})$ is the sum of single
puncture contributions. The general form of the spectrum including
the cases where nodes lie on $S$ has been computed in closed
form~\cite{ash2}.

The spectrum of the volume operator is also discrete~\cite{lee1,c3,loll1,ash22}. If we define
the volume operator $\hat V_\sigma(E)$ of a 3-dimensional region
$\sigma\subset \Sigma$ then non vanishing eigenstates
are given by spin-networks containing $n$-valent nodes in
$\sigma$ for $n > 3$. Volume is concentrated in nodes. For new results on the volume see
\cite{Bianchi:2011ub, Bianchi:2010gc}. Other geometric quantities have been 
considered in the quantization; in the particular in studies of coupling LQG to matter~\cite{Thiemann:1997rt} the introduction of a metric operator
is necessary. For another proposal of length operator see~\cite{Bianchi:2008es}.

\subsection{Quantum dynamics}

In contrast to the Gauss and vector constraints, the scalar
constraint does not have a simple geometrical meaning. This makes
its quantization more involved. Regularization choices have to be
made and the result is not unique. After Thiemann's first rigorous
quantization~\cite{th2} other well defined possibilities have been
found~\cite{c00,pul1,pul4}. This ambiguity affects dynamics
governed by
\begin{equation}\label{wd}
\hat {\cal S}\Psi=0.
\end{equation}

The difficulty in dealing with the scalar constraint is not
surprising. The vector constraint -- generating space
diffeomorphisms -- and the scalar constraint -- generating time
reparameterizations -- arise from the underlying 4-diffeomorphism
invariance of gravity. In the canonical formulation the 3+1
splitting breaks the manifest 4-dimensional symmetry. The price
paid is the complexity of the time re-parameterization constraint
$\cal S$. The situation is somewhat reminiscent of that in
standard quantum field theory where manifest Lorentz invariance is
lost in the Hamiltonian formulation\epubtkFootnote{There is however an
additional complication here: the canonical constraint algebra
does not reproduce the 4-diffeomorphism Lie algebra. This
complicates the geometrical meaning of $S$.}.

From this perspective, there has been growing interest in
approaching the problem of dynamics by defining a covariant
formulation of quantum gravity. The idea is that (as in the QFT
case) one can keep manifest 4-dimensional covariance in the path
integral formulation. The spin foam approach is an attempt to define
the path integral quantization of gravity using what we have learned
from LQG.

In standard quantum mechanics path integrals provide the solution
of dynamics as a device to compute the time evolution operator.
Similarly, in the generally covariant context it provides a tool
to find solutions to the constraint equations (this has been
emphasized formally in various places: in the case of gravity see
for example~\cite{hart1}, for a detailed discussion of this in the
context of quantum mechanics see~\cite{c6}). Recall discussion of Section~\ref{valin}.

Let us finish by stating some properties of $\hat S$ that do not
depend on the ambiguities mentioned above. One is the discovery
that smooth loop states naturally solve the scalar constraint
operator~\cite{jac,c8}. This set of states is clearly to small to
represent the physical Hilbert space (e.g., they span a zero
volume sector). However, this implies that $\hat S$ acts only on
spin network nodes. Its action modifies spin networks at nodes by
creating new links according to Figure~\ref{branch}\epubtkFootnote{This is
not the case in all the available definitions of the scalar
constraints as for example the one defined in~\cite{pul1,pul4}.}.
This is crucial in the construction of the spin foam approach of
the next section.

\epubtkImage{}{%
\begin{figure}[htbp]
\centerline{\hspace{0.5cm} \(\begin{array}{c}
\includegraphics[height=3cm]{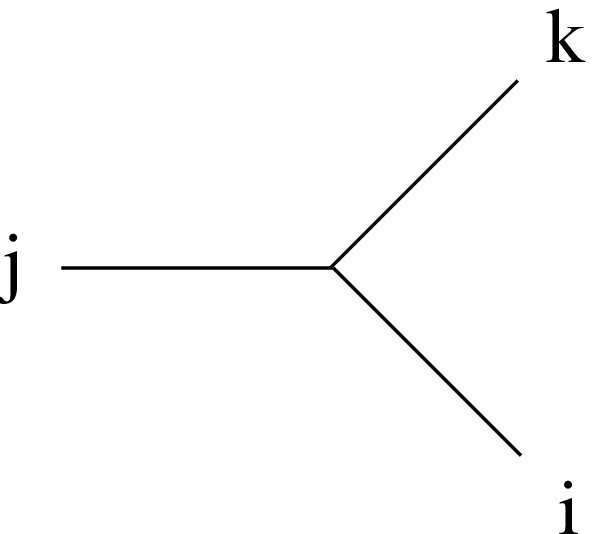}
\end{array}\ \   \rightarrow \ \
\begin{array}{c}
\includegraphics[height=3cm]{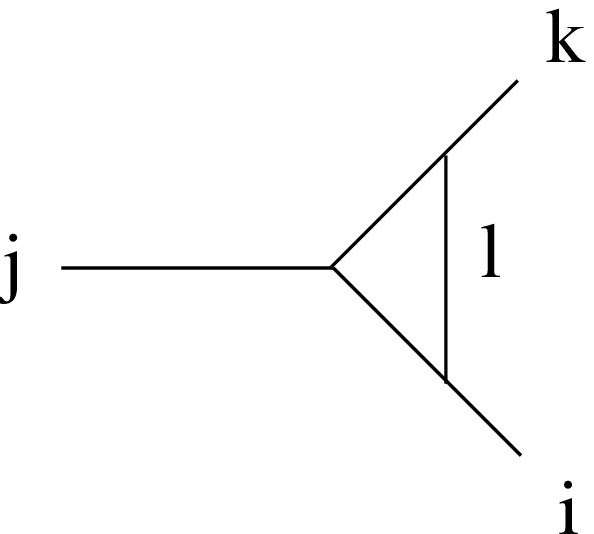}
\end{array} \) }
  \caption{A typical transition generated by the action of the scalar
    constraint}
  \label{branch}
\end{figure}}

\section{Spin Foams and the Path Integral for Gravity in a Nutshell}
\label{sec:intsf}

The possibility of defining quantum gravity using Feynman's path
integral approach has been considered since Misner~\cite{Misner:1957wq}. Given a
4-manifold $\cal M$ with boundaries $\Sigma_1$ and $\Sigma_2$, and
denoting by $G$ the space of metrics on ${\cal M}$, the transition
amplitude between $\left|\left[q_{ab}\right]\right>$ on $\Sigma_1$
and $\left|\left[q^{\prime}_{ab}\right]\right>$ on $\Sigma_2$ is
formally
\begin{equation}
\label{tarara}
\left<
\left[q^{}_{ab}\right]\right.\left|\right[q^{\prime}_{ab}\left]\right>
=\int \limits_{\left[g\right]} {\cal D}[g]\ \ e^{i S([g])},
\end{equation}
where the integration on the right is performed over all space-time
metrics up to $4$-diffeomorphisms $\left[g \right]\in G/{\vani
Diff({\cal M})}$ with fixed boundary values up to
$3$-diffeomorphisms $\left[q_{ab}\right]$,
$\left[q^{\prime}_{ab}\right]$, respectively.

There are various difficulties associated with (\ref{tarara}).
Technically there is the problem of defining the functional
integration over $\left[g\right]$ on the RHS. This is partially
because of the difficulties in defining infinite dimensional
functional integration beyond the perturbative framework. In
addition, there is the issue of having to deal with the space
$G/{\vani Diff({\cal M})}$, i.e., how to characterize the
diffeomorphism invariant information in the metric. This gauge
problem ($3$-diffeomorphisms) is also present in the definition of
the boundary data. There is no well defined notion of kinematical
state $\left|\left[q_{ab}\right]\right>$ as the notion of kinematical Hilbert
space in standard metric variables has never been defined.

The situation is different in the framework of loop quantum
gravity. The notion of quantum state of 3-geometry is rigorously
defined in terms of spin-network states. They carry the
diff-invariant information of the Riemannian structure of
$\Sigma$. In addition, and very importantly, these states are
intrinsically discrete (colored graphs on $\Sigma$) suggesting a
possible solution to the functional measure problem, i.e., the
possibility of constructing a notion of Feynman `path integral' in
a combinatorial manner involving sums over spin network
world sheets amplitudes. Heuristically, `4-geometries' are to be
represented by `histories' of quantum states of 3-geometries or
spin network states. These `histories' involve a series of
transitions between spin network states (Figure~\ref{3g}), and
define a foam-like structure (a `2-graph' or 2-complex) whose 
components inherit the spin representations from the
underlying spin networks. These spin network world sheets are the
so-called \emph{spin foams}.

The precise definition of spin foams was introduced by Baez
in~\cite{baez7} emphasizing their role as morphisms in the category
defined by spin networks\epubtkFootnote{The role of category theory
  for quantum gravity had been emphasized by Crane
  in~\cite{Crane:1993vs, Crane:1993vv, crane0}.}. A spin foam ${\cal
  F}: s\rightarrow s^{\prime}$, representing a transition from the
spin-network $s=(\gamma, \{ j_{\ell}\}, \{ \iota_n \})$ into
$s^{\prime}=(\gamma^{\prime}, \{ j_{\ell^{\prime}}\}, \{
\iota_{n^{\prime}}\})$, is defined by a $2$-complex  ${\cal J}$
bordered by the graphs of $\gamma$ and $\gamma^{\prime}$ respectively,
a collection of spins $\{ j_f\}$ associated with faces $f \in {\cal
  J}$ and a collection of intertwiners ${ \{ \iota_e}\}$ associated to
edges $e\in {\cal J}$. Both spins and intertwiners of exterior faces
and edges match the boundary values defined by the spin networks $s$
and $s^{\prime}$ respectively. Spin foams ${\cal F}: s\rightarrow
s^{\prime}$ and ${\cal F}^{\prime}: s^{\prime}\rightarrow
s^{\prime\prime}$ can be composed into ${\cal FF}^{\prime}: s
\rightarrow s^{\prime\prime}$ by gluing together the two corresponding
2-complexes at $s^{\prime}$. A spin foam model is an assignment of
amplitudes $A[{\cal F}]$ which is consistent with this composition
rule in the sense that
\begin{equation}\label{cobordism}
A[{\cal F F}^{\prime}]=A[{\cal F}]A[{\cal F}^{\prime}].
\end{equation}
Transition amplitudes between spin network states are defined by
\begin{equation}
\left<s,s^{\prime}\right>_{phys}=\sum_{{\cal F}: s\rightarrow
s^{\prime}} A[{\cal F}],
\end{equation}
where the notation anticipates the interpretation of such
amplitudes as defining the physical scalar product. The domain of
the previous sum is left unspecified at this stage. We shall
discuss this question further in Section~\ref{sci}. This last
equation is the spin foam counterpart of Equation~(\ref{tarara}).
This definition remains formal until we specify what the set of
allowed spin foams in the sum are and define the corresponding
amplitudes.

\epubtkImage{}{%
\begin{figure}[htbp]
\centerline{\hspace{0.5cm} \(\begin{array}{c}
\includegraphics[height=5cm]{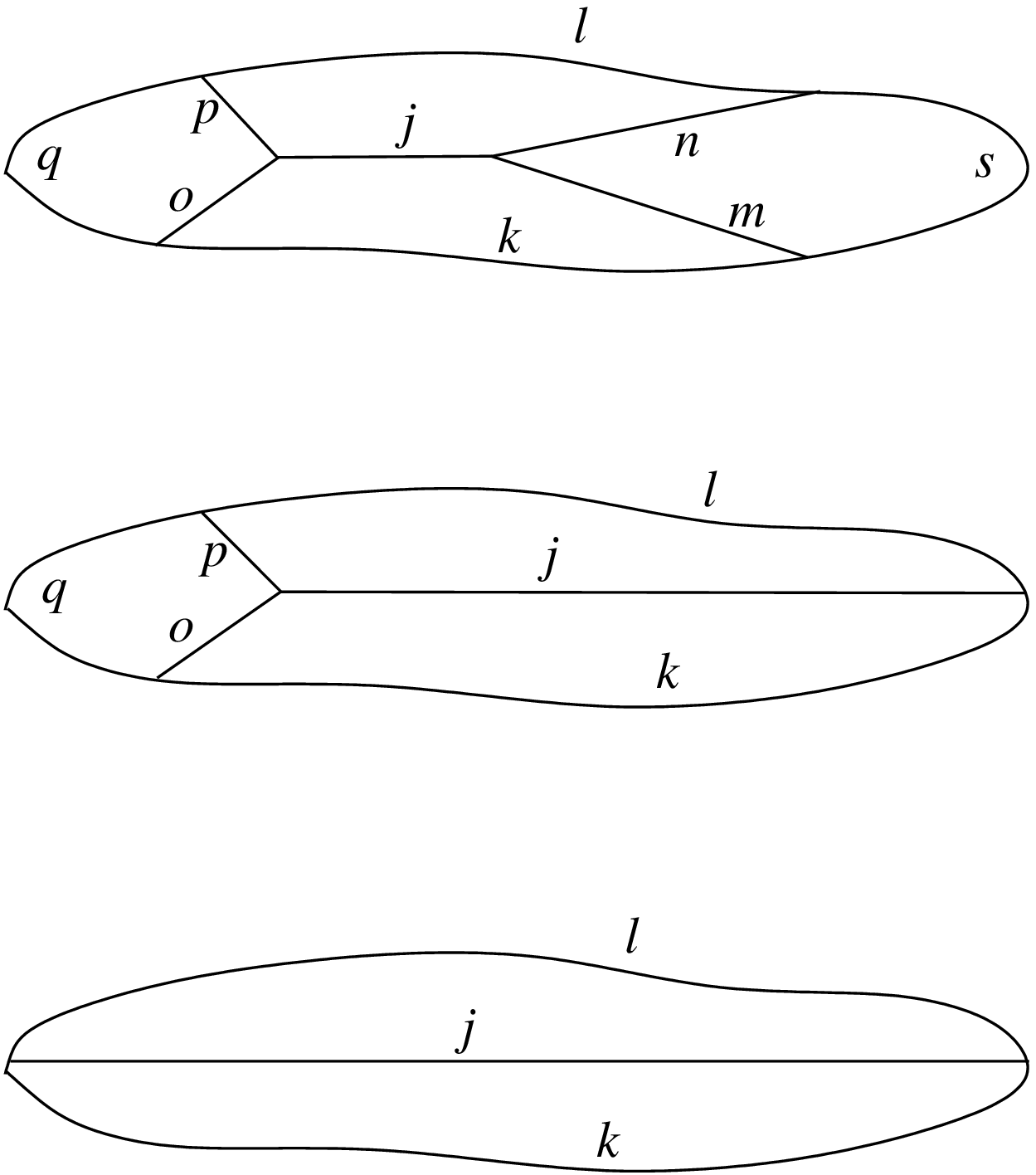}
\end{array}\ \   \rightarrow \ \
\begin{array}{c}
\includegraphics[height=5cm]{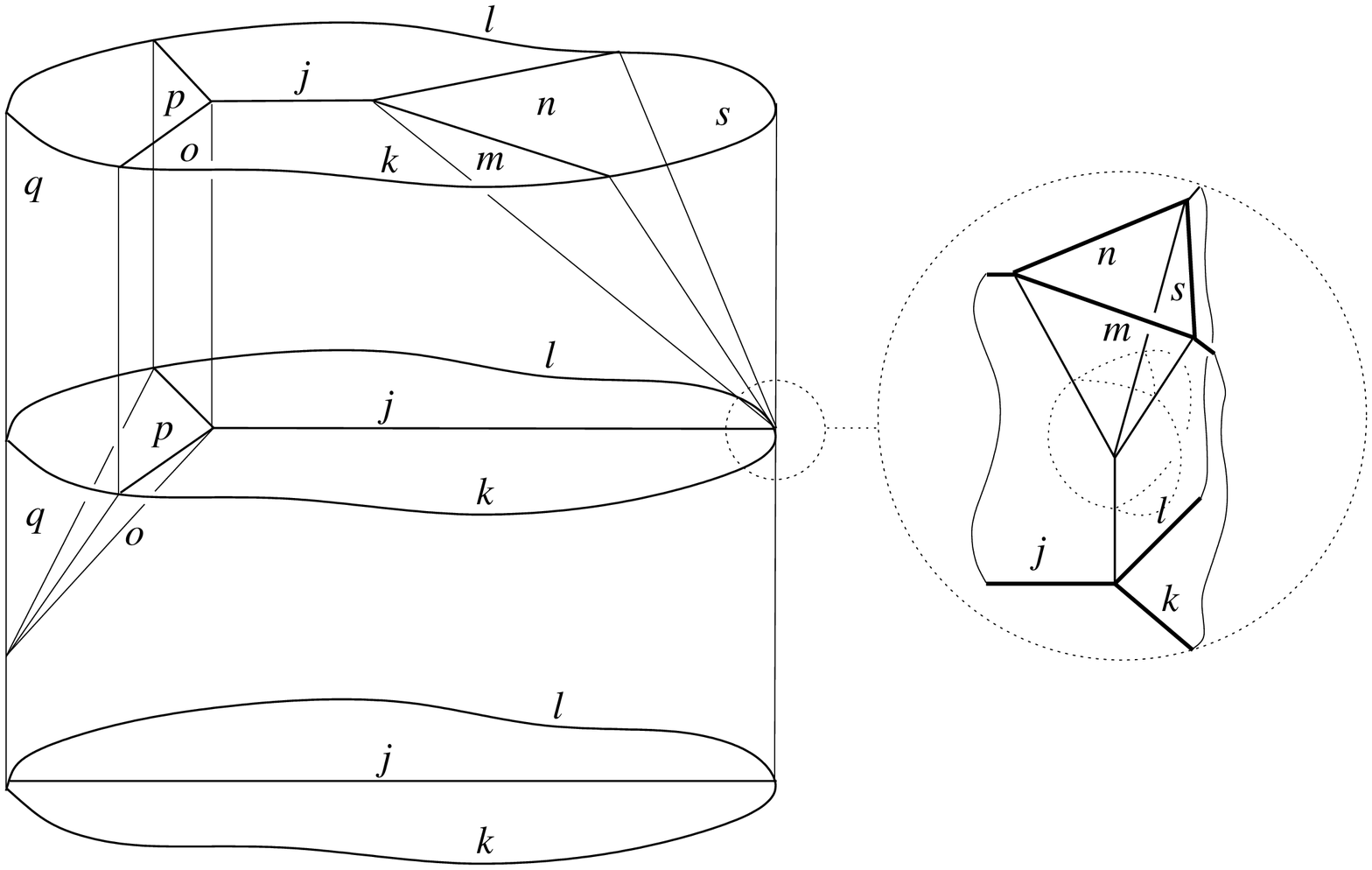}
\end{array} \) }
  \caption{A typical path in a path integral version of loop quantum
    gravity is given by a series of transitions through different
    spin-network states representing a state of 3-geometries. Nodes
    and links in the spin network evolve into 1-dimensional edges and
    faces. New links are created and spins are reassigned at vertexes
    (emphasized on the right). The `topological' structure is provided
    by the underlying 2-complex while the geometric degrees of freedom
    are encoded in the labeling of its elements with irreducible
    representations and intertwiners.}
  \label{3g}
\end{figure}}

The background-independent character of spin foams is manifest. The
2-complex can be thought of as representing `space-time' while
the boundary graphs as representing `space'. They do not carry
any geometrical information in contrast with the standard concept
of a lattice. Geometry is encoded in the spin labelings which
represent the degrees of freedom of the gravitational field.

In standard quantum mechanics the path integral is used to compute
the matrix elements of the evolution operator $U(t)$. It provides
in this way the solution for dynamics since for any kinematical
state $\Psi$ the state $U(t)\Psi$ is a solution to Schr\"odinger's
equation. Analogously, in a generally covariant theory the path
integral provides a device for constructing solutions to the
quantum constraints. Transition amplitudes represent the matrix
elements of the so-called generalized `projection' operator $P$
(i.e., $\left<s,s^{\prime}\right>_{phys}=\left<sP,s^{\prime}\right>$ recall the general discussion of Sections~\ref{valin}) such that $P\Psi$ is a
physical state for any kinematical state $\Psi$. As in the case of
the vector constraint the solutions of the scalar constraint
correspond to distributional states (zero is in the continuum part
of its spectrum). Therefore, ${\cal H}_{phys}$ is not a proper
subspace of $\cal H$ and the operator $P$ is not a projector
($P^2$ is ill defined)\epubtkFootnote{In the notation of the previous
section states in ${\cal H}_{phys}$ are elements of $Cyl^*$.}. In Section~\ref{sfm3d} we give an explicit example of this
construction.

\clearpage

\part{The new spin foam models for four dimensional gravity}
\label{thenew}

In this part we study the non-perturbative quantization of four
dimensional general relativity from the spin foam
perspective. Together with Part~\ref{cano}, his part of the article
form a  self contained body, thus can be studied completely
independently from the rest of the content of this review.

The new spin foam models for four dimensional quantum gravity are
introduced from a perspective that, in some aspects, is independent
from the one taken in the original works related to the so called EPRL
model~\cite{Engle:2007wy, Engle:2007qf, Engle:2007uq} as well as the
one used in the definition of the FK models~\cite{Freidel:2007py}. Our
starting point is the quantization of BF theory in the path integral
formulation, which leads to a well defined unambiguous state sum model
or topological field theory (see Section~\ref{BF}). The space of
histories of BF theory path integral will be constrained  to satisfy
the so-called linear simplicity constraints that reduce BF theory to
gravity.

In Section~\ref{BF} we review the quantization of BF theory and its
spin foam representation. In Section~\ref{eprl-r} we present the
Riemannian EPRL model together with various mathematical tools that
will be useful in the following sections. In Section~\ref{eprl-l} the
Lorentzian version of the EPRL model is reviewed. In Section~\ref{fk}
we present the FK model. In Section~\ref{BCM} we recall the definition
of the Barrett--Crane model.

Finally, there are various review works in the
literature~\cite{Rovelli:2011eq, Rovelli:2010bf} where the new models
are introduced from a rather minimalistic perspective by simply
postulating the defining amplitudes and deriving  their relation to
gravity a posteriori. Such choice is indeed quite advantageous  if the
novel simplicity of the new models is to be emphasised. In this
article we have chosen a complementary more constructive view. This
strategy allows for a systematic presentation of the ingredients that
go into the construction of the new models from the perspective of
continuum general relativity and BF theory. A possible drawback of
this choice is that most of the simplicial geometry intuitions used in
other derivations are almost completely avoided. We hope that this
lost will be compensated by the potential advantages of an alternative
viewpoint.

\clearpage

\section{Spinfoam Quantization of BF Theory}
\label{BF}

Here we follow the perspective of~\cite{baez5}.
Let $G$ be a compact group whose a Lie algebra $\frak g$ has an invariant 
inner product here denoted $\langle \rangle$, and $\cal M$ a ${\rm d}$-dimensional manifold.
Classical BF theory  is defined by the action
\begin{equation}\label{BFT}
S[{\rm B},\omega]=\int \limits_{\cal M}\langle {\rm B}\wedge {\rm F}(\omega) \rangle,
\end{equation}
where ${\rm B}$ is a $\frak g$ valued $({\rm d}-2)$-form,
$\omega$ is a connection on a $G$ principal bundle over
$\cal M$. The theory has no local excitations: all solutions of the equations of motion are locally related by gauge transformations. 
More precisely, the  gauge symmetries of the action are
the local $G$ gauge transformations
\begin{equation}\label{gauge1g}
\delta {\rm B} = \left[{\rm B},\alpha \right], \ \ \ \ \ \ \ \ \ \delta \omega
= {\rm d}_{\omega} \alpha,
\end{equation}
where $\alpha$ is a $\frak g$-valued 0-form, and the
`topological' gauge transformation
\begin{equation}\label{gauge2g}
\delta {\rm B} = {\rm d}_{\omega} \eta, \ \ \ \ \ \ \ \ \ \delta \omega =
0,
\end{equation}
where ${\rm d}_{\omega}$ denotes the covariant exterior derivative
and $\eta$ is a ${\frak g}$-valued 0-form. The first
invariance is manifest from the form of the action, while the
second is a consequence of the Bianchi identity, ${\rm d}_{
\omega}F(\omega)=0$. The gauge symmetries are so vast that all
the solutions to the equations of motion are locally pure gauge.
The theory has only global or topological degrees of freedom.

\vskip.2cm
\noindent \textbf{\emph{Remark:}} \emph{In the special case $G=SU(2)$ and ${\rm d}=3$ BF theory is (Riemannian) general relativity where the field ${\rm B}_a^i$ is given by the cotretrad frames $e_a^i$ of general relativity in the first order formalism. This simple example will be studied in more detail
in Section~\ref{sfm3d}. Another case of interest is $G=Spin(4)$  and $d=4$ as it will provide the basis
for the construction of the spinfoam models for four dimensional quantum (Riemannian) general relativity studied in Sections~\ref{eprl-r}, \ref{eprl-l}, \ref{fk} and \ref{BCM}.
The relationship with general relativity stems from the fact that constraining the field ${\rm B}_{ab}^{IJ}=\epsilon^{IJ}_{\ \ KL} e^K_ae^L_b$ in the action (\ref{BFT}) -- where $e_a^I$ is interpreted as the tetrad co-frame -- produces the action of general relativity in four dimensions.
 In the physically relevant cases of the above examples one needs to deal with non-compact groups -- $G=SL(2,\R)$ and $G=SL(2,\C)$ respectivelly. The  non-compacteness of the gauge group leads to certain infrared divergencies of transition amplitudes (infinite volume factors). We avoid such complications at this stage and concentrate on the compact $G$ case. The infinite volume divergences will be solved in the particular case of interest which is the spinfoam models for four dimensional Lorentzian gravity whose construction is reviewed in Section~\ref{eprl-l}.}
\vskip.2cm

For the moment we assume ${\cal M}$ to be a compact and
orientable.
The partition
function, ${\cal Z}$, is formally given by
\begin{equation}\label{zbfg}
{\cal Z}=\int  {\cal D}[{\rm B}] {\cal D}[\omega]\ \ \exp(i \int_{\va \cal M}
\langle {\rm B}\wedge F(\omega)\rangle).
\end{equation}
Formally integrating over the ${\rm B}$ field in (\ref{zbfg}) we
obtain
\begin{equation}\label{VAg}
{\cal Z}=\int {\cal D}[\omega] \ \ \delta \left(F(\omega)\right).
\end{equation}
The partition function ${\cal Z}$ corresponds to the `volume' of
the space of flat connections on $\cal M$.

In order to give a meaning to the formal expressions above, we
replace the ${\rm d}$-dimensional manifold ${\cal M}$ with an arbitrary
cellular decomposition $\Delta$. We also need the notion of the
associated dual 2-complex of $\Delta$ denoted by 
$\Delta^{\star}$. The dual 2-complex $\Delta^{\star}$ is a
combinatorial object defined by a set of vertices 
$v\in \Delta^{\star}$ (dual to d-cells in $\Delta$) edges 
$e\in \Delta^{\star}$ (dual to (d$-1$)-cells in $\Delta$) and faces $f\in \Delta^{\star}$ (dual to (d$-2$)-cells in $\Delta$).
In the case where $\Delta$ is a simplicial decomposition of $\cal M$ the structure of both $\Delta$ and $\Delta^{\star}$ is illustrated in Figures~\ref{cell2}, \ref{cell3}, and \ref{cell4}
in two, three, and four dimensions  respe1ctively.

\epubtkImage{}{%
\begin{figure}[htbp]
 \centerline{
   \includegraphics[height=5cm]{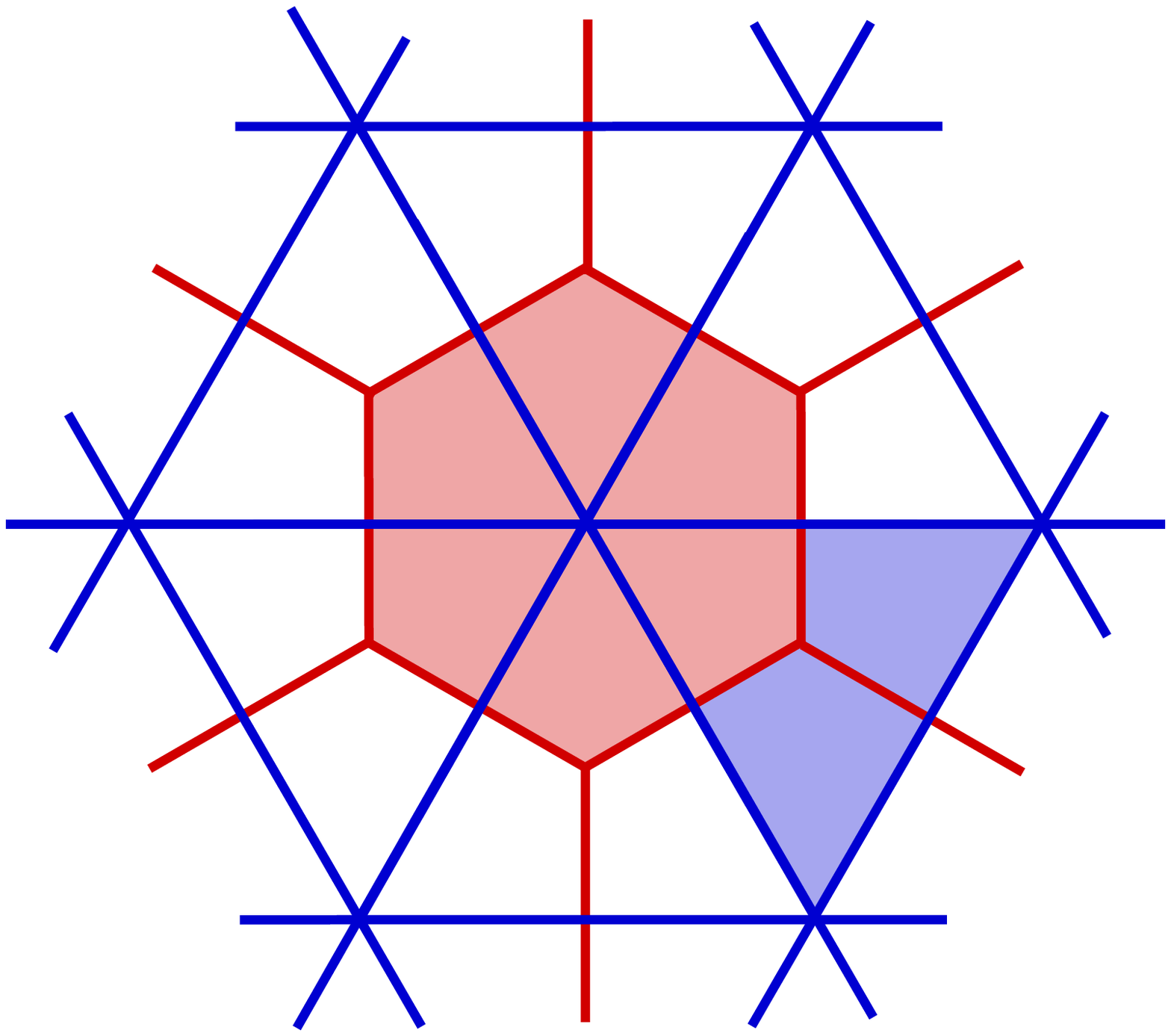}\qquad\qquad
   \includegraphics[height=5cm]{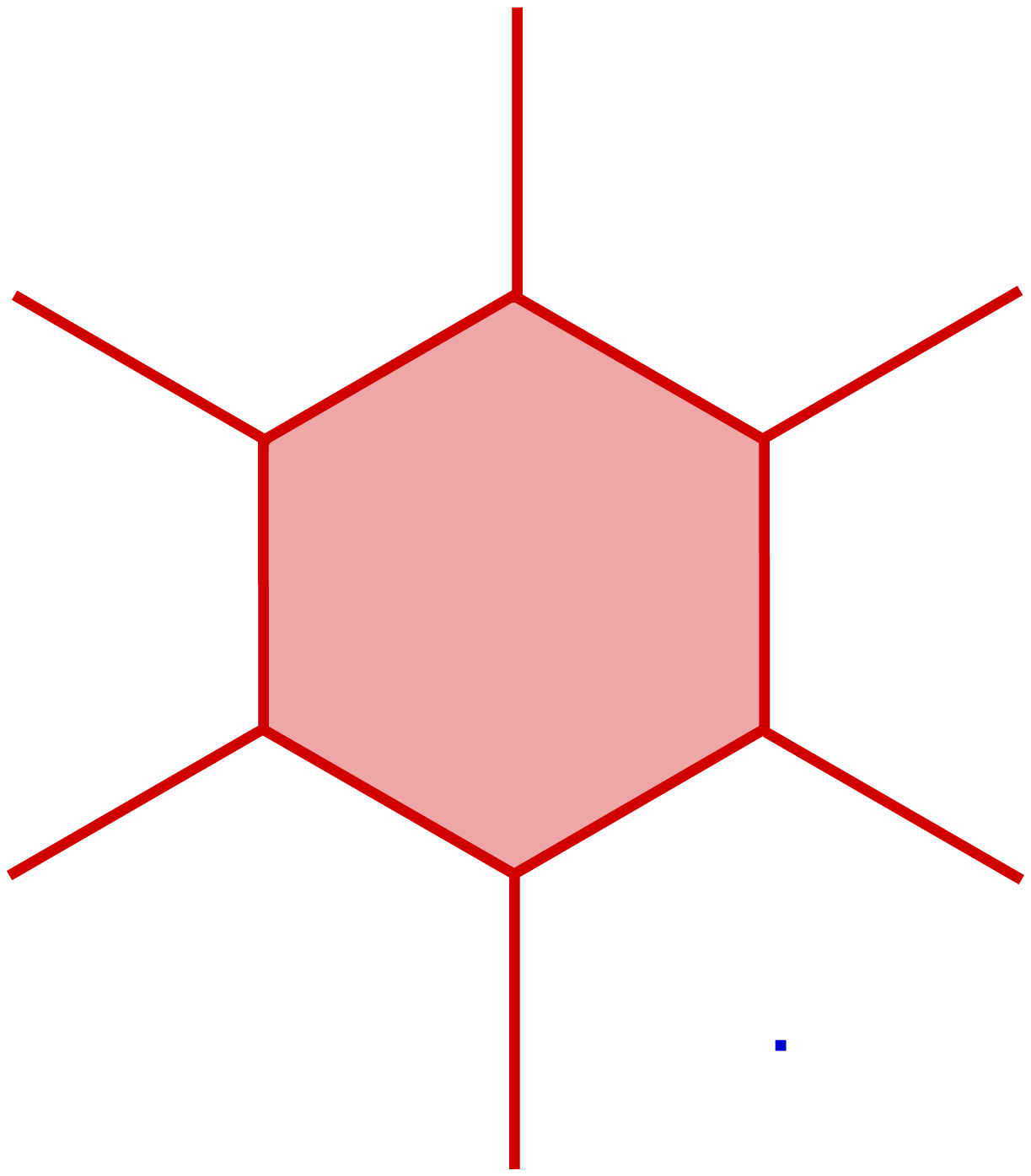}
 }
 \caption{\emph{Left:} a triangulation and its dual in two
   dimensions. \emph{Right:} the dual two complex; faces (shaded
   polygone) are dual to 0-simpleces in 2d.}
 \label{cell2}
\end{figure}}

\epubtkImage{}{%
\begin{figure}[htbp]
 \centerline{
   \includegraphics[height=5cm]{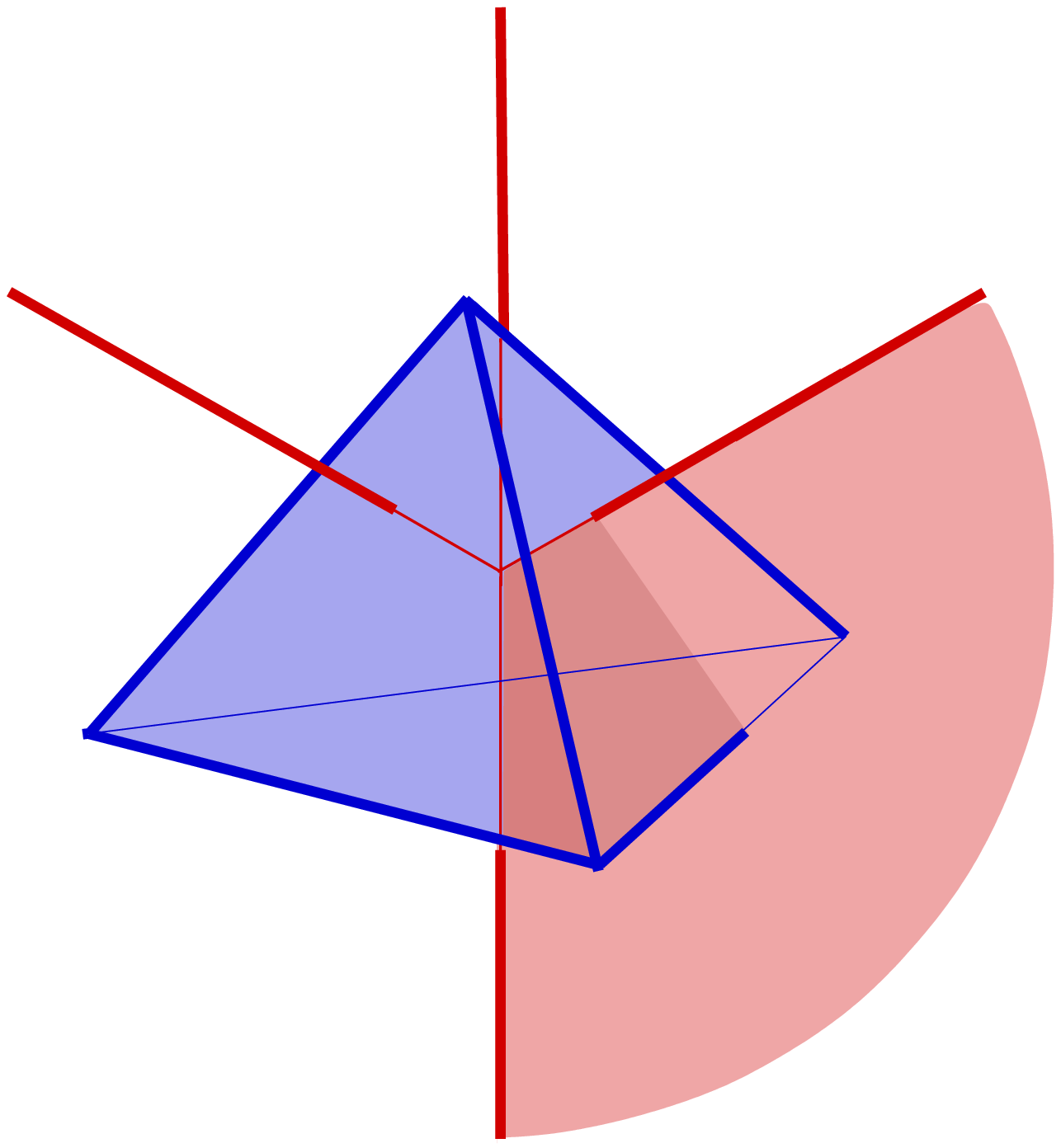}\qquad\qquad
   \includegraphics[height=5cm]{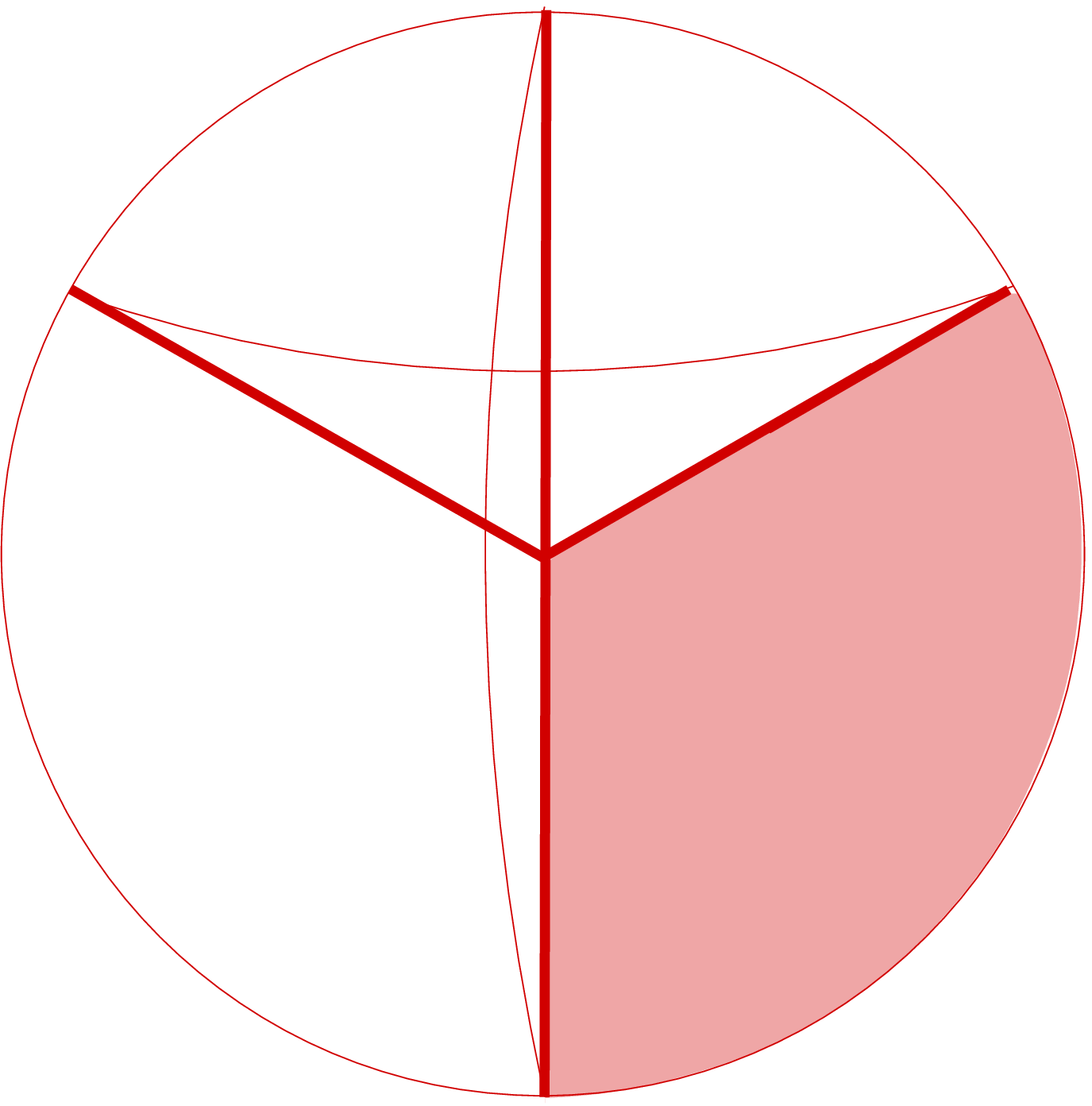}
 }
 \caption{\emph{Left:} a triangulation and its dual in three
  dimensions. \emph{Right:} the dual two complex; faces (shaded wedge)
  are dual to 1-simpleces in 3d.}
 \label{cell3}
\end{figure}}

\epubtkImage{}{%
\begin{figure}[htbp]
 \centerline{
   \includegraphics[height=5cm]{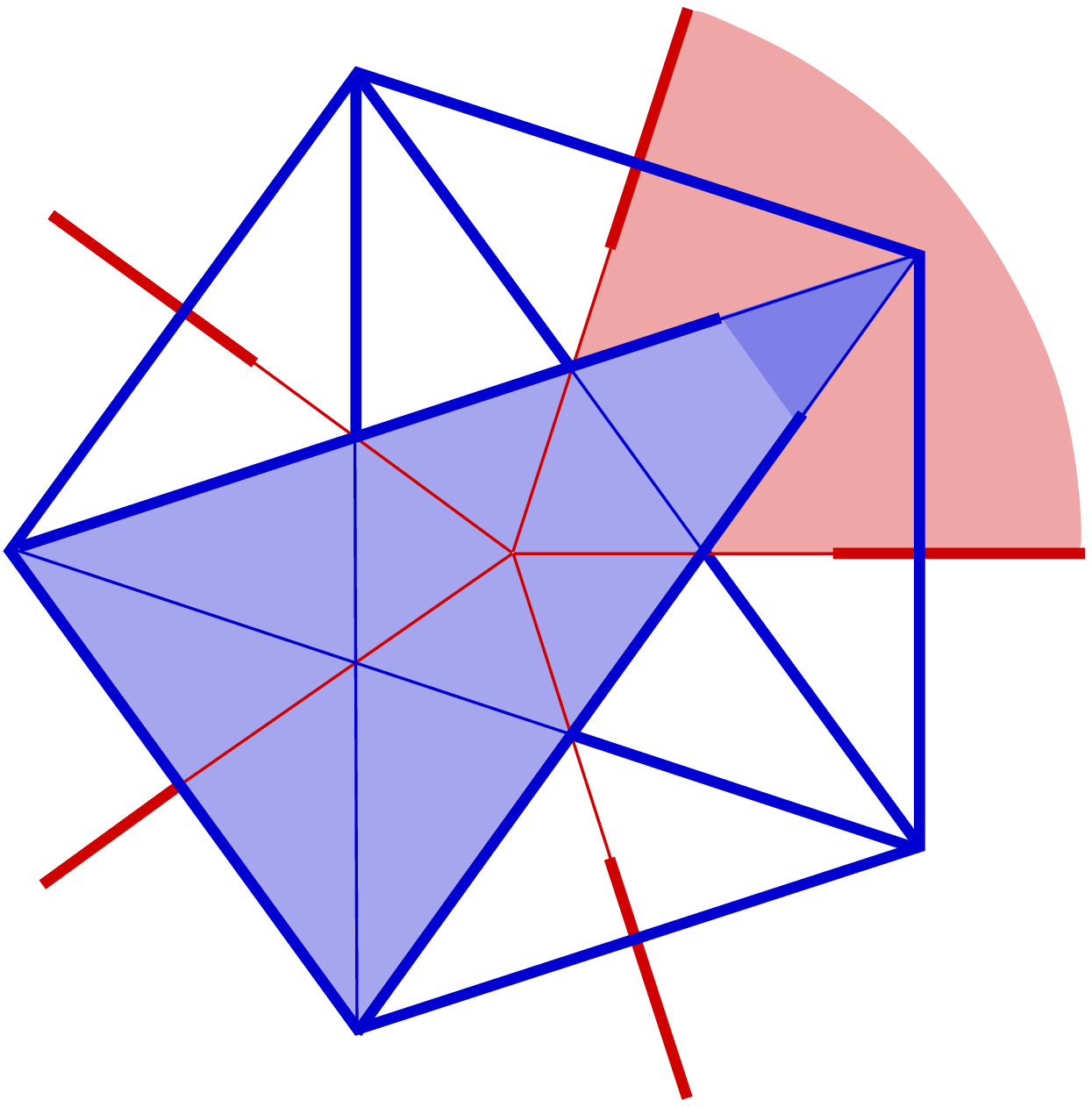}\qquad\qquad
   \includegraphics[height=5cm]{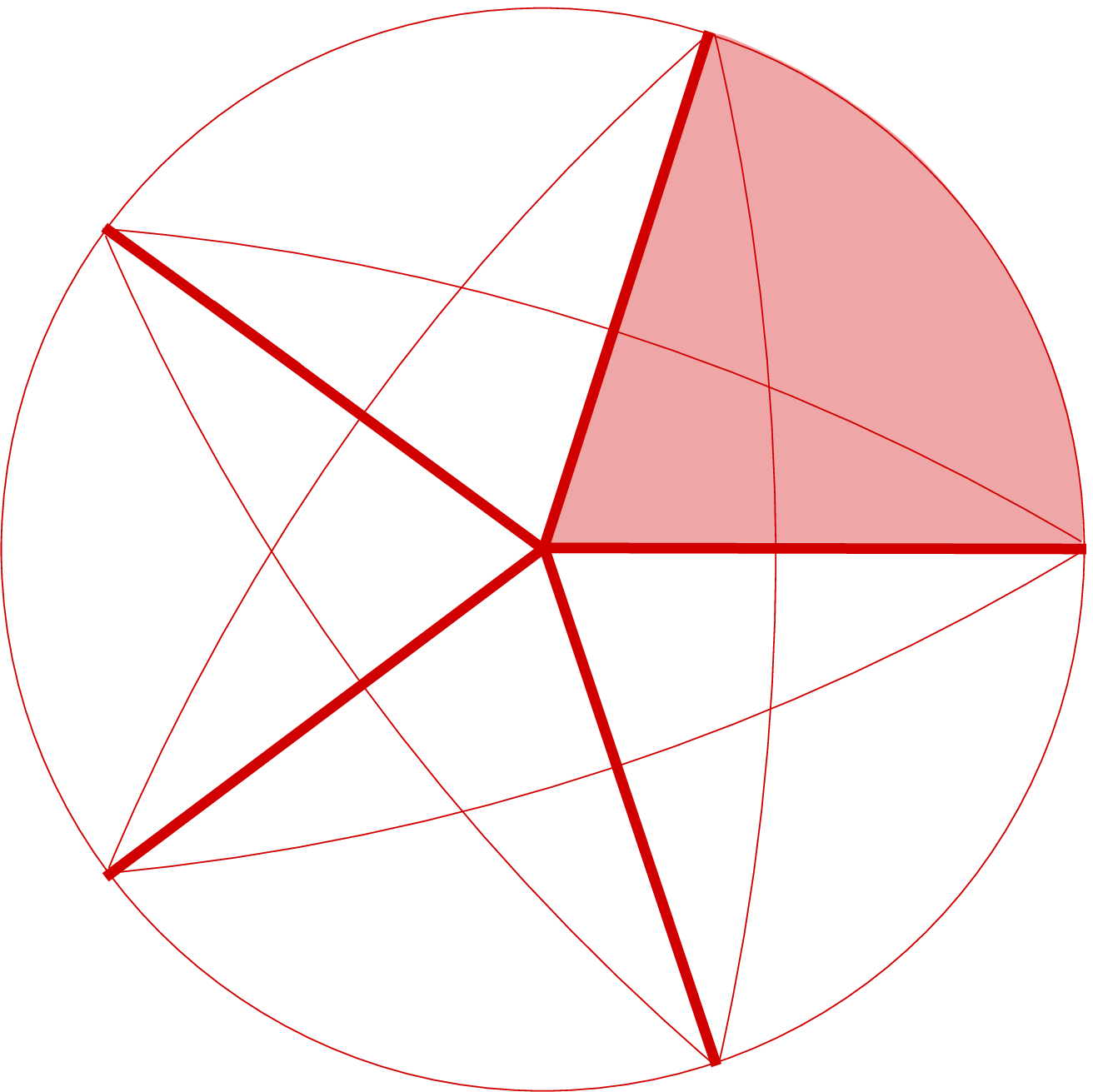}
 }
 \caption{\emph{Left:} a triangulation and its dual in four
  dimensions. \emph{Right:} the dual two complex; faces (shaded wedge)
  are dual to triangles in 4d. The shaded triangle dual to the shaded
  face is exhibited.}
 \label{cell4}
\end{figure}}

For simplicity we concentrate in the 
case when $\Delta$ is a triangulation. The field ${\rm B}$ is associated
with Lie algebra elements $B_f$ assigned to faces $f\in \Delta^{\star}$. We can think of it as
the integral of the (d$-2$)-form ${\rm B}$ on the (d$-2$)-cell dual to the face $f\in \Delta^{\star}$, namely \be\label{Bdisc}
B_{f}=\int\limits_{({\rm d}-2){\rm -cell}}\!\!\! {\rm B}.
\ee In other words
$B_f$ can be interpreted as the `smearing' of the continuous (d$-2$)-form ${\rm B}$ on
the (d$-2$)-cells in  $\Delta$. We use the one-to-one
correspondence between faces $f\in {\Delta}^{\star}$ and (d$-2$)-cells in  $\Delta$ to label the discretization of the ${\rm B}$ field $B_f$.
The connection $\omega$ is discretized by the assignment of group elements $g_e \in G$
to edges $e\in \Delta^{\star}$. One can think of the group elements $g_e$ as the holonomy of $\omega$ along 
$e\in \Delta^{\star}$, namely
\be
g_e={\rm  P} \exp (-\int_e \omega),
\ee
where the symbol $``{\rm  P} \exp"$ denotes the path-order-exponential that reminds us of the relationship of the holonomy with
the connection along the path $e\in \Delta^{\star}$.

With all this the discretized version of  the path integral (\ref{zbfg}) is
\begin{equation}\label{papart}Z(\Delta)= \int \prod_{e \in \Delta^{\star}}
dg_e \prod_{f \in \Delta^{\star}} dB_f \ e^{iB_f U_f} = \int \prod_{e \in \Delta^{\star}}
dg_e \prod_{f \in \Delta^{\star}} \delta(g_{e_1} \cdots g_{e_n}),
\end{equation}
where $U_f=g_{e_1} \cdots g_{e_n}$ denotes the holonomy around faces, and 
the second equation is the result of the ${\rm B}$ integration: it can be thus regarded as the analog of (\ref{VAg}) \footnote{It is important to point out that the integration over the algebra valued $B$ field does not exactly give the group delta function. For instance, in the simple case where $G=SU(2)$ with $B\in su(2)$ integration leads to the $SO(3)$ delta distribution (which only contains integer spin representations in the mode expansion (\ref{deltaPW})). Generally, one ignores this fact and uses the $G$-delta distribution in the models found in the literature.}. 
The integration measure $dB_f$ is the standard Lebesgue measure while the integration 
in the group variables is done in terms of the invariant measure in $G$ (which is the unique Haar measure when $G$ is compact).
For given $h\in G$ and test function $F(g)$ the invariance property reads as follows
\be\label{invariance}
\int dg F(g)=\int dg F(g^{-1})=\int dg F(gh)=\int dg F(hg) 
\ee
The 
Peter-Weyl's theorem provides a useful formula or the Dirac delta distribution appearing in (\ref{papart}), namely
\be\label{deltaPW}
\delta(g)=\sum_{\rho} d_{\rho} {\rm Tr}[\rho(g)],\ee where $\rho$ are irreducible unitary representations of $G$. From the previous expression one obtains
\begin{equation}\label{coloring4}
{\cal Z}(\Delta)=\sum \limits_{{\cal C}:\{\rho\} \rightarrow \{
f\}} \int \ \prod_{e \in \Delta^{\star}} dg_e \ \prod_{f
\in \Delta^{\star}} {\rm d}_{\rho_f} \ {\rm
Tr}\left[\rho_f(g^1_e\dots g^{\va N}_e)\right].
\end{equation}
Integration over the
connection can be performed as follows. 
In a triangulation $\Delta$, the edges $e\in \Delta^{\star}$ bound
precisely $\rm d$ different faces; therefore, the $g_e$'s in (\ref{coloring4})
appear in $\rm d$ different traces. The relevant
formula is
\begin{equation}\label{4dp}
P^{e}_{inv}(\rho_1,\cdots, \rho_{\rm d}):= \int dg_{e}\ {\rho_1(g_{e})}\otimes \rho_2(g_{e}) \otimes \cdots \otimes \rho_{\rm d}(g_{e}).
\end{equation}
For compact $G$ it is easy to prove using the invariance (and normalization) of the the integration measure (\ref{invariance}) that $P^{e}_{inv}=(P^{e}_{inv})^2$ i
s the projector onto ${\rm Inv}[\rho_1\otimes \rho_2 \otimes \cdots
\otimes \rho_{\rm d}]$. In this way the spin foam amplitudes of $SO(4)$ BF theory reduce to 
\begin{eqnarray}\label{bf4} Z_{BF}(\Delta)=\sum \limits_{ {\cal
C}_f:\{f\} \rightarrow \rho_f }  \ \prod_{f \in \Delta^{\star}} {\rm d}_{\rho_f}
\prod_{e \in {\Delta^{\star}}} P^{e}_{inv}(\rho_1,\cdots, \rho_{\rm d}).
\end{eqnarray}
In other words, the $BF$ amplitude associated to a two complex $\Delta^{\star}$ is simply given by 
the sum over all possible assignments of irreducible representations of $G$ to faces of the number obtained by the natural contraction of the network of projectors $P^e_{inv}$ according to the 
pattern provided defined by the two-complex 
$\Delta^{\star}$. 

There is a nice graphical representation of the partition function of BF theory that will be very useful for some calculations.
On the one hand, using this graphical notation  one can easily prove the discretization independence of the BF amplitudes. On the other hand this graphical notation will simplify the presentation of the new spin foam models of quantum gravity that will be considered in the following sections. This useful notation was introduced by Oeckl~\cite{Oeckl:2005rh,  Oeckl:2000hs} and used in~\cite{Girelli:2001wr} to give a general prove of the discretization independence of the BF partition function and the Turaev-Viro invariants for their definition on general cellular decompositions.

Let us try to present this notation in more detail:
The idea is to represent each representation matrix appearing in (\ref{coloring4}) by a line (called a wire) labeled by an irreducible representation, and integrations on the group by a box (called a cable). The traces in equation (\ref{coloring4}) imply that there is a wire, labelled by the representation $\rho_f$, winding around each face  $f\in \Delta^{\star}$. In addition, there is a cable (integration on the group) associated with each edge $e\in \Delta^{\star}$.  As in (\ref{bf4}), there is a projector $P^{e}_{inv}$ is the projector into ${\rm Inv}[\rho_1\otimes \rho_2 \otimes \cdots \otimes \rho_{\rm d}]$ associated to each edge. This will be represented by a cable with $\rm d$ wires as shown in (\ref{cabled}). Such graphical representation allows for a simple diagrammatic expression of the BF quantum amplitudes. 
\be P^{e}_{inv}(\rho_1,\rho_2, \rho_3,\cdots, \rho_{\rm d })\ \equiv \!\!\!\!\!\!\!\!
\psfrag{a}{\!\!\!\!\!\!\!\!\!\!\!\!\!\!\!\!\!\!\!\!\!\!\!\!\!\!\!\!\!\!\!$$}
\psfrag{x}{\!\!$\rho_1$}\psfrag{y}{\!\!$\rho_2$}\psfrag{z}{\!\!$\rho_3$}\psfrag{w}{$\cdots$}\psfrag{u}{\!\!$\rho_{\rm d}$}
\begin{array}{c}
\includegraphics[width=3cm]{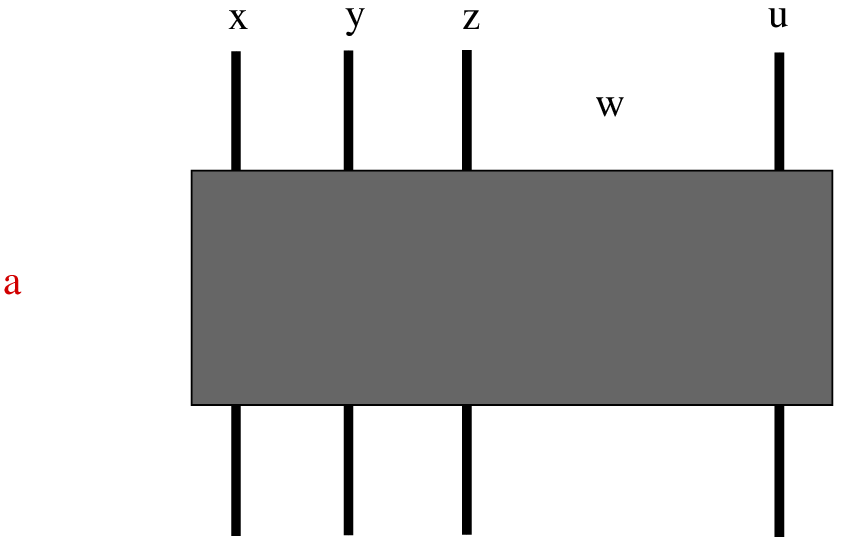}
\end{array}
\label{cabled}
\ee
We can now express the product of projection operators defining the BF quantum amplitudes in equation (\ref{bf4}) in an entirely  graphical way.  
Let us show this in 2, 3 and 4 dimensions. The generalization to arbitrary dimension is obvious.

According to Figure~(\ref{cell2}) in two dimensions edges $e\in \Delta^{\star}$ are shared by two faces. This means that there are two representation matrices associated to the group element $g_e$ and hence two wires labelled by two in principle different $\rho_f$. The endpoint of the open wires are connected to the neighbouring 
vertices and form close loops around each face (due to the trace in (\ref{coloring4})). There is a cable with two wires on each $e\in \Delta^{\star}$. The BF amplitude is obtained by plugging together  all the cables  in the way dictated by the dual 2-complex $\Delta^{\star}$.  In two dimensions the result is 
\be
Z_{BF}(\Delta)=\sum \limits_{ {\cal
C}_f:\{f\} \rightarrow \rho_f }  \ \prod\limits_{f \in \Delta^{\star}} {\rm d}_{\rho}  \begin{array}{c}\psfrag{a}{$\rho_{1}$}
\psfrag{b}{$\rho_{2}$}
\psfrag{c}{$\rho_{3}$}
\includegraphics[width=2.2cm]{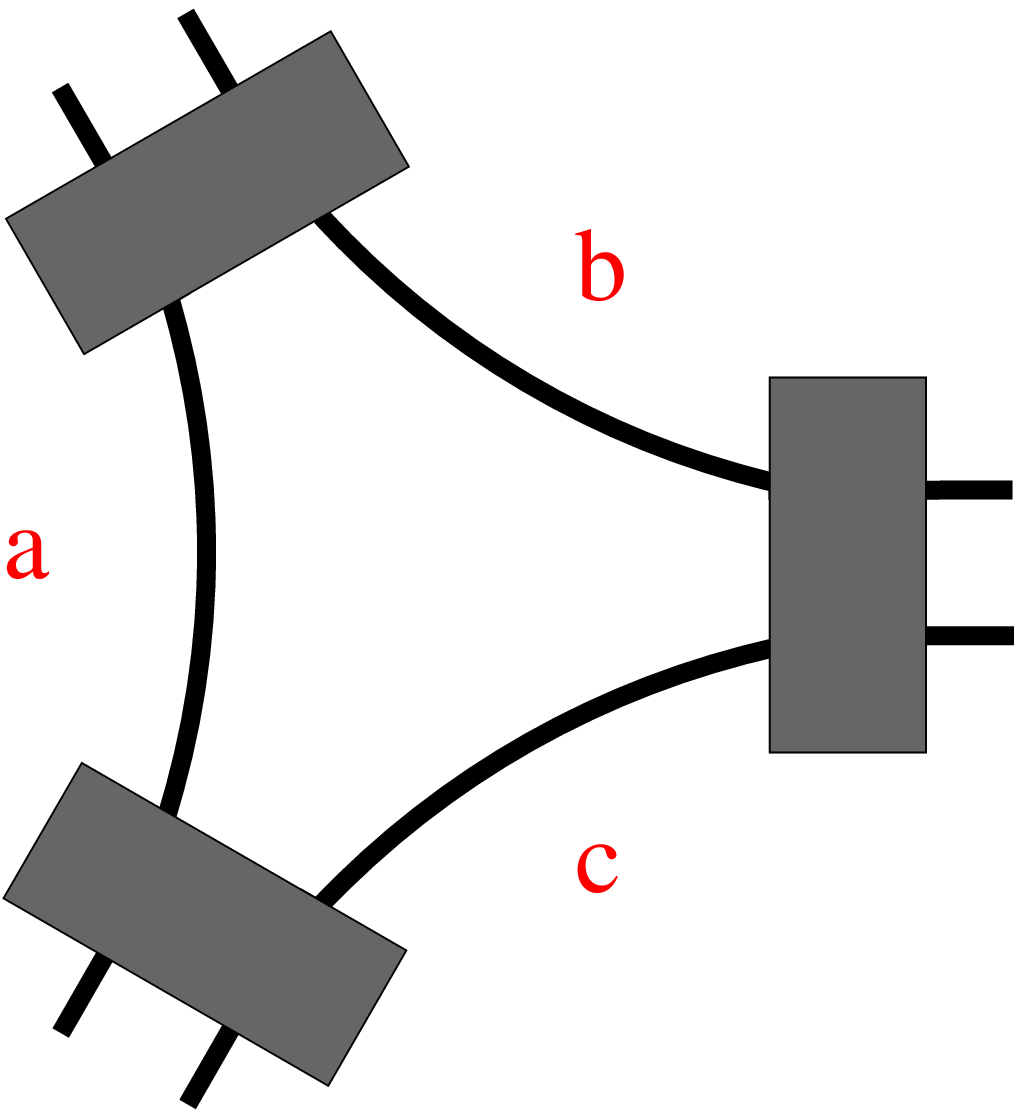}
\end{array},
\label{cabled-2d}
\ee
where it is understood that the open ends of the wires are connected to neighbouring vertices according to the combinatorial structure provided by the cellular 2-complex $\Delta^{\star}$. 
In other words, the graph on the right hand side of the previous equation is just the basic building block of a closed cable-wire diagram involving the whole complex $\Delta$.

The construction is now obvious in three dimensions. The only difference is that now three faces share each edge and hence cables have three wires labelled by three different representations. The wires are connected according to the structure provided by the dual two complex $\Delta^{\star}$ (see now Figure~\ref{cell3}) so that a close wire-loop is obtained for each face $f\in \Delta^{\star}$. The result is
\be
Z_{BF}(\Delta)=\sum \limits_{ {\cal
C}_f:\{f\} \rightarrow \rho_f }  \ \prod\limits_{f \in \Delta^{\star}} {\rm d}_{\rho}
 \begin{array}{c}\psfrag{a}{$\rho_{1}$}
\psfrag{b}{$\rho_{2}$}
\psfrag{c}{$\rho_{3}$}
\psfrag{d}{$\rho_{4}$}
\psfrag{e}{$\rho_{5}$}
\psfrag{f}{$\rho_{6}$}
\includegraphics[width=3.5cm]{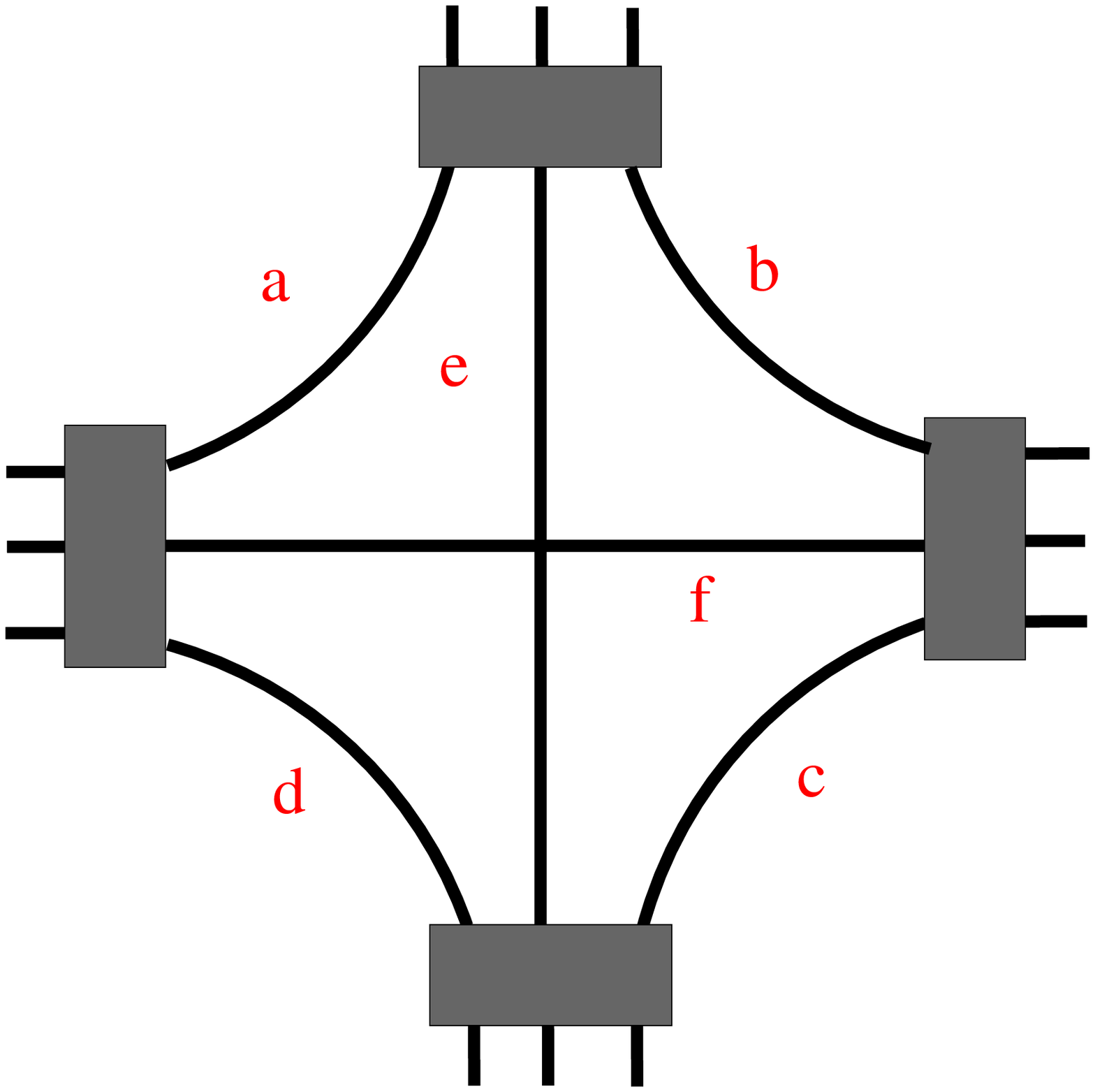}
\end{array}.
 \label{cabled-3d}
\ee
At each vertex we have six wires corresponding to the six faces $f\in \Delta^{\star}$ sharing a vertex $v\in \Delta^{\star}$.  As in the 2d case, the cable-diagram diagram on the right is only a piece of a 
global diagram embracing the whole complex $\Delta$.

Finally, in four dimensions the construction follows the same lines. Now edges are shared by four faces; each cable has now four wires. The
cable wire diagram giving the BF amplitude is disctated by the combinatorics of the dual two complex $\Delta^{\star}$. From Figure~\ref{cell4} we get 
\be
Z_{BF}(\Delta)=\sum \limits_{ {\cal
C}_f:\{f\} \rightarrow \rho_f }  \ \prod\limits_{f \in \Delta^{\star}} {\rm d}_{\rho}
\begin{array}{c}\psfrag{a}{$\rho_{1}$}
\psfrag{b}{$\rho_{2}$}
\psfrag{c}{$\rho_{3}$}
\psfrag{d}{$\rho_{4}$}
\psfrag{f}{$\rho_{5}$}
\psfrag{g}{$\rho_{6}$}
\psfrag{h}{$\rho_{7}$}
\psfrag{i}{$\rho_{8}$}
\psfrag{j}{$\rho_{9}$}
\psfrag{k}{$\rho_{10}$}
\includegraphics[width=5cm]{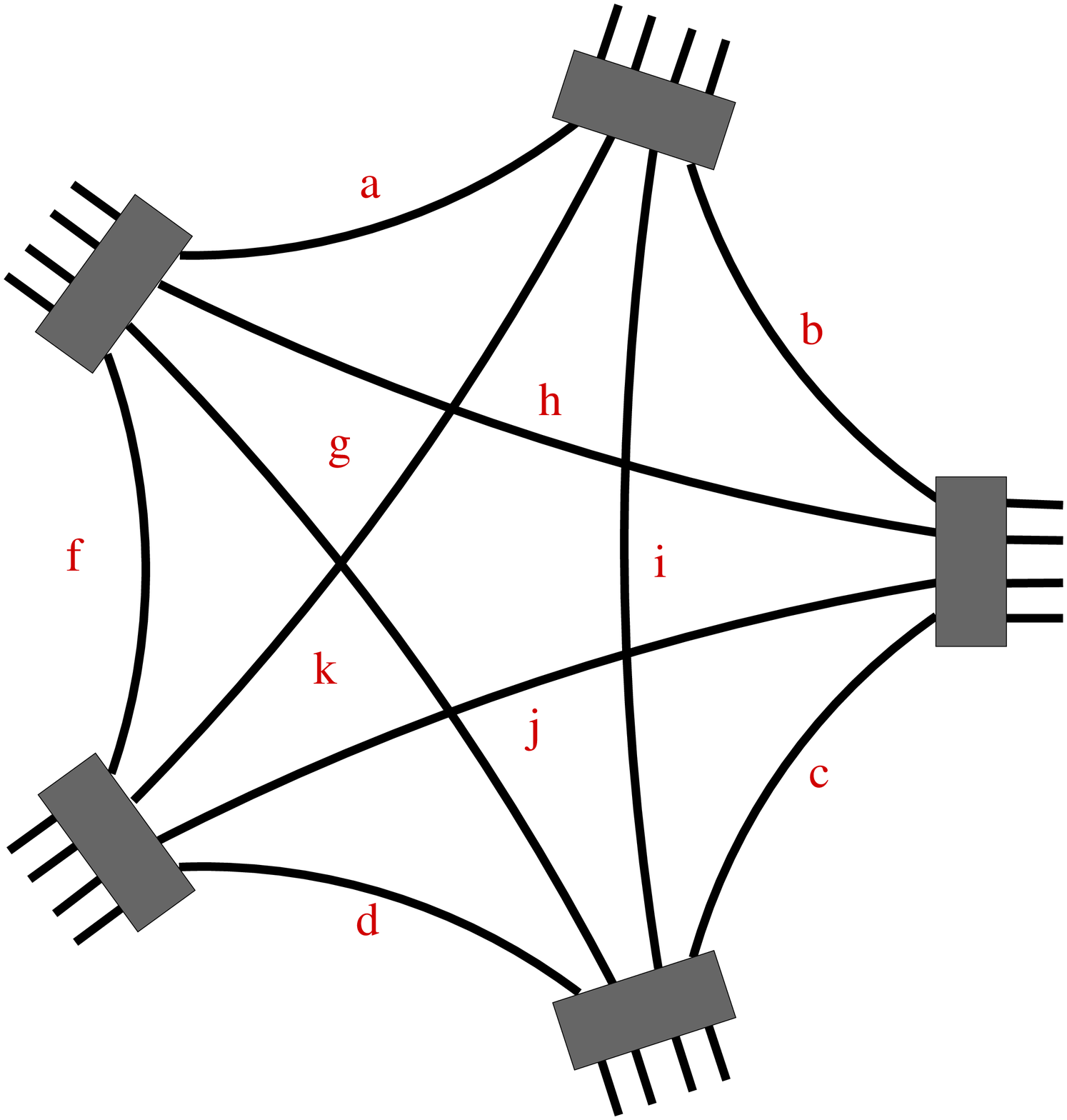}
\end{array}
\label{cabled-4d}.
\ee
 The 10 wires corresponding to the 10 faces $f\in \Delta^{\star}$ sharing a vertex $v\in \Delta^{\star}$ are connected to the neighbouring  vertices through the 5 cables (representing the projectors in (\ref{bf4}) and Figure~\ref{cabled}) associated to the 5 edges  $e\in \Delta^{\star}$ sharing the vertex $v\in \Delta^{\star}$.

\subsection{Special Cases}

We end this section with some simple examples of BF theory that will be useful in following applications.

\subsubsection{$SU(2)$ BF theory in 2d: the simplest topological model}

The amplitude of BF theory in two dimensions can be  entirely worked out due to the simplicity of the projectors $P^{e}_{inv}$ in this case.
For concreteness here we take the special case $G=SU(2)$ as this structure group will play an important case in the gravity context, while  the techniques used here will provide a nice  warming-up exercise for what follows.
Irreducible representations of $SU(2)$ are labelled by spins (half-integers): from now on we set $\rho_f=j_{f}\in \N/2$. In two dimensions, the relevant projectors are 
$P^{e}_{inv}:j_{f}\otimes j_{f'}\to{\rm Inv} [j_{f}\otimes j_{f'}]$ which in turn are non trivial if and only if $j_f=j_{f'}^{*}$. In our graphical notation the projector is
\be
P_{inv}^e(j,j^{\prime})=
\begin{array}{c}
\psfrag{x}{$j$}\psfrag{y}{$j'$}
\includegraphics[height=1.5cm]{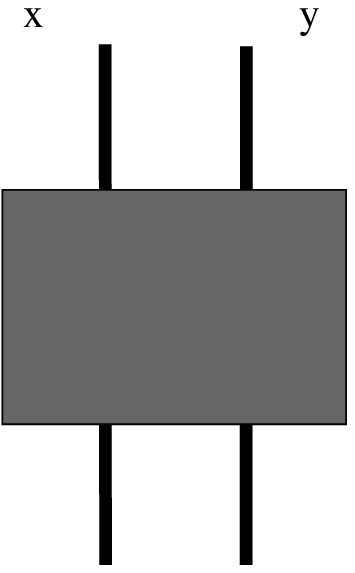}
\end{array}=\frac{\delta_{j,j'}}{(2j+1)}
\begin{array}{c}\psfrag{a}{$$}\psfrag{x}{$j$}\psfrag{y}{$j'$}
\includegraphics[height=1.5cm]{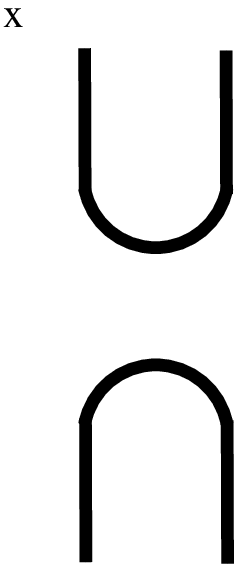}
\end{array},
\label{p2}\ee
where the lines on the right hand side denote the identity representation matrices in the representation $j$.
Replacing this in the expression of the BF amplitude given in equation (\ref{cabled-2d}) one gets
\be
Z^{2d}_{BF}(\Delta)=\sum \limits_{j }  \ \prod\limits_{f \in \Delta^{\star}} (2j+1)
\prod\limits_{e \in {\Delta^{\star}}} \frac{1}{(2j+1)}
\begin{array}{c}
\psfrag{a}{$j$}
\includegraphics[width=1.5cm]{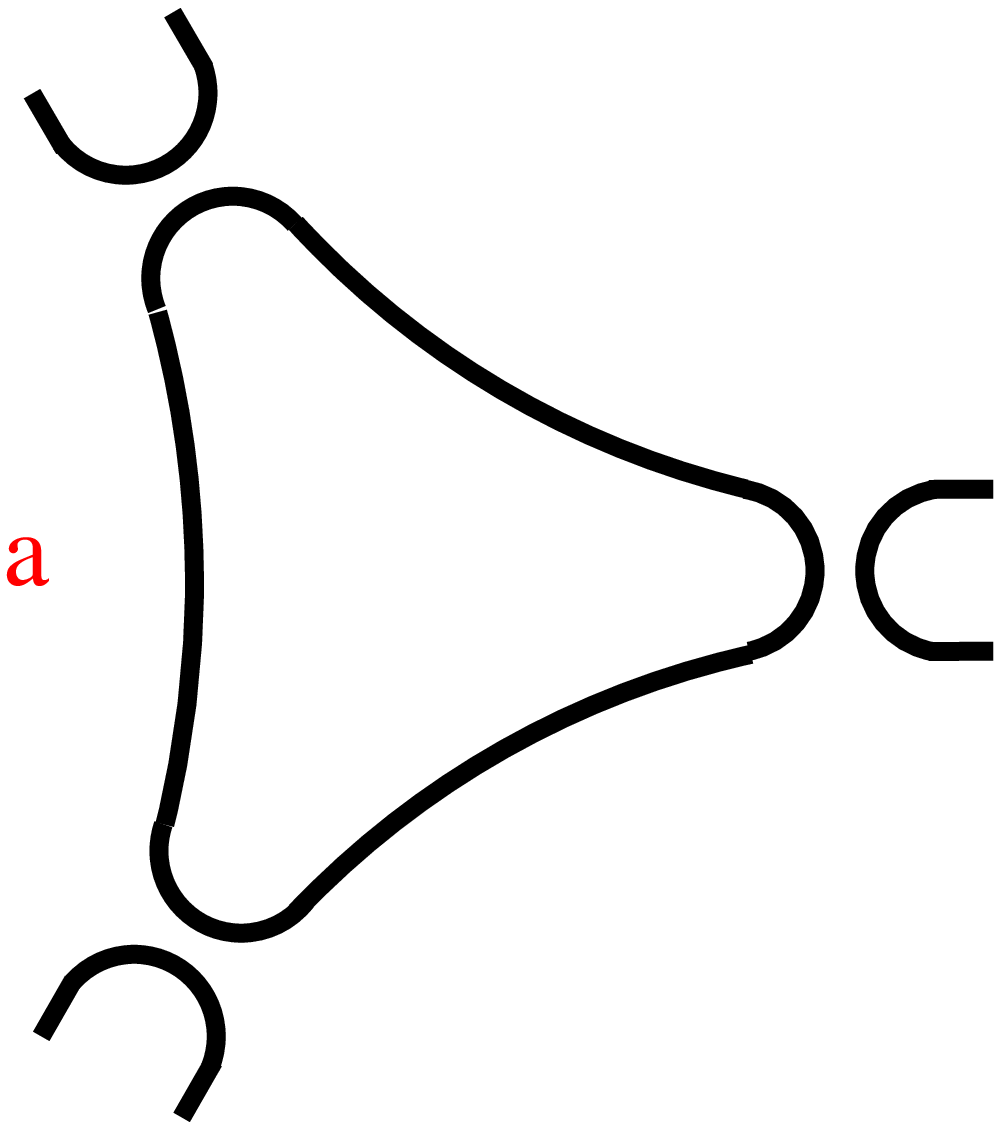}
\end{array},
\label{cabled-2d-su2}
\ee
where the sum over $j_f$ in (\ref{cabled-2d}) has collapsed to a single sum as according to (\ref{p2}).  Spin foam amplitudes vanish unless all the spins assigned to faces are equal, i.e. $j_f=j$ for all $f\in \Delta^{\star}$.
In addition to a factor ${\rm d}_{j}=(2j+1)$ for each face $f\in \Delta^{\star}$ one gets a factor $1/(2j+1)$ for each edge $e\in \Delta^{\star}$ from the projectors
(see \ref{p2}). Finally, for each vertex $v\in \Delta^{\star}$ one obtains a factor $(2j+1)$ coming from the remaining closed loop that is left over once (\ref{p2}) is used in all the edges associated with $v$ (see (\ref{cabled-2d-su2})). 
Finally, if we denote $N_{f}$, $N_e$, and $N_v$ the number of faces, the number of edges, and the number of vertices in $\Delta^{\star}$ respectively; then it follows that the BF partition function becomes
\begin{eqnarray}\label{bf4} Z^{2{\rm d}}_{BF}(\Delta)=\sum \limits_{j}  (2j+1)^{N_f-N_e+N_v}=\sum \limits_{j}  (2j+1)^{\chi},
\end{eqnarray}
where $\chi$ is the Euler character of the two dimensional manifold $\cal M$. In this simple two dimensional example one discovers explicitly two important properties
of the partition function (and quantum amplitudes) of BF theory:  first, quantum amplitudes are independent of  the regulating triangulation used in their definition, and second their discretization independent value  depend only on 
global properties of the spacetime defining topological invariants of the manifold $\cal M$.  One can genuinly write
\be
Z^{2{\rm d}}_{BF}({\cal M})=Z^{2{\rm d}}_{BF}(\Delta).
\ee

\subsubsection*{Extra remarks on two dimensional BF theory}

Two dimensional BF theory has been used as the basic theory in an attempt to define a manifold independent 
model of QFT in~\cite{Livine:2003kn}. It is also related to gravity in two dimensions  in two ways: on the one hand it is  equivalent to the so-called 
Jackiw-Teitelboim model~\cite{JTMODEL1, JTMODEL2}, on the other hand it is related to usual 2d gravity via constraints in a way similar to the one exploited in four dimensions
(see next section).  The first relationship has been used in the canonical quantization of the Jackiw-Teitelboim model  in~\cite{Constantinidis:2008ty}.
The second relationship has been explored in~\cite{Oriti:2004qk}

\subsubsection{$SU(2)$ BF theory and 3d Riemannian gravity}
\label{BF3}

In three dimensions edges $e\in \Delta^{\star}$ are shared by three faces and thus the relevant projector is represented by a cable with three wires.
In the case $G=SU(2)$ the singlet component ${\rm Inv}[\j_1\otimes j_2\otimes j_3]$  is one dimensional when it is non-trivial. Therefore, we can write the projector as 
\be
\begin{array}{c}\psfrag{z}{$m$}
\psfrag{a}{$=$}
\psfrag{x}{$j$}\psfrag{y}{$k$}
\includegraphics[height=2cm]{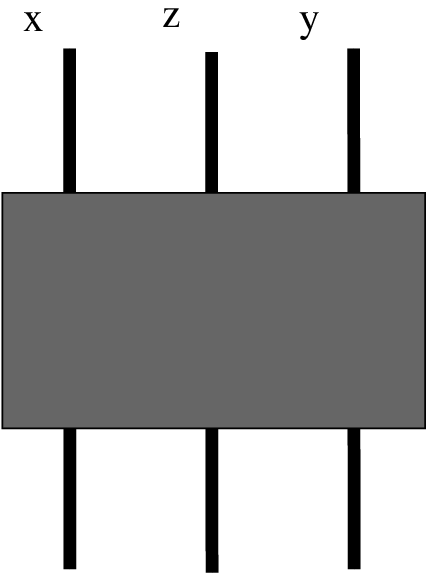}
\end{array}
\begin{array}{c}\psfrag{z}{$m$}
\psfrag{a}{$=$}
\psfrag{x}{$j$}\psfrag{y}{$k$}
\includegraphics[height=2cm]{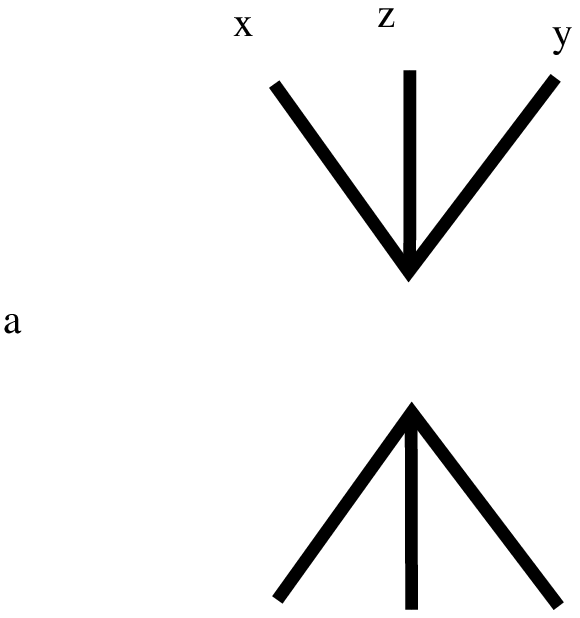}
\end{array},
\label{cab-3d}
\ee 
 where the three valent open graphs on the right hand side represent the unique normalized invariant vector in $j\otimes m \otimes k$. 
 Now we can play the same game as in the previous example. Replacing the expression of the projectors in equation (\ref{cabled-3d}) we get
\be
\ \ \ \ \ \ \ \ \ \ \ \ \ \ \ \begin{array}{c}\psfrag{x}{\!\!\!\!\!\!\!\!\!\!\!\!\!\!\!\!\!\!\!\!\!\!\!\!\!\!\!\!\!\!\!\!\!\!\!\!\!\!\!\!\!\!\!\!\!\!\!\!\!\!\!\!\!\!\!\!\!\!\!\!\!\!\!$Z_{BF}(\Delta)=\sum \limits_{ {\cal
C}_f:\{f\} \rightarrow \rho_f }  \ \prod\limits_{f \in \Delta^{\star}} {\rm d}_{\rho}
$}
\psfrag{a}{$j_1$}
\psfrag{b}{$j_2$}\psfrag{c}{$j_3$}\psfrag{d}{$j_4$}\psfrag{e}{$j_5$}\psfrag{f}{$j_6$}
\includegraphics[width=6cm]{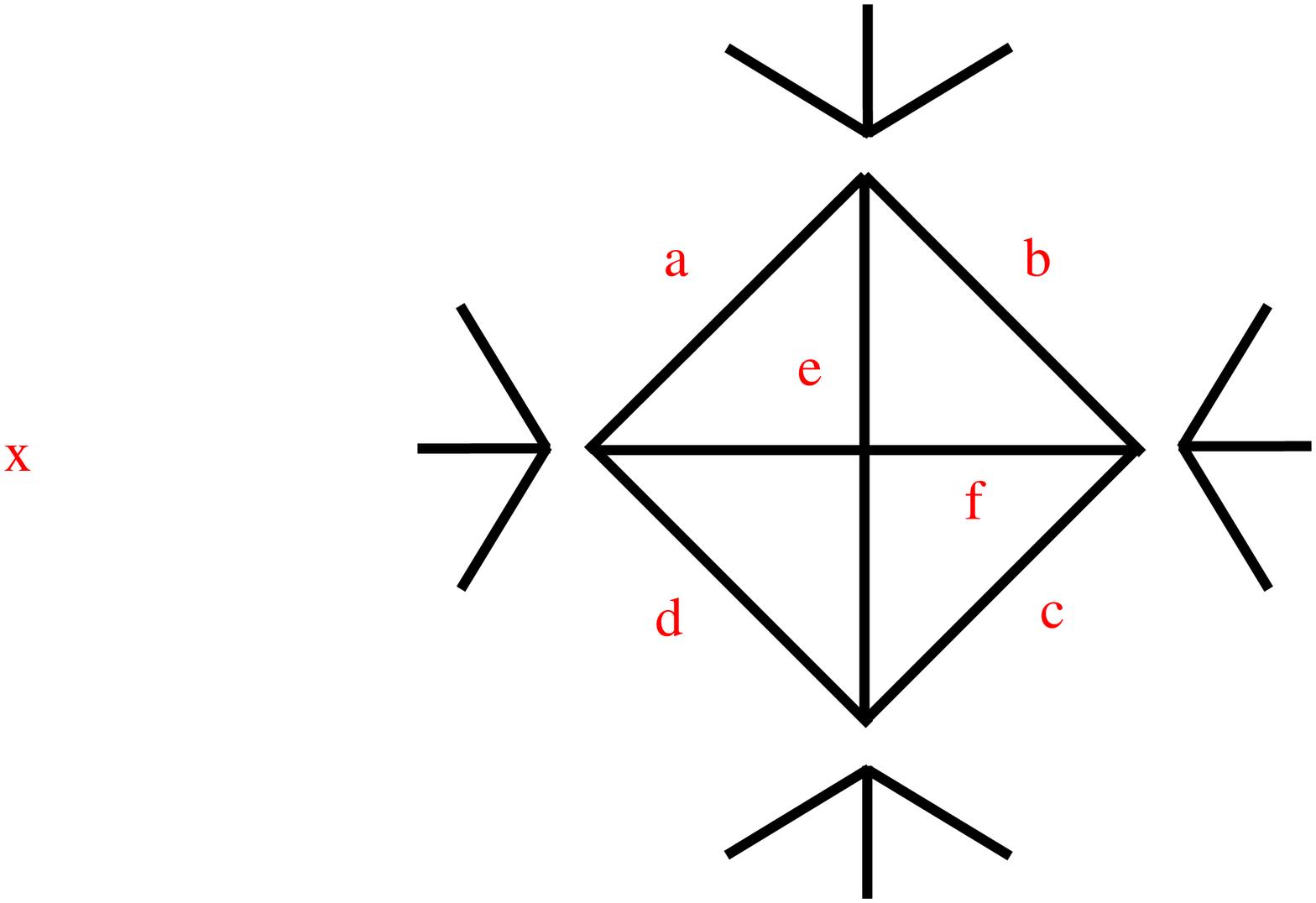}
\end{array}.
\label{cabled-3d-1}
\ee
Hence, the amplitude can be written as a sum over spins $j_f$ associated to vertices of a product of face and vertex amplitudes, namely 
\be
\ \ \ \ \ \ \ \ \ \begin{array}{c}
\psfrag{x}{\!\!\!\!\!\!\!\!\!\!\!\!\!\!\!\!\!\!\!\!\!\!\!\!\!\!\!\!\!\!\!\!\!\!\!\!\!\!\!\!\!\!\!\!\!\!\!\!\!\!\!\!\!\!\!\!\!\!\!\!$Z_{BF}(\Delta)=\sum \limits_{ {\cal
C}_f:\{f\} \rightarrow \rho_f }  \ \prod\limits_{f \in \Delta^{\star}} {\rm d}_{\rho}
\prod\limits_{v \in {\Delta^{\star}}}$}
\psfrag{a}{$j_1$}
\psfrag{b}{$j_2$}\psfrag{c}{$j_3$}\psfrag{d}{$j_4$}\psfrag{e}{$j_5$}\psfrag{f}{$j_6$}
\includegraphics[width=5.5cm]{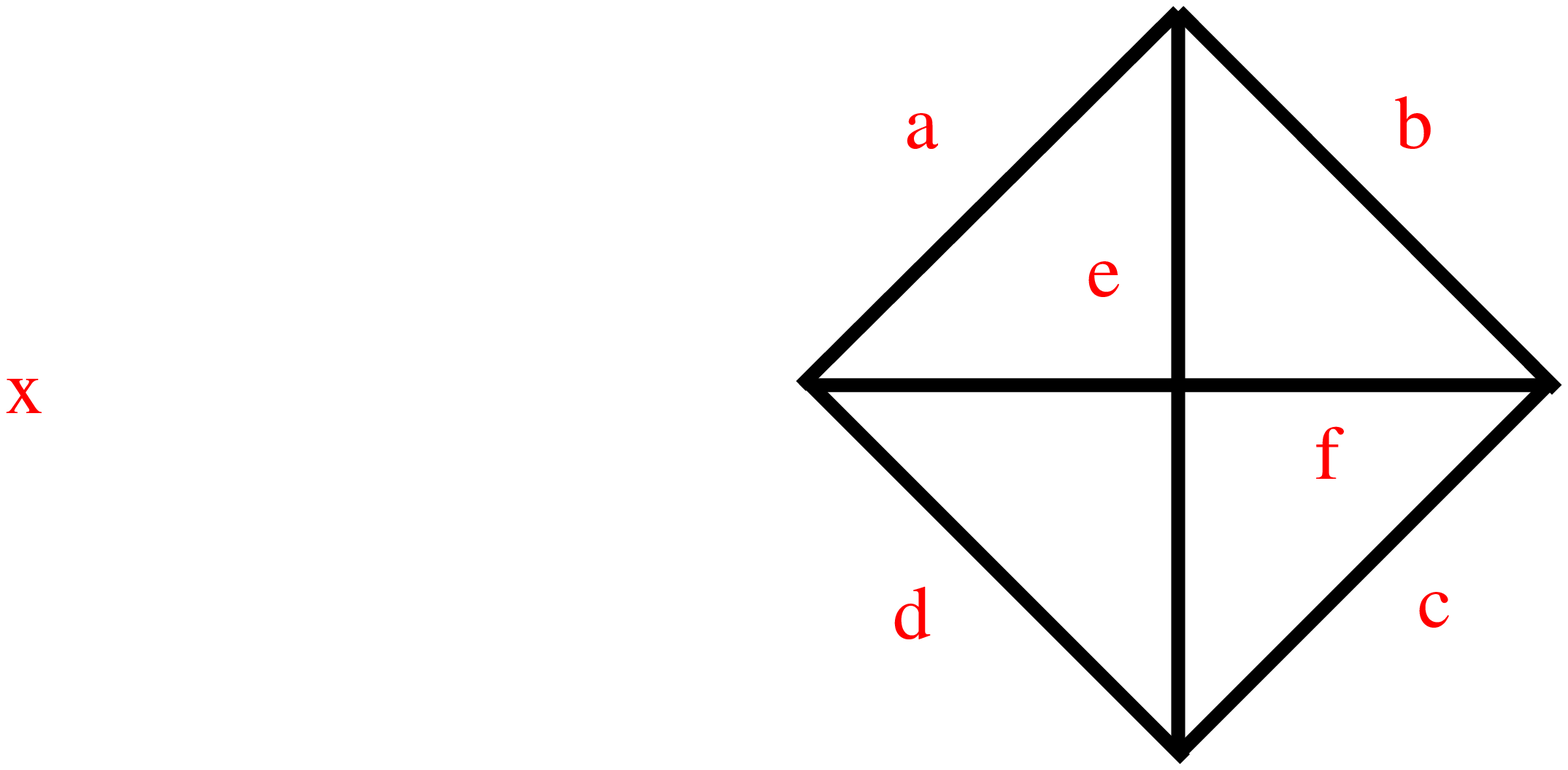}
\end{array}.
\label{cabled-3d-3}
\ee
The vertex amplitude is a $6j$-symbol and depends on six spins labelling  the corresponding faces associated with a vertex.

\subsubsection*{Extra remarks on three dimensional BF theory}

Three dimensional BF theory and the spin foam quantization presented above is intimately related to classical and 
quantum gravity in three dimensions (for a classic reference see~\cite{carlip}). We will discuss this relationship in detail in Part~\ref{part3d}.
The state sum as presented above matches the quantum amplitudes first proposed by Ponzano and Regge in the 1960s  based on their 
discovery of the asymptotic expressions of the 6j symbols~\cite{ponza} and is 
often referred to as the Ponzano-Regge model. Divergences in the above formal expression require regularization.
We will see in Part~\ref{part3d} that natural regularizations are available and that the model is well defined~\cite{Barrett:2008wh, Noui:2004iy, frei8}.
For a detailed study of the divergence structure of the model see~\cite{Bonzom:2010ar, Bonzom:2010zh, Bonzom:2011br, Bonzom:2012mb}. 
The quantum deformed version of the above amplitudes  lead to the so called
Turaev-Viro model~\cite{TV}  which is expected to correspond to the quantization of three dimensional Riemannian gravity in the 
presence of a non vanishing positive cosmological constant. For the definition of observables in the latter context as well as in the analog four dimensional analog see~\cite{Barrett:2004im}.

The topological character of BF theory can be preserved by the coupling of the theory with topological defects that
play the role of point particles. In the spin foam literature this has been considered form the canonical perspective in~\cite{Noui:2004jb, a21}
and from the covariant perspective extensively by Freidel and Louapre~\cite{Freidel:2004vi}. These theories have been shown by Freidel and Livine to be 
dual, in a suitable sense, to certain non-commutative fields theories in three dimensions~\cite{Freidel:2005bb, Freidel:2005me}.

Concerning coupling BF theory with non topological matter see~\cite{Fairbairn:2006dn, Dowdall:2010ej} for the case of fermionic matter, and~\cite{Speziale:2007mt} for gauge fields.
A more radical perspective for the definition of matter in 3d gravity is taken in~\cite{Fairbairn:2007sv}. For three dimensional supersymmetric  BF theory models see~\cite{Livine:2003hn, Baccetti:2010xd}

Recursion relations
for the 6j vertex amplitudes have been investigated in ~\cite{Bonzom:2011jh, Dupuis:2009qw}. They provide a tool for studying dynamics in spin foams of 3d gravity and might be useful in higher dimensions~\cite{Bonzom:2009zd}.

\subsubsection{$SU(2)$ BF theory: The Oouguri model}

In four dimensions the structure is essentially the same. Now edges are shared by four representations. In the case of $G=SU(2)$ the vector space ${\rm Inv}[j_1 \otimes j_2\otimes j_3\otimes j_4]$ has a dimension that is generically greater than one. Consequently, if one wants to write the BF quantum partition function as a sum over representations of products of faces and vertex amplitudes (the spin foam representation) one has to express the projectors in terms of a basis of invariant vectors in  $j_1 \otimes j_2\otimes j_3\otimes j_4$.  Graphically, this can be done as follows
\be
\begin{array}{c}\psfrag{x}{$\rho_1$}
\psfrag{y}{$\rho_2$}
\psfrag{z}{$\rho_3$}
\psfrag{w}{$\rho_4$}
\psfrag{h}{ $ $}
\psfrag{a}{$j^{}_1$}
\psfrag{b}{$j^{}_2$}
\psfrag{c}{$j^{}_3$}
\psfrag{d}{$j^{}_4$}
\psfrag{A}{$j^{+}_1$}
\psfrag{B}{$j^{+}_2$}
\psfrag{C}{$j^{+}_3$}
\psfrag{D}{$j^{+}_4$}
\psfrag{im}{$\iota^{}$}
\psfrag{ip}{$\iota^{+}$}
\includegraphics[height=2cm]{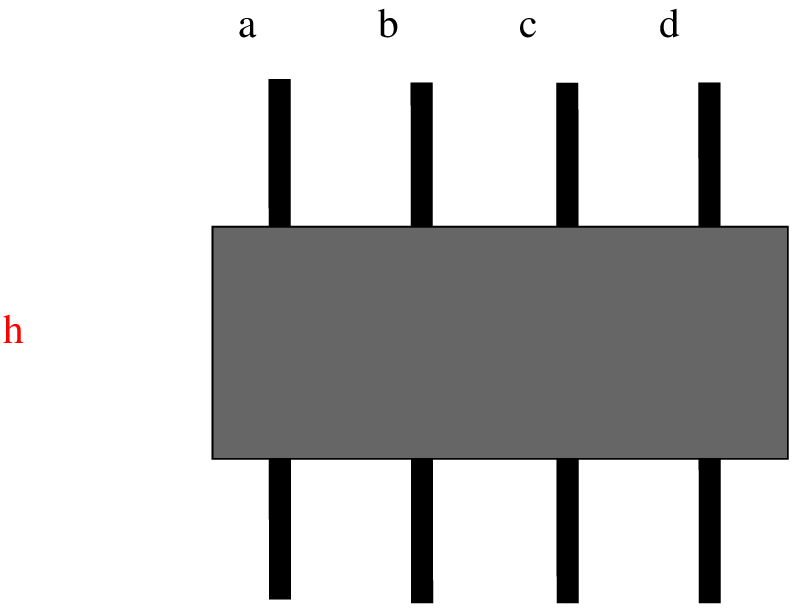}
\end{array}
={\sum\limits_{\iota^{}}}
\begin{array}{c}\psfrag{x}{$\rho_1$}
\psfrag{y}{$\rho_2$}
\psfrag{z}{$\rho_3$}
\psfrag{w}{$\rho_4$}
\psfrag{h}{ $ $}
\psfrag{a}{$j^{}_1$}
\psfrag{b}{$j^{}_2$}
\psfrag{c}{$j^{}_3$}
\psfrag{d}{$j^{}_4$}
\psfrag{A}{$j^{+}_1$}
\psfrag{B}{$j^{+}_2$}
\psfrag{C}{$j^{+}_3$}
\psfrag{D}{$j^{+}_4$}
\psfrag{im}{$\iota^{}$}
\psfrag{ip}{$\iota^{+}$}
\psfrag{g}{$$}
\includegraphics[height=2cm]{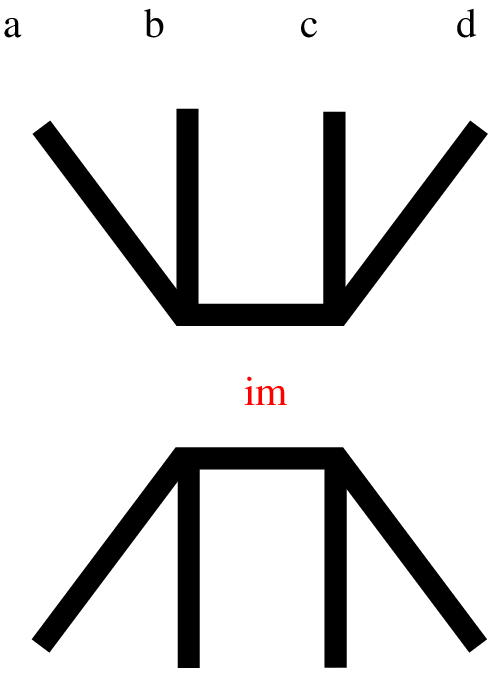}
\end{array},
\label{cab-3d}
\ee
where the four-valent diagrams on the right hand side represent elements of an orthonormal  basis of invariant  vectors in $j_1\otimes j_2\otimes j_3\otimes j_4$ and its dual, which are labelled by a half integer $\iota$. As in the previous lower dimensional cases we replace the previous equation in the expression (\ref{cabled-4d}) to obtain  
\be
Z_{BF}(\Delta)=\sum \limits_{ {\cal
C}_f:\{f\} \rightarrow j_f }  \ \prod\limits_{f \in \Delta^{\star}} {\rm d}_{j_f}\sum \limits_{ {\cal
C}_e:\{e\} \rightarrow \iota_e } 
\begin{array}{c}\psfrag{a}{$\iota_1$}
\psfrag{b}{$\iota_2$}
\psfrag{c}{$\iota_3$}
\psfrag{d}{$\iota_4$}
\psfrag{e}{$\iota_5$}
\psfrag{A}{$j_1$}
\psfrag{B}{$j_2$}
\psfrag{C}{$j_3$}
\psfrag{D}{$j_4$}
\psfrag{E}{$j_5$}
\psfrag{F}{$j_6$}
\psfrag{G}{$j_7$}
\psfrag{H}{$j_8$}
\psfrag{I}{$j_9$}
\psfrag{J}{$j_{10}$}
\includegraphics[height=6cm]{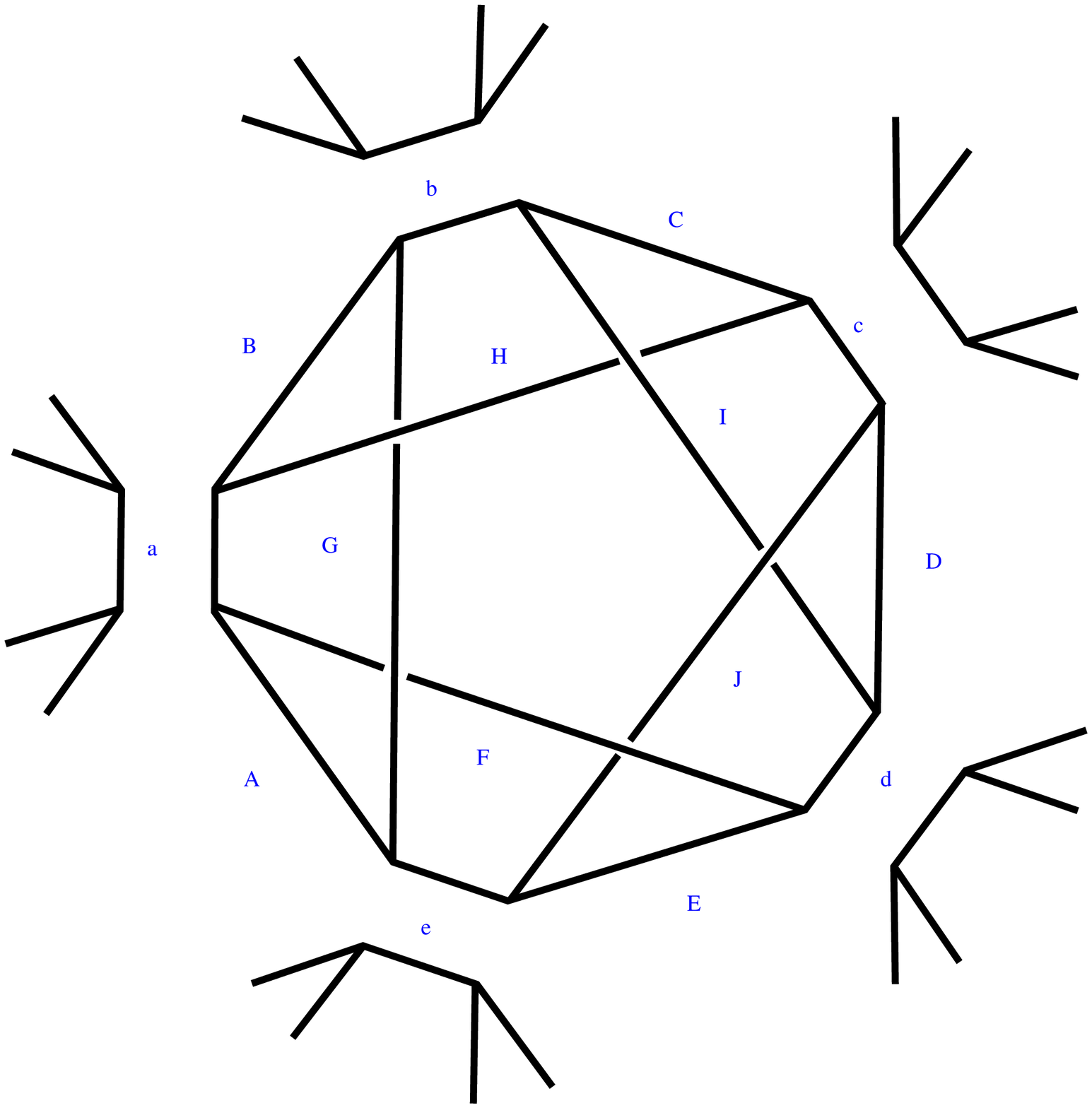}
\end{array}, 
\label{BFsu2-1}
\ee
which can be re-written in the spinfoam representation as
\be
Z_{BF}(\Delta)=\sum \limits_{ {\cal
C}_f:\{f\} \rightarrow j_f }  \ \prod\limits_{f \in \Delta^{\star}} {\rm d}_{j_f} \sum \limits_{ {\cal
C}_e:\{e\} \rightarrow \iota_e } \ \prod\limits_{v\in \Delta^{\star}} 
 \begin{array}{r}\psfrag{a}{$\iota_1$}
\psfrag{b}{$\iota_2$}
\psfrag{c}{$\iota_3$}
\psfrag{d}{$\iota_4$}
\psfrag{e}{$\iota_5$}
\psfrag{A}{$j_1$}
\psfrag{B}{$j_2$}
\psfrag{C}{$j_3$}
\psfrag{D}{$j_4$}
\psfrag{E}{$j_5$}
\psfrag{F}{$j_6$}
\psfrag{G}{$j_7$}
\psfrag{H}{$j_8$}
\psfrag{I}{$j_9$}
\psfrag{J}{$j_{10}$}
\includegraphics[height=4cm]{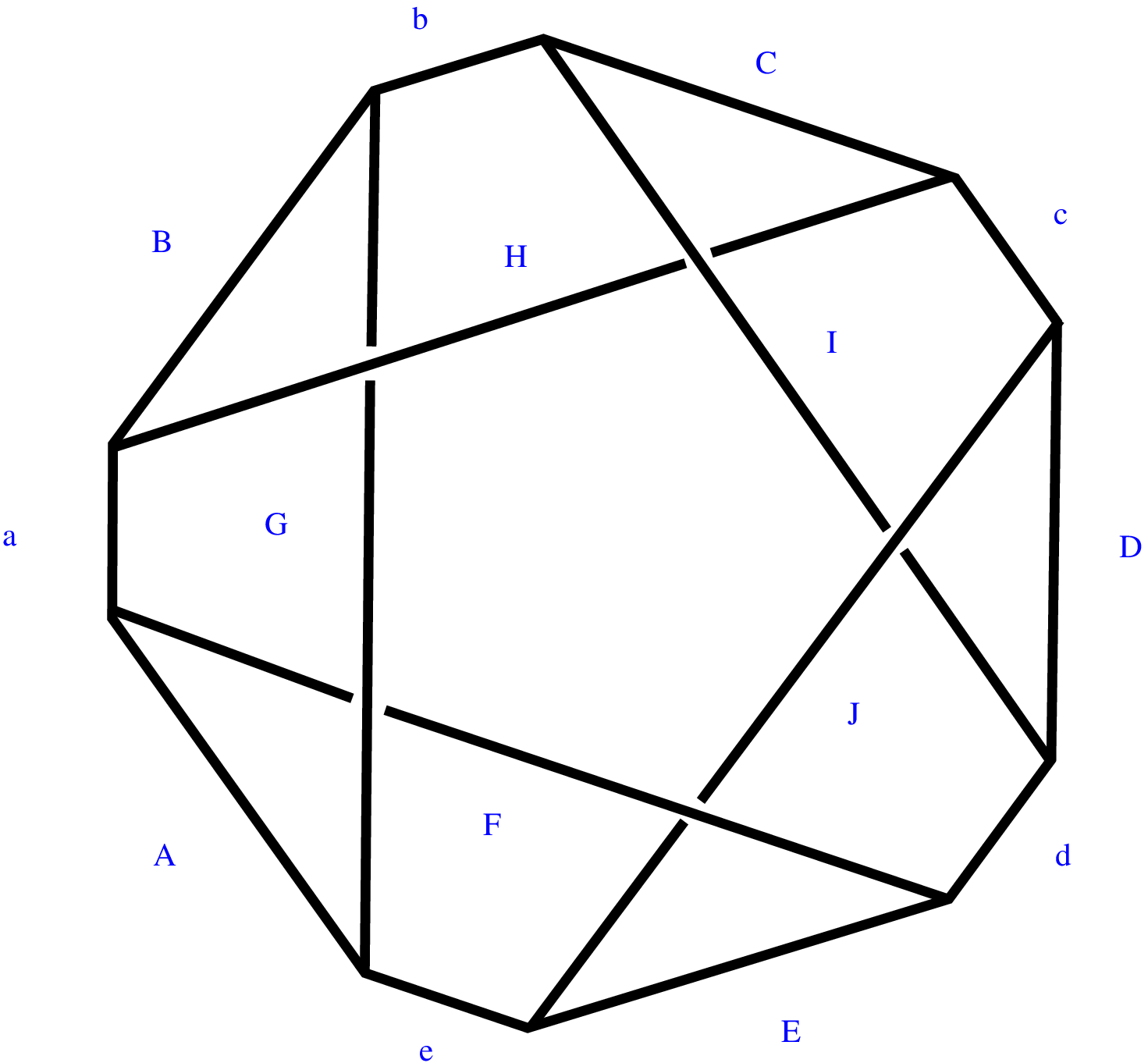}.
\end{array}\label{BFSU2}
\ee
The vertex amplitude in the previous expression is often called a 15j-symbol.

\subsubsection{$SU(2)\times SU(2)$ BF theory: a starting point for 4d Riemannian gravity}

For completeness we now present the BF quantum amplitudes in the case $G=SU(2)\times SU(2)$. This special case is of fundamental importance in the construction of the gravity models presented in the following  sections. The construction mimics that of the previous section in all details. In fact the product form of the structure group implies the simple relationship $Z_{BF}(SU(2)\times SU(2))=Z_{BF}(SU(2))^2$. Nevertheless, it is important for us to present this example in explicit form as it will provide the graphical notation that is needed to introduce the gravity models in a simple manner. As in the previous cases the spin foam representation of the BF partition function follows from expressing the projectors in (\ref{cabled-4d}) in the an orthonormal basis of intertwiners. From the previous example and the product form of the structure group we have  
\be
\begin{array}{c}
\psfrag{x}{$\rho_1$}
\psfrag{y}{$\rho_2$}
\psfrag{z}{$\rho_3$}
\psfrag{w}{$\rho_4$}
\psfrag{h}{ $=$}
\psfrag{a}{$j^{-}_1$}
\psfrag{b}{$j^{-}_2$}
\psfrag{c}{$j^{-}_3$}
\psfrag{d}{$j^{-}_4$}
\psfrag{A}{$j^{+}_1$}
\psfrag{B}{$j^{+}_2$}
\psfrag{C}{$j^{+}_3$}
\psfrag{D}{$j^{+}_4$}
\psfrag{im}{$\iota^{-}$}
\psfrag{ip}{$\iota^{+}$}
\psfrag{g}{$={\LARGE \sum\limits_{\iota^{-}\iota^{+}}}$}
\includegraphics[height=2cm]{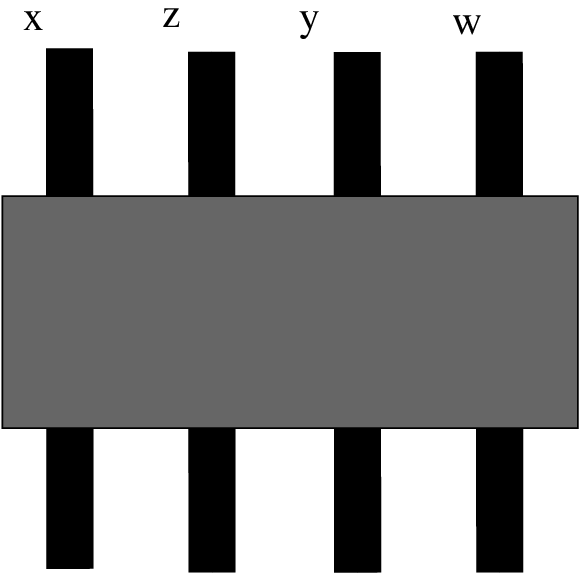}
\end{array}=
\begin{array}{c}
\psfrag{x}{$\rho_1$}
\psfrag{y}{$\rho_2$}
\psfrag{z}{$\rho_3$}
\psfrag{w}{$\rho_4$}
\psfrag{h}{ $=$}
\psfrag{a}{$j^{-}_1$}
\psfrag{b}{$j^{-}_2$}
\psfrag{c}{$j^{-}_3$}
\psfrag{d}{$j^{-}_4$}
\psfrag{A}{$j^{+}_1$}
\psfrag{B}{$j^{+}_2$}
\psfrag{C}{$j^{+}_3$}
\psfrag{D}{$j^{+}_4$}
\psfrag{im}{$\iota^{-}$}
\psfrag{ip}{$\iota^{+}$}
\psfrag{g}{$={\LARGE \sum\limits_{\iota^{-}\iota^{+}}}$}
\includegraphics[height=2cm]{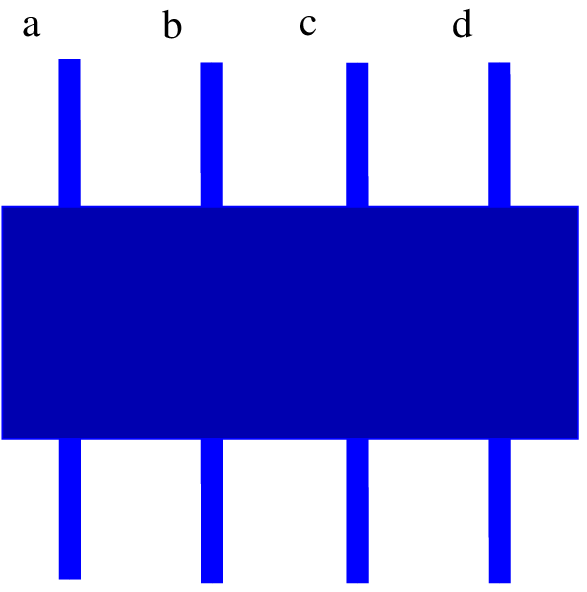}
\end{array}
\begin{array}{c}
\psfrag{x}{$\rho_1$}
\psfrag{y}{$\rho_2$}
\psfrag{z}{$\rho_3$}
\psfrag{w}{$\rho_4$}
\psfrag{h}{ $=$}
\psfrag{a}{$j^{-}_1$}
\psfrag{b}{$j^{-}_2$}
\psfrag{c}{$j^{-}_3$}
\psfrag{d}{$j^{-}_4$}
\psfrag{A}{$j^{+}_1$}
\psfrag{B}{$j^{+}_2$}
\psfrag{C}{$j^{+}_3$}
\psfrag{D}{$j^{+}_4$}
\psfrag{im}{$\iota^{-}$}
\psfrag{ip}{$\iota^{+}$}
\psfrag{g}{$={\LARGE \sum\limits_{\iota^{-}\iota^{+}}}$}
\includegraphics[height=2cm]{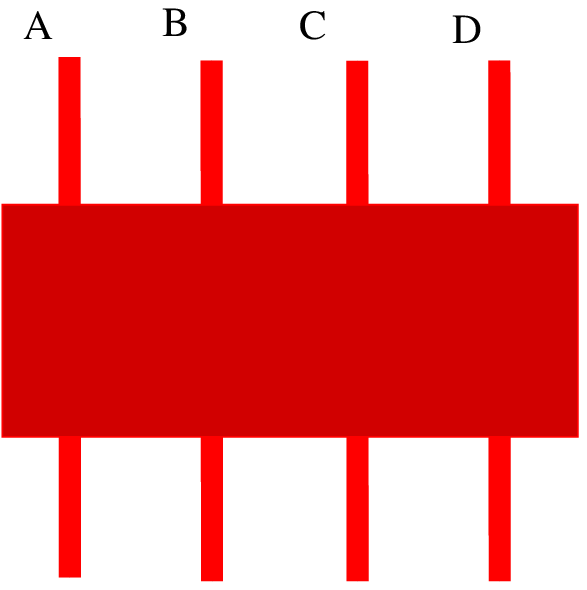}
\end{array}=\sum\limits_{\iota^{-}\iota^{+}}
\begin{array}{c}
\psfrag{x}{$\rho_1$}
\psfrag{y}{$\rho_2$}
\psfrag{z}{$\rho_3$}
\psfrag{w}{$\rho_4$}
\psfrag{h}{ $=$}
\psfrag{a}{$j^{-}_1$}
\psfrag{b}{$j^{-}_2$}
\psfrag{c}{$j^{-}_3$}
\psfrag{d}{$j^{-}_4$}
\psfrag{A}{$j^{+}_1$}
\psfrag{B}{$j^{+}_2$}
\psfrag{C}{$j^{+}_3$}
\psfrag{D}{$j^{+}_4$}
\psfrag{im}{$\iota^{-}$}
\psfrag{ip}{$\iota^{+}$}
\psfrag{g}{$={\LARGE }$}
\includegraphics[height=2cm]{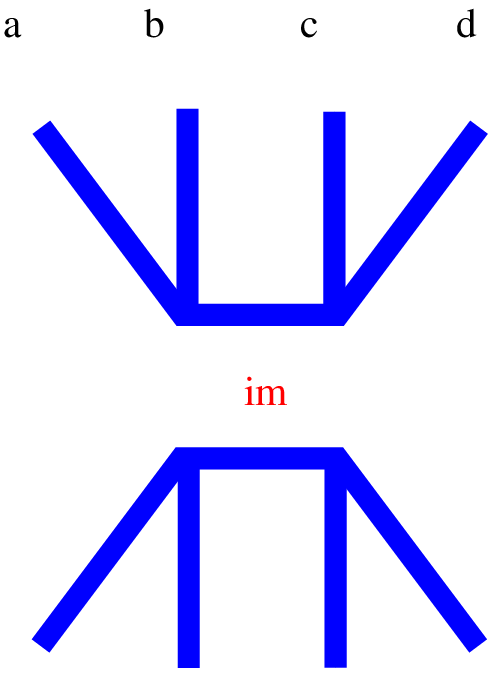}
\end{array}\ \ \ \ 
\begin{array}{c}
\psfrag{x}{$\rho_1$}
\psfrag{y}{$\rho_2$}
\psfrag{z}{$\rho_3$}
\psfrag{w}{$\rho_4$}
\psfrag{h}{ $=$}
\psfrag{a}{$j^{-}_1$}
\psfrag{b}{$j^{-}_2$}
\psfrag{c}{$j^{-}_3$}
\psfrag{d}{$j^{-}_4$}
\psfrag{A}{$j^{+}_1$}
\psfrag{B}{$j^{+}_2$}
\psfrag{C}{$j^{+}_3$}
\psfrag{D}{$j^{+}_4$}
\psfrag{im}{$\iota^{-}$}
\psfrag{ip}{$\iota^{+}$}
\psfrag{g}{$={\LARGE \sum\limits_{\iota^{-}\iota^{+}}}$}
\includegraphics[height=2cm]{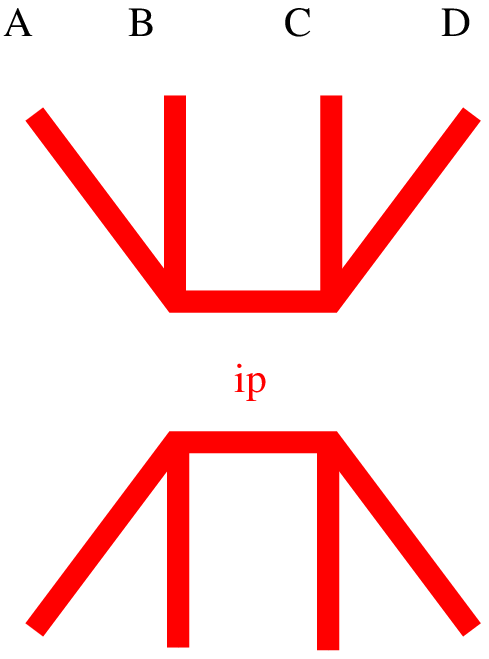},
\end{array}
\label{cab-44d}
\ee 
 where $\rho_f=j_f^{-}\otimes j_f^+$,  and $j_f^{\pm}$ and $\iota^{\pm}$ are half integers labelling left and right representations of $SU(2)$ that 
 defined the irreducible unitary representations of $G=SU(2)\times SU(2)$. Accordingly, when replacing the previous expression in (\ref{cabled-4d}) one gets 
\be\label{bf-so4}
Z_{BF}(\Delta)=\sum \limits_{ {\cal
C}_f:\{f\} \rightarrow \rho_f }  \ \prod\limits_{f \in \Delta^{\star}} {\rm d}_{j_f^{-}}{\rm d}_{j_f^{+}}
\begin{array}{c}
\includegraphics[width=5cm]{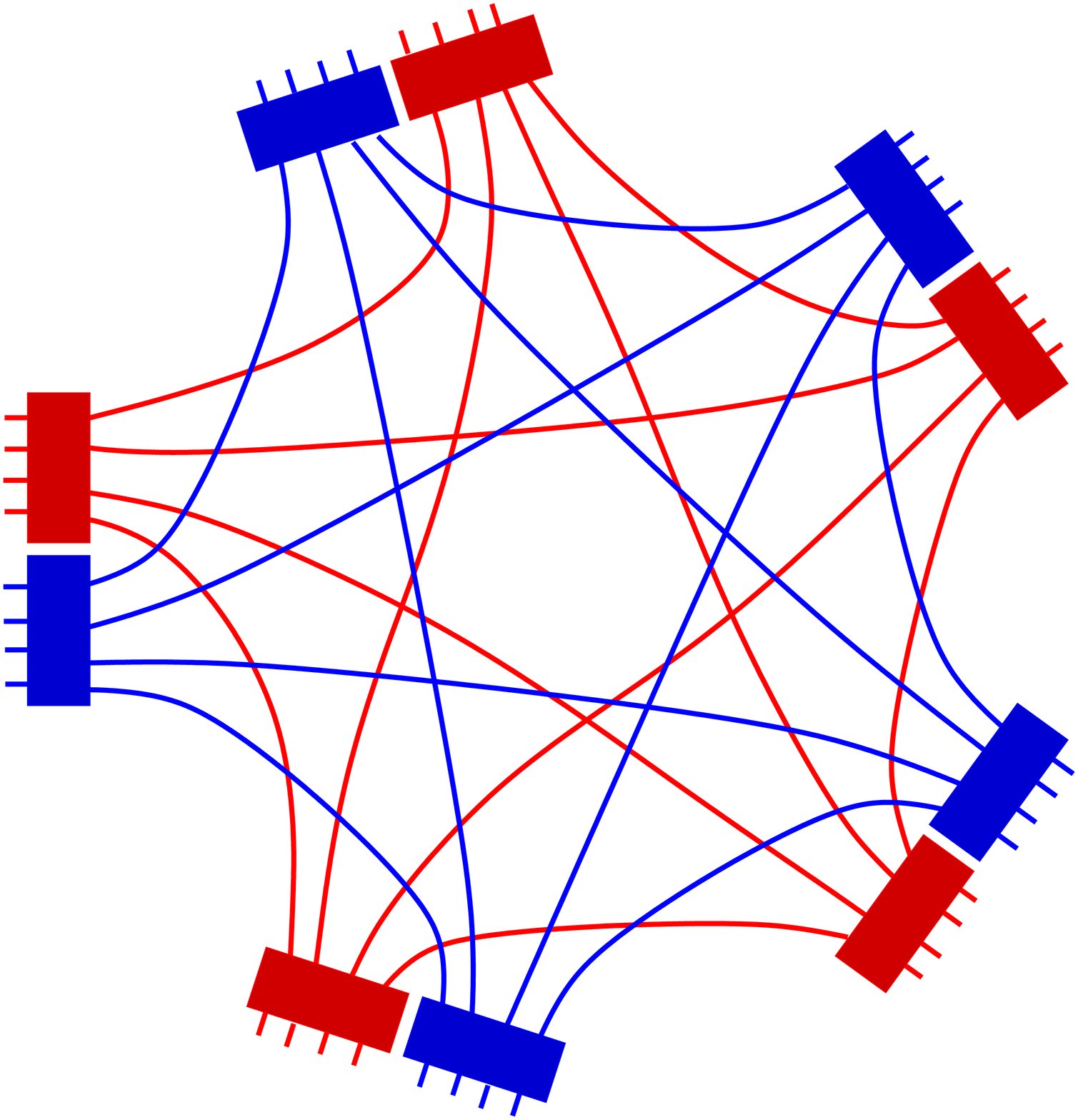}
\end{array},
\ee and equivalently
\be
Z_{BF}(\Delta)=\sum \limits_{ {\cal
C}_f:\{f\} \rightarrow \rho_f }  \ \prod\limits_{f \in \Delta^{\star}} {\rm d}_{j_f^{-}}{\rm d}_{j_f^{+}} \sum \limits_{ {\cal
C}_e:\{e\} \rightarrow \iota_e } 
\begin{array}{c}
\includegraphics[width=5cm]{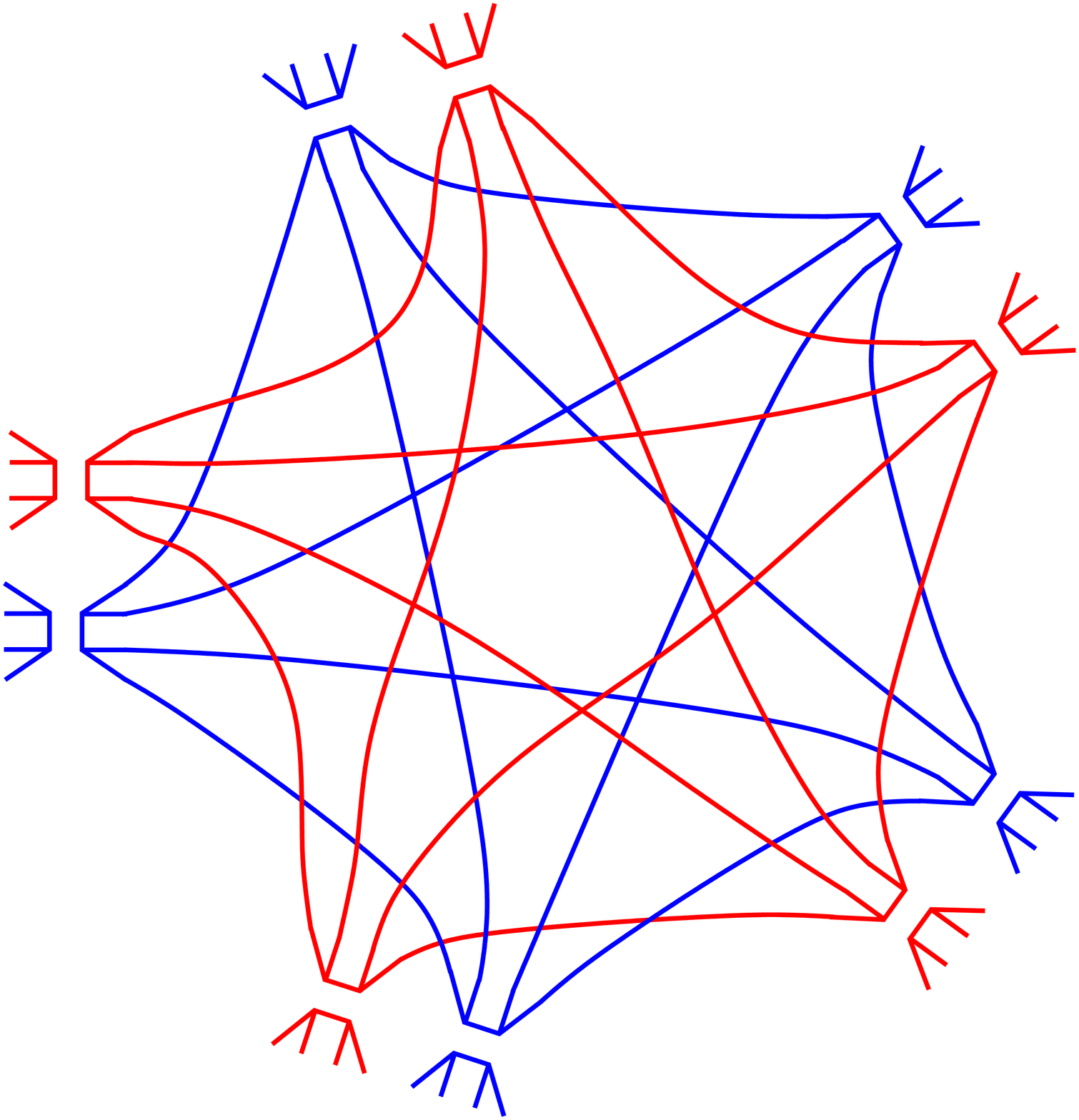}
\end{array}
\ee
from which we obtain finally the spin foam representation of the $SU(2)\times SU(2)$ partition function as a product of two $SU(2)$ amplitudes, namely
\ba &&
Z_{BF}(\Delta)=\sum \limits_{ {\cal
C}_f:\{f\} \rightarrow \rho_f }  \ \prod\limits_{f \in \Delta^{\star}} {\rm d}_{j_f^{-}}{\rm d}_{j_f^{+}} \sum \limits_{ {\cal
C}_e:\{e\} \rightarrow \iota_e } \ \prod\limits_{v\in \Delta^{\star}} \n \\ && \ \ \ \ \ \ \ \ \  \ \ \ \ \ \ \  \ \ \ \ \ \
\begin{array}{c}
\psfrag{a}{$\iota^{-}_1$}
\psfrag{b}{$\iota^{-}_2$}
\psfrag{c}{$\iota^{-}_3$}
\psfrag{d}{$\iota^{-}_4$}
\psfrag{e}{$\iota^{-}_5$}
\psfrag{ap}{$\iota^{+}_1$}
\psfrag{bp}{$\iota^{+}_2$}
\psfrag{cp}{$\iota^{+}_3$}
\psfrag{dp}{$\iota^{+}_4$}
\psfrag{ep}{$\iota^{+}_5$}
\psfrag{A}{$j^{-}_1$}
\psfrag{B}{$j^{-}_2$}
\psfrag{C}{$j^{-}_3$}
\psfrag{D}{$j^{-}_4$}
\psfrag{E}{$j^{-}_5$}
\psfrag{F}{$j^{-}_6$}
\psfrag{G}{$j^{-}_7$}
\psfrag{H}{$j^{-}_8$}
\psfrag{I}{$j^{-}_9$}
\psfrag{J}{$j^{-}_{10}$}
\psfrag{Ap}{$j^{+}_1$}
\psfrag{Bp}{$j^{+}_2$}
\psfrag{Cp}{$j^{+}_3$}
\psfrag{Dp}{$j^{+}_4$}
\psfrag{Ep}{$j^{+}_5$}
\psfrag{Fp}{$j^{+}_6$}
\psfrag{Gp}{$j^{+}_7$}
\psfrag{Hp}{$j^{+}_8$}
\psfrag{Ip}{$j^{+}_9$}
\psfrag{Jp}{$j^{+}_{10}$}
\includegraphics[height=4cm]{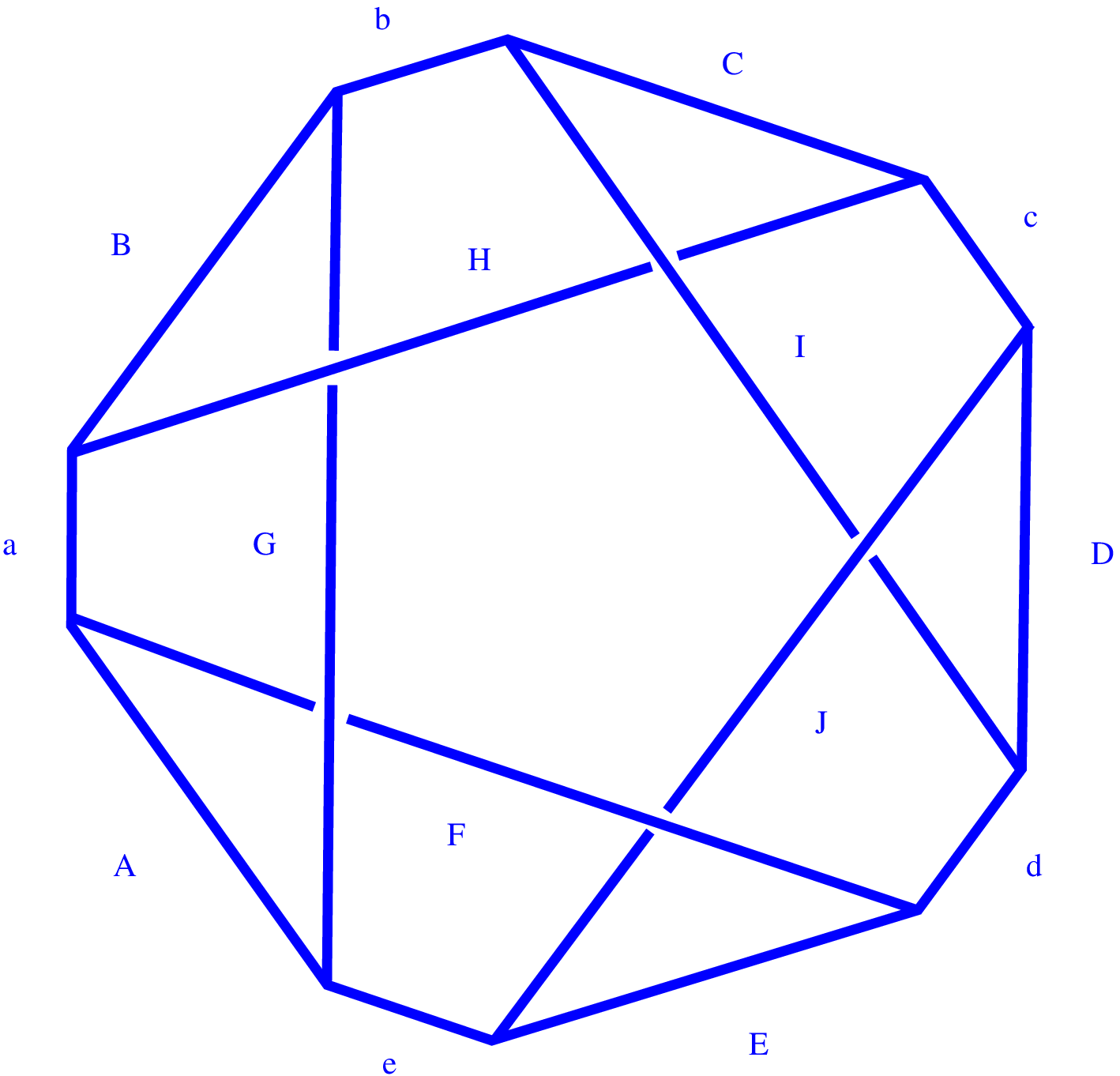}
\end{array}\ \  \ \
\begin{array}{c}
\psfrag{Q}{\!\!\!\!\!\!\!\!\!\!\!\!\!\!\!\!\!\!\!\!\!\!\!\!\!\!\!\!\!\!\!\!\!\!\!\!\!\!\!\!\!\!\!\!\!\!\!\!\!\!\!\!\!\!\!\!\!\!\!\!\!\!\!\!\!\!\!\!\!\!\!\!\!\!\!\!\!\!\!\!\!\!\!\!\!\!\!\!\!\!\!\!\!\!\!\!\!\!\!\!\!\!$Z_{BF}(\Delta)=\sum \limits_{ {\cal
C}_f:\{f\} \rightarrow \rho_f }  \ \prod\limits_{f \in \Delta^{\star}} {\rm d}_{j_f^{-}}{\rm d}_{j_f^{+}} \sum \limits_{ {\cal
C}_e:\{e\} \rightarrow \iota_e } \ \prod\limits_{v\in \Delta^{\star}} $}
\psfrag{a}{$\iota^{-}_1$}
\psfrag{b}{$\iota^{-}_2$}
\psfrag{c}{$\iota^{-}_3$}
\psfrag{d}{$\iota^{-}_4$}
\psfrag{e}{$\iota^{-}_5$}
\psfrag{ap}{$\iota^{+}_1$}
\psfrag{bp}{$\iota^{+}_2$}
\psfrag{cp}{$\iota^{+}_3$}
\psfrag{dp}{$\iota^{+}_4$}
\psfrag{ep}{$\iota^{+}_5$}
\psfrag{A}{$j^{-}_1$}
\psfrag{B}{$j^{-}_2$}
\psfrag{C}{$j^{-}_3$}
\psfrag{D}{$j^{-}_4$}
\psfrag{E}{$j^{-}_5$}
\psfrag{F}{$j^{-}_6$}
\psfrag{G}{$j^{-}_7$}
\psfrag{H}{$j^{-}_8$}
\psfrag{I}{$j^{-}_9$}
\psfrag{J}{$j^{-}_{10}$}
\psfrag{Ap}{$j^{+}_1$}
\psfrag{Bp}{$j^{+}_2$}
\psfrag{Cp}{$j^{+}_3$}
\psfrag{Dp}{$j^{+}_4$}
\psfrag{Ep}{$j^{+}_5$}
\psfrag{Fp}{$j^{+}_6$}
\psfrag{Gp}{$j^{+}_7$}
\psfrag{Hp}{$j^{+}_8$}
\psfrag{Ip}{$j^{+}_9$}
\psfrag{Jp}{$j^{+}_{10}$}
\includegraphics[height=4cm]{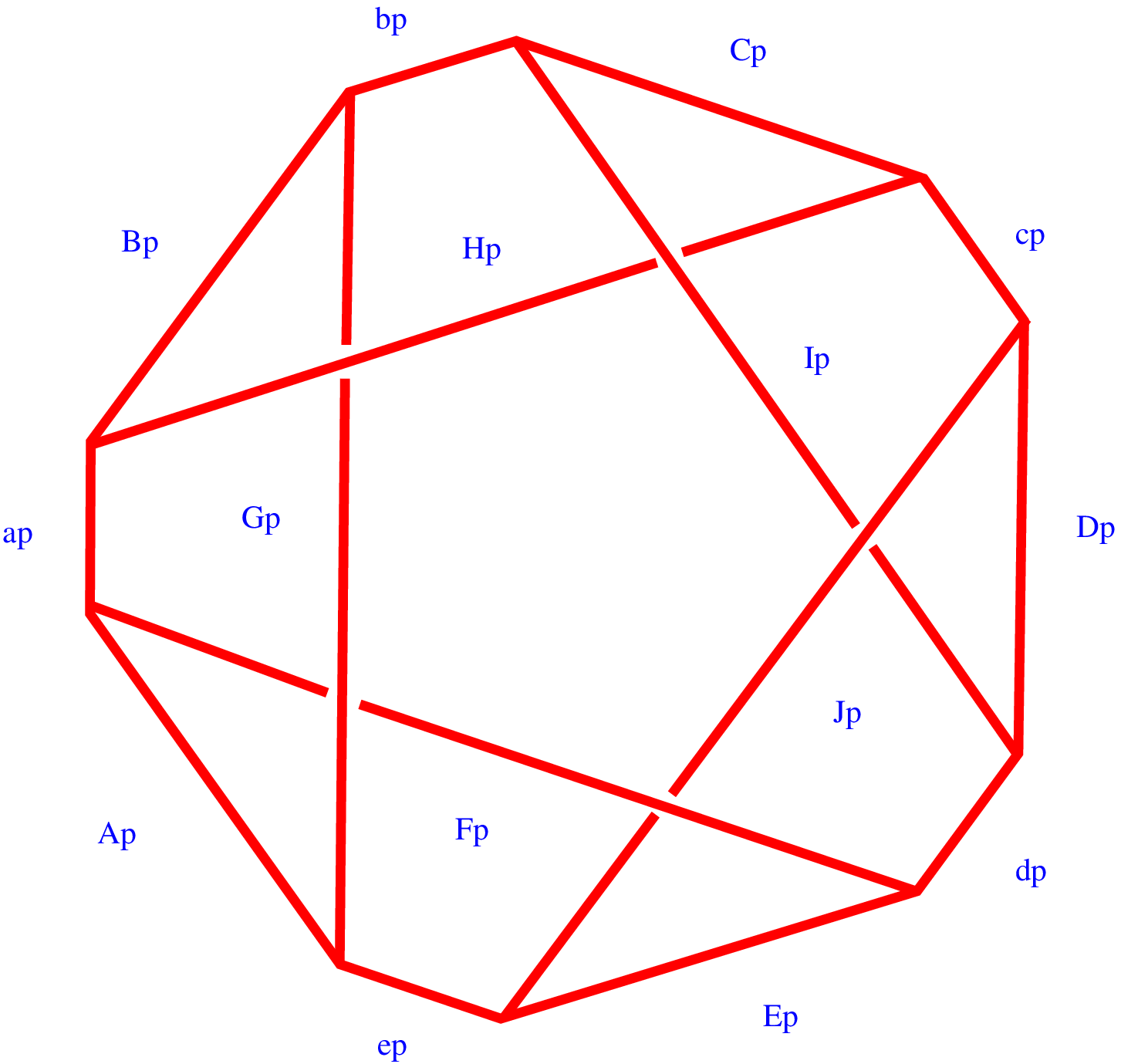}
\end{array}
\label{BF4V-b}
\ea

\subsubsection*{Extra remarks on four dimensional BF theory}

The state sum (\ref{coloring4}) is generically divergent (due to the
gauge freedom analogous to (\ref{gauge2g})). A regularized version
defined in terms of $SU_q(2)\times SU_q(2)$ was introduced by
Crane and Yetter ~\cite{crane0, crane00}. As in three dimensions, if an appropriate regularization of bubble divergences is provided,
(\ref{coloring4}) is topologically invariant and the spin foam path
integral is discretization independent.

As in the three dimensional case BF theory in any dimensions can be coupled to topological defects~\cite{Baez:2006sa}. In the four dimensional case
defects are string-like~\cite{Fairbairn:2007fb} and can carry extra degrees of freedom such as topological Yang-Mills fields~\cite{Montesinos:2007dc}.
The possibility that quantum gravity could be defined directly form these simple kind of topological theories has also been considered outside spin foams~\cite{tHooft:2008kk} 
 (for which the UV problem described in the introduction is absent) 
is attractive and should, in my view, be considered further. 

It is also possible to introduce one dimensional particles in four dimensional BF theory and gravity as shown in~\cite{Freidel:2006hv}.

\subsection{The coherent states representation}
\label{cohecohe}

In this section we introduce the coherent state representation of  the $SU(2)$ and $Spin(4)$  path integral of BF theory. 
This will be particularly important for the definition of the models defined by 
Freidel and Krasnov in~\cite{Freidel:2007py} that we will address in Section~\ref{fk} as well as in the semiclassical analysis of the new models reported in Section~\ref{semiclas}.
The relevance of such representation for spin foams was first emphasized  by 
Livine and Speziale in~\cite{Livine:2007vk}.

\subsubsection{Coherent states}

Coherent states associated to the representation theory of a compact group have been studied by Thiemann and collaborators~\cite{Thiemann:2000zf, Thiemann:2000bw, Sahlmann:2001nv, Thiemann:2000bw, Thiemann:2000ca, Thiemann:2000bx, Thiemann:2000by, Thiemann:2002vj, Bahr:2007xa, Bahr:2007xn, Flori:2008nw}  see also~\cite{Bianchi:2009ky}.
Their importance for the new spin foam models was put forward by Livine and Speziale in~\cite{Livine:2007vk} where the emphasis is put on coherent states of intertwiners or the so-called quantum tetrahedron (see also~\cite{Conrady:2009px}). Here we follow the presentation of~\cite{Freidel:2007py}.

In order to built coherent states for $\Spin(4)$ we start by introducing them in the case of $SU(2)$. Starting from the representation space $\sH_j$ of dimension $\rd_j\equiv 2j+1$ one can write
the resolution of the identity in tems of the canonical orthonormal basis $|j,m\rangle$ as
\be\label{ident-m}
1_j = \sum_m |j,m \rangle \langle j,m|,
\ee
where $-j\le m\le j$. There exists an over complete basis $|j,g \rangle \in \sH_{j}$ labelled by $g\in SU(2)$ such that \be\label{ident-coherent}
1_j=
\rd_j \int\limits_{{\rm SU}(2)} dg \, |j,g \rangle \langle j,g|,
\ee
The states  $|j,g \rangle \in \sH_{j}$ are $SU(2)$ coherent states defined by the action of the group on maximum weight
states $|j,j\rangle$ (themselves coherent), namely 
\be\label{def}
|j,g \rangle \equiv g |j,j\rangle = \sum_m  |j,m \rangle D^j_{mj}(g),
\ee
where $D^j_{mj}(g)$ are the matrix elements of the unitary representations in the $|j,m\rangle$ (Wigner matrices).
Equation (\ref{ident-coherent}) follows from the orthonormality of unitary representation matrix elements, namely
\be
 \rd_j \int_{{\rm SU}(2)} dg \, |j,g \rangle \langle j,g|,= \rd_j \sum_{mm'} |j,m \rangle \langle j,m'|
\int_{{\rm SU}(2)} dg \, D^j_{mj}(g) \overline{D^j_{m'j}(g)}= \sum_m |j,m \rangle \langle j,m| ,
\ee
where in the last equality we have used the orthonormality of the matrix elements graphically represented in
(\ref{p2}). The decomposition of the identity (\ref{ident-coherent}) can be expressed as en integral on the two-sphere of directions $S^2=SU(2)/U(1)$
 by noticing that $D^j_{mj}(g)$ and $D^j_{mj}(gh)$ differ only by a phase for 
any group element $h$ from a suitable $U(1)\subset SU(2)$. Thus one has
\be\label{patacu}
1_j = \rd_j \int_{S^2}  dn \, |j,n \rangle \langle j,n|,
\ee
where  $n\in S^2$ is integrated with the invariant measure of the sphere.
The states $|j,n\rangle$ form (an over-complete) basis in $\sH_{j}$. $SU(2)$ coherent states 
have the usual semiclassical properties. Indeed if one considers the operators $J^i$ 
generators of $su(2)$ one has
\be \label{geo}
\langle j,n|\hat{J}^{i}|j,n \rangle  = 
j \,n^{i},
\ee
where $n^i$ is the corresponding three dimensional unit vector for $n\in S^2$. 
The fluctuations of  $\hat{J}^{2}$ are also minimal with $\Delta J^{2} =\hbar^2 j$, where we have restored $\hbar$ for clarity.
The fluctuations go to zero in the limit $\hbar\to 0$ and $j\to\infty$ while $\hbar j$ is kept constant. This kind of limit will be used often as a notion of semiclassical limit in spin foams. 
The state  $|j,n\rangle$ is a semiclassical state describing a vector in $\mathbb{R}^{3}$ of length $j$ and of direction 
$n$. It will convenient to introduce the following graphical notation for equation (\ref{patacu})
\be\label{patacul}
\begin{array}{c}\psfrag{a}{$\ \ j$}
\includegraphics[height=2cm]{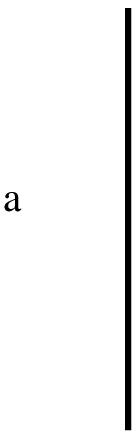}
\end{array}={\rm d}_{j}\,\int\limits_{S^2} dn \
\begin{array}{c}\psfrag{a}{$j$}\psfrag{b}{$\!n$}
\includegraphics[height=2cm]{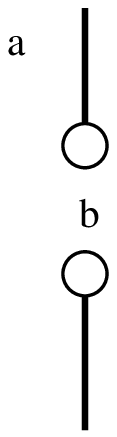}
\end{array}
\ee
Finally, an important property of $SU(2)$ coherent states stemming from the fact that \[|j,j\rangle=|{\van \frac{1}{2}},{\van \frac{1}{2}}\rangle|{\van \frac{1}{2}},{\van \frac{1}{2}}\rangle\cdots |{\van \frac{1}{2}},{\van \frac{1}{2}}\rangle\equiv |{\van \frac{1}{2}},{\van \frac{1}{2}}\rangle^{\otimes 2j}\]
is that 
\be\label{exp}
|j,n\rangle=|{\van \frac{1}{2}},n \rangle^{\otimes 2j}.
\ee
The above property will be of key importance in constructing 
effective discrete actions for spin foam models. In particular, it will play a central role in the study of the semiclassical limit of 
the EPRLand FK modesl studied in Sections~\ref{eprl-r}, \ref{eprl-l}, and \ref{fk}. In the following subsection we provide an example for $Spin(4)$ BF theory.

\subsubsection{$Spin(4)$ BF theory: amplitudes in the coherent state basis}

Here we study the coherent states representation of the path integral for $Spin(4)$ BF theory. The construction presented here can be extended to more general cases. The present case is however of particular importance for the
study of gravity models presented in  Sections~\ref{eprl-r}, \ref{eprl-l}, and \ref{fk}. With the introductions of coherent states one achieved the most difficult part of the work. In order to express the 
$Spin(4)$ BF amplitude in the coherent state representation one simply inserts a resolution of the identity in the form (\ref{patacu}) on each and every wire connecting neighbouring vertices in the expression (\ref{bf-so4})
for the BF amplitudes. The result is 
\ba\label{bf-cohe}
&& Z_{BF}(\Delta)=\sum \limits_{ {\cal
C}_f:\{f\} \rightarrow \rho_f }  \ \prod\limits_{f \in \Delta^{\star}} {\rm d}_{j_f^{-}}{\rm d}_{j_f^{+}} \n \\ && \int \prod_{e\in  \in \Delta^{\star}} {\rm d}_{j_{ef}^{-}}{\rm d}_{j_{ef}^{+}} dn^{-}_{ef}dn^{+}_{ef}
\begin{array}{c}\psfrag{w}{$$}
\psfrag{a}{\!\!\,$\va n^{\va -}_{ 1}$}
\psfrag{A}{\!\!\,$\va n^+_1$}
\psfrag{b}{\!\!\,$\va n^-_2$}
\psfrag{B}{\!\!\,$\va n^+_2$}
\psfrag{c}{\!\!\,$\va n^-_3$}
\psfrag{C}{\!\!\,$\va n^+_3$}
\psfrag{d}{\!\!\,$\va n^-_4$}
\psfrag{D}{\!\!\,$\va n^+_4$}
\includegraphics[width=7cm]{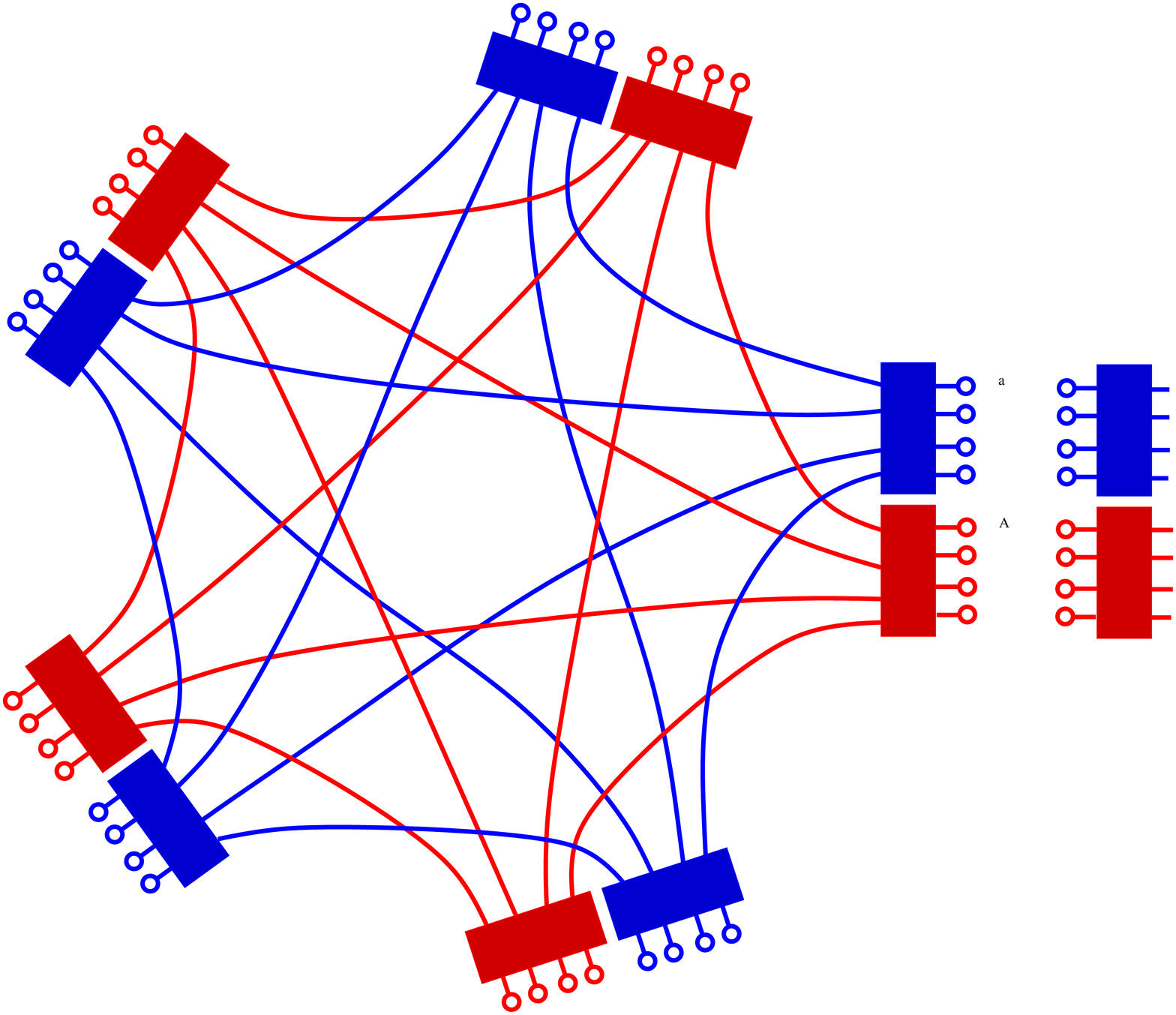}
\end{array},
\ea
where we have explicitly written the $n_{\pm}\in S^2$ integration variables only on a single cable. One observes that there is one $n_{\pm}\in S^2$ per each wire coming out at an edge $e\in\Delta^{\star}$; as 
wires are in one-to-one correspondence with faces $f\in \Delta^{\star}$ the integration variables  $n^{\pm}_{ef}\in S^2$ are labelled by an edge and face subindex. 
In order to get an expression of the BF path integral in terms of an affective action we restore at this stage the explicit group integrations represented by the 
boxes in the previous equation.  One gets,
\ba&&\!\!\!\!\!\!\!\!\!\!\!\!\!\!\!\!\!\!\!\!\!\!\!\!\!\!\!
Z_{BF}(\Delta)=\sum \limits_{ {\cal
C}_f:\{f\} \rightarrow \rho_f }  \ \prod\limits_{f \in \Delta^{\star}} {\rm d}_{j_f^{-}}{\rm d}_{j_f^{+}} \int \prod_{e\in   \Delta^{\star}} {\rm d}_{j_{ef}^{-}}{\rm d}_{j_{ef}^{+}} dn^{-}_{ef}dn^{+}_{ef} \n \\  \ \ \ \ \ \ \ \ \  \ 
&& \prod_{v\in \Delta^{\star}} \prod_{e,e'\in v} dg^{-}_{ev}dg^{+}_{ev} \ (\langle n^{-}_{ef} |(g^{-})^{-1}_{ev} g^{-}_{e'v}| n^-_{e'f}\rangle)^{2j^{-}_{f}}(\langle n^{+}_{ef} |(g^{+})^{-1}_{ev} g^{+}_{e'v}| n^+_{e'f}\rangle)^{2j^{+}_{f}},
\ea
where we have used the coherent states property (\ref{exp}),  and   $| n^{\pm}\rangle$ is a simplified notation for $|{\van \frac{1}{2}}, n^{\pm}\rangle$. 
Also we denote by $g^{\pm}_{ev}$ the group elements associated to the five boxes for each given vertex (see eq. (\ref{bf-cohe})).
The previous equation can be finally written as  
\ba
\label{cspig}Z_{BF}(\Delta)=\sum \limits_{ {\cal
C}_f:\{f\} \rightarrow \rho_f }  \ \prod\limits_{f \in \Delta^{\star}} {\rm d}_{j_f^{-}}{\rm d}_{j_f^{+}} \int \prod_{e\in  \Delta^{\star}} {\rm d}_{j_{ef}^{-}}{\rm d}_{j_{ef}^{+}} dn^{-}_{ef}dn^{+}_{ef} dg^{-}_{ev}dg^{+}_{ev}
\ \exp{(S^{d}_{j^{\pm},\bn^{\pm}}[g^{\pm}])}, \ea
where the discrete action \be\label{discrete-action-g}
S^{d}_{j^{\pm},\bn^{\pm}}[g^{\pm}]=\sum_{v\in\Delta^{\star}} S^v_{j_v,\bn_v}[g^{\pm}]\ee with 
\be
\label{v-action}
S^v_{j,\bn}[g] = \sum\limits^{5}_{a < b=1} 2j_{ab} \ln \, \la n_{ab}| g^{-1}_a g_b| \, n_{ba} \ra,
\ee
and the indices $a,b$ label the five edges of a given vertex. The previous expression is exactly equal to the form (\ref{coloring4}) of the BF amplitude. In the case of the gravity models studied in what follows,
the coherent state path integral representation will be the basic tool for the study of the semiclassical limit of the models and the relationship with Regge discrete formulation of general relativity.

\section{The Riemannian EPRL model}
\label{eprl-r}
 
In this section we introduce the Riemannian version of the
Engle--Pereira--Rovelli--Livine (EPRL) \cite{Engle:2007uq,
  Engle:2007wy}. The section is organized as follows. The relevant
$Spin(4)$ representation theory is introduced in
Section~\ref{rept}. In Section~\ref{7-2} we present and discuss the
linear simplicity constraints -- classically equivalent to the
Plebanski constraints -- and discuss their implementation in the
quantum theory. In Section~\ref{path-pre} we introduce the EPRL model
of Riemannian gravity. In Section~\ref{cuadratiqui} we prove the
validity of the quadratic Plebanski constraints -- reducing BF theory
to general relativity -- directly in the spin foam representation. In
Section~\ref{anofree} we discuss a certain modification of the EPRL
model. In Section~\ref{7-6} we present the coherent state
representation of the EPRL model which is essential for the
semiclassical analysis of Section~\ref{semiclas}. The material of this
section will also allow us to describe the construction of the closely
related (although derived from a different logic) Riemannian FK
constructed in~\cite{Freidel:2007py}. The idea that linear simplicity
constraints are more convenient for dealing with the constraints that
reduce BF theory to gravity was pointed out by Freidel and Krasnov in
this last reference. We have shown in Part~\ref{cano} that they arise
directly from the canonical analysis of the Plebanski formulation of
gravity.

\subsection{Representation theory of $Spin(4)$ and the canonical
  basis}
\label{rept}

In order to introduce the definition of the EPRL model  we need to briefly review the representation 
theory of $Spin(4)$. The group $Spin(4)=SU(2)\times SU(2)$ with Lie algebra $spin(4)=su(2)\oplus su(2)$ can explicitly be described in terms of generators $J^i_{\pm}$ such that 
\be\label{lieso4}
[J^i_{\pm},J^{j}_{\pm}]=\epsilon^{ij}_{\ \ k} J^{k}_{\pm}.
\ee 
Unitary irreducible representations $\sH_{j^+,j^-}$ of $Spin(4)$ are given by the product of unitary irreducible represetantions of $SU(2)$, i.e., they are labelled by 
two half-integers $j^{\pm}$. A standard basis in $\sH_{j^+,j^-}$ is given by vectors $|j^+,j^-,m^+,m^-\rangle$ of eigenstates of the Casimirs $C_1=J_+^2+J_-^2$ and $C_2=J_+^2-J_-^2$ and the components $J_+^3$ and $J_-^3$, explicitly
\ba\label{casi}
&& C_1 |j^+,j^-,m^+,m^-\rangle=(j_+(j_++1)+j_-(j_-+1))\, |j^+,j^-,m^+,m^-\rangle\n\\
&& C_2 |j^+,j^-,m^+,m^-\rangle=(j_+(j_++1)-j_-(j_-+1))\, |j^+,j^-,m^+,m^-\rangle\n\\
&& J^3_{\pm} |j^+,j^-,m^+,m^-\rangle=m^{\pm}\, |j^+,j^-,m^+,m^-\rangle. 
\ea
The definition of the EPRL model  requires the introduction of an (arbitrary)
 subgroup $SU(2)\subset Spin(4)$. This subgroup will be shown, \emph{a posteriori} , to be the internal gauge group 
 of the gravitational phase space in connection variables (appearing in the classical canonical study of Section~\ref{canolys} and leading to the quantization of Section~\ref{lqg}).  In the quantum theory, the representation theory of this $SU(2)$ subgroup will be hence important. This importance will soon emerge as apparent from  the imposition of the constraints that define the EPRL as well as the FK model of Section~\ref{fk}.
 
 The link between the unitary representations of $spin(4)$ and those of $su(2)$ comes from the fact that the former can be expressed as a direct sum of the latter according to 
 \be\label{spin4su2g}
 \sH_{j^+,j^-}=\bigoplus\limits_{j=|j^{+} - j^{-}|}^{j^{+} + j^{-}} \sH_{j}.\ee
 As the unitary irreducible representations of the subgroup $SU(2)\in Spin(4)$ are essential in understanding the link of the EPRL model and the operator canonical formulation of LQG it will be convenient to express the action of the generators of the Lie algebra $spin(4)$ in a basis adapted to the above equation. In order to do this we first notice that the Lie algebra (\ref{lieso4}) can be equivalently characterized in terms of the generators of a rotation subgroup $L^i$ and the remaining ``\emph{boost}'' generators $K^i$ as follows
\ba&& \n
[L_3,L_{\pm}]=\pm \ L_{\pm} \ \ \ \ \ [L_+,L_{-}]=2\ L_{3} \\
&& \n
[L_+,K_{+}]= [L_-,K_{-}]=[L_3,K_{3}]=0 \\
&& \n
[K_3,L_{\pm}]=\pm \ K_{\pm} \ \ \ \ \ [L_\pm,K_{\mp}]=\pm 2\ K_{3}\ \ \ \ \ [L_3,K_{\pm}]=\pm \ K_{\pm} \\
&& [K_3,K_{\pm}]=\pm \ L_{\pm} \ \ \ \ \ [K_+,K_{-}]= 2\ L_{3},
\ea
where $K_{\pm}=K^1 \pm i K^2$ and $L_{\pm}=L^1 {\pm}i L^2$ respectively.
The action of the previous generators in the basis $ |j^+,j^-,j ,m\rangle$ can be shown to be
\ba
&& L^3 |j^+,j^-,j,m\rangle = m |j^+,j^-,j,m\rangle, \nonumber \\
&& L^+ |j^+,j^-,j,m\rangle = \sqrt{(j+m+1)(j-m)} |j^+,j^-,j,m+1\rangle, \nonumber \\
&&  L^- |j^+,j^-,j,m\rangle = \sqrt{(j+m)(j-m+1)} |j^+,j^-,j,m-1\rangle, \nonumber \\
&&  K^3 |j^+,j^-,j,m\rangle = \n \\
&& \alpha_j\sqrt{j^2-m^2} |j^+,j^-,j-1,m\rangle+ \gamma_j m |j^+,j^-,j,m\rangle 
-\alpha_{j+1}\sqrt{(j+1)^2-m^2} |j^+,j^-,j+1,m\rangle,\nonumber \\
&&  K^+ |j^+,j^-,j,m\rangle = \alpha_j\sqrt{(j-m)(j-m-1)}
|j^+,j^-,j-1,m+1\rangle \nonumber \\
&& + \gamma_j\sqrt{(j-m)(j+m+1)}|j^+,j^-,j,m+1\rangle \nonumber
\\ && +\alpha_{j+1}\sqrt{(j+m+1)(j+m+2)} |j^+,j^-,j+1,m+1\rangle,\nonumber \\
&& K^- |j^+,j^-,j,m\rangle = -\alpha_j\sqrt{(j+m)(j+m-1)}
|j^+,j^-,j-1,m-1\rangle
\nonumber \\ && + \gamma_j\sqrt{(j+m)(j-m+1)} |j^+,j^-,j,m-1\rangle \nonumber \\
&& -\alpha_{j+1}\sqrt{(j-m+1)(j-m+2)}|j^+,j^-,j+1,m-1\rangle,
\label{operadores eigenedos}
\ea
where 
\be
\gamma_{j}=\frac{j^+(j^++1)-j^-(j^-+1)}{j(j+1)}
\ \ \ \ \ \ \ \ \ \ 
\alpha_{j}=\sqrt{\frac{(j^2-(j^++j^-+1)^2)(j^2-(j^+-j^-)^2)}{j^2(4j^2-1)}}.
\ee
The previous equations will be important in what follows: they will allow for the characterisation of the solutions of the quantum simplicity constraints in a direct manner. 
This concludes the review of the representation theory that is necessary for the definition of the EPRL model.

\subsection{The linear simplicity constraints}\label{7-2}

As first shown in \cite{Freidel:2007py}, the  quadratic simplicity constraints (\ref{ito}) -- and more precisely in their dual version presented below (\ref{dual}) -- are equivalent in the discrete setting to 
the linear constraint on each face of a given tetrahedron
\be\label{constrainty-e}
D_{f}^i=L_{f}^i-\frac{1}{\gamma} K_{f}^i\approx 0,
\ee
where the label $f$ makes reference to a face $f\in \Delta^{\star}$, and where (very importantly) the subgroup $SU(2)\subset Spin(4)$ that is necessary for the definition of the above 
constraints is chosen arbitrarily at each tetrahedron, equivalent on each edge $e\in \Delta^{\star}$. Such choice of the rotation subgroup is the precise analog of the time gauge in the canonical analysis of Section
\ref{canocal}. Moreover,  the linear simplicity constraint above is the simplicial version of the canonical primary constraint (\ref{piripipi}).  The EPRL model is defined by imposing the previous constraints as operator equations on the Hilbert spaces defined by the unitary irreducible representations of $Spin(4)$ that take part in the state-sum of BF theory.  We will show in Section~\ref{cuadratiqui} that the models constructed on the requirement of
a suitable imposition of the linear constraints (\ref{constrainty-e}) satisfy the usual quadratic Plebanski constraints in the path integral formulation (up to quantum corrections which are neglected in the usual semiclassical limit).

It will be useful in what follows to write the above constraint in its equivalent form
\be\label{jojo}
D_{f}^i=(1-\gamma)J^{+i}_{f}-(1+{\gamma}) J_{f}^{-i}\approx 0,
\ee 

 From the commutation relations (\ref{lieso4}) of previous section we can easily compute the
commutator of the previous tetrahedron constraints and conclude that in fact it does not close, namely
\ba\label{algebry}
[D_{f}^i,D_{f'}^j]&=&\delta_{f f'}  \epsilon^{ij}_{\ \, k} \left[(1+\frac{1}{\gamma^2}) L_{f}^k-\frac{2}{\gamma} K_e^k\right]=\n \\
&=&2 \delta_{e e'}  \epsilon^{ij}_{\ \, k} D^k+\delta_{e e'} \frac{1-\gamma^2}{\gamma^2} \epsilon^{ij}_{\ \, k}  L_{f}^k.
\ea
The previous commutation relations imply  that the constraint algebra is not closed and cannot therefore be imposed as operator equations on the states summed over in the BF partition function in general. There are two interesting exceptions to the previous statement: 
\begin{enumerate}
\item The first one is to take $\gamma=\pm 1$ in which case the constraint (\ref{constrainty}) can be imposed strongly. More precisely, for the plus sign  it 
reduces to the condition $J^i_{f+}=0$ (or equivalently for the other sign $J^i_{f-}=0$) which amounts to setting all the left quantum numbers in the BF state-sum for the group $Spin(4)$ to zero. 
It is immediate to see that if we impose such restriction on the $Spin(4)$ BF amplitude  (\ref{BF4V-b}) it simply reduces to the amplitude (\ref{BFSU2}) defining $SU(2)$ BF theory. Thus, this possibility does not lead to an 
acceptable model for quantum gravity in 4d.
\item The second possibility is to work in the sector where $L^i_f=0$. We will show later that this corresponds to the 
famous Barret-Crane model \cite{BC2} with the various limitations discussed in Section~\ref{BCM}.
\end{enumerate}
The EPRL model is obtained by restricting the representations appearing in the expression of the BF partition 
function so that at each tetrahedron the linear constraints (\ref{constrainty}) the strongest possible way that is compatible with the uncertainties 
relations stemming from 
(\ref{algebry}). In addition one would  add  the requirement that the state-space of tetrahedra is compatible with the state-space 
of the analogous excitation in the canonical context of LQG so that arbitrary states in the kinematical state of LQG have non trivial 
amplitudes in the model.

\subsection*{On the weak imposition of the linear simplicity constraints}

We now discuss the weak imposition of the linear simplicity constraints in the quantum framework.
There are essentially three views in the literature: two of them, discussed below, concerns directly the  way the 
EPRL model has been usually introduced. The third possibility is the semiclassical view based on the coherent state representation leading to the FK model (see Section~\ref{fk}).

\subsubsection{Riemannian model: The Gupta--Bleuler criterion}

Due to the fact that the constraints $D^{i}_f$ do not form a closed (first class) algebra in the generic case one needs to 
devise a weaker sense in which they are to be imposed. One possibility is to consider the Gupta--Bleuler criterion 
consisting in selecting a suitable class of states for which the matrix elements on $D_{f}^i$ vanish.  In this respect one notices that
if we chose the subspace  $\sH_{j}\subset\sH_{j^{+},j^{-}}$ one has
\ba && \n
  \langle j^+,j^-,j,q|D^{3}_f |j^+,j^-,j,m\rangle=  \delta_{q,m}m (1-\frac{\gamma_{j}}{\gamma})\\ && \n
  \langle j^+,j^-,j,q|D^{\pm}_f |j^+,j^-,j,m\rangle=  \delta_{q\pm 1,m}\sqrt{(j\pm m+1)(j\mp m)}(1-\frac{\gamma_{j}}{\gamma}).
\ea
One immediately observes that matrix elements of the linear constraints vanish in this subclass if one can chose
\be\gamma_{j}=\frac{j^+(j^++1)-j^-(j^-+1)}{j(j+1)}=\gamma\label{coky}
\ee
There are two cases:
\begin{enumerate}
\item {\bf Case $\gamma<1$: } Following \cite{Ding:2009jq}, in this case one restricts the $Spin(4)$ representations to \be j^{\pm}=(1\pm \gamma)\frac{j}{2}, \ee which amounts to choosing the \emph{maximum weight} component $j=j^++j^-$ in the expansion (\ref{spin4su2}). 
Simple algebra shows that condition (\ref{coky}) is met. There are indeed other solutions \cite{Ding:2010fw} of the Gupta--Bleuler criterion in this case. 
 \item {\bf Case $\gamma>1$: } In this case \cite{Alexandrov:2010pg} one restricts the $Spin(4)$ representations to \be j^{\pm}=( \gamma\pm 1)\frac{j}{2} +\frac{\gamma-1}{2}, \ee which amounts to choosing the \emph{minimum weight} component $j=j^+-j^-$ in the expansion (\ref{spin4su2}). This choice of $j^{\pm}$ is the solution to condition (\ref{coky}). 
\end{enumerate}
\subsubsection{Riemannian model:
The Master constraint criterion}
Another criterion for weak imposition can be developed by studying the spectrum of the Master constraint $M_f=D_f\cdot D_f$. Strong imposition of the constraints $D_f^i$ would amount to
looking for the kernel of the master constraint $M_f$. However, generically the positive operator associated with the master constraint does not contain the zero eigenvalue in the spectrum due to the open nature of the constraint algebra (\ref{algebry}).  
The proposal of \cite{Engle:2007mu}, that we follow here,  is to look for the minimum eigenvalue among spaces $\sH_j\in \sH_{j^{+},j^{-}}$. Explicitly 
 \be
 M_f|\psi>=m_{j^{\pm},j}|\psi>,
 \ee
 where
\ba  m_{j^{\pm},j}=(1-\gamma)^2 j_+(j_++1)+(1+\gamma^2) j_-(j_-+1)-(1-\gamma^2)[j(j+1)- j_+(j_++1)-j_-(j_-+1)]\n
\ea
There are two cases:
\begin{enumerate}
\item {\bf Case $\gamma<1$: }   In this case the minimum eigenvalue is obtained for \be j^{\pm}=(1\pm \gamma)j/2, \ee i.e., one has $$m_{j^{\pm},j}\ge m_j$$ where $m_{j}=\hbar^2 (1-\gamma^2)j$ where we have restored the explicit dependence on $\hbar$  so that it is apparent that 
the selected eigenvalue vanishes in the  the semiclassical limit $\hbar \to 0$ and $k\rightarrow \infty$ with 
$\hbar k=$constant.  
 \item {\bf Case $\gamma>1$: } In this case the minimum eigenvalue is obtained for \be j^{\pm}=(\gamma \pm 1)j/2, \ee for which $m_{j}=\hbar^2 (\gamma^2-1)j$ and and thus vanishes in the  the semiclassical limit $\hbar \to 0$ and $k\rightarrow \infty$ with 
$\hbar k=$constant.  
 \end{enumerate}

The master constraint criterion will be used here as the basic justification for the definition of the EPRL model. Notice however, that the master constraint is only invariant under the subgroup $SU(2)\subset Spin(4)$
chosen at each tetrahedron. Its lack of Lorentz invariance is clear from the fact that its validity is justified from the canonical analysis in the time gauge. Nevertheless, 
as shown in \cite{Rovelli:2010ed}, one can also write down a criterion (equivalent to the master constraint one) in a manifestly $Spin(4)$ invariant way. This reflects the arbitrariness in the choice of the internal subgroup $SU(2)\subset Spin(4)$ and 
implies the consistency of the time gauge. In order to get a gauge invariant constraint one starts from the  master constraint and  uses the $D^i_f=0$ classically to rewrite it entirely in terms of the $Spin(4)$ Casimirs, namely 
\ba\n
M_f &=& (1-\gamma)^2J^2_++(1+\gamma)^2 J_-^2-2(1-\gamma^2) J_+\cdot J_{-}\\
&\approx&(1+\gamma^2) C_2 - 2 C_1 \gamma,
\ea
where $C_1$ and $C_2$ are the two $spin(4)$ Casimirs given in equation (\ref{casi}). The Lorentz invariant version of the master constraint is hence
\be
M^{\va LI}_f=(1+\gamma^2) C_2 - 2 C_1 \gamma=0
\ee
One can verify for the cases $\gamma<1$ and $\gamma>1$ the Lorentz invariant criterion is just equivalent to the master constraint analysis above.

\subsubsection{Riemannian model: Restrictions on the Immirzi
  parameter}

We have just shown how the states in $\sH_{j^+,j^-}$ can be restricted in order to satisfy the linear simplicity constraints according to different criteria.
From all these the EPRL model is defined most simply using the Master constraint criterion. In all cases the restriction takes the form of the selection of a subspace
$\sH_{j}\subset \sH_{j^+,j^-}$ given by unitary irreducible representations of a subgroup $SU(2)\in Spin(4)$. As we will show in what follows this implies that the boundary states of the spin foam models
defined by the EPRL construction are in correspondence to $SU(2)$ spin network states: the states of the kinematical Hilbert space  of LQG reviewed in Part~\ref{cano}. 
This is very nice as it suggests that the spin foam amplitudes can be interpreted as dynamical transition amplitudes of the canonical quantum theory.
Now in order for this to be so, one would like arbitrary spins $j\in \N/2$ to be admissible in the state sum. We shall see now that this imposes restrictions on the possible values
of the Immirzi parameter. However, these restrictions seem so far of little physical importance as they are a special feature of the Riemannian
 model that is absent in its Lorentzian relative.
As above there are two distinct cases:
 \begin{enumerate}
\item {\bf Case $\gamma<1$: }  If one wants the spin  $j\in \N/2$ to be arbitrary then the only possibilities are $\gamma=0$ or $\gamma=1$. Rational $\gamma$ values would restrict the spins $j$ to a subclass in $\N/2$. This restriction is not natural from the viewpoint of LQG. ITs relevance if any remains misterious at this stage.  

\item {\bf Case $\gamma>1$: } In this case $\gamma\in \N$ would allow any $j\in \N/2$. Rational choices of $\gamma$ have the same effect as in the previous case.
\end{enumerate}

\subsubsection{Riemannian model: Overview of the solutions of the
  linear simplicity constraints}
\label{resume-r}

We have reviewed different criteria given in the literature for the definition of the subset of quantum states that are selected in the construction of the EPRL model.
From all the criteria the Master constraint one is the most direct and clear-cut. For the Riemannian
 model the result is:
\begin{enumerate}
\item {\bf Case $\gamma<1$: }   \be j^{\pm}=(1\pm \gamma)j/2, \ee
 \item {\bf Case $\gamma>1$: }  \be j^{\pm}=(\gamma \pm 1)j/2, \ee 
\end{enumerate}
where $j\in \N/2$ is arbitrary -- except from the requirement that $j^{\pm}\in \N/2$ which depends of the value of the Immirzi parameter according to the previous section.
From now on we denote the subset of admissible representation of $Spin(4)$ as\be
\sK_{\gamma}\subset {\rm Irrep}(Spin(4)).
\label{kkr}
\ee
It follows from the discussion in the previous sections that for  states $\Psi\in\sK_{\gamma}$ satisfy the constraints (\ref{constrainty}) in the following  semiclassical sense:
\be\label{needed}
( K^i_f-\gamma  L_f^i) \Psi=\sO_{sc},
\ee 
where the symbol $\sO_{sc}$ (order semiclassical) denotes a quantity that vanishes in limit $\hbar\to 0$, $j\to \infty$ with $\hbar j=$constant. Equivalently one has 
\be
[(1-\gamma) J_+^i-(1+\gamma) J_-^i] \Psi=\sO_{sc}.
\ee 
This last equation can be written graphically as:
\be
\begin{array}{c}
\psfrag{x}{$\!\!\!\!\!\!\!\!\!\!-(1+\gamma)$}
\psfrag{y}{$\!\!\!\!\!\!\!\!\!\!+(1-\gamma)$}
\psfrag{z}{$=\sO_{sc}$}
\psfrag{j}{$k$}
\psfrag{jp}{$j_{-}$}
\psfrag{jm}{$j_{+}$} 
\includegraphics[height=3cm]{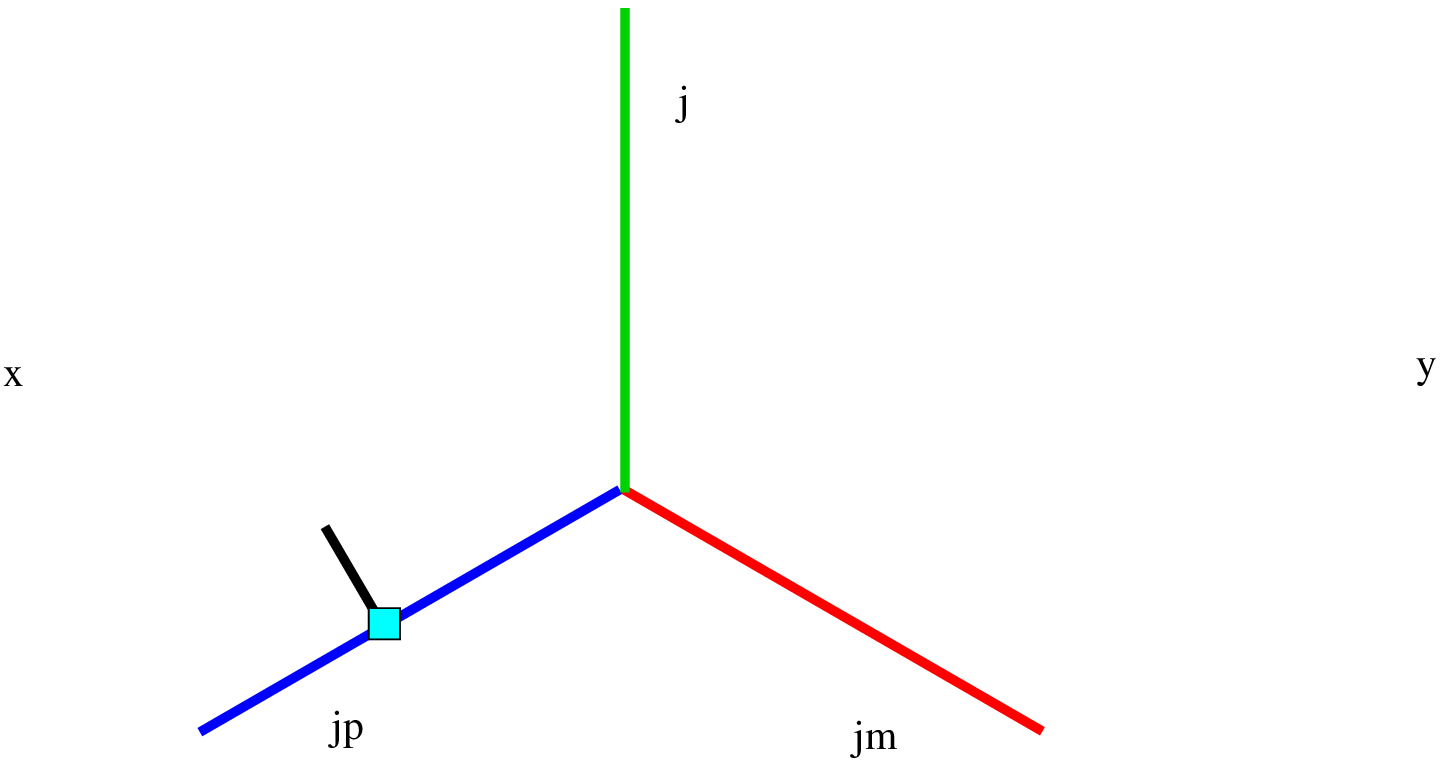} \includegraphics[height=3cm]{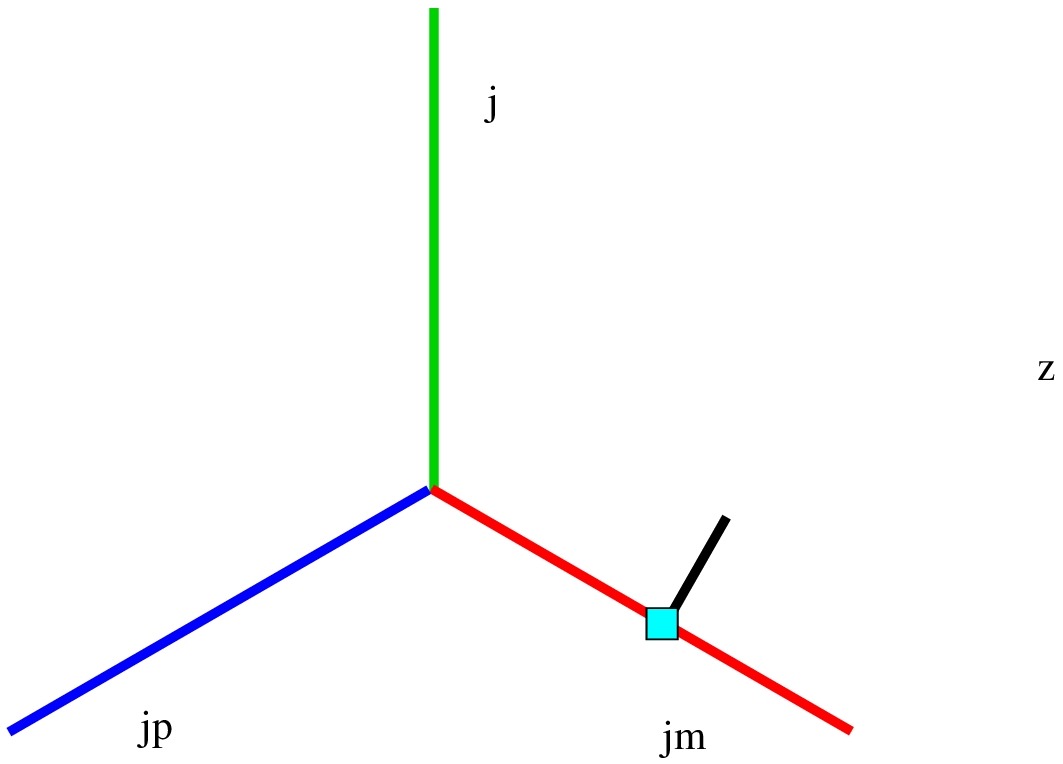}
\end{array}
\label{trival}\ee
The previous equation will be of great importance in the graphical calculus that will alow us to show that the 
linear constraint imposed here at the level of states imply the vanishing of the quadratic Plebanski constraints (\ref{ito}) and 
their fluctuations, computed in the path integral sense,  in the appropriate large spin semiclassical limit.

\subsection{Presentation of the EPRL amplitude}
\label{path-pre}

Here we complete the definition of the EPRL models by imposing the linear constraints on the BF amplitudes constructed
in Section~\ref{BF}. We will also show that the path-integral expectation value of the Plebanski constraints~(\ref{ito}), as well as their fluctuations, 
vanish in a suitable semiclassical sense.  This shows that the EPRL model can be considered as a lattice definition of the a quantum gravity theory.

We start with the Riemannian model for which a straightforward graphical notation is available. The first step is the translation of Equation~(\ref{spin4su2g}) in terms of the graphical 
notation introduced in Section~\ref{BF}. Concretely, for $\gamma<1$ one has $j^{\pm}=(1\pm\gamma) j/2\in \sK_{\gamma}$---as defined in (\ref{kkr})---becomes
\vskip.5cm
\be
\begin{array}{c}
\psfrag{j}{$\alpha$}
\psfrag{jp}{$\!\!\!\!\!\!\!\!\!\!\!\!\!\!\!\!\!\!(1-\gamma) \frac{j}{2}$}
\psfrag{jm}{$(1+\gamma) \frac{j}{2}$} 
\includegraphics[height=2.5cm]{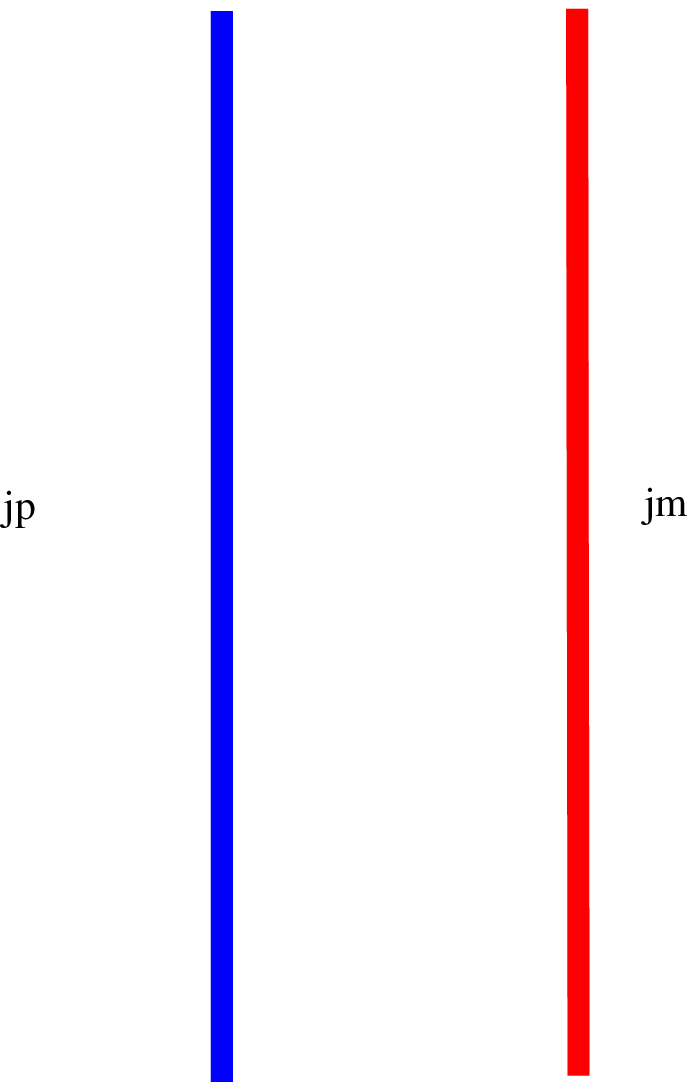}\end{array} \ \ \ \ \ \ \ \ \ \ \ \ =\bigoplus \limits_{\alpha=\gamma j}^{j} \ \ \ \ \ \ \begin{array}{c} \psfrag{j}{$\alpha$}
\psfrag{jp}{$\!\!\!\!\!\!(1-\gamma) \frac{j}{2}$}
\psfrag{jm}{$(1+\gamma) \frac{j}{2}$}  \includegraphics[height=2.5cm]{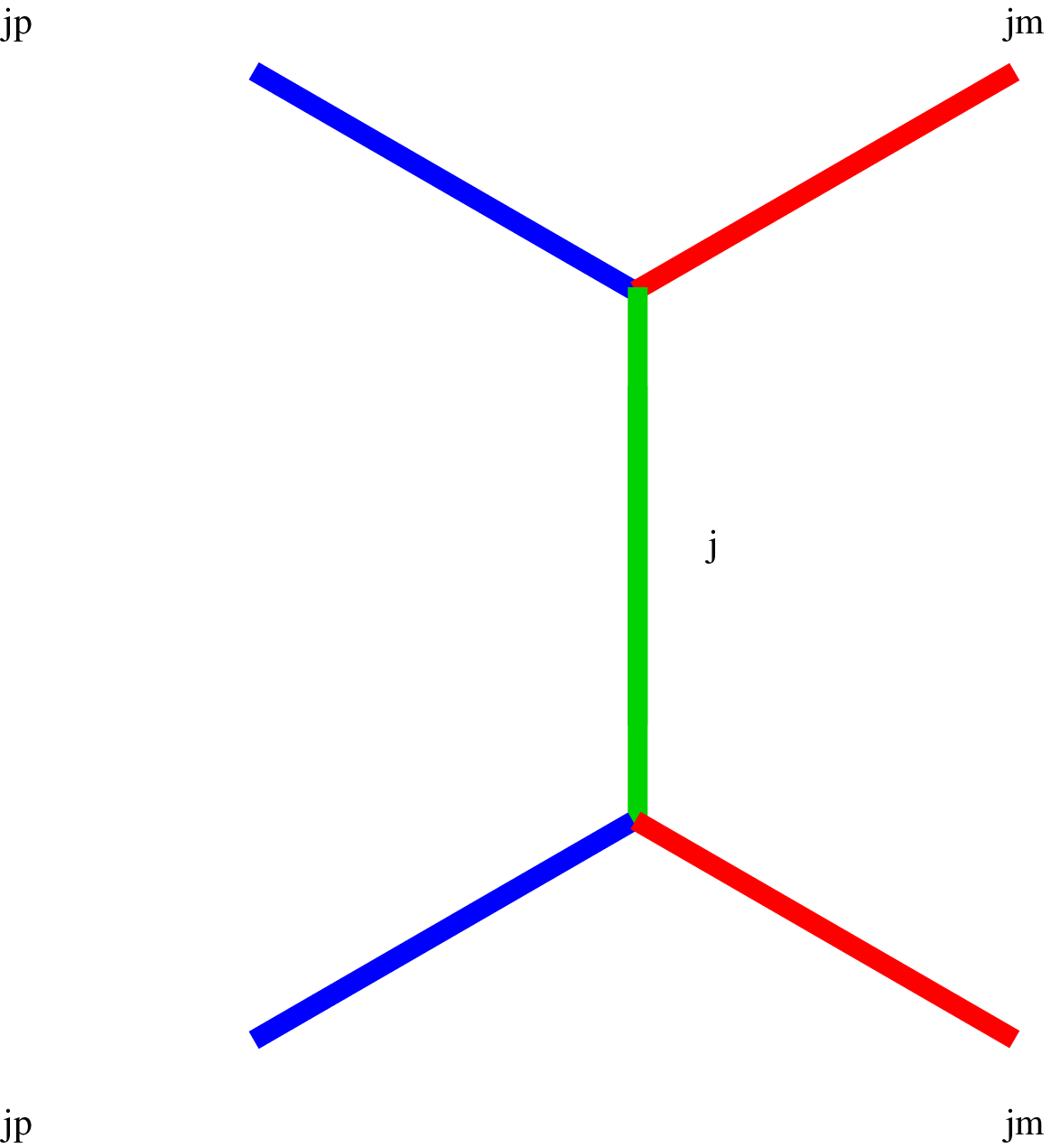}
\end{array}
\label{trivalis}\ee
For $\gamma>1$ we have 
\vskip.5cm
\be
\begin{array}{c}
\psfrag{j}{$\alpha$}
\psfrag{jp}{$\!\!\!\!\!\!\!\!\!\!\!\!\!\!\!\!\!\!(\gamma-1) \frac{j}{2}$}
\psfrag{jm}{$(1+\gamma) \frac{j}{2}$} 
\includegraphics[height=2.5cm]{travalisto}\end{array} \ \ \ \ \ \ \ \ \ \ \ \ =\bigoplus \limits_{\alpha= j}^{\gamma j} \ \ \ \ \ \ \begin{array}{c} \psfrag{j}{$\alpha$}
\psfrag{jp}{$\!\!\!\!\!\!(\gamma-1) \frac{j}{2}$}
\psfrag{jm}{$(1+\gamma) \frac{j}{2}$}  \includegraphics[height=2.5cm]{travalis}
\end{array}
\label{trivalisto}\ee
\vskip.5cm
The implementation of the linear constraints of Section (\ref{7-2}) consist in restricting the representations $\rho_f$ of $Spin(4)$ appearing in the state sum amplitudes of BF theory as written in Equation (\ref{bf-so4})
to the subclass $\rho_f\in \sK_{\gamma} \subset {\rm Irrep}(Spin (4))$, defined above,  while projecting to the highest weight term in (\ref{trivalis}) for $\gamma<1$. For $\gamma>1$ one must take the minimum weight term in  (\ref{trivalisto}) .
The action of this projection will be denoted $\sY_{j}:\sH_{(1+\gamma)j/2,|(1-\gamma)|j/2}\to \sH_{j}$, graphically
\be\label{Y}
\sY_{j}\left[\ \ \ \ \ \ \ \ 
\begin{array}{c}\psfrag{j}{$\alpha$}
\psfrag{jp}{$\!\!\!\!\!\!\!\!\!\!\!\!\!\!\!\!\!\! |\gamma-1| \frac{j}{2}$}
\psfrag{jm}{$(1+\gamma) \frac{j}{2}$} 
\includegraphics[width=1cm]{travalisto}
\end{array}\ \ \ \ \ \ \ \ \ \right]=
\begin{array}{c} \psfrag{j}{$j$}
\psfrag{jp}{$ $}
\psfrag{jm}{$ $}  \includegraphics[height=2cm]{travalis}
\end{array}.
\ee 
Explicitly, one takes the expression of the BF partition function (\ref{bf4}) and modifies it by replacing the projector
 $P^e_{inv}(\rho_1,\cdots, \rho_4)$ with $\rho_1,\cdots \rho_4\in \sK_{\gamma}$ by a new object 
 \be\label{regalo} P_{eprl}^e(j_1,\cdots, j_4) \equiv P^e_{inv} (\rho_{1}\cdots \rho_4)( \sY_{j_1}\otimes\cdots\otimes\sY_{j_4} )P^e_{inv} (\rho_{1}\cdots \rho_4)\ee with $j_1, \cdots j_4\in \N/2$
implementing the linear constraints described in the previous section. Graphically the modification of BF theory that produces the EPRL model corresponds to the replacement  \be\label{eprl-projection}
P_{inv}^e(\rho_1\cdots\rho_4)=
\begin{array}{c}\psfrag{a}{$ $}\psfrag{b}{$ $}\psfrag{c}{$ $}\psfrag{d}{$ $}
\includegraphics[height=1cm]{cable-4d-1}
\end{array}\begin{array}{c}\psfrag{A}{$ $}\psfrag{B}{$ $}\psfrag{C}{$ $}\psfrag{D}{$ $}
\includegraphics[height=1cm]{cable-4d-2}
\end{array} \ \ \ \ \ \begin{array}{c} \includegraphics[width=1cm]{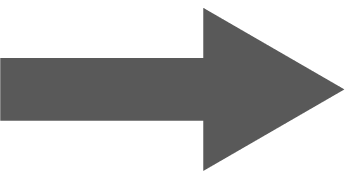}\end{array}
\ \ \ \ \  P_{eprl}^{e}(j_1\cdots j_4)=
\begin{array}{c}
\includegraphics[width=1.4cm]{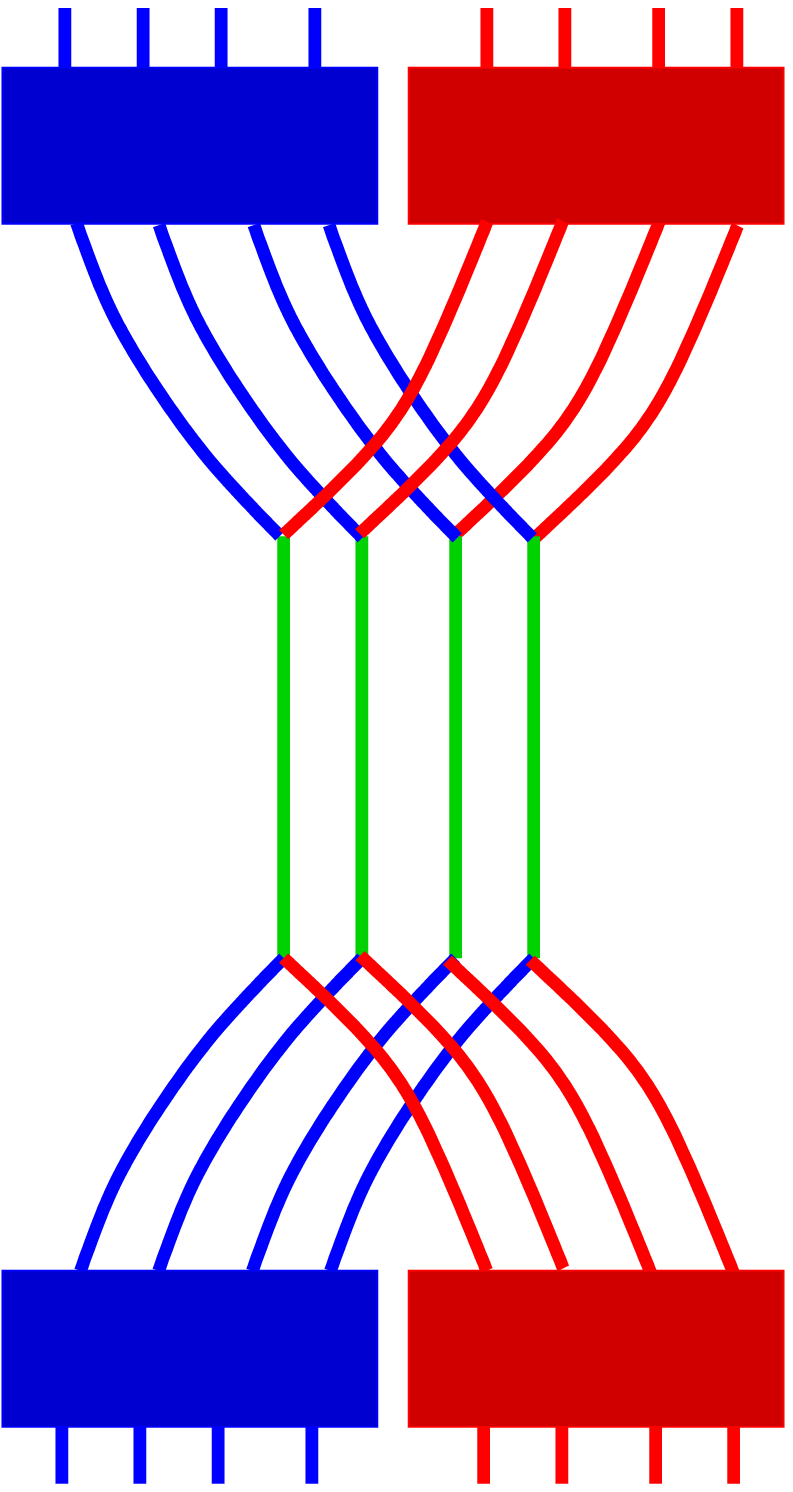}
\end{array}
\ee 
on the expression (\ref{bf-so4}), 
where we have dropped the representation labels from the figure for simplicity.  We have done the operation (\ref{Y}) on each an every  of the four pairs of representations. The $Spin(4)$ integrations represented by the two boxes at the top and bottom of the previous graphical expression restore the full $Spin(4)$ invariance as the projection (\ref{Y}) breaks this latter symmetry for being based on the selection of a special subgroup $SU(2)\subset Spin(4)$ in its definition (see section \ref{anofree} for an important implication). One should simply keep in mind that green wires in the previous 
two equations and in what follows are labeled by arbitrary spins $j$ (which are being summed over in the expression of the amplitude (\ref{eprl-so4})), while 
red and blue wires are labelled by $j^{+}=(1+\gamma)j/2$ and $j^{-}=|1-\gamma|j/2$ respectively. With this (\ref{bf-so4}) is modified to
\ba\label{eprl-so4}\n
Z^{E}_{eprl}(\Delta)&=&\sum \limits_{ \rho_f \in \sK}  \ \prod\limits_{f \in \Delta^{\star}} {\rm d}_{|1-\gamma|\frac{j}{2}}{\rm d}_{(1+\gamma)\frac{j}{2}}
\prod_{e} P^{e}_{eprl}(j_1,\cdots,j_4)=\\
&=&\sum \limits_{ \rho_f \in \sK}  \ \prod\limits_{f \in \Delta^{\star}} {\rm d}_{|1-\gamma|\frac{j}{2}}{\rm d}_{(1+\gamma)\frac{j}{2}}
\begin{array}{c}
\includegraphics[width=5cm]{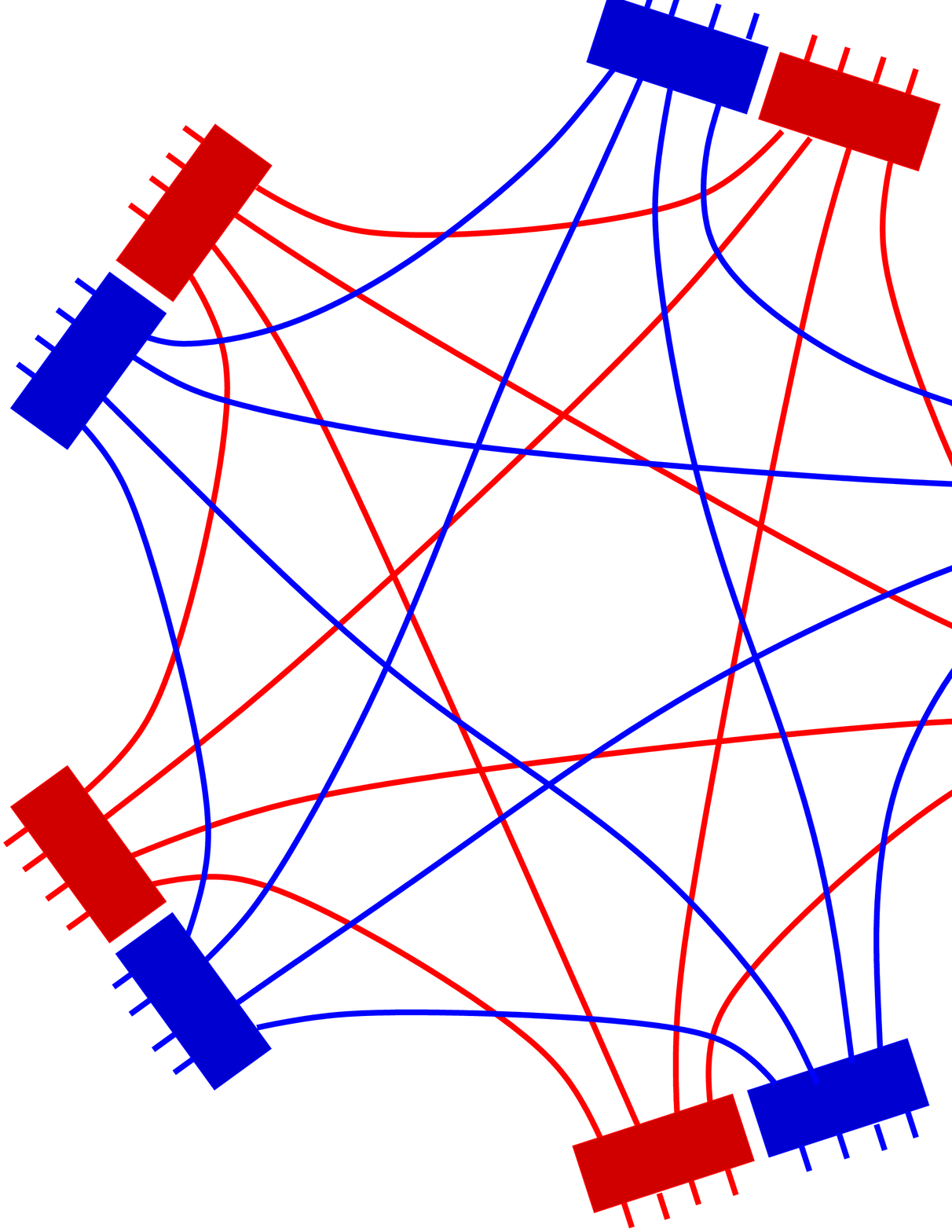}
\end{array},
\ea
The previous expression is defines the EPRL model amplitude. 

\subsubsection{The spin foam representation of the EPRL amplitude}

Now we will work out the spin foam representation of the EPRL amplitude which  at this stage will take no much more effort than the
derivation of the spin foam representation for $Spin(4)$ BF theory as we went from equation (\ref{bf-so4}) to (\ref{BF4V-b}) 
in Section \ref{BF}. The first step is given in the following equation
\ba  \begin{array}{c}
\psfrag{w}{$$}
\includegraphics[width=6cm]{eprl3} \end{array}&=& \begin{array}{c}\psfrag{w}{$$} \includegraphics[width=6cm]{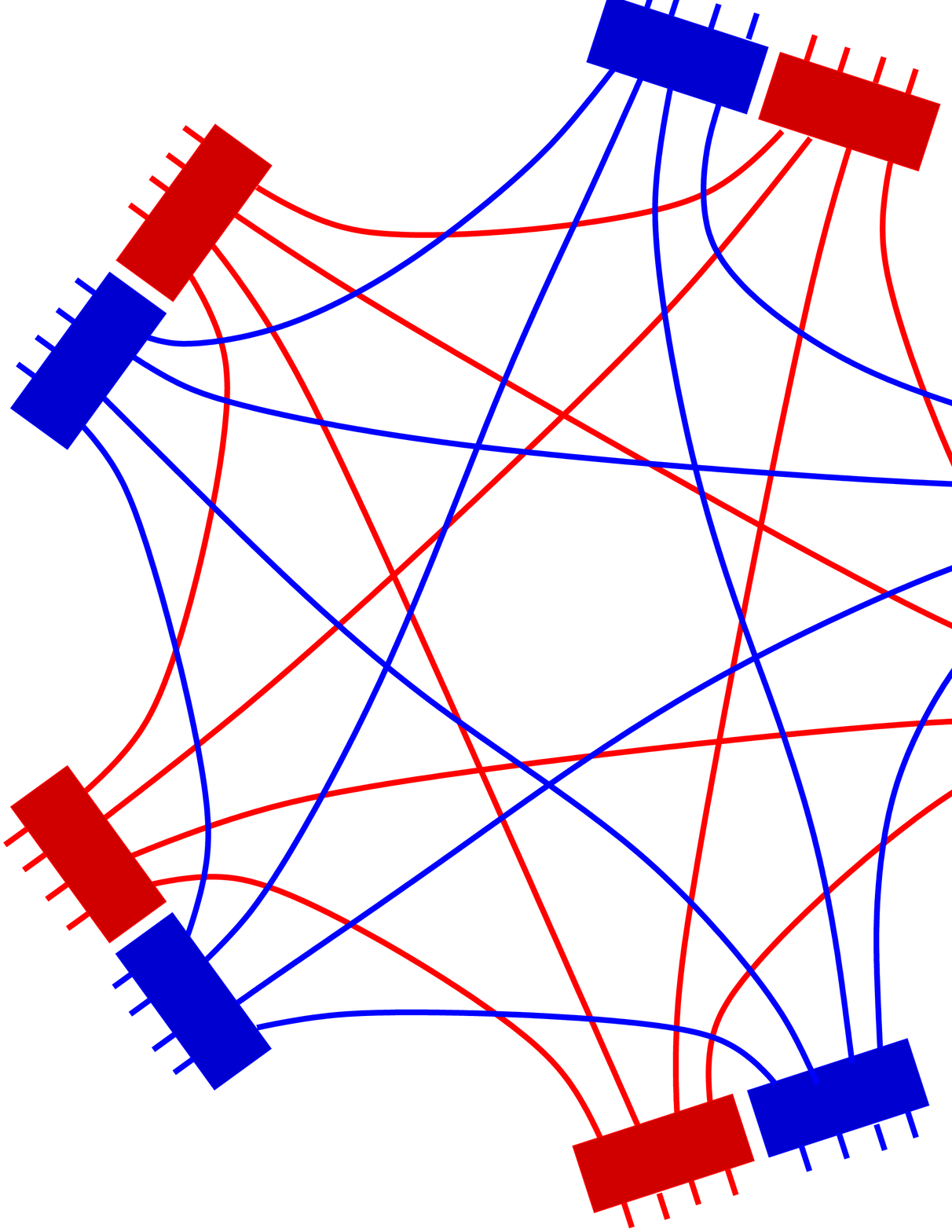}\label{dosf}
\end{array} =\n \\ 
&=& \sum_{\iota }\begin{array}{c}
\psfrag{a}{$\van \iota$}
\psfrag{b}{$\! \van \bar \iota$}
\psfrag{w}{}
\includegraphics[width=6cm]{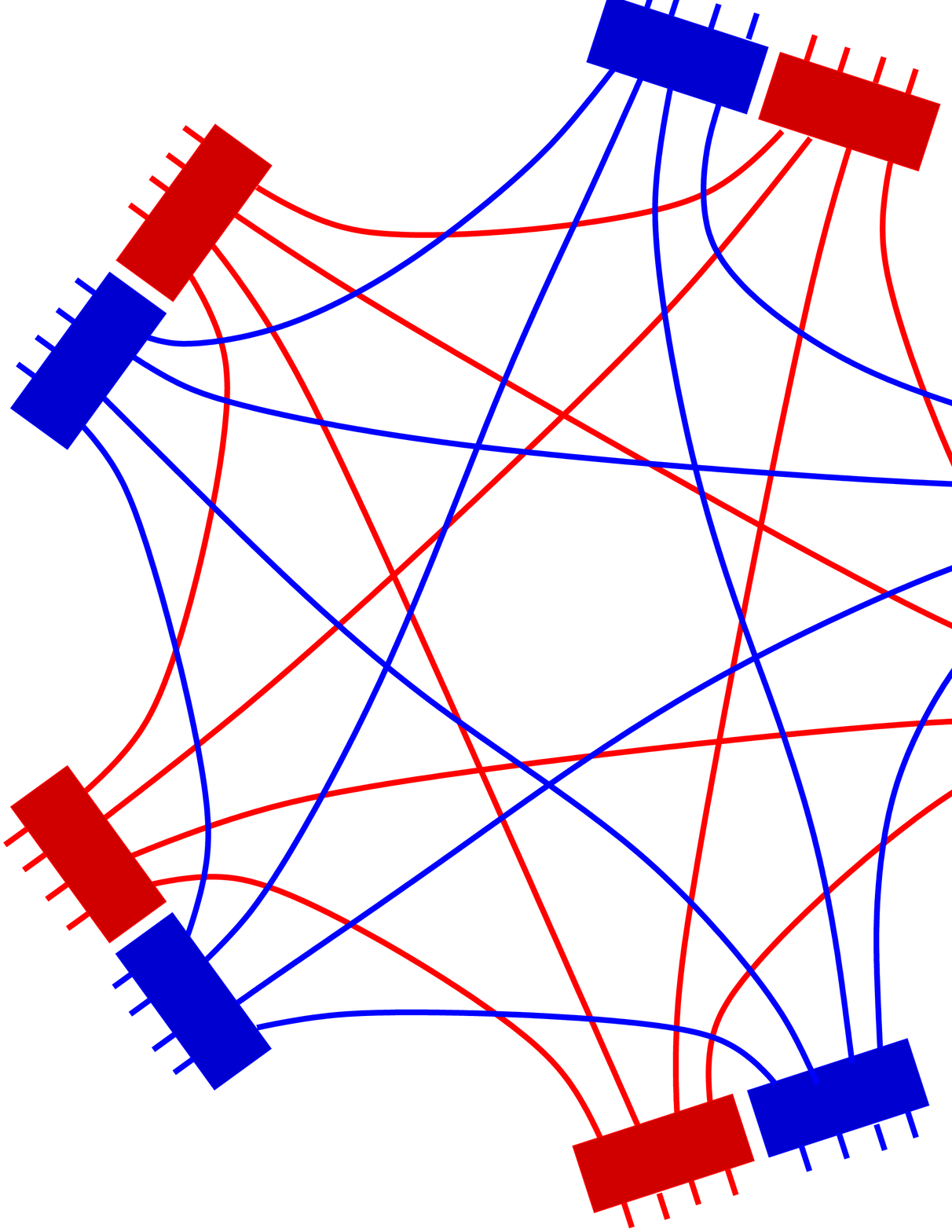}
\end{array}
\label{dosf}
\ea
which follows basically from the invariance of the Haar measure (\ref{invariance})(in the last line we have used (\ref{cab-3d})). More presicely, the integration of the subgroup $SU(2)\in Spin(4)$, represented by the green box on the right,
can be absorbed by suitable redefinition of the integration on the right and left copies of $SU(2)$, represented by the red and blue boxes respectively.
With this we can already write the spin foam representation of the EPRL model, namely  
\be\label{eprl-sf}
Z^{E}_{eprl}(\Delta)=\sum \limits_{ j_f} \sum_{\iota_e}  \\ \prod\limits_{f \in \Delta^{\star}} {\rm d}_{|1-\gamma|\frac{j}{2}}{\rm d}_{(1+\gamma)\frac{j}{2}} 
\prod_{v\in \Delta^{\star}}\begin{array}{c}\psfrag{a}{$\van \iota_1$}
\psfrag{b}{$\van  \iota_2$}
\psfrag{c}{$\van  \iota_3$}
\psfrag{d}{$\van  \iota_4$}
\psfrag{e}{$\van  \iota_5$}
\includegraphics[width=5cm]{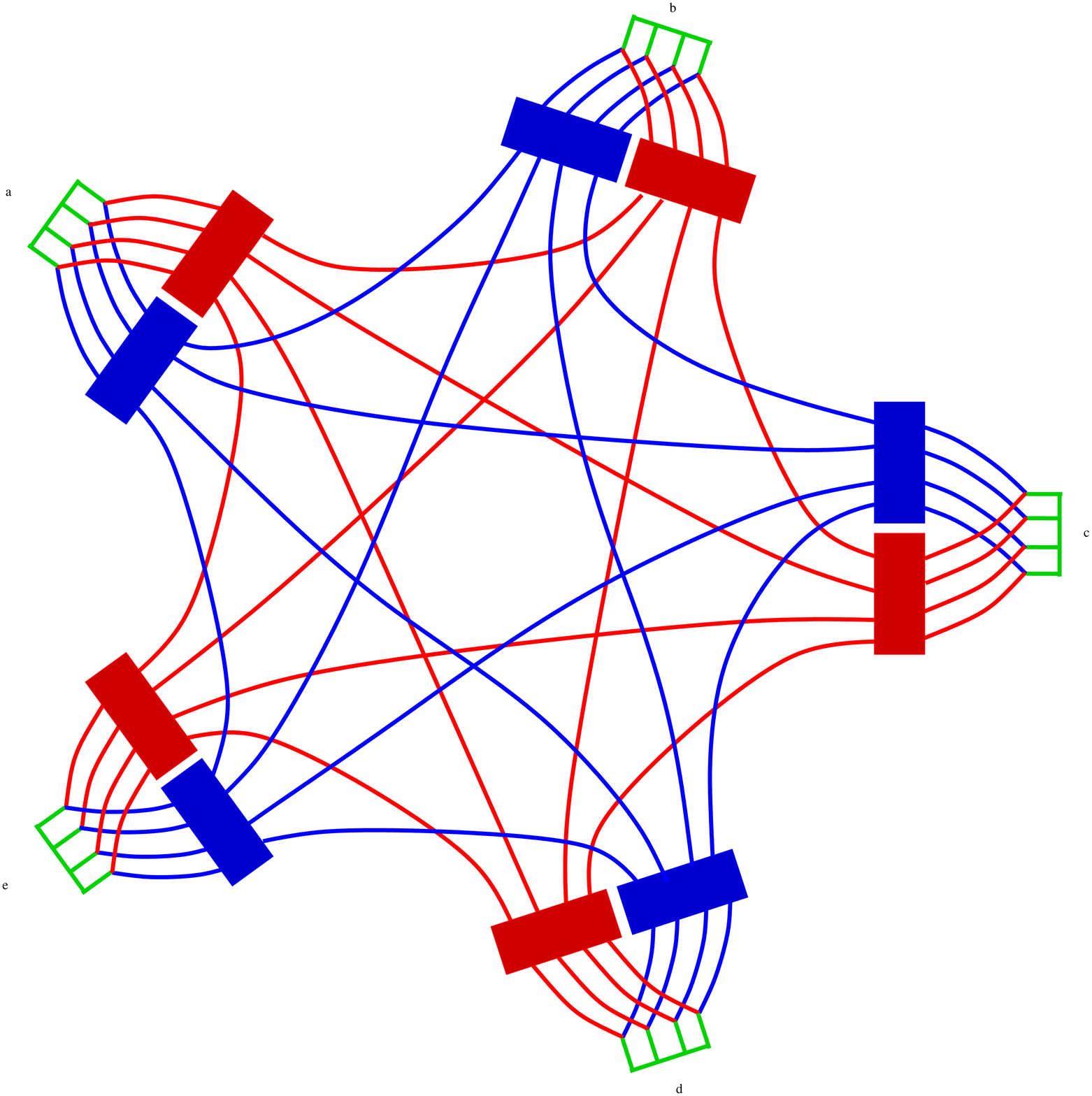}
\end{array},
\ee 
where the vertex amplitude (graphically represented) depends on the 10 spins $j$ associated to the face-wires and the 5 intertwiners associated to the five edges (tetrahedra).
As in previous equations we have left the spin labels of wires implicit for notational simplicity.
We can write the previous spin foam amplitude in another form by  integrating out all the projectors (boxes) explicitly.  Using, (\ref{cab-3d})  we get \be
\begin{array}{c}
\psfrag{w}{$= \sum\limits_{\iota_{\va +}\iota_{\va -}\iota} $}
\psfrag{x}{$\!\!\!\!\!\!\!\!
 \vani \iota_{\va +}\, \ \ \ \bar\iota_{\va +}$}
\psfrag{z}{$\!\!\!\!\!\!\!\! \vani \iota_{\va -}\,\ \ \ \bar\iota_{\va-}$}
\psfrag{y}{$\!\!\, \vani \iota\, \bar\iota$}
\includegraphics[width=6cm]{eprl}\ \ \ \ \  \includegraphics[width=6cm]{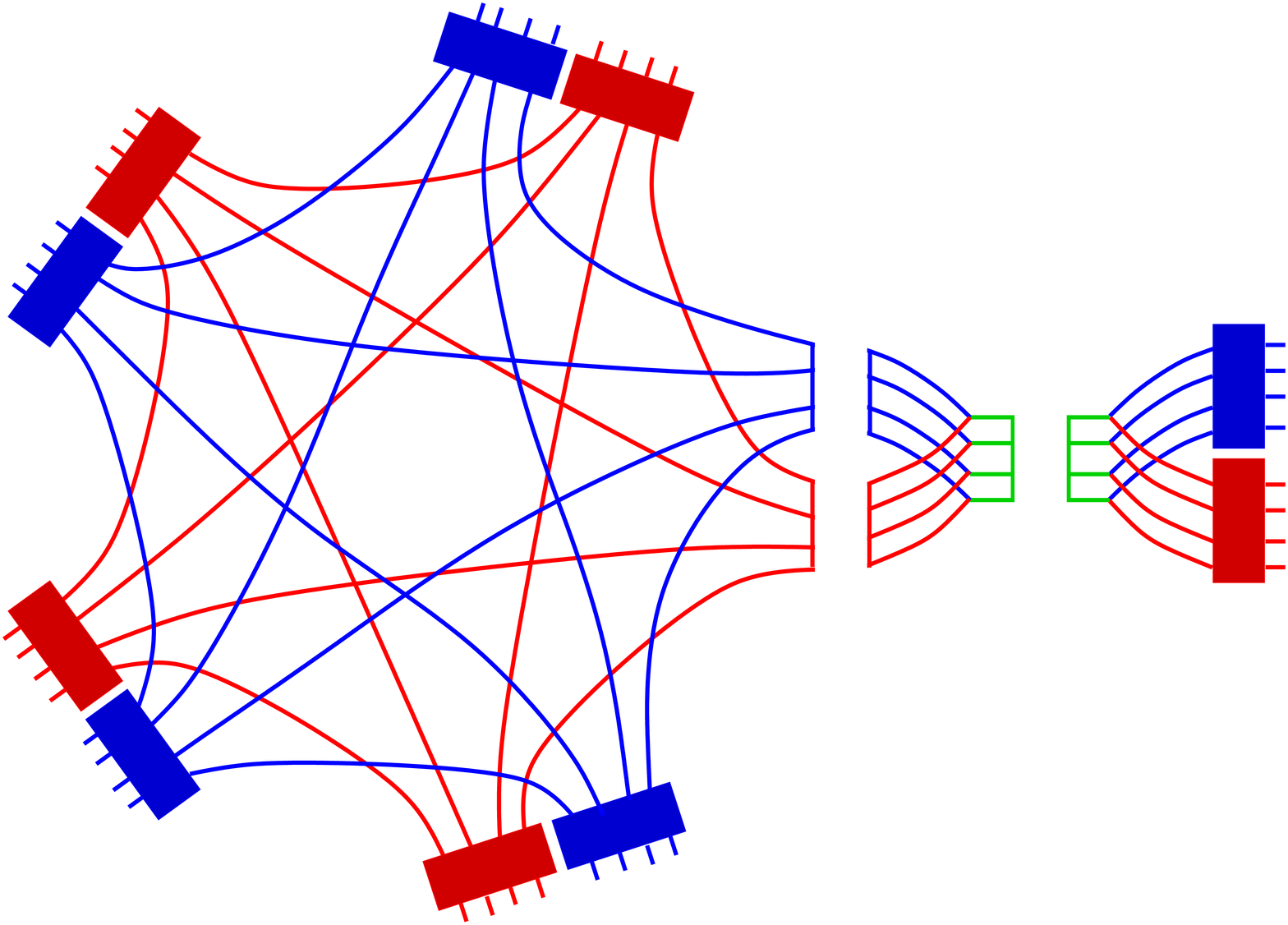}
\end{array}
\label{unof}
\ee
thus replacing this in (\ref{eprl-so4}) we get 
\ba 
&&
Z^{E}_{eprl}(\Delta)=\sum \limits_{ j_f }  \ \prod\limits_{f \in \Delta^{\star}}   {\rm d}_{|\gamma-1|\frac{j}{2}}{\rm d}_{(\gamma+1)\frac{j}{2}} \sum \limits_{\iota_e } \ \prod\limits_{v\in \Delta^{\star}}  \\ &&   \sum_{\iota^{-}_1\cdots \iota^-_5}\sum_{\iota^{+}_1\cdots \iota^+_5} \prod\limits_{a=1}^{5} f^{\iota_a}_{\iota^{-}_a,\iota^{+}_{a}}
\begin{array}{c}
\psfrag{a}{$\van \iota^{-}_1$}
\psfrag{b}{$\van \iota^{-}_2$}
\psfrag{c}{$\van \iota^{-}_3$}
\psfrag{d}{$\van \iota^{-}_4$}
\psfrag{e}{$\van \iota^{-}_5$}
\psfrag{A}{$\!\!\!\!\!\!\!\!\!\!\van |1-\gamma|\frac{j_1}{2}$}
\psfrag{B}{$\!\!\!\!\!\!\!\!\!\! \van |1-\gamma|\frac{j_2}{2}$}
\psfrag{C}{$\!\!\!\!\!\!\!\!\!\!\!\! \van |1-\gamma|\frac{j_3}{2}$}
\psfrag{D}{$\!\!\!\!\!\!\!\!\!\!\!\! \van |1-\gamma|\frac{j_4}{2}$}
\psfrag{E}{$\!\!\!\!\!\!\!\!\!\!\!\! \van |1-\gamma|\frac{j_5}{2}$}
\psfrag{F}{$\!\!\!\!\!\!\!\! \van |1-\gamma|\frac{j_6}{2}$}
\psfrag{G}{$\!\!\!\!\!\!\!\! \van |1-\gamma|\frac{j_7}{2}$}
\psfrag{H}{$\!\!\!\!\!\!\!\! \van |1-\gamma|\frac{j_8}{2}$}
\psfrag{I}{$\!\!\!\!\!\!\!\!\!\!\!\! \van |1-\gamma|\frac{j_9}{2}$}
\psfrag{J}{$\!\!\!\!\!\!\!\!\!\!\!\! \van |1-\gamma|\frac{j_{10}}{2}$}
\includegraphics[height=4.7cm]{BF4V-g}
\end{array}\ \ \ \ \ 
\begin{array}{c}
\psfrag{ap}{$\van \iota^{+}_1$}
\psfrag{bp}{$\van \iota^{+}_2$}
\psfrag{cp}{$\van \iota^{+}_3$}
\psfrag{dp}{$\van \iota^{+}_4$}
\psfrag{ep}{$\van \iota^{+}_5$}
\psfrag{Ap}{$\!\!\!\!\!\!\!\! \van |1+\gamma|\frac{j_1}{2}$}
\psfrag{Bp}{$\!\!\!\!\!\!\!\!\!\! \van |1+\gamma|\frac{j_2}{2}$}
\psfrag{Cp}{$\!\!\!\!\!\!\!\!\!\!\!\! \van |1+\gamma|\frac{j_3}{2}$}
\psfrag{Dp}{$\!\!\!\!\!\!\!\!\!\!\!\! \van |1+\gamma|\frac{j_4}{2}$}
\psfrag{Ep}{$\!\!\!\!\!\!\!\!\!\!\!\! \van |1+\gamma|\frac{j_5}{2}$}
\psfrag{Fp}{$\!\!\!\!\!\!\!\! \van |1+\gamma|\frac{j_6}{2}$}
\psfrag{Gp}{$\!\!\!\!\!\!\!\! \van |1+\gamma|\frac{j_7}{2}$}
\psfrag{Hp}{$\!\!\!\!\!\!\!\! \van |1+\gamma|\frac{j_8}{2}$}
\psfrag{Ip}{$\!\!\!\!\!\!\!\!\!\!\!\! \van |1+\gamma|\frac{j_9}{2}$}
\psfrag{Jp}{$\!\!\!\!\!\!\!\!\!\!\!\! \van |1+\gamma|\frac{j_{10}}{2}$}
\includegraphics[height=4.7cm]{BF4V-b}
\end{array}\n 
\label{SF-eprl}
\ea
where the coefficients $f^{\iota}_{\iota^+\iota^{-}}$ are the so-called fusion coefficients
which appear in their graphical form already in (\ref{unof}), more explicitly
\be f^{\iota}_{\iota^+\iota^{-}}(j_1,\cdots,j_4)=
\begin{array}{c}
\psfrag{x}{$ \van \iota_{ +}$}
\psfrag{z}{$\van  \iota_{\va -}$}
\psfrag{y}{$\van \!\!\, \iota$}
\psfrag{a}{$\van |1-\gamma| \frac{j_1}{2}$}
\psfrag{b}{$\van |1-\gamma| \frac{j_2}{2}$}
\psfrag{c}{$\van |1-\gamma| \frac{j_3}{2}$}
\psfrag{d}{$\van |1-\gamma| \frac{j_4}{2}$}
\psfrag{ap}{$\van |1+\gamma| \frac{j_1}{2}$}
\psfrag{bp}{$\van |1+\gamma| \frac{j_2}{2}$}
\psfrag{cp}{$\van |1+\gamma| \frac{j_3}{2}$}
\psfrag{dp}{$\van |1+\gamma| \frac{j_4}{2}$}
\psfrag{A}{$\van {j_1}$}
\psfrag{B}{$\van {j_2}$}
\psfrag{C}{$\van {j_3}$}
\psfrag{D}{$\van {j_4}$}
\includegraphics[height=5cm]{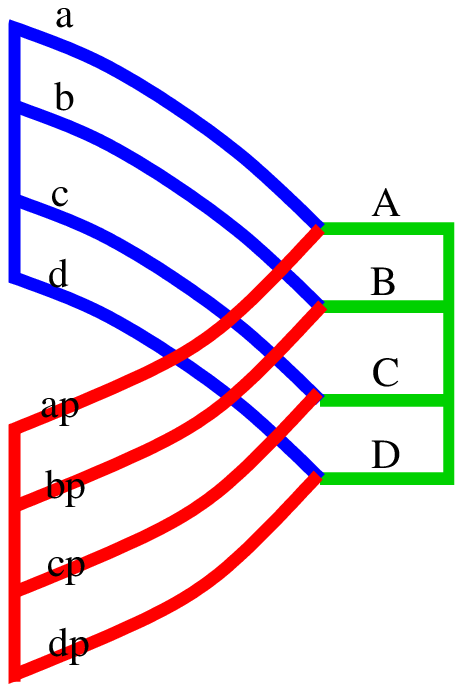}
\end{array}
\ee
The previous is the form of the EPRL model as derived in \cite{Engle:2007wy}.

\subsection{Proof of validity of the Plebanski constraints}
\label{cuadratiqui}

In this section we prove that the quadratic constraints are satisfied in the sense 
that their path integral expectation value and fluctuation vanish in the appropriate 
semiclassical limit.

\subsubsection{The dual version of the constraints}

In this section we rewrite the Plebanski constraints (\ref{ito}) in an equivalent version where spacetime indices are traded with 
internal Lorentz indices, namely 
\be \epsilon_{IJKL} B^{IJ}_{\mu\nu}B^{KL}_{\rho\sigma}-e \
 \epsilon_{\mu\nu\rho\sigma}\approx 0. \label{dual}\ee
The constraints in this form are more suitable for the translation into the 
discrete formulation. More precisely, according to (\ref{Bdisc}), the smooth fields $B_{\mu\nu}^{IJ}$ is now associated with
the discrete quantities $B_{\va {\rm triangles}}^{IJ}$, or equivalently $B^{IJ}_{f}$ as, we recall, faces $f\in \Delta^{\star}$ are in one-to-one correspondence to 
triangles in four dimensions.  The constraints (\ref{dual}) are local constraints valid at every spacetime point. 
In the discrete setting, spacetime points are represented by four-simplexes or (more addapted to our discussion) vertices $v\in \Delta^{\star}$.
With all this the constraints (\ref{dual}) are discretized as follows:
\be\label{2s}
\mbox{\bf Triangle (or diagonal) constraints: \ \ \ \ \ \ \ $ \epsilon_{IJKL} B^{IJ}_{f}B^{KL}_{f}=0$,}
\ee
for all  $f\in v$, i.e., for each and every face of the 10 possible faces touching the vertex $v$.
\be\label{3s}
\mbox{\bf Tetrahedron constraints: \ \ \ \ \ \ \  $\epsilon_{IJKL} B^{IJ}_{f}B^{KL}_{f'}=0$,}
\ee
for all $f,f'\in v$ such that they are dual to triangls sharing a one-simplex, i.e., belonging to the same tetrahedron out of the five possible ones. 
\be\label{4s}
\mbox{\bf 4-simplex constraints: \ \ \ \ \ \ \  $\epsilon_{IJKL} B^{IJ}_{f}B^{KL}_{\bar f}=e_v$,}
\ee
for any pair of faces $f,\bar f\in v$ that are dual to triangles sharing a single point. The last constraint will require a more detailed discussion. At this point let us point out that the 
constraint (\ref{4s}) is interpreted as a definition of the four volume $e_v$ of the four-simplex. The constraint requires that such definition be consistent, i.e., the true condition is
\be\label{4strue}
\epsilon_{IJKL} B^{IJ}_{f}B^{KL}_{\bar f}=\epsilon_{IJKL} B^{IJ}_{f'}B^{KL}_{\bar f'}=\epsilon_{IJKL} B^{IJ}_{f''}B^{KL}_{\bar f''}=\cdots=e_v
\ee
for all five different possible pairs of $f$ and $\bar f$ in a four simplex, and where we assume the pairs $f$-$\bar f$ are ordered 
in agreement with the orientation of the complex $\Delta^{\star}$.

\subsubsection{The path integral expectation value of the Plebanski constraints}

Here we prove that the Plebanski constraint are satisfied by the EPRL amplitudes in the 
path integral expectation value sense.

\subsubsection*{The triangle constraints:} 

We start from the simplest case:  the triangle (or diagonal) constraints (\ref{2s}).
We choose a face $f\in v$ (dual to a triangle) in the cable-wire-diagram of Equation (\ref{eprl-so4}).
This amounts to choosing a pair of wires (right and left representations) connecting two nodes in the vertex cable wire diagram. 
The two nodes are dual to the two tetrahedra -- in the four simplex dual to the vertex -- sharing the chosen triangle.
From  (\ref{defidifi}) one can show that
\be
 \epsilon_{IJKL} B^{IJ}_{f}B^{KL}_{f}\propto (1+\gamma)^2 J_{f}^{-}\cdot J_{f}^{-} -(1-\gamma)^2 J_{f}^{+}\cdot J_{f}^{+},
\ee
where $J_{f}^{\pm}$ denotes the self-dual and anti-self-dual parts of $\Pi^{IJ}_{f}$.  
The path integral expectation value of the triangle constraint is then
\ba \label{pipo}
 && \langle (1+\gamma)^2 J_{f}^{-}\cdot J_{f}^{-} -(1-\gamma)^2 J_{f}^{+}\cdot J_{f}^{+}\rangle \propto \\ 
&& \n (1+\gamma)^2 \begin{array}{c}
\includegraphics[width=4.5cm]{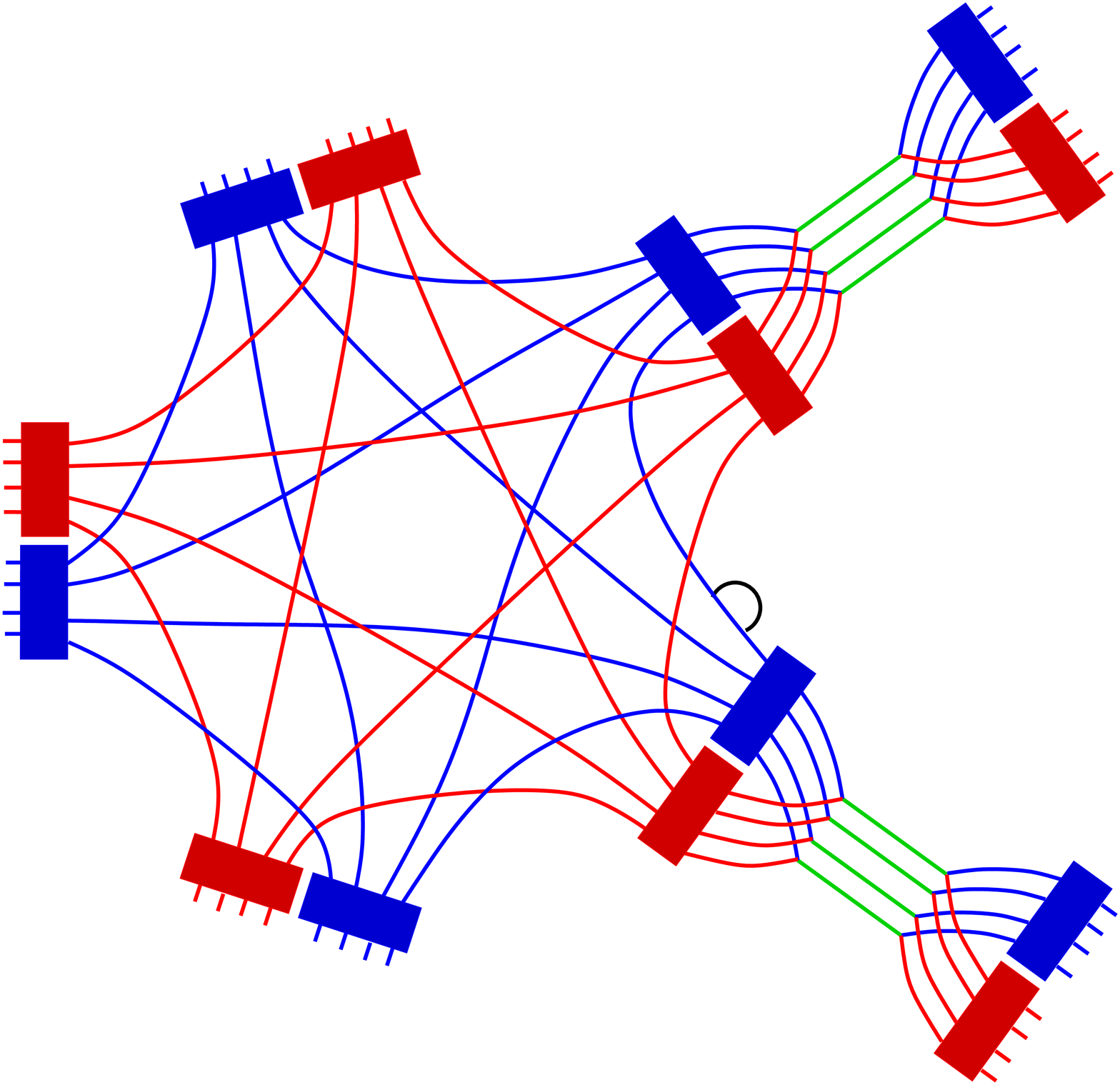}
\end{array}
-(1-\gamma)^2
\begin{array}{c}
\includegraphics[width=4.5cm]{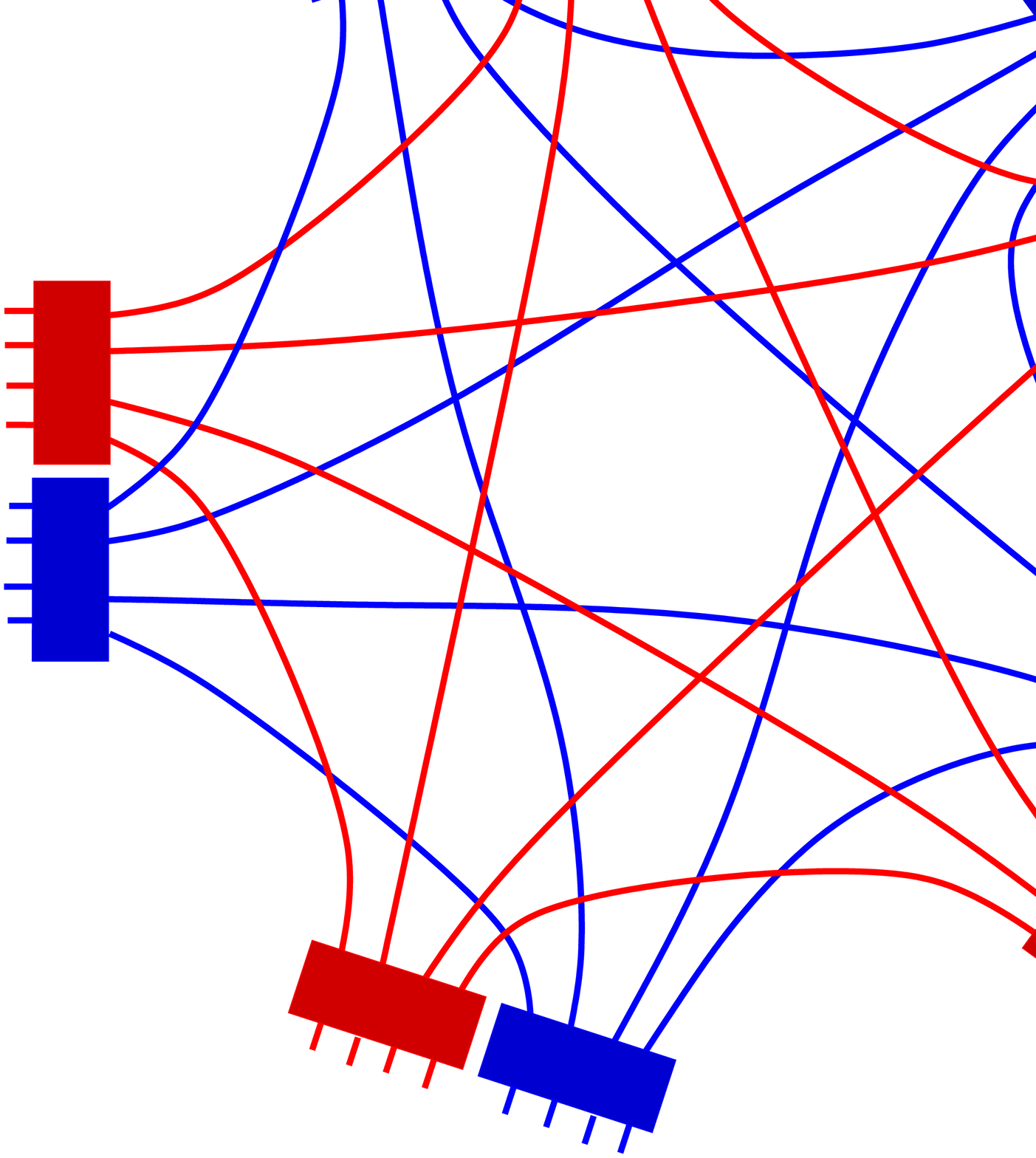}
\end{array}=\sO_{sc},
\ea
where the double graspings on the anti-self-dual (blue) wire and the self-dual (red) wire represent the action of the Casimirs
$J_{f}^{-}\cdot J_{f}^{-}$ and $J_{f}^{+}\cdot J_{f}^{+}$ on the cable-wire diagram of the corresponding vertex. Direct evaluation
shows that the previous diagram is proportional to $\hbar^2 j_f$ which vanishes in the semiclassical limit $\hbar\to 0$, $j\to \infty$
with $\hbar j=$constant. We use the notation already adopted in (\ref{needed}) and call such quantity $\sO_{sc}$. This concludes the proof that the triangle Plebanski constraints 
are satisfied in the semiclassical sense.

 \subsubsection*{The tetrahedra constraints:} 
 
The proof of the validity of the tetrahedra constraints (\ref{3s}). In this case we also have
\be
(1+\gamma)^2 \begin{array}{c}
\includegraphics[width=4.5cm]{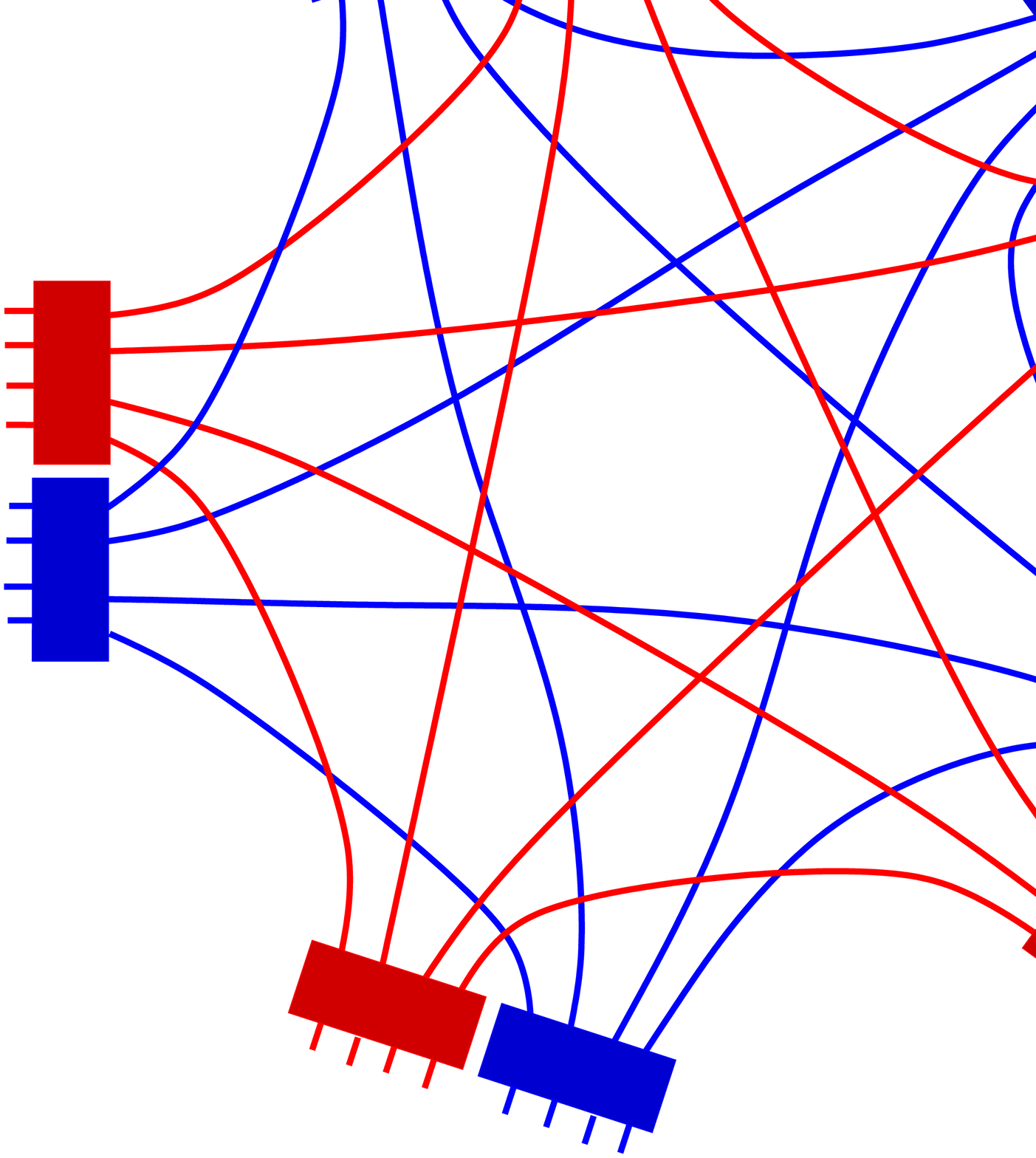}
\end{array}
-(1-\gamma)^2
\begin{array}{c}
\includegraphics[width=4.5cm]{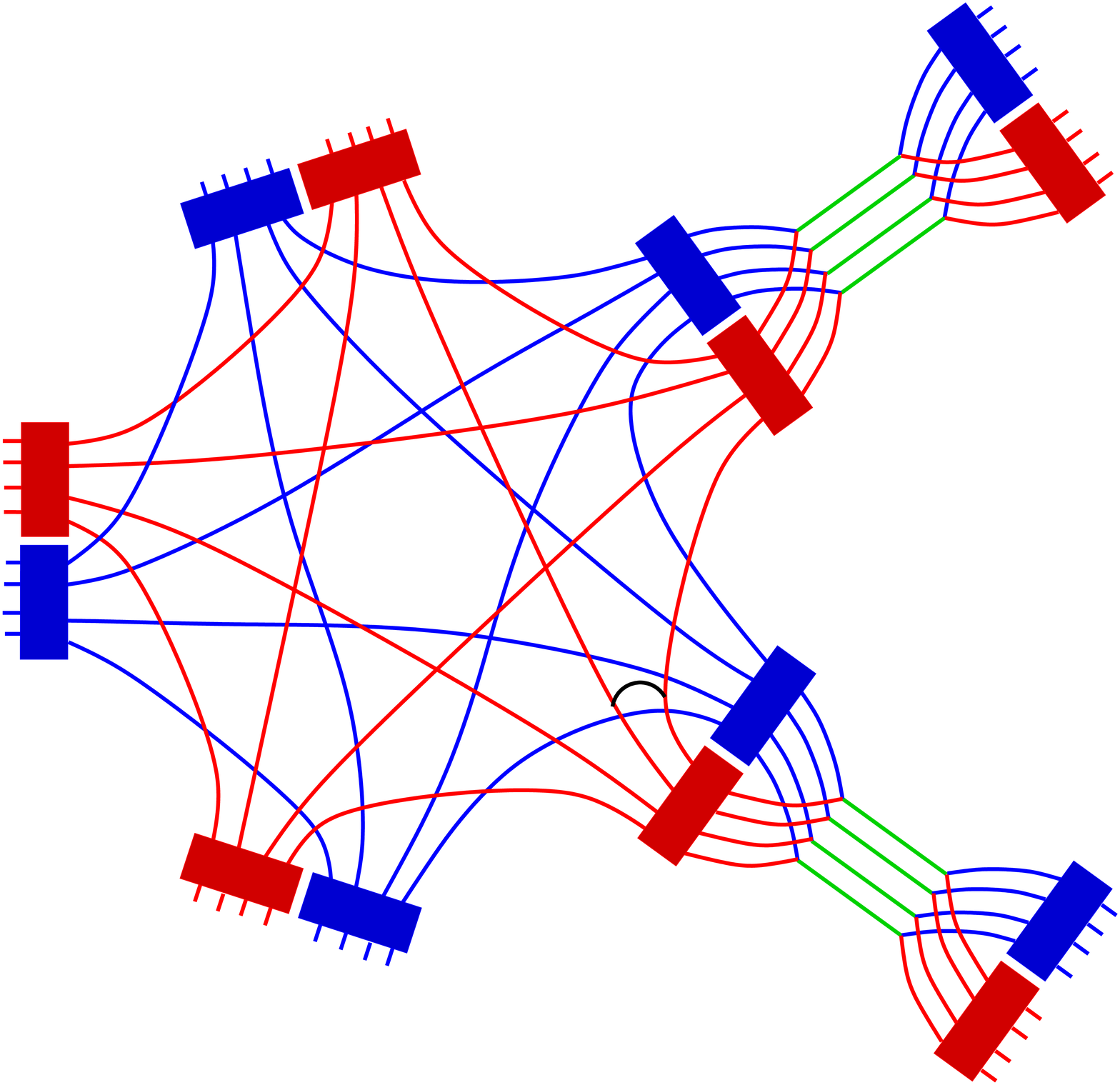}
\end{array}=\sO_{sc}.\label{tetra}
\ee
where we have chosen an arbitrary pair of faces. 
In order to prove this let us develop the term on the right. The result follows from 
\ba &&
\begin{array}{c}
\includegraphics[width=2cm]{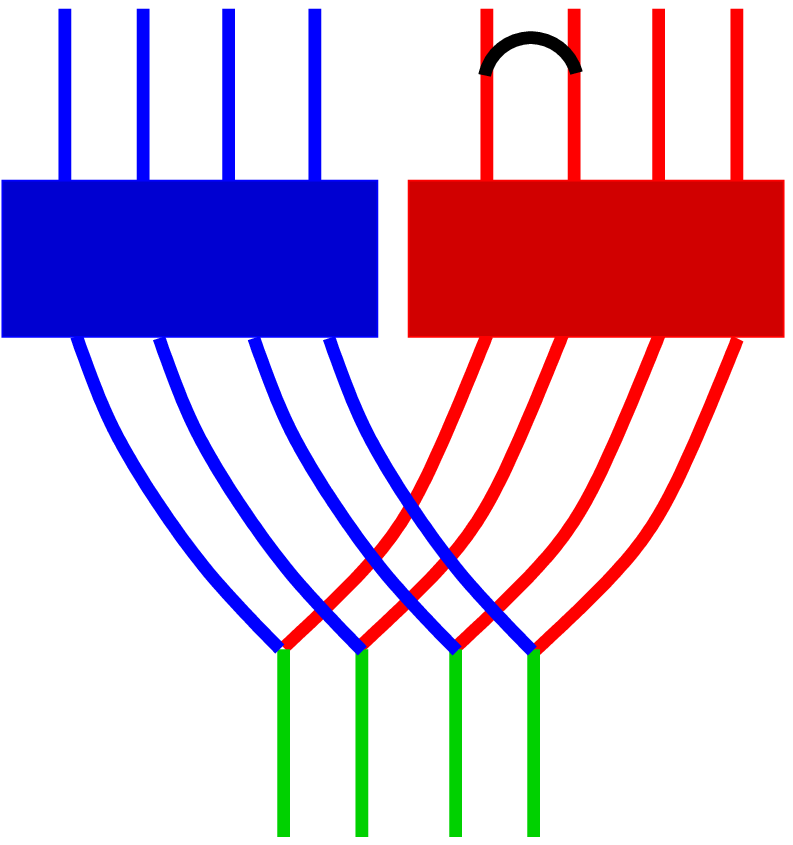}
\end{array}
=
\begin{array}{c}
\includegraphics[width=2cm]{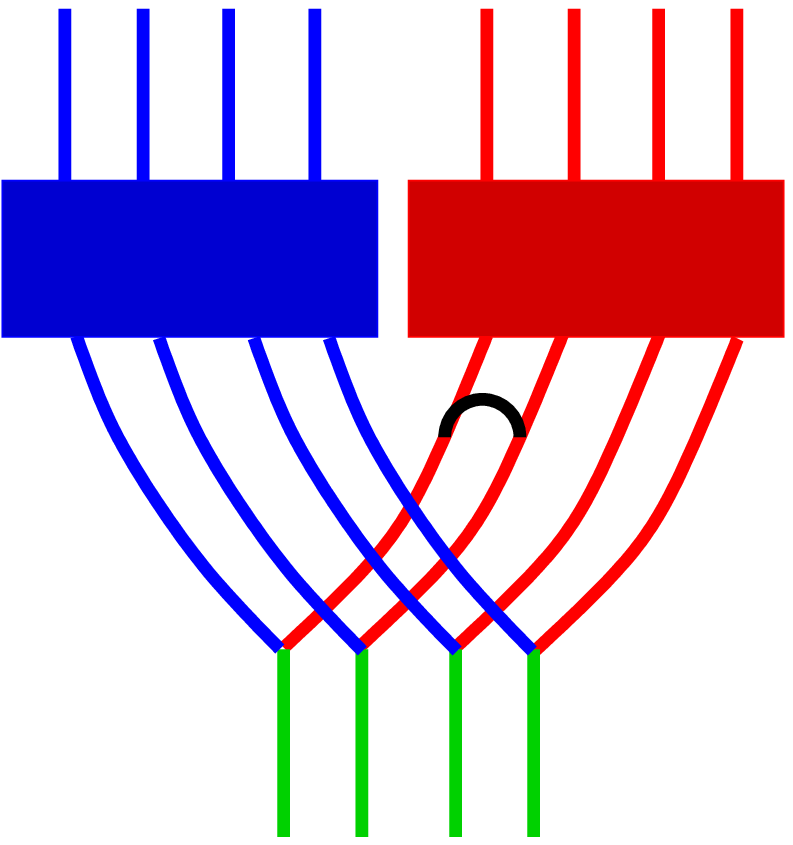}
\end{array}=\n \\
&& =\frac{(1+\gamma)}{|1-\gamma|}
\begin{array}{c}
\includegraphics[width=2cm]{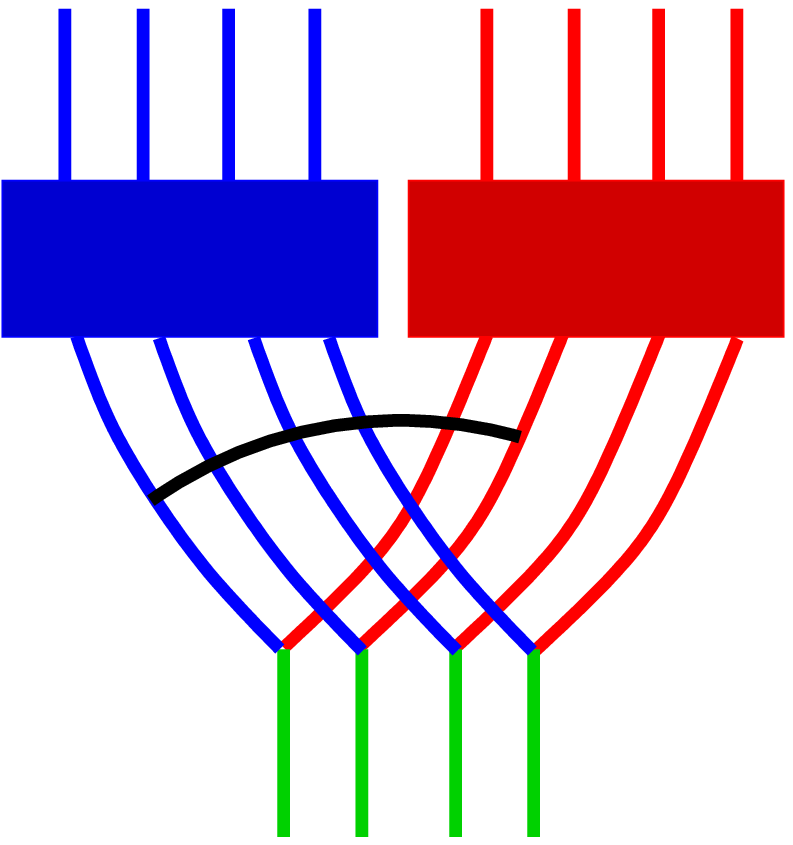}
\end{array}+\sO_{sc}
=\frac{(1+\gamma)^2}{(1-\gamma)^2}
\begin{array}{c}
\includegraphics[width=2cm]{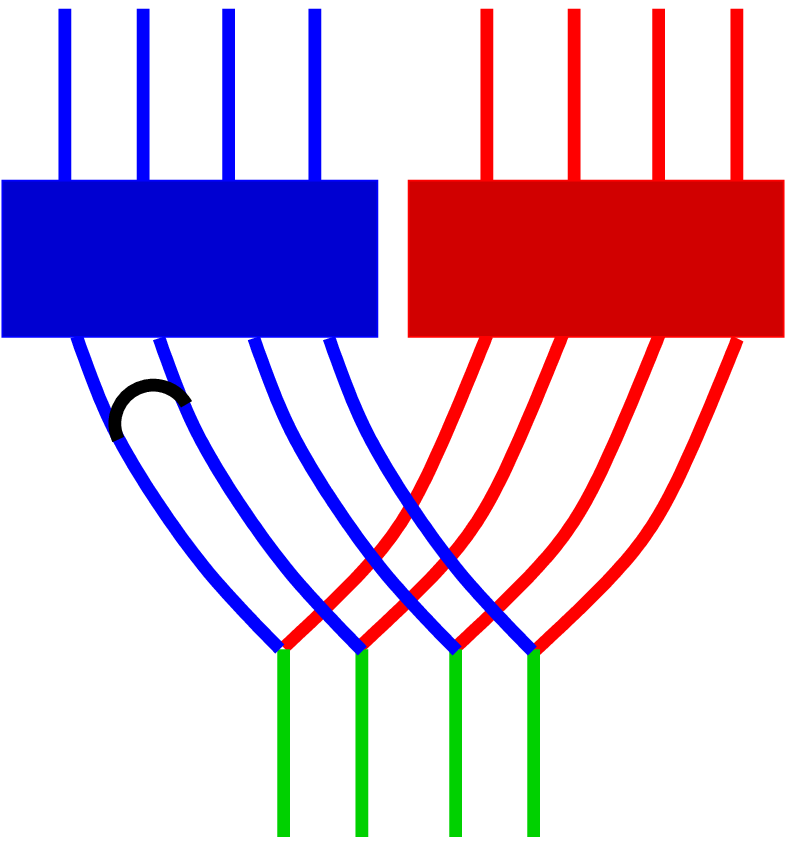}
\end{array}+\sO_{sc}=\n \\ &&=\frac{(1+\gamma)^2}{(1-\gamma)^2} \begin{array}{c}
\includegraphics[width=2cm]{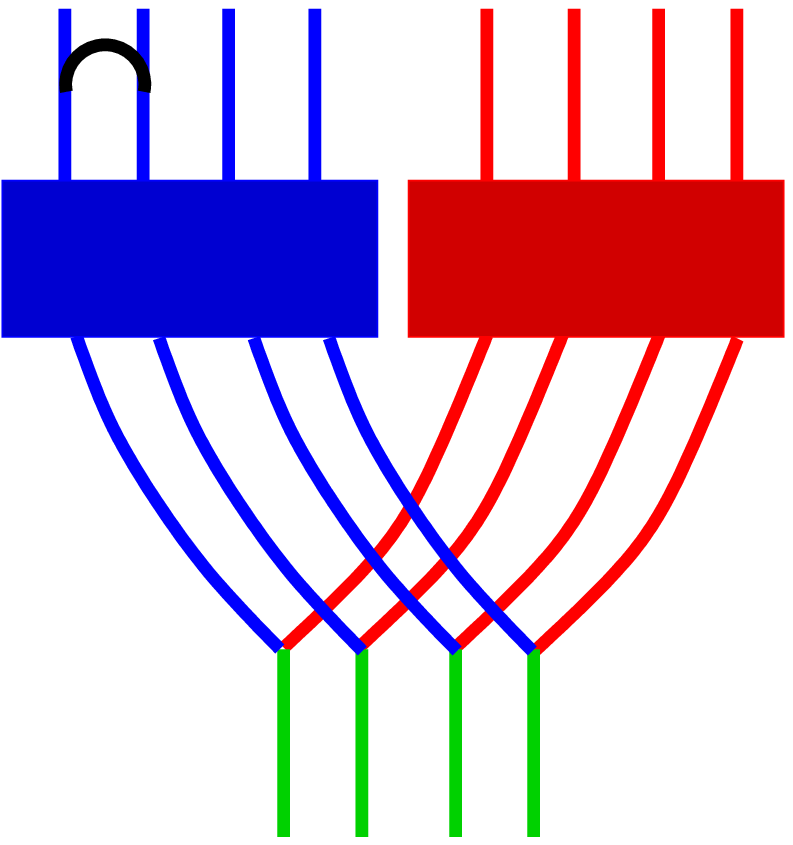}
\end{array}+\sO_{sc}, \label{pipisi}
\ea
where in the first line we have used the fact that the double grasping can be shifted through the group integration (due to gauge invariance (\ref{invariance})), and
in the first and second terms on the second line we have used Equation (\ref{trival}) to move the graspings on self-dual wires to the corresponding anti-self-dual wires. 
Equation (\ref{tetra}) follows immediately from the pervious one; the argument works in the same way for any other pair of faces.
Notice that the first equality in Equation (\ref{pipisi}) implies that we can view the Plebanski constraint as applied in the frame of the tetrahedron as well as in a Lorentz invariant 
framework (the double grasping defines an intertwiner operator commuting with the projection $P^e_{inv}$ represented by the box). An analogous statement also holds 
for the triangle constraints (\ref{pipo}).

\subsubsection*{The 4-simplex constraints}

Now we show the validity of the four simplex constraints in their form (\ref{4strue}). As we show below, this last set of constraints 
follow from the $Spin(4)$ gauge invariance of the EPRL node (i.e., the validity of the Gauss law) plus the validity of the tetrahedra constraints (\ref{3s}).
Gauge invariance of the node  takes the following form in graphical notation
\be
\begin{array}{c}
\includegraphics[width=2cm]{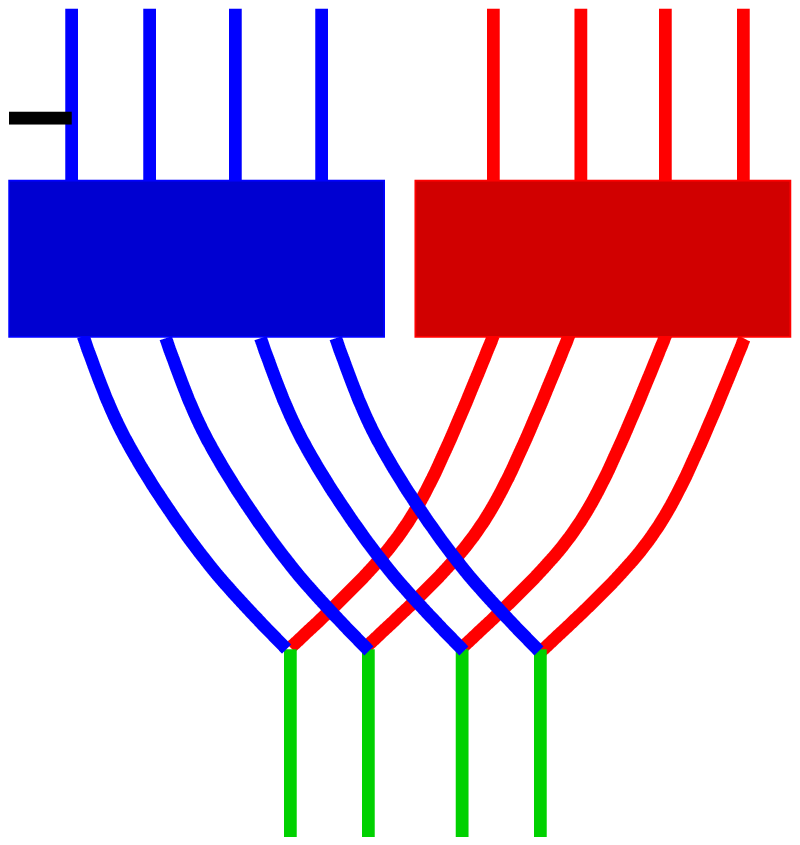}
\end{array}
+
\begin{array}{c}
\includegraphics[width=2cm]{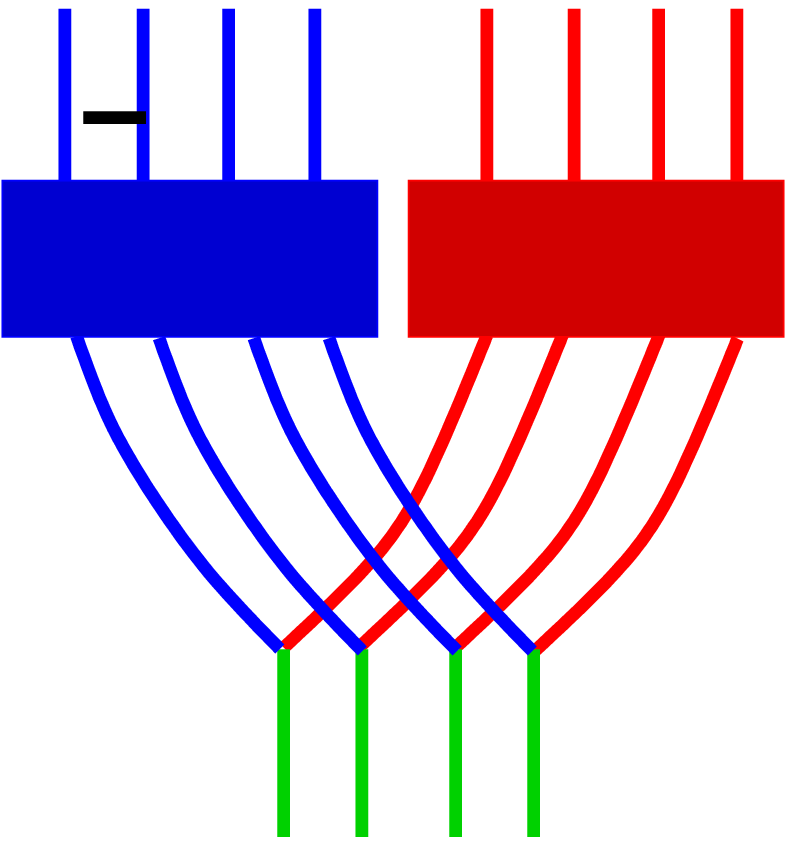}
\end{array}+\begin{array}{c}
\includegraphics[width=2cm]{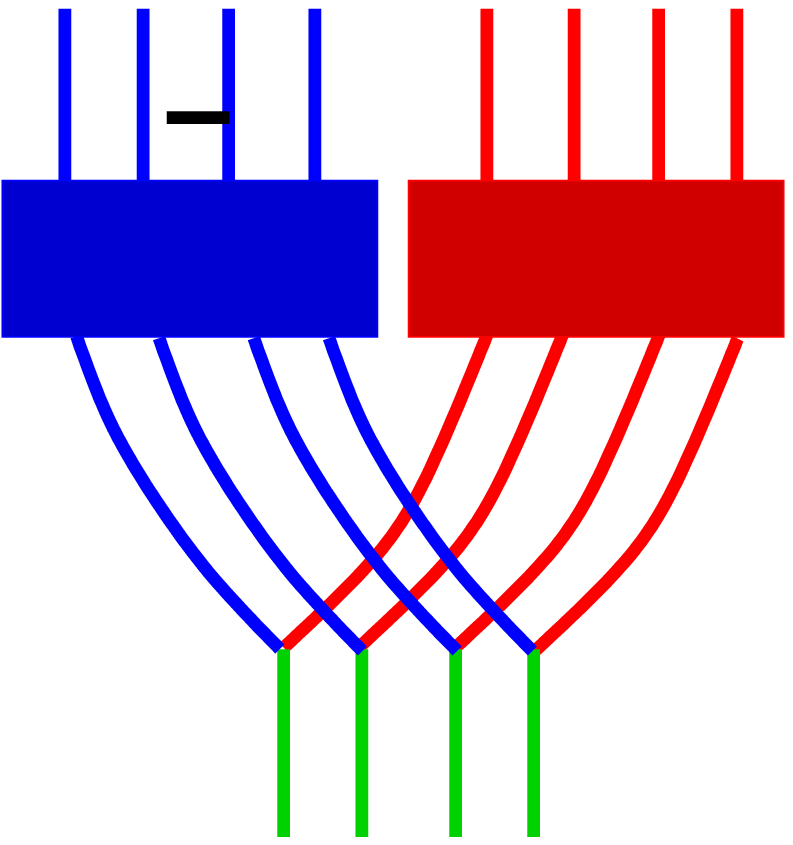}
\end{array}+\begin{array}{c}
\includegraphics[width=2cm]{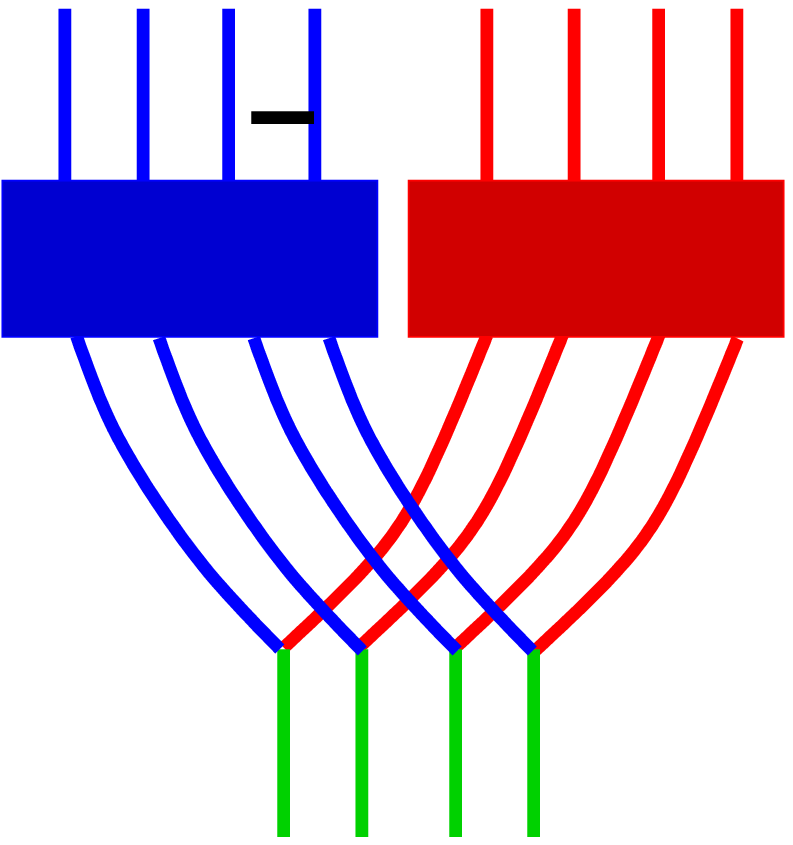}
\end{array}=0,\label{eprl-gauss}
\ee
where the above equation represents the gauge invariance under infinitesimal left $SU(2)$ rotations.
An analogous equation with insertions on the right is also valid. The validity of the previous equation can 
again be related to the invariance of the Haar measure used in the integration on the gauge group that 
defines the boxes (\ref{invariance}).

Now we chose an arbitrary pair $f$ and $\bar f$ (where, recall, $\bar f$ is one of the three possible faces whose dual triangle only shares a point with the 
corresponding to $f$) and will show how the four volumen $e_v$ defined by it equals the one defined by any other admissible pair. The first  step is to show that
we get the same result using the pair $f$-$\bar f$ and $f$-$\bar{\bar f}$, where $\bar{\bar f}$ is another of the three admissible faces opposite to $f$. 
The full result follows from applying the same procedure iteratively to reach any admissible pair. It will be obvious from the treatment given below that this is possible.
Thus, for a given pair of admissible faces we have
\ba && \n
\!\!\!\!\!\!\!\!\!\!\!\!e_v=(1+\gamma)^2 \begin{array}{c}
\includegraphics[width=4cm]{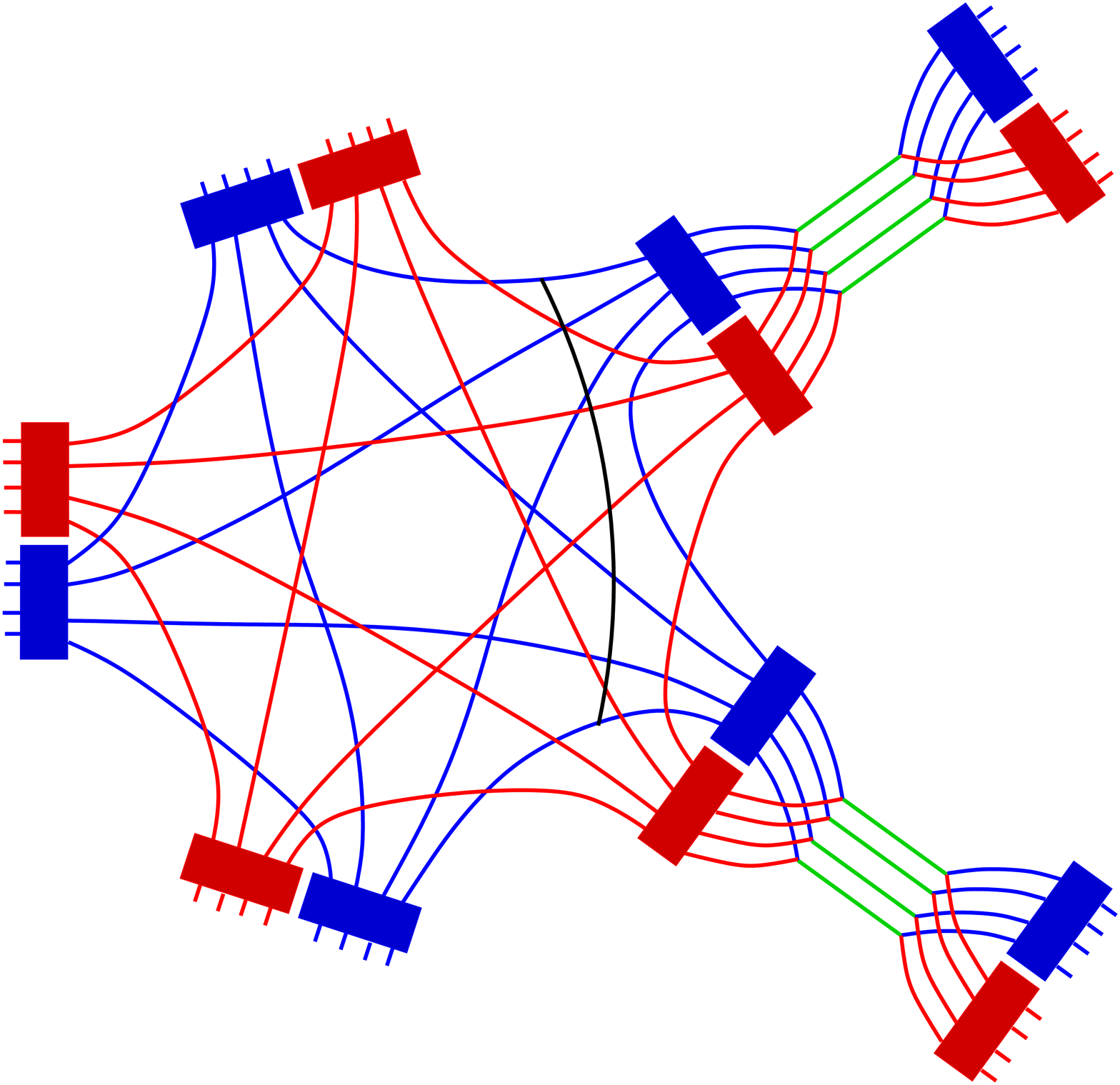}
\end{array}
-(1-\gamma)^2
\begin{array}{c}
\includegraphics[width=4cm]{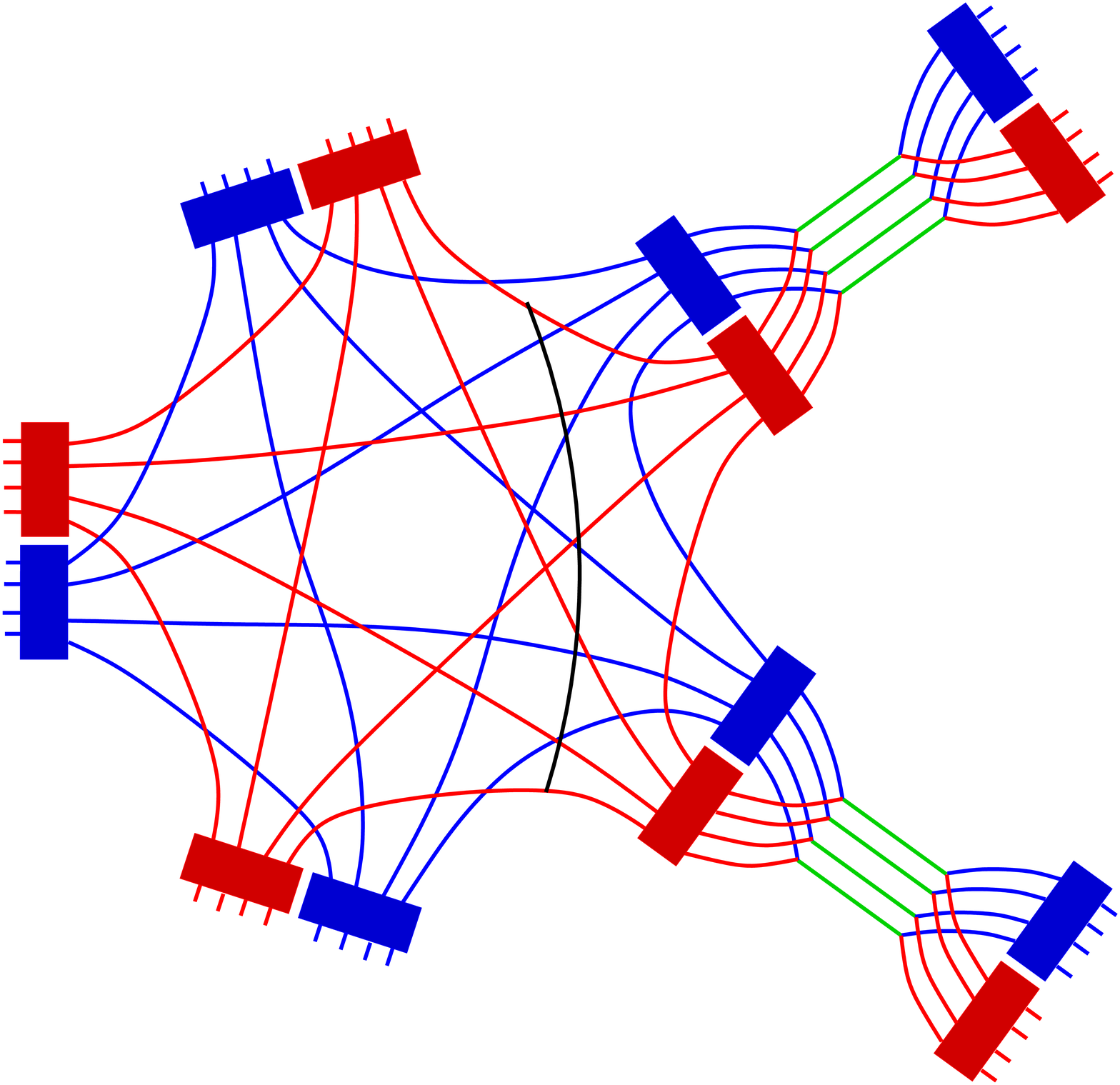}
\end{array} =\\ 
&& \!\!\!\!\!\!\!\!\!\!\!\! -(1+\gamma)^2\left[ \begin{array}{c}
\includegraphics[width=4cm]{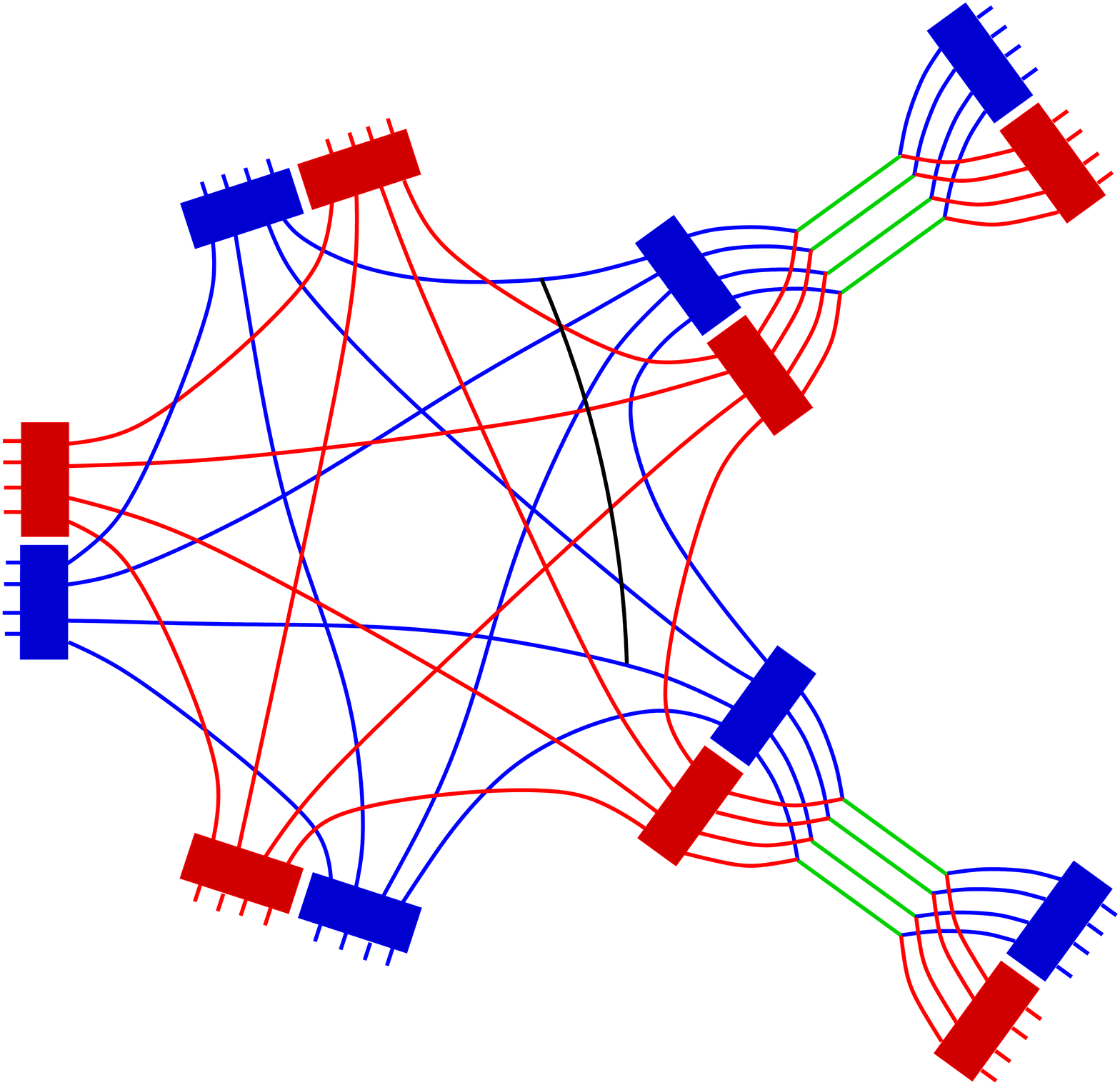}
\end{array}+\begin{array}{c}
\includegraphics[width=4cm]{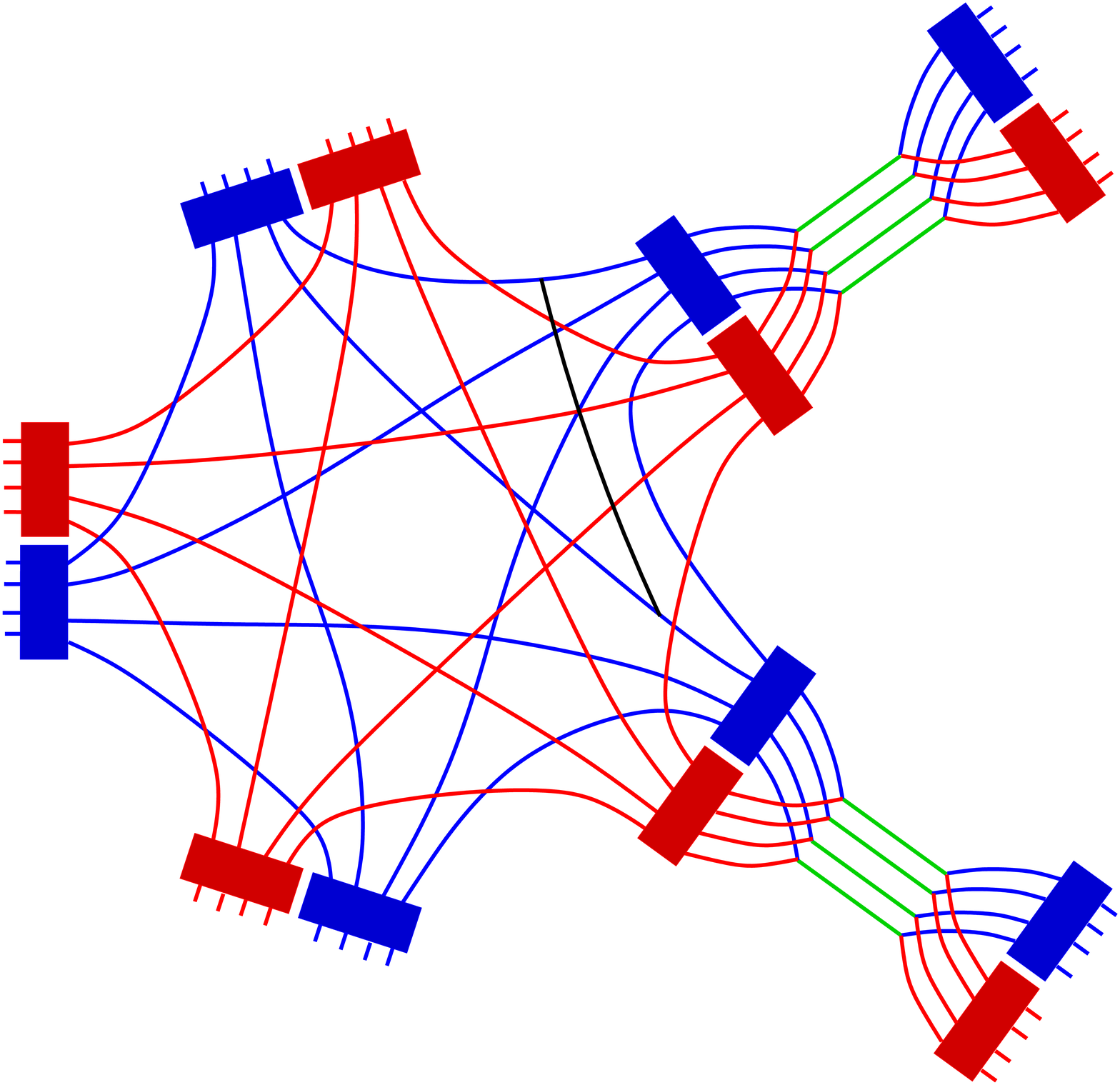}
\end{array}+\begin{array}{c}
\includegraphics[width=4cm]{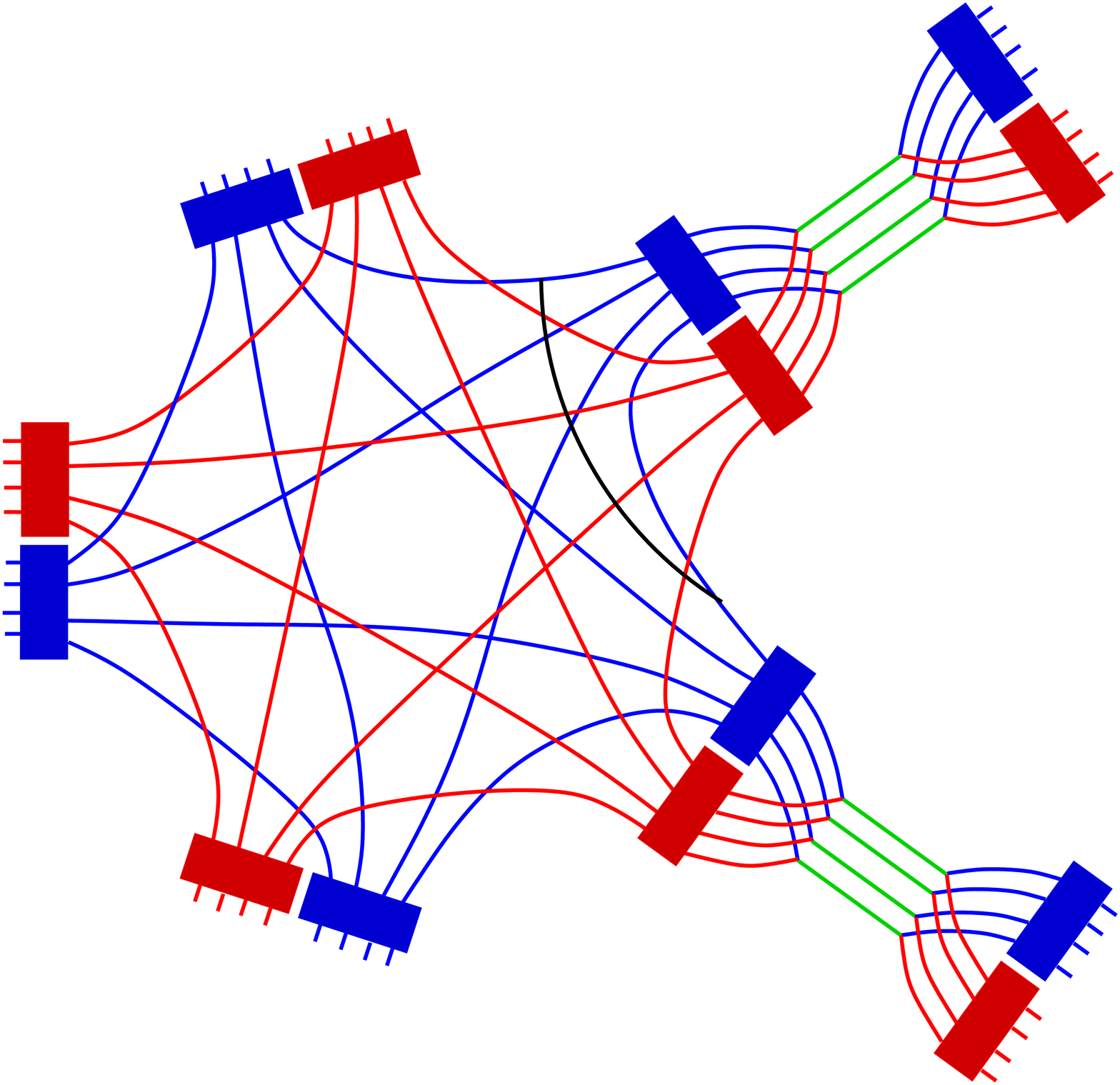}
\end{array}\right]+\n \\
&& \!\!\!\!\!\!\!\!\!\!\!\! +
(1-\gamma)^2\left[ \begin{array}{c}
\includegraphics[width=4cm]{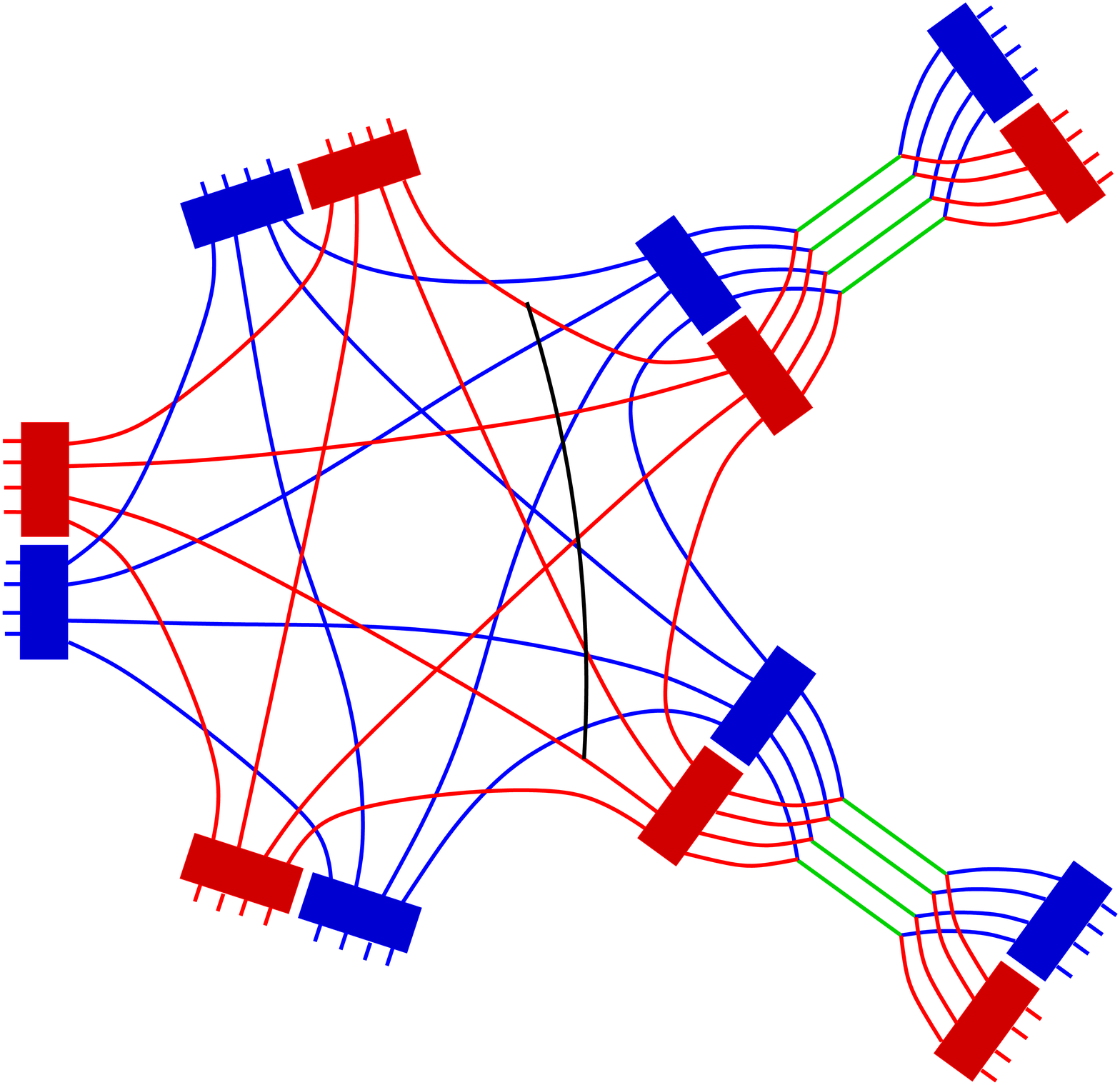}
\end{array}+\begin{array}{c}
\includegraphics[width=4cm]{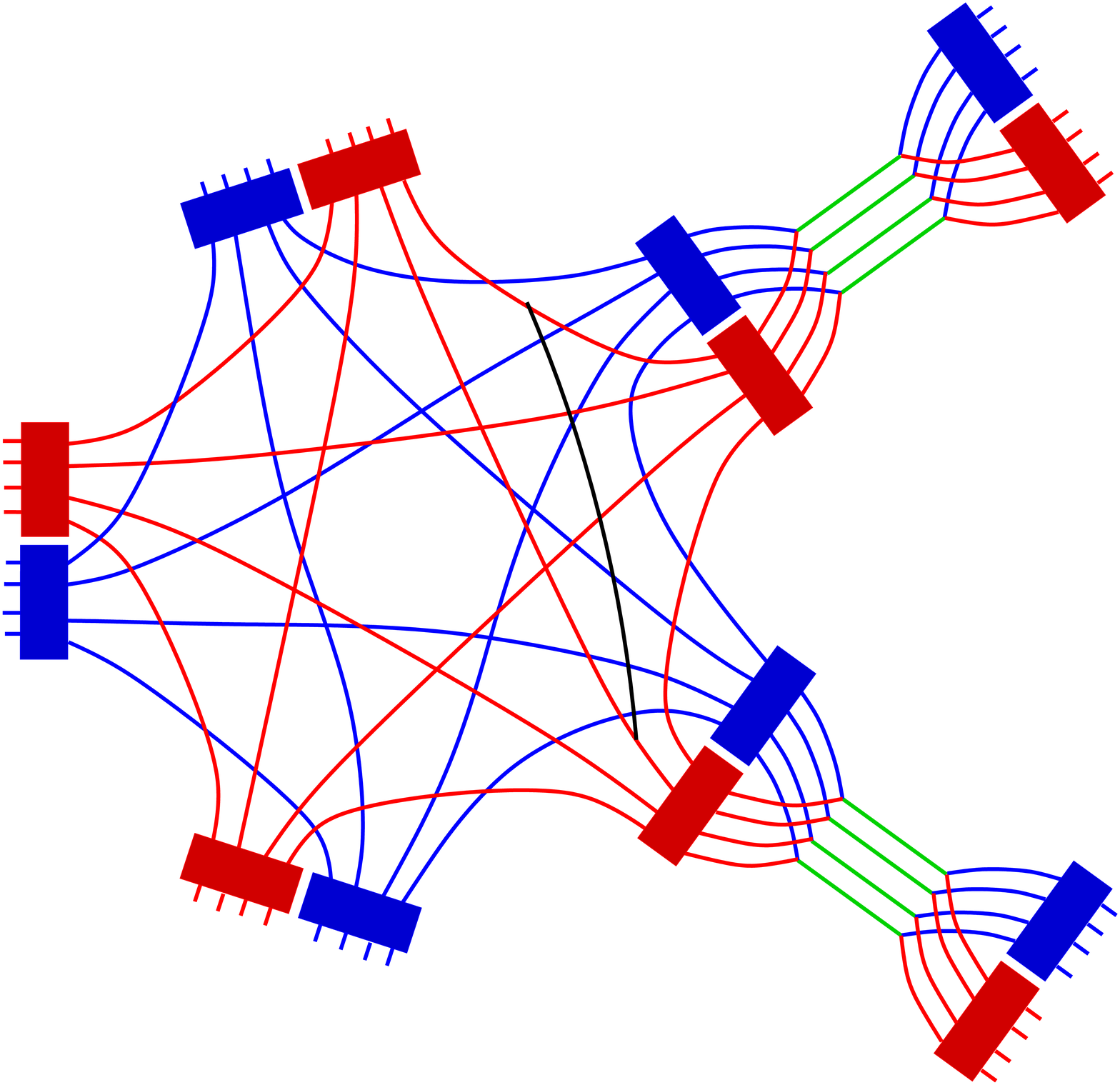}
\end{array}+\begin{array}{c}
\includegraphics[width=4cm]{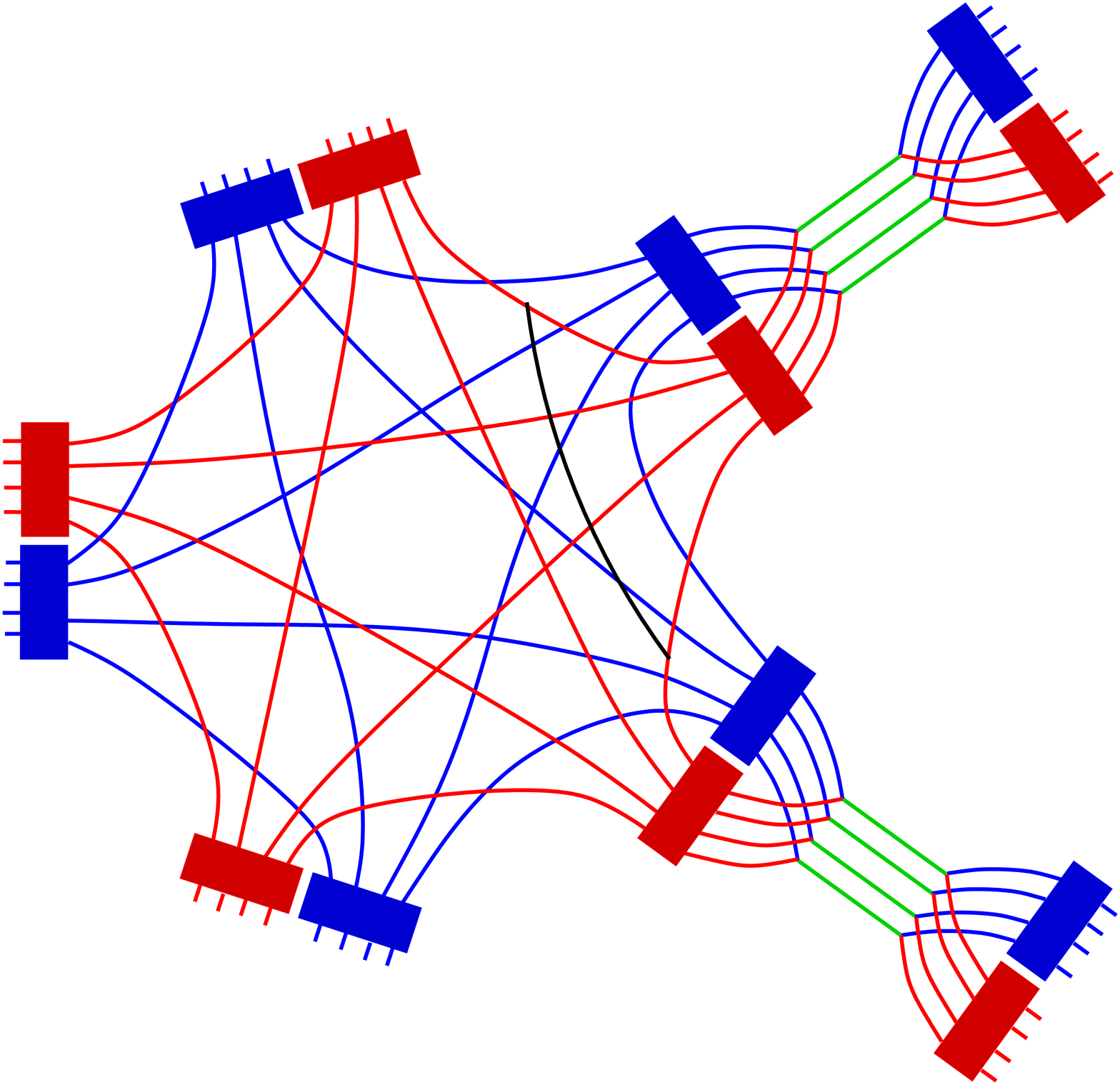}
\end{array}\right]=\n \\ 
&& \!\!\!\!\!\!\!\!\!\!\!\!
-(1+\gamma)^2 \begin{array}{c}
\includegraphics[width=4cm]{5-1}
\end{array}
+(1-\gamma)^2
\begin{array}{c}
\includegraphics[width=4cm]{6-1}
\end{array} +\sO_{sc},
\ea
where going from the first line to the second and third lines we have simply used (\ref{eprl-gauss}) on the bottom graspings on the right and left wires.
The last line results from the validity of (\ref{3s}): notice that the second terms in the second and third lines add up to $\sO_{sc}$ as well as the third terms in the second and third line. There is an overall minus sign which amounts for an orientation factor. It should be clear that we can apply the same procedure to arrive at any admissible pair. 

\subsection{Modifications of the EPRL model}\label{anofree}

\subsubsection{$P_{eprl}$ is not a projector}

Let us study in a bit more detail the object $P_{eprl}^e(j_1,\cdots, j_4)$. We see that it is made of two ingredients. The first one is the 
projection to the maximum  weight subspace $\sH_{j}$ for $\gamma>1$  in the decomposition of $\sH_{j^+,j^-}$  for $j^{\pm}=(1\pm\gamma)j/2$ ($j^{\pm}=(\gamma\pm 1)j/2$ for $\gamma>1$) 
in terms of irreducible representations of an arbitrarily chosen $SU(2)$ subgroup of $Spin(4)$. The second ingredient is to eliminate the dependence on the choice of subgroup by 
group averaging with respect to the full gauge group $Spin(4)$. This is diagramaticaly represented in (\ref{eprl-projection}). However $P_{eprl}^e(j_1,\cdots, j_4)$ is not a projector, namely
\be
P_{eprl}^e(j_1,\cdots, j_4)^2\not=P_{eprl}^e(j_1,\cdots, j_4).
\ee
Technically this follows from (\ref{regalo}) and the fact that  
 \be [P^e_{inv} (\rho_{1}\cdots \rho_4),( \sY_{j_1}\otimes\cdots\otimes\sY_{j_4} )]\not=0\ee 
i.e., the projection imposing the linear constraints (defined on the frame of a tetrahedrom or edge) and the $Spin(4)$ (or Lorentz) group 
averaging -- rendering the result gauge invariant -- do not commute. The fact the  $P_{eprl}^e(j_1,\cdots, j_4)$  is not a projection operator has important 
consequences in the mathematical structure of the model:
\begin{enumerate}
\item 
From (\ref{eprl-so4}) one can  immediately obtain the following expression for the EPRL amplitude
\be\label{eprl-p}
Z_{eprl}(\Delta)=\sum \limits_{ \rho_f \in \sK}  \ \prod\limits_{f \in \Delta^{\star}} {\rm d}_{|1-\gamma|\frac{j}{2}}{\rm d}_{(1+\gamma)\frac{j}{2}}
\prod \limits_{e} P_{eprl}^e(j_1,\cdots, j_4).
\ee 
This expression has the formal structure of expression (\ref{bf4}) for BF theory. The formal similarity however is broken by the fact that  $P_{eprl}^e(j_1,\cdots, j_4)$ is not a projection operator.
From the formal perspective is the possibility that the amplitudes  be defined in term of a network of projectors (as in BF theory) might provide an interesting structure that might be of relevance in the definition of a discretization independent model (see discussion in Part \ref{sci}). On the contrary, the failure of $P_{eprl}^e(j_1,\cdots, j_4)$ to be a projector may lead, in my opinion, to difficulties in the limit where the
complex $\Delta$ is refined: the increasing of the number of edges might produce either trivial or divergent amplitudes \footnote{This is obviously not clear from the form of
(\ref{eprl-p}). We are extrapolating the properties of $(P_{eprl}^e)^{N}$ for large $N$ to those of the amplitude (\ref{eprl-p}) in the large number of edges limit implied by the continuum limit.}. 

\item Another difficulty associated with $P_{eprl}^e(j_1,\cdots, j_4)^2\not=P_{eprl}^e(j_1,\cdots, j_4)$ is the failure of the amplitudes of the EPRL model, as defined here, to be consistent with
the abstract notion of spin foams as defined in \cite{baez7}. This is a point of crucial importance under current discussion in the community. The point is that the cellular decomposition $\Delta$
has no physical meaning and is to be interpreted as a subsidiary regulating structure to be removed when computing physical quantities. Spin foams (as defined in Section \ref{sec:intsf}) can fit in different ways on a given 
$\Delta$, yet any of these different embeddings represent the same physical process (like the same gravitational field in different coordinates). Consistency requires the spin foam amplitudes to be independent of the embedding, i.e., well defined on the equivalence classes of spin foams as defined by Baez in \cite{baez7} (the importance of these consistency requirements was emphasized in \cite{myo}). The amplitude (\ref{eprl-p}) fails this requirement due to  $P_{eprl}^e(j_1,\cdots, j_4)^2\not=P_{eprl}^e(j_1,\cdots, j_4)$. 

\end{enumerate}
\subsubsection{The Warsaw proposal}

The above difficulties have a simple solution in the Riemannian case. As proposed in \cite{Bahr:2010bs, Kaminski:2009cc} one can obtain a consistent modification of the EPRL model 
by replacing $P^e_{eprl}$ in (\ref{eprl-p}) by a genuine projector $P^{e}_{w}$, graphically
\ba  && \n P_{eprl}^{e}(j_1\cdots j_4)=\!\!\!\!\!
\begin{array}{c}
\includegraphics[height=3cm]{proj-eprl}
\end{array}\!\!\!\!\! =\sum_{\alpha}\!\!\!\!\! \begin{array}{c}\psfrag{a}{$\! \van \alpha$}\psfrag{b}{$\!\van \alpha$}
\includegraphics[height=3cm]{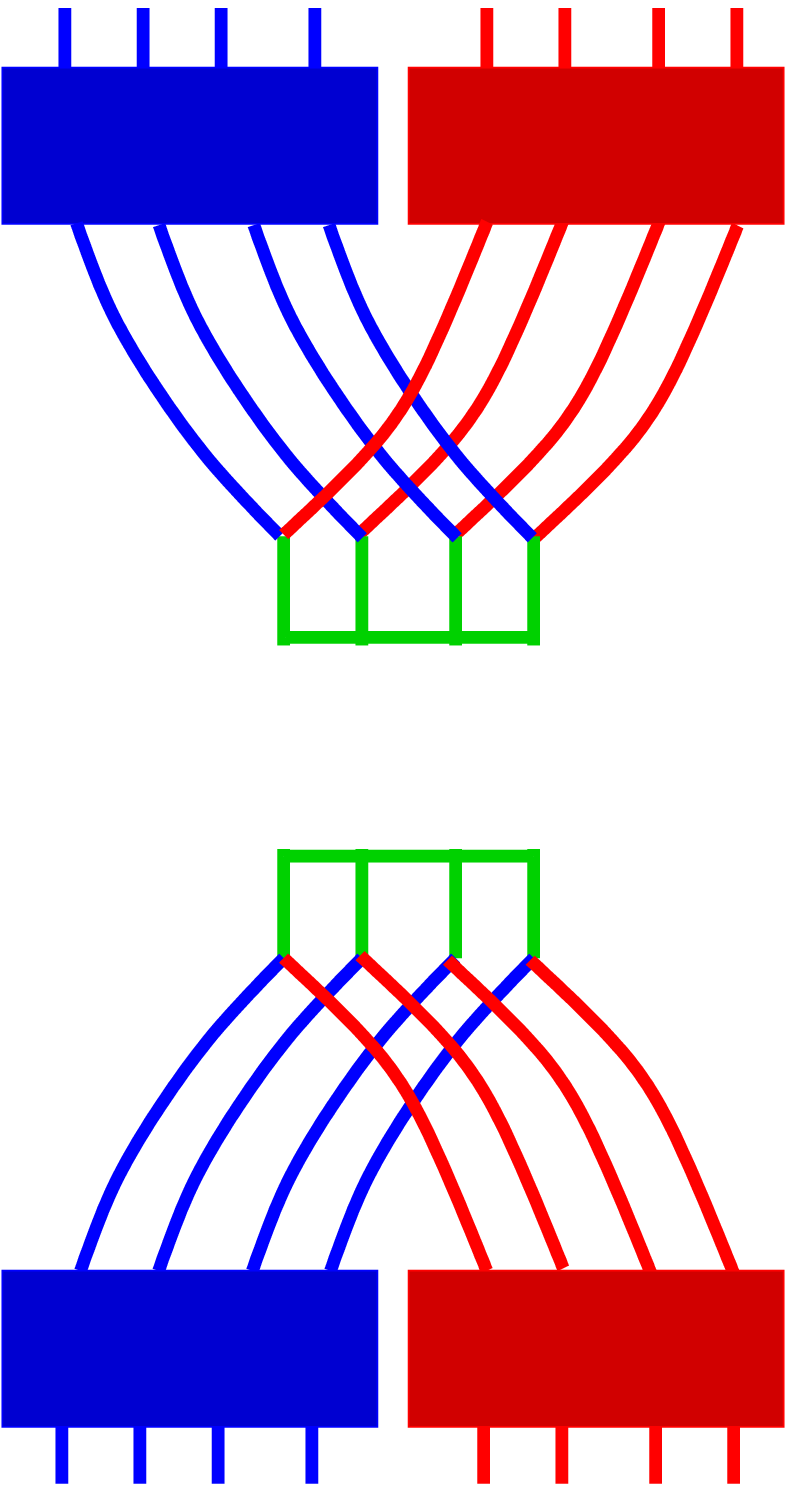}
\end{array}\!\!  \\ &&  \ \ \ \ \ \ \ \ \ \ \ \ \ \ \ \ \ \ \ \ \ \ \  \begin{array}{c} \includegraphics[width=1cm]{arrowbig}\end{array}
\ \   P_{w}^{e}(j_1\cdots j_4)=\sum_{\alpha\beta}   g^e_{\alpha \beta}  
\!\!\!\!\! \begin{array}{c}\psfrag{a}{$\! \van \alpha$}\psfrag{b}{$\! \van \beta$}
\includegraphics[height=3cm]{warsaw-eprl}
\end{array},
\ea
where
\be   \n g^e_{\alpha \beta}=\left(\sum_{\iota^+\iota^-} f^{\alpha}_{\iota^+\iota^-} f^{\beta}_{\iota^+\iota^-}\right)^{-1}  
\!\!\!\!\! ={\rm Inverse}\left( \begin{array}{c}\psfrag{a}{$\!\!\!\! \van \alpha$}\psfrag{b}{$\van \beta$}
\includegraphics[height=1.5cm]{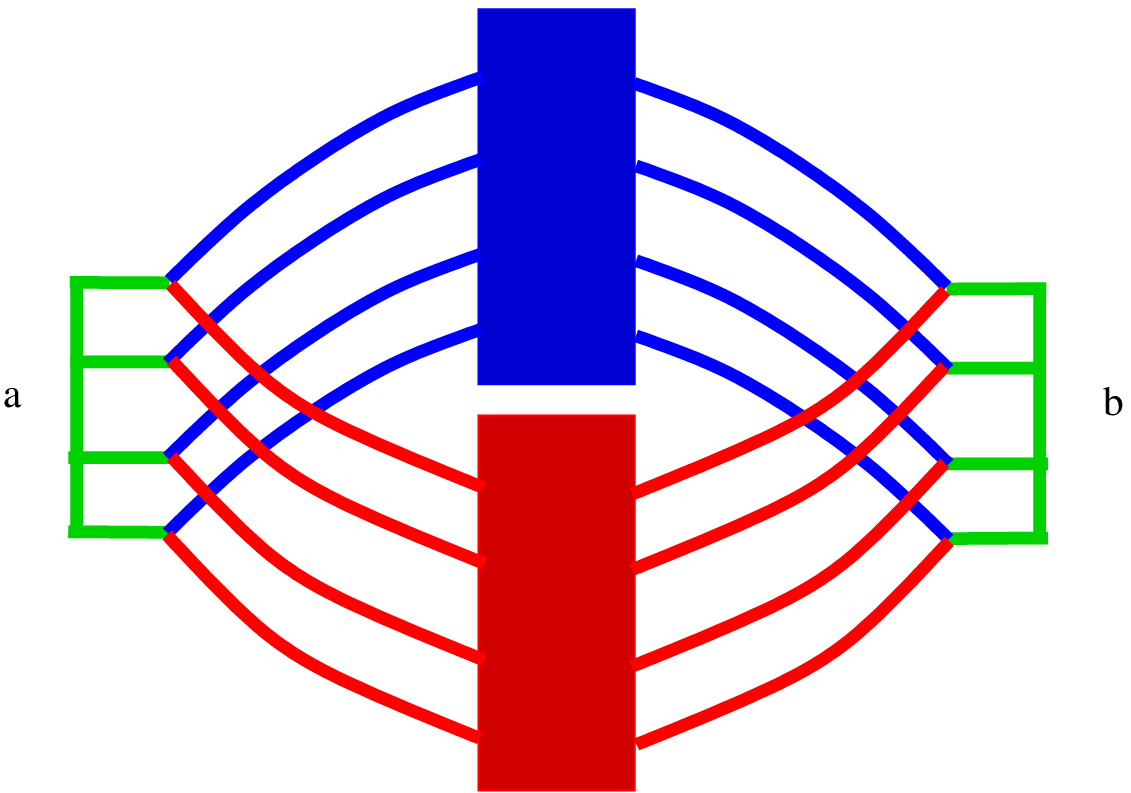}
\end{array}\right).
\ee 
It is easy to check that by construction \be
(P_{w}^{e}(j_1\cdots j_4))^2=P_{w}^{e}(j_1\cdots j_4).
\ee
The variant of the EPRL model proposed in \cite{Bahr:2010bs, Kaminski:2009cc} takes then the form
\ba
\label{eprl-warsaw}  Z_{eprl}(\Delta)&=&\sum \limits_{ j_f}  \ \prod\limits_{f \in \Delta^{\star}} {\rm d}_{|1-\gamma|\frac{j}{2}}{\rm d}_{(1+\gamma)\frac{j}{2}}
\prod \limits_{e} P_{w}^e(j_1,\cdots, j_4)\\
&=&\sum \limits_{ j_f }\sum \limits_{ \iota_{ev} }  \ \prod\limits_{f \in \Delta^{\star}} {\rm d}_{|1-\gamma|\frac{j}{2}}{\rm d}_{(1+\gamma)\frac{j}{2}} \prod_{e\in \Delta^{\star}} g^e_{\iota^e_{v_s}\iota^e_{v_t}} \prod_{v\in \Delta^{\star}}\begin{array}{c}\psfrag{a}{$\van \iota^1_{v}$}
\psfrag{b}{$\van  \iota^2_{v}$}
\psfrag{c}{$\van  \iota^3_{v}$}
\psfrag{d}{$\van  \iota^4_{v}$}
\psfrag{e}{$\van  \iota^5_{v}$}
\includegraphics[width=5cm]{eprl-vertex}
\end{array}. \n \ea
Thus in the modified EPRL model edges $e\in \Delta^{\star}$ are assigned pairs of intertwiner quantum numbers $\iota^e_{v_s}$ and $\iota^e_{v_t}$  
and an edge amplitude given by the matrix elements $g^e_{\iota^e_{v_s},\iota^{e}_{v_t}}$  (where $v_s$ and $v_t$ stand for the source and target vertices of the given oriented edge).  
The fact that edges are not assigned a single quantum number is not really significative; one could go to a basis of normalized eigenstates of $P^e_{w}$ and rewrite
the modified model above as a spin foam model where edges are assigned a single (basis element) quantum number.  As the nature of such basis and the quantum 
geometric interpretation of its elements is not clear at this stage, it seems simpler to represent the amplitudes of the  modified model in the above form.

The advantages of the modified model are important,; however, a generalization of the above modification of the EPRL model in the Lorentzian case is still lacking.
Notice that this modification does not interfere with the results on the semiclassical limit (to leading order) as reviewed in Section \ref{semiclas}. The reason is that the 
matrix elements $g^e_{\alpha\beta}\to \delta_{\alpha\beta}$ in that limit \cite{Alesci:2008un}.

\subsection{The coherent states representation}
\label{7-6}

We have written the amplitude defining the EPRL model by constraining the 
state sum of BF theory.  For semiclassical studies that we will review in 
Section \ref{semiclas} it is convenient to express the EPRL amplitude in terms of the 
coherent states basis. The importance of coherent states in spin foam models
was put forward in \cite{Livine:2007vk} and explicitly used to re-derive the EPRL model in \cite{Livine:2007ya}. The coherent state technology was used by Freidel and Krasnov
in \cite{Freidel:2007py} to introduce a new kind of spin foam models for gravity: the FK models.
In some cases the FK model is equivalent to the EPRL model; we will review this in detail in 
Section \ref{fk}. 

The coherent state representation of the EPRL model is obtained by replacing (\ref{patacul}) in each of the intermediate 
$SU(2)$ (green) wires in the expression (\ref{eprl-so4}) of the EPRL amplitudes, namely 
\ba && \begin{array}{c}
\psfrag{w}{$$}
\psfrag{a}{$n_{\va 1}$}
\psfrag{A}{$\va n_1$}
\psfrag{b}{$\va n_2$}
\psfrag{B}{$\va n_2$}
\psfrag{c}{$\va n_3$}
\psfrag{C}{$\va n_3$}
\psfrag{d}{$\va n_4$}
\psfrag{D}{$\va n_4$}
\includegraphics[width=6cm]{eprl3}\end{array}=\n \\ && \ \ \ \ \ \ \ \ \ \ \ \ \ \ \ \ \ \ \ \ \ \ \ \ \ \ \ =\int\limits_{[S^2]^4} \prod\limits_{\va I=1}^{\va 4}{\rm d}_{j_I}  dn_{\va I}  \begin{array}{c}\psfrag{w}{$$}
\psfrag{a}{$\va n_1$}
\psfrag{A}{$\va n_1$}
\psfrag{b}{$\va n_2$}
\psfrag{B}{$\va n_2$}
\psfrag{c}{$\va n_3$}
\psfrag{C}{$\va n_3$}
\psfrag{d}{$\va n_4$}
\psfrag{D}{$\va n_4$}
\includegraphics[width=6cm]{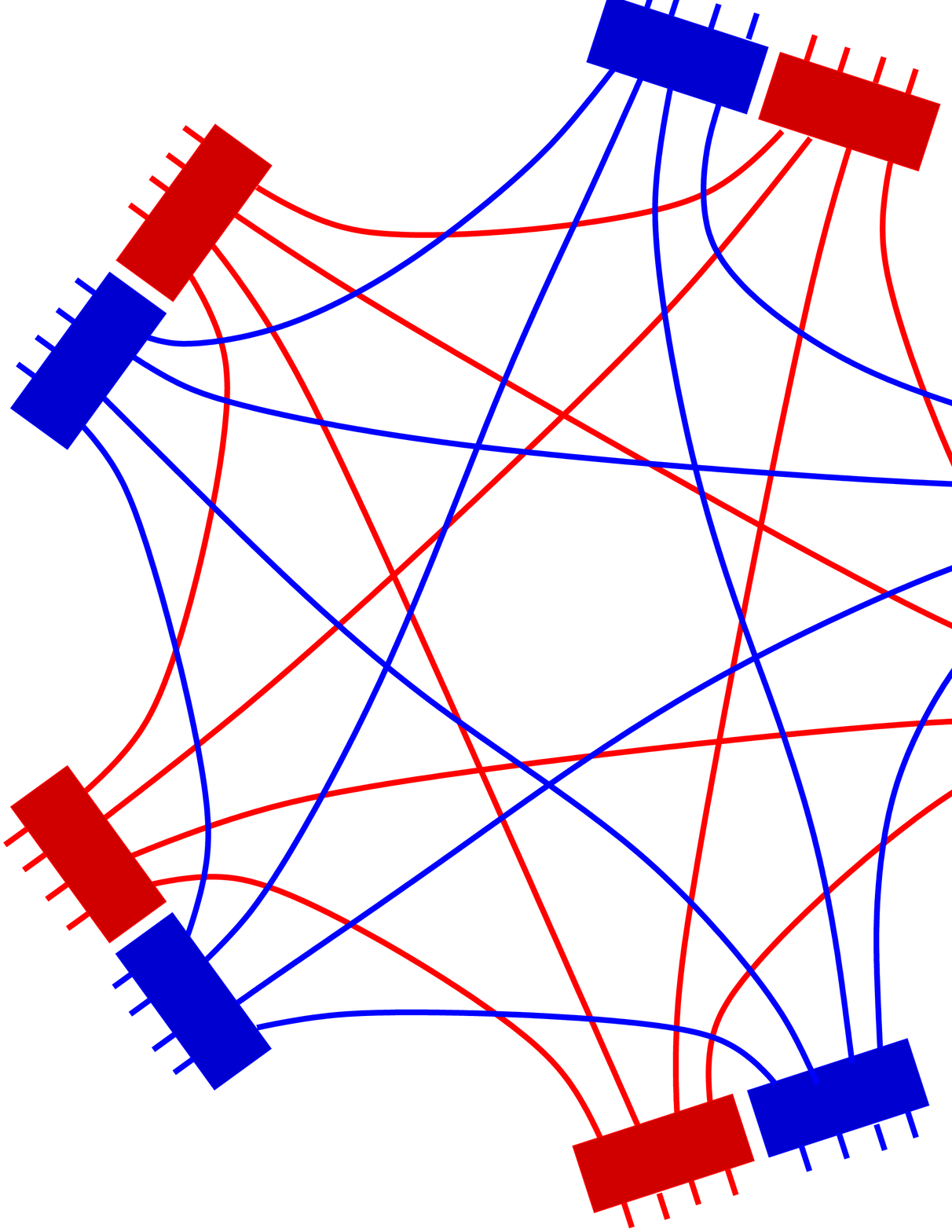}\end{array}\label{tresf}
\ea
\subsubsection*{The case $\gamma<1$}
In this case the coherent state 
property (\ref{exp}) implies 
\be\begin{array}{c}
\psfrag{w}{$=\int\limits_{[S^3]^4} \prod\limits_{\va I=1}^{\va 4} dn_{\va I}$}
\psfrag{a}{$\va n_1$}
\psfrag{A}{$\va n_1$}
\psfrag{b}{$\va n_2$}
\psfrag{B}{$\va n_2$}
\psfrag{c}{$\va n_3$}
\psfrag{C}{$\va n_3$}
\psfrag{d}{$\va n_4$}
\psfrag{D}{$\va n_4$}
\psfrag{x}{$$}
\includegraphics[width=6.5cm]{eprl4}\end{array}= \begin{array}{c}\psfrag{w}{$=\int\limits_{[S^3]^4} \prod\limits_{\va I=1}^{\va 4} dn_{\va I}$}
\psfrag{a}{$\va n_1$}
\psfrag{A}{$\va n_1$}
\psfrag{b}{$\va n_2$}
\psfrag{B}{$\va n_2$}
\psfrag{c}{$\va n_3$}
\psfrag{C}{$\va n_3$}
\psfrag{d}{$\va n_4$}
\psfrag{D}{$\va n_4$}
\psfrag{x}{$$}
\includegraphics[width=6.5cm]{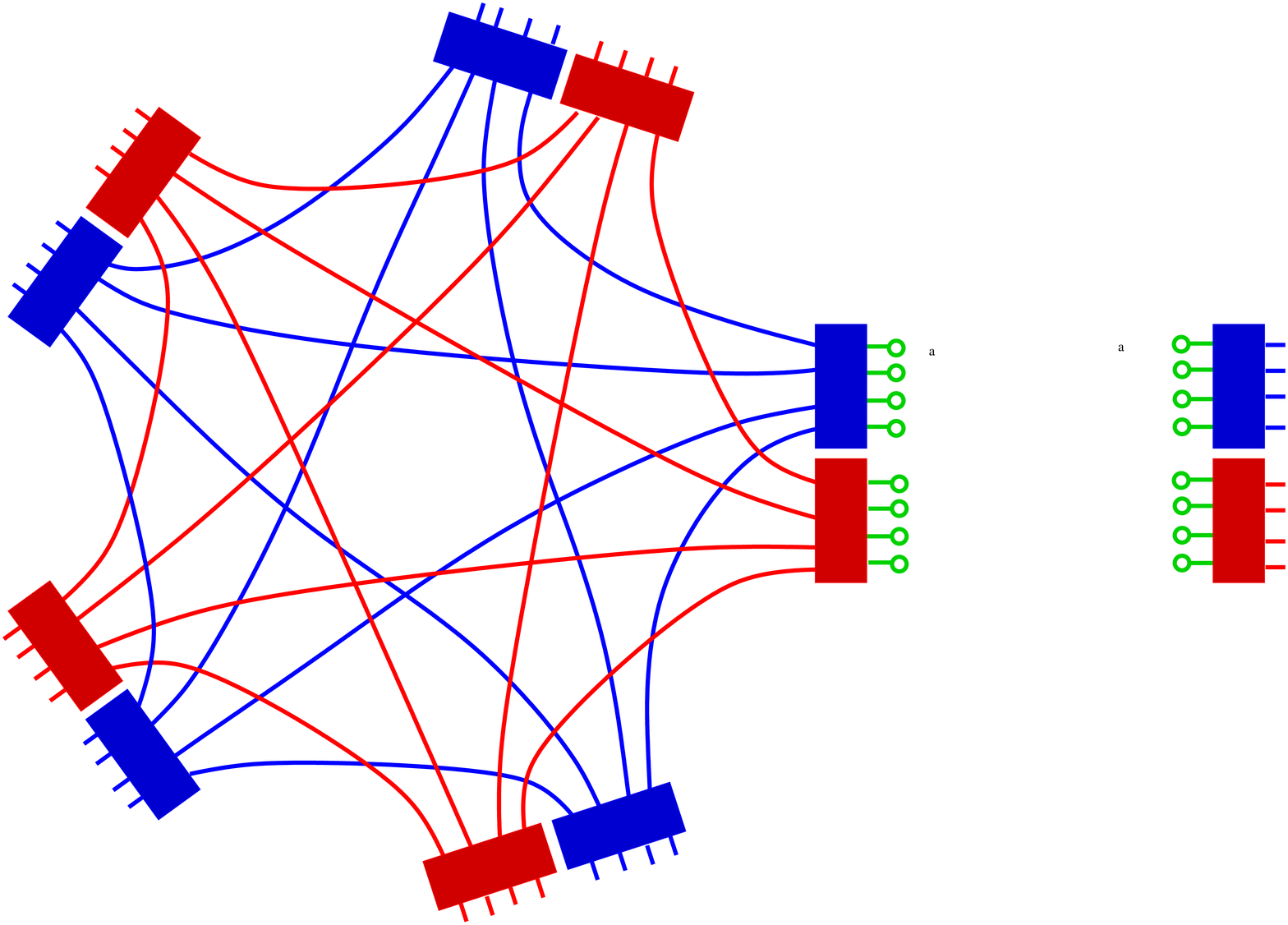}\end{array}
\label{cuatrofy},\ee
where we used in the last line the fact that for $\gamma<1$ the representations $j$ of the subgroup $SU(2)\in Spin(4)$  are maximum weight, i.e., $j=j^++j^-$.
Doing this at each edge we get
\ba\label{eprl-cohe}
&& Z^{E}_{eprl}(\Delta)=\sum \limits_{j_f}  \ \prod\limits_{f \in \Delta^{\star}} {\rm d}_{j_f^{-}}{\rm d}_{j_f^{+}} \n \\ && \int \prod_{e\in  \in \Delta^{\star}} {\rm d}_{j_{ef}} dn_{ef}
\begin{array}{c}\psfrag{w}{$$}
\psfrag{a}{$\va n_{ 1}$}
\psfrag{A}{$\va n_1$}
\psfrag{b}{$\va n_2$}
\psfrag{B}{$\va n_2$}
\psfrag{c}{$\va n_3$}
\psfrag{C}{$\va n_3$}
\psfrag{d}{$\va n_4$}
\psfrag{D}{$\va n_4$}
\includegraphics[width=7cm]{coherent-v}
\end{array},
\ea
where we have explicitly written the $n\in S^2$ integration variables on a single cable. The expression above is very similar to the coherent states representation of $Spin(4)$ BF theory given in Equation (\ref{bf-cohe}). In fact one would get the above
expression if one would start form the expression (\ref{bf-cohe}) and would set $n^{+}_{ef}=n^{-}_{ef}=n_{ef}$ while dropping for example all the sphere integrations corresponding to the $n^{+}_{ef}$ (or equivalently $n^{-}_{ef}$).  
Moreover, by construction the coherent states participating in the previous amplitude satisfy the linear constraints (\ref{jojo}) in expectation values, namely
\ba\n 
\langle j, n_{ef} | D^i_f| j, n_{ef}\rangle &=&\\ &=& 
\langle j, n_{ef} | (1-\gamma) J^{+i}_ f+ (1+\gamma) J^{-i}_f  |j, n_{ef}\rangle=0.
\ea
Thus the coherent states participating in the above representation of the EPRL amplitudes solve the linear simplicity constraints in the usual semiclassical sense.
The same manipulations leading to (\ref{discrete-action}) in Section \ref{BF} lead to a discrete effective action for the EPRL model, namely
\ba
\label{eprl-cohe}Z^{\va \gamma<1}_{eprl}=\sum \limits_{ j_f }  \ \prod\limits_{f \in \Delta^{\star}} {\rm d}_{(1-\gamma)\frac{j_f}{2}}{\rm d}_{(1+\gamma)\frac{j_f}{2}} \int \prod_{e\in  \Delta^{\star}} {\rm d}_{j_{ef}} dn_{ef} dg^{-}_{ev}dg^{+}_{ev}
\ \exp{(S^{\va \gamma<1}_{j^{\pm},\bn}[g^{\pm}])}, \ea
where the discrete action \be\label{discrete-action}
S^{\va \gamma<1}_{j^{\pm},\bn}[g^{\pm}]=\sum_{v\in\Delta^{\star}} (S^v_{(1-\gamma)\frac{j_f}{2},\bn}[g^{-}]+S^v_{(1+\gamma)\frac{j_f}{2},\bn}[g^{+}])\ee with 
\be
\label{v-action}
S^v_{j,\bn}[g] = \sum\limits^{5}_{a < b=1} 2j_{ab} \ln \, \la n_{ab}| g^{-1}_a g_b| \, n_{ba} \ra,
\ee
and the indices $a,b$ label the five edges of a given vertex. The previous expression is exactly equal to the form (\ref{coloring4}) of the BF amplitude. In the case of the gravity models presented here,
the coherent state path integral representation (analogous to (\ref{cspig})) will be the basic tool for the study of the semiclassical limit of the models and the relationship with Regge discrete formulation of general relativity. 

\subsubsection*{The case $\gamma>1$}

The case $\gamma>1$ is more complicated \cite{Barrett:2009gg}. The reason is that the step (\ref{cuatrofy}) directly leading  to the discrete action in the previous case is no longer valid as the
representations of the subgroup $SU(2)\in Spin(4)$ are now minimum instead of maximum weight. However, the representations $j^{+}=j^{-}+j$ are maximum weight. We can therefore insert 
coherent states resolution of the identity on the right representations and get:
\ba && \n \begin{array}{c}
\psfrag{w}{$=\int\limits_{[S^3]^4} \prod\limits_{\va I=1}^{\va 4} dn_{\va I}$}
\psfrag{a}{$\va n_1$}
\psfrag{A}{$\va n_1$}
\psfrag{b}{$\va n_2$}
\psfrag{B}{$\va n_2$}
\psfrag{c}{$\va n_3$}
\psfrag{C}{$\va n_3$}
\psfrag{d}{$\va n_4$}
\psfrag{D}{$\va n_4$}
\psfrag{x}{$$}
\includegraphics[width=2.7cm]{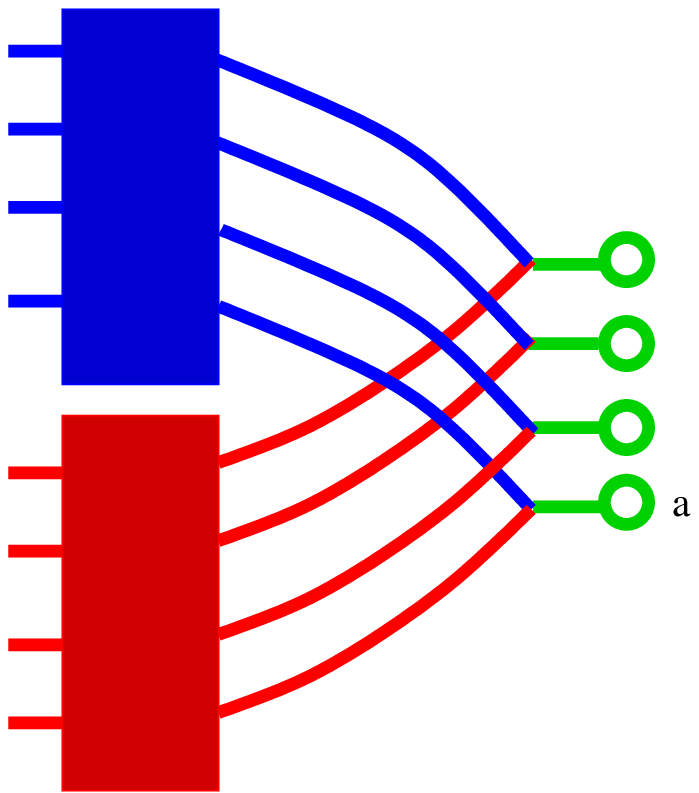}\end{array}= \int\limits_{[S^3]^4} \prod \limits_{\va I=1}^{\va 4} {\rm d}_{(1+\gamma)\frac{j_{I}}{2}}\ dm_{\va I} \begin{array}{c}
\psfrag{w}{$=\int\limits_{[S^3]^4} \prod\limits_{\va I=1}^{\va 4} dn_{\va I}$}
\psfrag{m1}{$\va m_1$}
\psfrag{m2}{$\va m_2$}
\psfrag{m3}{$\va m_3$}
\psfrag{m4}{$\va m_4$}
\psfrag{a}{$\va n_1$}
\psfrag{A}{$\va n_1$}
\psfrag{b}{$\va n_2$}
\psfrag{B}{$\va n_2$}
\psfrag{c}{$\va n_3$}
\psfrag{C}{$\va n_3$}
\psfrag{d}{$\va n_4$}
\psfrag{D}{$\va n_4$}
\psfrag{x}{$$}
\includegraphics[width=2.7cm]{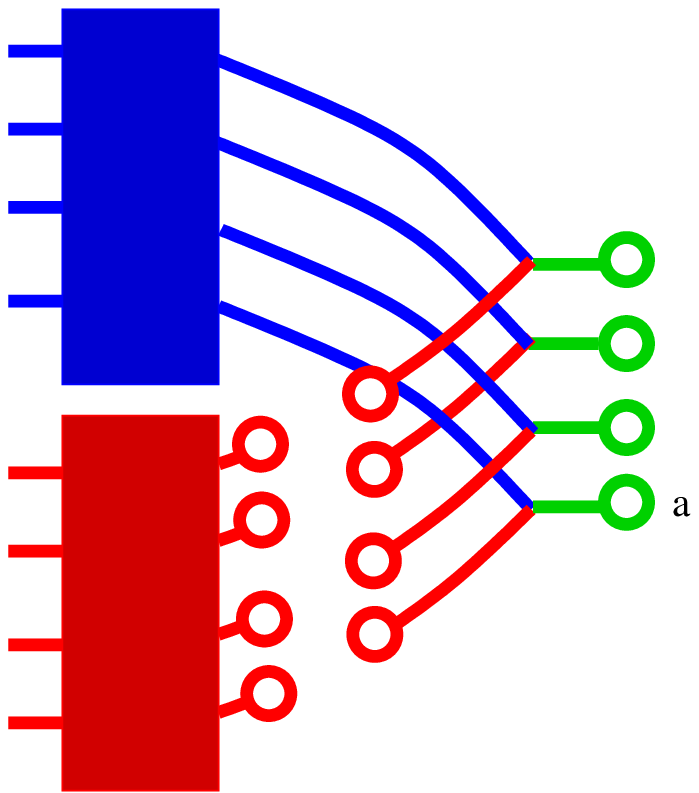}\end{array}=\\ && =\int\limits_{[S^3]^4} \prod \limits_{\va I=1}^{\va 4} {\rm d}_{(1+\gamma)\frac{j_{I}}{2}}\ dm_{\va I}  \begin{array}{c}
\psfrag{w}{$=\int\limits_{[S^3]^4} \prod\limits_{\va I=1}^{\va 4} dn_{\va I}$}
\psfrag{m1}{$\va m_1$}
\psfrag{m2}{$\va m_2$}
\psfrag{m3}{$\va m_3$}
\psfrag{m4}{$\va m_4$}
\psfrag{a}{$\va n_1$}
\psfrag{A}{$\va n_1$}
\psfrag{b}{$\va n_2$}
\psfrag{B}{$\va n_2$}
\psfrag{c}{$\va n_3$}
\psfrag{C}{$\va n_3$}
\psfrag{d}{$\va n_4$}
\psfrag{D}{$\va n_4$}
\psfrag{x}{$$}
\includegraphics[width=2.7cm]{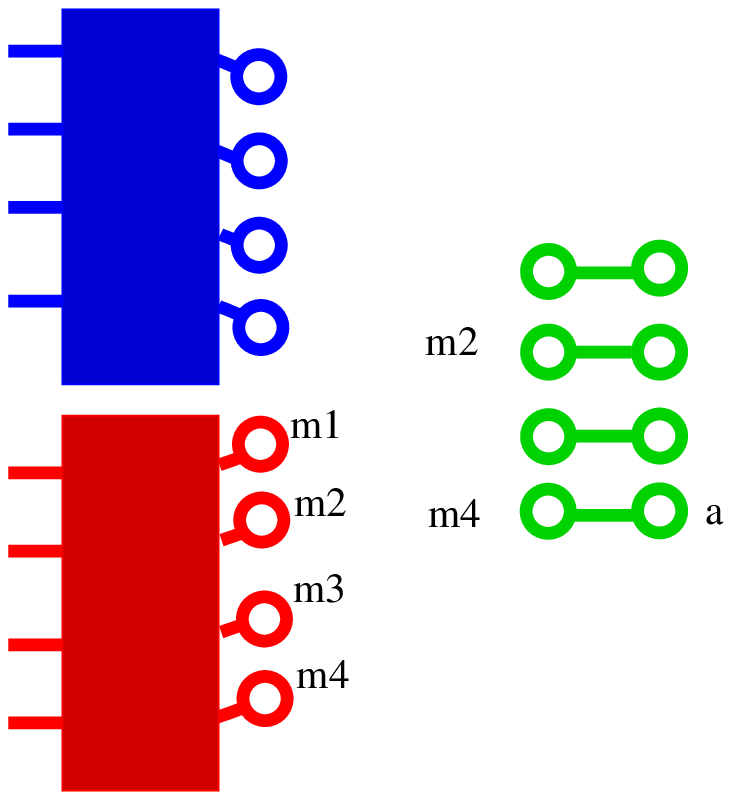}\end{array}
\label{cuat},\ea
where we are representing the relevant part of the diagram appearing in equation (\ref{tresf}). In the last line we have used that $j^+=j+j^-$ (i.e. maximum weight), and the graphical notation
$\begin{array}{c}\psfrag{a}{$\! \!\! \, m$}
\psfrag{b}{$n$}
\includegraphics[width=1cm]{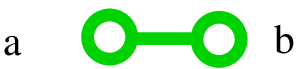}
\end{array}\equiv \langle m |n\rangle$ as it follows from our previous conventions.
With all this, one gets
\ba
\label{eprl-cohe-g}&& Z^{\va \gamma> 1}_{eprl}= \\ \n &&\sum \limits_{ j_f }  \ \prod\limits_{f \in \Delta^{\star}} {\rm d}_{(1-\gamma)\frac{j_f}{2}}{\rm d}_{(1+\gamma)\frac{j_f}{2}} \int \prod_{e\in  \Delta^{\star}} {\rm d}_{j_{ef}} {\rm d}_{(1+\gamma)\frac{j_{ef}}{2}} dn_{ef} dm_{ef} dg^{-}_{ev}dg^{+}_{ev}
\ \exp{(S^{\va \gamma>1}_{j^{\pm},\bn,\bm} [g^{\pm}])}, \ea
where the discrete action \be\label{discrete-action-g<}
S^{\va \gamma>1}_{j^{\pm},\bn,\bm}[g^{\pm}]=\sum_{v\in\Delta^{\star}} S^v_{j^{\pm},\bn,\bm}[g^{\pm}]\ee with
\ba\label{gamma-g}
S^v_{j^{\pm},\bn,\bm}[g^{\pm}]=\sum_{1\le a<b\le 5} j_{ab} (1+\gamma)\log( \langle m_{ab} |g^{+}_{ab}|m_{ba} \rangle)+j_{ab} (\gamma-1) \log(\langle m_{ab} |g^{-}_{ab}|m_{ba}\rangle)+\n \\ 
\ \ \ \ \ \ \ \ \ \ \ \ \ \ \ \ \ \ \ \ \ +2j_{ab} \left(\log(\langle n_{ab} |m_{ab} \rangle)+\log(\langle m_{ba} |n_{ba} \rangle)\right).
\ea

\subsection{Some additional remarks}

It is important to point out that the commutation relations of basic fields---reflecting the simple algebraic structure of $spin(4)$---used here is the one induced by the canonical analysis 
of BF theory presented previously. The presence of constraints generally modifies canonical commutation relations in particular in the presence of second class constraints. 
For some investigation of the issue in the context of the EPRL and FK  models see \cite{Alexandrov:2010pg}. 
In \cite{Alexandrov:2008da} it is pointed out that the presence of secondary constraints in the canonical analysis of Plebanski action should translated in additional constraints
in the holonomies of the spin foam models here considered (see also \cite{Alexandrov:2007pq}). A possible view is  that the simplicity constraints are here imposed  {\em for all times} and thus
secondary constraints should be imposed automatically. 

There are alternative derivations of the models presented in the previous sections. In particular 
one can derive them from a strict Lagrangean approach of Plebanski's action. Such viewpoint is taken in 
\cite{Bonzom:2009hw, Bonzom:2009wm, Bonzom:2008ru}.  
The path integral formulation of Plebansky theory using commuting $B $-fields was studied in \cite{Han:2010rb}, where it is shown that only in the appropriate semiclassical limit
the amplitudes coincide with the ones presented in the previous sections (this is just another indication that the construction of the models have a certain semiclassical input; see below). 
The spin foam quantization of the  Holst formulation of gravity via cubulations was investigated in 
\cite{Baratin:2008du}.
The simplicity constraints can also be studied from the perspective of the $U(N)$ formulation of quantum geometry 
 \cite{Dupuis:2010iq, Dupuis:2011fz, Dupuis:2011dh} which also has been explored in the Lorentzian sector \cite{Dupuis:2011wy}. Such $U(N)$ treatment is related to previous work  which has been extended to a completely new perspective on quantum geometry 
 with possible advantageous features \cite{Borja:2010rc, Livine:2011gp}. For additional discussion on the simplicity constraints see \cite{Dittrich:2010ey}.

\section{The Lorentzian EPRL Model: Representation Theory and Simplicity Constraints} 
\label{eprl-l}

In this section we introduce some elements of $SL(2,\C)$
representation theory and show how the so called linear simplicity
constraints are solved in the Lorentzian EPRL
model~\cite{Pereira:2007nh, Engle:2008ev, Engle:2007wy}.

\subsection{Representation theory of $SL(2,\C)$ and the canonical basis}

Unitary irreducible representations $\sH_{\rho,k}$ of $SL(2,\C)$ are infinite dimensional and are labelled by 
a positive real number $p \in \R^+$  and a half-integer $k\in \N/2$. The  two Casimirs are $C_1=\frac{1}{2}J_{IJ}J^{IJ}=L^2-K^2$ and $C_2=\frac{1}{2}{}^\star \!J_{IJ}J^{IJ}=K\cdot L$ where $L^i$ are the generators of an arbitrary rotation subgroup and
$K^i$ are the generators of the corresponding boosts. 
The Casimirs act on $ |p,k\rangle\in \sH_{p,k}$ as follows 
\ba\label{casiL}
&& C_1 |p,k\rangle=(k^2-p^2-1)\, |p,k\rangle\n\\
&& C_2  |p,k\rangle=2pk\,  |p,k\rangle. 
\ea
For detail on the representation theory of $SL(2,\C)$ see \cite{ruhl, gelfand, gelfand2}.
As in the Riemannian
 case, the definition of the EPRL model  requires the introduction of an (arbitrary)
 subgroup $SU(2)\subset SL(2,\C)$. This subgroup correspond to the internal gauge group 
 of the gravitational phase space in connection variables in the time gauge.  In the quantum theory, the representation theory of this $SU(2)$ subgroup will be hence important. This importance will soon emerge as apparent from  the imposition of the constraints that define the EPRL.  The link between the unitary representations of $SL(2,\C)$ and those of $SU(2)$ is given by  the decomposition 
 \be\label{spin4su2}
 \sH_{p,k}=\bigoplus \limits_{j=k}^{\infty} \sH_{j}.\ee
 As the unitary irreducible representations of the subgroup $SU(2)\in SL(2,\C)$ are essential in understanding the link of the EPRL model and the operator canonical formulation of LQG it will be convenient to express the action of the generators of the Lie algebra $sl(2,\C)$ in a basis adapted to the above equation. In order to do this we first notice that the Lie algebra (\ref{lieso4}) can be equivalently characterized in terms of the generators of the rotation subgroup $L^i$ and the remaining {\em boost} generators $K^i$ as follows
\ba&& \n
[L_3,L_{\pm}]=\pm \ L_{\pm} \ \ \ \ \ [L_+,L_{-}]=2\ L_{3} \\
&& \n
[L_+,K_{+}]= [L_-,K_{-}]=[L_3,K_{3}]=0 \\
&& \n
[K_3,L_{\pm}]=\pm \ K_{\pm} \ \ \ \ \ [L_\pm,K_{\mp}]=\pm 2\ K_{3}\ \ \ \ \ [L_3,K_{\pm}]=\pm \ K_{\pm} \\
&& [K_3,K_{\pm}]=\mp \ L_{\pm} \ \ \ \ \ [K_+,K_{-}]= -2\ L_{3},
\ea
where $K_{\pm}=K^1 \pm i K^2$ and $L_{\pm}=L^1 {\pm}i L^2$ respectively.
The action of the previous generators in the basis $ |p,k,j ,m\rangle$ can be shown to be \cite{gelfand2}
\ba
&& L^3 |p,k,j,m\rangle = m |p,k,j,m\rangle, \nonumber \\
&& L^+ |p,k,j,m\rangle = \sqrt{(j+m+1)(j-m)} |p,k,j,m+1\rangle, \nonumber \\
&&  L^- |p,k,j,m\rangle = \sqrt{(j+m)(j-m+1)} |p,k,j,m-1\rangle, \nonumber \\
&&  K^3 |p,k,j,m\rangle =  \alpha_j\sqrt{j^2-m^2} |p,k,j-1,m\rangle+ \gamma_j m |p,k,j,m\rangle 
-\alpha_{j+1}\sqrt{(j+1)^2-m^2} |p,k,j+1,m\rangle,\nonumber \\
&&  K^+ |p,k,j,m\rangle = \alpha_j\sqrt{(j-m)(j-m-1)}
|p,k,j-1,m+1\rangle \nonumber \\
&& + \gamma_j\sqrt{(j-m)(j+m+1)}|p,k,j,m+1\rangle \nonumber
\\ && +\alpha_{j+1}\sqrt{(j+m+1)(j+m+2)} |p,k,j+1,m+1\rangle,\nonumber \\
&& K^- |p,k,j,m\rangle = -\alpha_j\sqrt{(j+m)(j+m-1)}
|p,k,j-1,m-1\rangle
\nonumber \\ && + \gamma_j\sqrt{(j+m)(j-m+1)} |p,k,j,m-1\rangle \nonumber \\
&& -\alpha_{j+1}\sqrt{(j-m+1)(j-m+2)}|p,k,j+1,m-1\rangle,
\label{operadores eigenedos}
\ea
where 
\be
\gamma_{j}=\frac{kp}{j(j+1)}
\ \ \ \ \ \ \ \ \ \ 
\alpha_{j}=i\sqrt{\frac{(j^2-k^2)(j^2+p^2)}{j^2(4j^2-1)}}
\ee
This concludes the review of the representation theory that is necessary for the definition of the EPRL model.

\subsection{Lorentzian model: the linear simplicity constraints}\label{reshe}

As in the Riemannian case it can be shown as  in Section (\ref{cuadratiqui}) that the quadratic simplicity constraints (\ref{dual}) are implied, in the discrete setting, by 
the linear constraint on each face
\be\label{constrainty}
D_{f}^i=L_{f}^i-\frac{1}{\gamma} K_{f}^i\approx 0,
\ee
where the label $f$ makes reference to a face $f\in \Delta^{\star}$, and where (very importantly) the subgroup $SU(2)\subset SL(2,\C)$ that is necessary for the definition of the above 
constraints is chosen arbitrarily at each tetrahedron, equivalent on each edge $e\in \Delta^{\star}$.
The commutation relations of the previous tetrahedron constraint are
\ba\label{algebry}
[D_{f}^i,D_{f'}^j]&=&\delta_{f f'}  \epsilon^{ij}_{\ \, k} \left[(1-\frac{1}{\gamma^2}) L_{f}^k-\frac{2}{\gamma} K_e^k\right]=\n \\
&=&2 \delta_{e e'}  \epsilon^{ij}_{\ \, k} D^k-\delta_{e e'} \frac{\gamma^2+1}{\gamma^2} \epsilon^{ij}_{\ \, k}  L_{f}^k.
\ea
The previous commutation relation implies that the constraint algebra is not closed and cannot therefore be imposed as operator equations of the states summed over in the BF partition function in general. There are two interesting exceptions: 
\begin{enumerate}
\item The first one is to take $\gamma=\pm i$ in which case the constraint (\ref{constrainty}) could in principle be imposed strongly.  The self-dual sector is however not
present in $\sH_{p,k}$.

\item The second possibility is to work in the sector where $L^i_e=0$. Which, as in the Riemannian
 case, it corresponds to the Barret-Crane model \cite{BC1} with the various limitations discussed in Section \ref{BCM}.
\end{enumerate}
The EPRL model is obtained by restricting the representations appearing in the expression of the BF partition 
function so that at each tetrahedron the linear constraints (\ref{constrainty}) the strongest possible way that is compatible with the uncertainties 
relations stemming from 
(\ref{algebry}). In addition one would  add  the requirement that the state-space of tetrahedra is compatible with the state-space 
of the analogous excitation in the canonical context of LQG so that arbitrary states in the kinematical state of LQG have non trivial 
amplitudes in the model.

\subsection*{On the weak imposition of the linear simplicity constraints}

We now discuss the weak imposition of the linear simplicity constraints in the quantum framework.
There are essentially three views in the literature: two of them, discussed below, concerns directly the  way the 
EPRL model has been usually introduced. The third possibility is the semiclassical view based on the coherent state representation leading to the FK model (see section \ref{fk}).

\subsubsection{Lorentzian model:
The Gupta-Bleuler criterion}

As in the Riemannian
 case, the fact that the constraints $D^{i}_f$ do not form a closed (first class) algebra in the generic case 
requires a weaker sense in which the constraints are imposed. One possibility is to consider the Gupta-Bleuler criterion 
consisting in selecting a suitable class of states for which the matrix elements on $D_{f}^i$ vanish.  In this respect one notices that
if we chose the subspace  $\sH_{j}\subset\sH_{p,k}$ one has
\ba && \n
  \langle p,k,j,q|D^{3}_f |p,k,j,m\rangle=  \delta_{q,m}m (1-\frac{\gamma_{j}}{\gamma})\\ && \n
  \langle p,k,j,q|D^{\pm}_f |p,k,j,m\rangle=  \delta_{q\pm 1,m}\sqrt{(j\pm m+1)(j\mp m)}(1-\frac{\gamma_{j}}{\gamma}).
\ea
One immediately observes that matrix elements of the linear constraints vanish in this subclass if one can choose
\be\gamma_{j}=\frac{kp}{j(j+1)}=\gamma\label{coky}
\ee
Now the situation simplifies with respect to the Riemannian
 case as there is only one single case. The linear simplicity constraint is satisfied as long as $ p=\gamma j(j+1)/k$. 
Therefore, one has an integer worth of possibilities. In particular, given $j$ one has $k=j-r$ for all possible half-integers such that $k>0$
\cite{Ding:2010fw}. As we will see in what follows the best choice from the point of view of the Master constraint criterion is $k=j$ from which it follows
\be
p=\gamma (j+1).
\ee

\subsubsection{Lorentzian model: The Master constraint criterion}

Another criterion for weak imposition can be developed by studying the spectrum of the Master constraint \ba\n M_f=D_f\cdot D_f&=&L^2-\frac{1}{\gamma} 2 L \cdot K+\frac{1}{\gamma^2}K^2\\ &=& \frac{1}{\gamma^2}[ (1+{\gamma^2}) L^2- 2{\gamma} C_2 - C_1].\ea Strong imposition of the constraints $D_f^i$ would amount to
looking for the kernel of the master constraint $M_f$. However, generically the positive operator associated with the master constraint does not have the zero eigenvalue in the spectrum due to the open nature of the constraint algebra (\ref{algebry}).  
The proposal of \cite{Engle:2007mu} is to look for the minimum eigenvalue among spaces $\sH_j\in \sH_{p,k}$. Explicitly 
 \be
 M_f|\psi>=m_{p,k,j} |\psi>,
 \ee
 where
\ba  m_{p,k,j}=\frac{1}{\gamma^2}[p^2 +1-k^2+(1+\gamma^2) j(j +1)-2\gamma k p].\n
\ea
For given $j$ the minimum eigenvalue is obtained when $k=j$ and $p=\gamma j$ in which case one gets $m_{p,k,j}\ge m_j$ for $m_j=\hbar^2 (1+\gamma^2) j+1$. 
Again we see that the important condition \be
M_f |\gamma j,j,j,m\rangle=\sO_{sc}
\ee 
holds (recall the notion $\sO_{sc}$ from Section \ref{resume-r}). Interestingly, the states that satisfy the previous semiclassical criterion satisfy the Gupta-Bleuler criterion not exactly and only in the large spin sense.
However, the advantage of the master constraint criterion is clear from the point of view of cylindrical consistency as the choice $p=\gamma (j+1)$ obtained 
in the previous subsection would imply that $j=0$ spin network links would be have a non trivial amplitude in the path integral defined in Section \ref{path-pre}. Such 
thing would be in sharp conflict with diffeomorphism invariance in the model. 

Therefore, as in the Riemannian
 case we will use the master constraint criterion as the basic ingredient in the definition of the EPRL model. One can write a Lorentz invariant constraint \cite{Rovelli:2010ed}  
from the master constraint and using the $D^i_f=0$ classically. This leads to the following  Lorentz invariant condition in terms of the $SL(2,\C)$ Casimirs 
\ba
M^{\va LI}_f =(1-\gamma^2) C_2 - 2 C_1 \gamma,
\ea
where $C_1$ and $C_2$ are the two $sl(2,\C)$ Casimirs given in equation (\ref{casiL}). Surprisingly, the conditions $p=\gamma j$ and $k=j$ minimizing the
master constraint solve its Lorentz invariant version strongly! This resolves the cylindrical consistency issue pointed out above. More recently, it has been shown that
the constraint solution $p=\gamma j$ and $k=j$ also follows naturally from a spinor formulation of the simplicity constraints \cite{Wieland:2011ru, Dupuis:2011wy, Livine:2011vk}.

\subsubsection{Lorentzian model: Restrictions on the Immirzi parameter}

Unlike the Riemannian model, all values of spin $j$ are allowed, and
no restrictions on the Immirzi parameter arise.

\subsubsection{Lorentzian model: Overview of the solutions of the
  linear simplicity constraints}
\label{resume-l}

In the Lorentzian model one restricts the $SL(2,\C)$ representations of BF theory to those satisfying \be p=\gamma j \ \ \ k=j\ee
for $j\in \N/2$. From now on we denote the subset of admissible representation of $SL(2,\C)$ as \be\label{kkl}
\sK_{\gamma}\subset {\rm Irrep}(SL(2,\C))
\ee
 The admissible quantum states $\Psi$ are elements of the subspace $\sH_{j}\subset \sH_{\gamma j,j}$ (i.e., minimum weight states) satisfy the constraints (\ref{constrainty}) in the following  emiclassical sense:
\be
( K^i_f-\gamma  L_f^i) \Psi=\sO_{sc},
\ee 
where as in the Riemannian case the symbol $\sO_{sc}$ (order semiclassical) 
denotes a quantity that vanishes in limit $\hbar\to 0$, $j\to \infty$ with $\hbar j=$constant.

\subsection{Presentation of the EPRL Lorentzian model}
\label{alala}

As briefly discussed in Section \ref{eprl-l}, unitary irreducible representations of $SL(2,\C)$ are infinite dimensional and labelled by a  positive real number $p\in \R^{+}$ and a half-integer $k\in \N/2$. These representation are the ones that intervene in the harmonic analysis of square integrable functions of $SL(2,\C)$ \cite{gelfand}. Consequently, 
one has an explicit expression of the delta function distribution (defined on such test function), namely
\be
\delta(g)=\sum_{k}\int_{\R^+} dp\   (p^2+k^2) \ \sum_{j,m} D^{p,k}_{jmjm}(g)
\ee
 where $D^{p,k}_{jmj'm'}(g)$  with $j\ge k$ and $j\ge m \ge-j$ (similarly for the primed indices) are the matrix elements of the unitary representations $p-k$ in the so-called canonical basis \cite{ruhl}. One can use the previous expression the Lorentzian version of Equation (\ref{coloring4}) in order to introduce a formal definition of the BF amplitudes, which now would involve integration of the continuous labels $p_f$ in addition of sums over discrete quantum numbers such as $k$, $j$ and $m$.
 The Lorentzian version of the EPRL model can be obtained from the imposition of the linear simplicity constraints, discussed in Section \ref{reshe}, to this formal expression.
As the continuum labels $p_f$ are restricted to $p_f=\gamma j_f$ the Lorentzian EPRL model becomes a state-sum model as its Riemannian relative.
Using the following graphical notation 
\be
D^{p,k}_{jmj'm'}(g)=\ \ \ \begin{array}{c}\psfrag{a}{$\van p$}\psfrag{b}{$\van k$}\psfrag{c}{$\van \!\!\! j',m'$}\psfrag{d}{$\van \!\!\!\!\!\!\!j,m$}
\includegraphics[width=2.5cm]{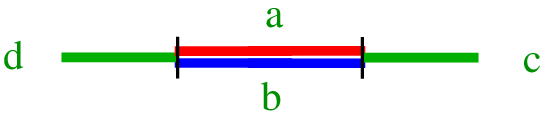}
\end{array}
\ee
the amplitude is
\ba\label{eprl-sl2c}\n
Z^{L}_{eprl}(\Delta)= \sum \limits_{ j_f}  \ \prod\limits_{f \in \Delta^{\star}} (1+\gamma^2) j_f^2 \begin{array}{c}
\includegraphics[width=5cm]{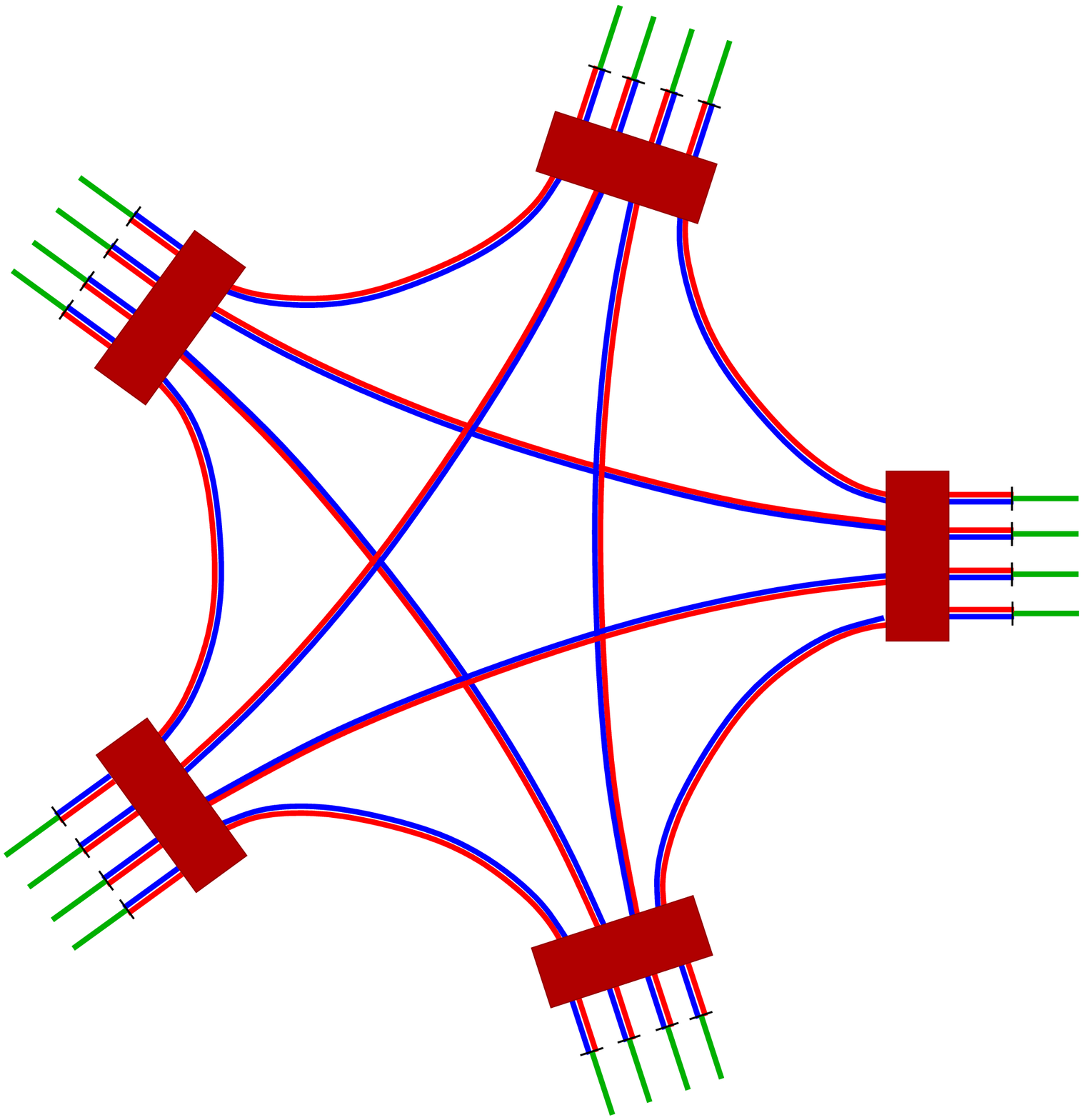}
\end{array},
\ea
where the boxes now represent $SL(2,\C)$ integrations with the invariant measure. The previous amplitude is equivalent to the its spin foam representation
\ba\label{eprl-sf-sl2c}\n
Z^{L}_{eprl}(\Delta)= \sum \limits_{ j_f}\sum \limits_{ \iota_e}  \ \prod\limits_{f \in \Delta^{\star}} (1+\gamma^2) j_f^2 \prod_{v\in \Delta^{\star}} \begin{array}{c}\psfrag{a}{$\iota_1$}\psfrag{b}{$\iota_2$}\psfrag{c}{$\iota_3$}\psfrag{d}{$\iota_4$}\psfrag{e}{$\iota_5$}
\includegraphics[width=5cm]{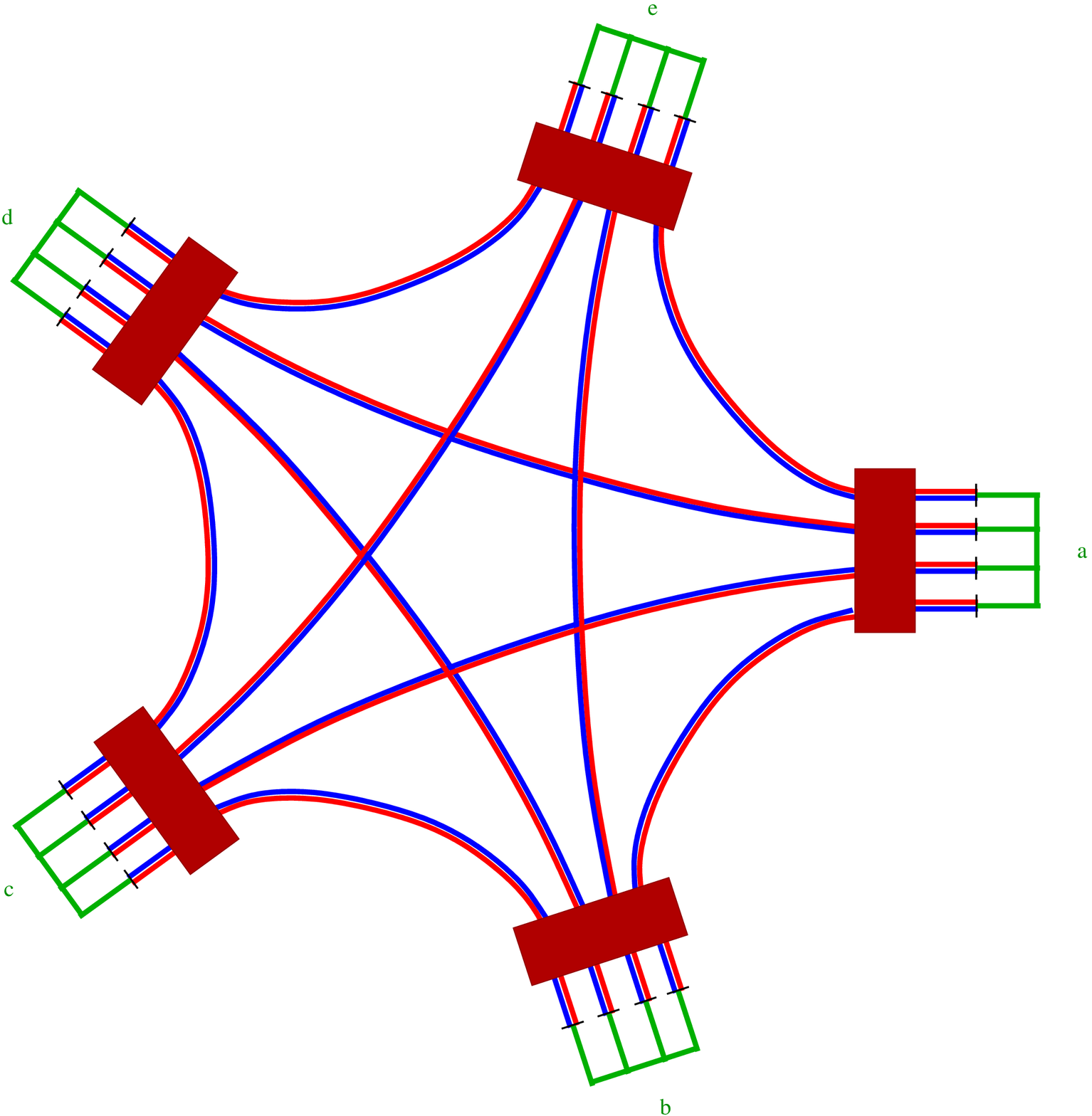}
\end{array},
\ea
The vertex amplitude is divergent due to the presence of a redundant integration over $SL(2,\C)$, it becomes finite by dropping an arbitrary integration, i.e. removing any of the 5 boxes in the vertex expression \cite{Engle:2008ev}.

\subsection{The coherent state representation}

It is immediate to obtain the coherent states representation of the Lorentzian models. As in the Riemannian case, one simply inserts resolution of the identities 
(\ref{ident-coherent}) on the intermediate $SU(2)$  (green) wires in (\ref{eprl-sl2c}) from where it results 
\ba\label{eprl-cohe-l}
&& Z^{L}_{eprl}(\Delta)=\sum \limits_{j_f}  \ \prod\limits_{f \in \Delta^{\star}} (1+\gamma^2) j^2 \n \\ && \int \prod_{e\in  \in \Delta^{\star}} {\rm d}_{j_{ef}} dn_{ef}
\begin{array}{c}\psfrag{w}{$$}
\psfrag{a}{$\va n_{ 1}$}
\psfrag{A}{$\va n_1$}
\psfrag{b}{$\va n_2$}
\psfrag{B}{$\va n_2$}
\psfrag{c}{$\va n_3$}
\psfrag{C}{$\va n_3$}
\psfrag{d}{$\va n_4$}
\psfrag{D}{$\va n_4$}
\includegraphics[width=7cm]{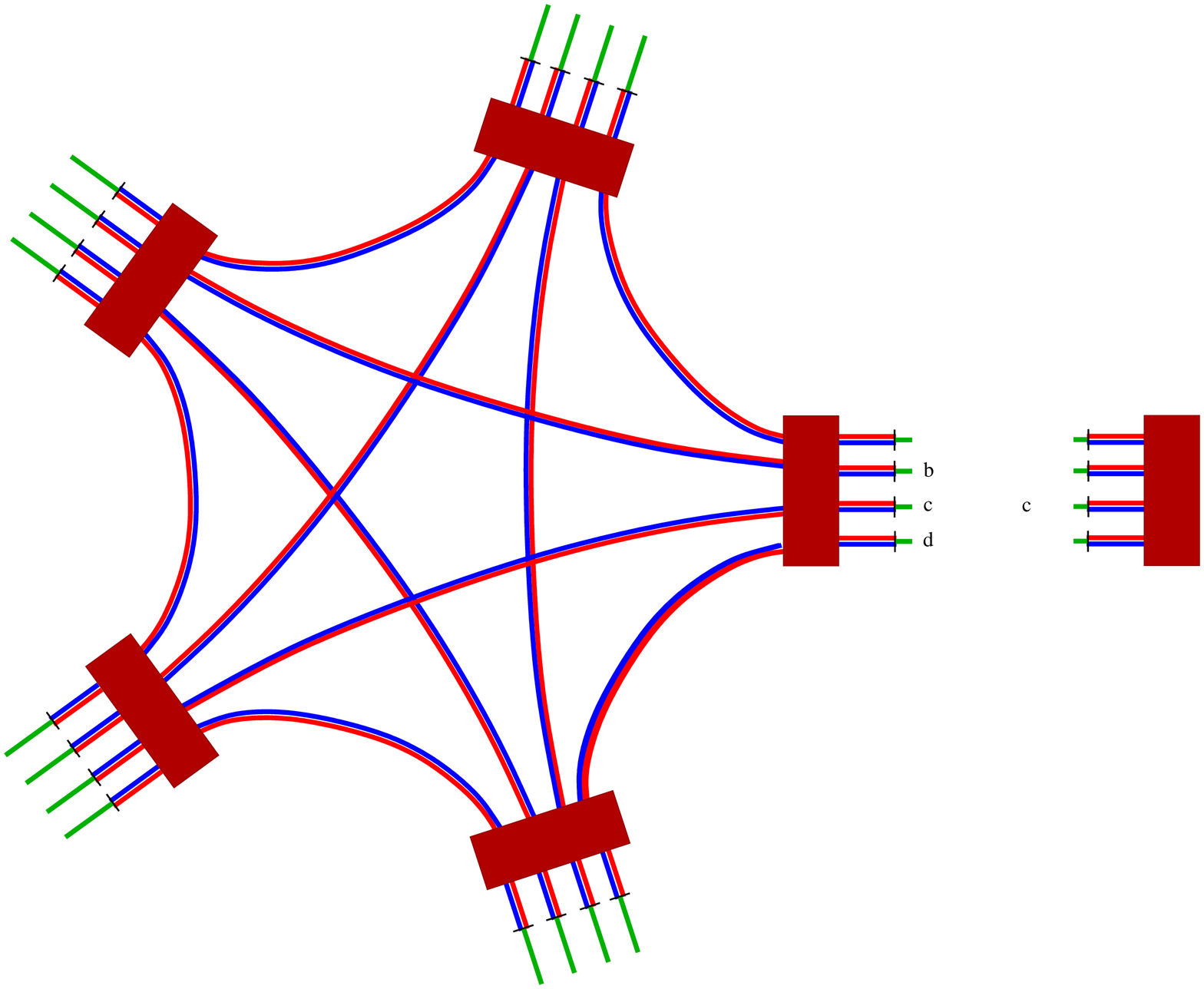}
\end{array},
\ea

%

\section{The Freidel--Krasnov Model}
\label{fk}

Shortly after the appearance of the paper \cite{Engle:2007uq}, Freidel and Krasnov \cite{Freidel:2007py} introduced a set of new spin foam models 
for four dimensional gravity using the coherent state basis of the quantum tetrahedron of Livine and Speziale \cite{Livine:2007vk}.
The idea is to impose the linearized simplicity constraints (\ref{constrainty-e}) directly as a semiclassical condition on the coherent state basis.
As we have seen above, coherent states are quantum states of the right and left tetrahedra in BF theory which have a clear-cut semiclassical interpretation 
through their property  (\ref{geo}).  We have also seen that the imposition of the linear constraints (\ref{constrainty-e}) {\em a la} EPRL is in essence semiclassical as they are strictly 
valid only in the large spin limit. In the FK approach one simply accept from the starting point that, due to their property of not defining set that is closed 
under commutation relations, the Plebansky  are to be imposed semiclassically. One defines new models by restricting 
the set of coherent states entering in the coherent state representation of $Spin(4)$ BF theory (\ref{bf-cohe}) to those that satisfy condition 
(\ref{constrainty-e}) in expectation values.  They also emphasize how the model \cite{Engle:2007uq} corresponds indeed to the sector $\gamma=\infty$ which 
has been shown to be topological \cite{Liu:2009em}.  

\subsubsection*{The case $\gamma<1$} For $\gamma<1$ the vertex amplitude is identical to the EPRL model. This is apparent in the coherent state expression of the EPRL model (\ref{eprl-cohe}).
Thus we have
\ba\label{FK<}&& 
Z_{fk}^{\va \gamma<1}(\Delta)=\sum \limits_{ j_f}   \ \prod\limits_{f \in \Delta^{\star}} {\rm d}_{|1-\gamma|\frac{j}{2}}{\rm d}_{(1+\gamma)\frac{j}{2}} 
\n \\ && \ \ \ \ \ \ \ \ \ \ \ \ \  \prod_{e\in \Delta^\star} \int {\rm d}_{(1+\gamma)\frac{j}{2}}{\rm d}_{(\gamma-1)\frac{j_{ef}}{2}} dn_{ef} 
\begin{array}{c}\psfrag{w}{$=\int\limits_{[S^3]^4} \prod\limits_{\va I=1}^{\va 4} dn_{\va I}$}
\psfrag{a}{$\va n_1$}
\psfrag{A}{$\va n_1$}
\psfrag{b}{$\va n_2$}
\psfrag{B}{$\va n_2$}
\psfrag{c}{$\va n_3$}
\psfrag{C}{$\va n_3$}
\psfrag{d}{$\va n_4$}
\psfrag{D}{$\va n_4$}
\psfrag{x}{$$}
\includegraphics[width=6.5cm]{eprl5}\end{array}.
\ea 
From the previous expression we conclude that the vertex amplitudes of the FK and EPRL model coincide for $\gamma<1$
\be
A_{v\ fk}^{\va \gamma<1}=A_{v \ eprl}^{\va \gamma<1}.
\ee
Notice however that different weights are assigned to edges in the FK model.  This is due to the fact that one is restricting  the $Spin(4)$ resolution of identity
in the coherent basis in the previous expression, while in the EPRL model the coherent state resolution of the identity is used for $SU(2)$ representations.
This difference is important and has to do with the still un-settled discussion concerning the measure in the path integral representation (see Section \ref{medida} and references therein).

\subsubsection*{The case $\gamma>1$} 
For the case $\gamma>1$ the FK amplitude is given by
\ba\label{FK>}&& 
Z_{fk}^{\va \gamma>1}(\Delta)=\sum \limits_{ j_f}   \ \prod\limits_{f \in \Delta^{\star}} {\rm d}_{|1-\gamma|\frac{j}{2}}{\rm d}_{(1+\gamma)\frac{j}{2}} 
\n \\ && \ \ \ \ \ \ \ \ \ \ \ \ \  \prod_{e\in \Delta^\star} \int {\rm d}_{(1+\gamma)\frac{j}{2}}{\rm d}_{(\gamma-1)\frac{j_{ef}}{2}} dn_{ef} 
\begin{array}{c}\psfrag{w}{$=\int\limits_{[S^3]^4} \prod\limits_{\va I=1}^{\va 4} dn_{\va I}$}
\psfrag{a}{$\va n_1$}
\psfrag{A}{$\va -n_1$}
\psfrag{b}{$\va n_2$}
\psfrag{B}{$\va -n_2$}
\psfrag{c}{$\va n_3$}
\psfrag{C}{$\va -n_3$}
\psfrag{d}{$\va n_4$}
\psfrag{D}{$\va -n_4$}
\psfrag{x}{$$}
\includegraphics[width=6.5cm]{eprl5}\end{array}.
\ea 
The study of the coherent state representation of the FK model for $\gamma>1$,  and comparison with  equation (\ref{cuat}) for the EPRL model, clearly shows the difference between the two models in this regime. 
\ba
\label{fk-cohe}Z^{\va \gamma}_{fk}=\sum \limits_{ j_f }  \ \prod\limits_{f \in \Delta^{\star}} {\rm d}_{(1-\gamma)\frac{j_f}{2}}{\rm d}_{(1+\gamma)\frac{j_f}{2}} \int \prod_{e\in  \Delta^{\star}} {\rm d}_{|1-\gamma|\frac{j_{ef}}{2}} {\rm d}_{(1+\gamma)\frac{j_{ef}}{2}} dn_{ef} dg^{-}_{ev}dg^{+}_{ev}
\ \exp{(S^{\va fk\ \gamma}_{j^{\pm},\bn}[g^{\pm}])}, \ea
where the discrete action \be\label{fk-discrete-action}
S^{\va fk\ \gamma}_{j^{\pm},\bn}[g^{\pm}]=\sum_{v\in\Delta^{\star}} (S^v_{(1-\gamma)\frac{j_f}{2},\bn}[g^{-}]+S^v_{(1+\gamma)\frac{j_f}{2},s(\gamma) \bn}[g^{+}]), \ee where $s(\gamma)={\rm sign}(1-\gamma)$ and  
\be
\label{fk-v-action}
S^v_{j,\bn}[g] = \sum\limits^{5}_{a < b=1} 2j_{ab} \ln \, \la n_{ab}| g^{-1}_a g_b| \, n_{ba} \ra,
\ee
with the indices $a,b$ labelling the five edges of a given vertex.

\section{The Barrett--Crane Model}
\label{BCM}

The Barrett--Crane models \cite{BC1, BC2} can be easily obtained in the present framework as a suitable limiting case $\gamma\to \infty$ of the previous constructions.
One way to get it is to take the limit $\gamma\to \infty$ of the linear simplicity constraint (\ref{constrainty-e}) at the classical level. In this way the constraint becomes
\be
D^i=L_{f}^i \approx 0
\ee
with which the constraint algebra closes. One can impose the constraints strongly in this case. The solution is obtained by setting $j_{f}=0$ in all faces at the level of the amplitude (\ref{eprl-so4}).
An immediate consequence of this is that $j_{f}^{+}=j_{f}^-$ and all tetrahedra are assigned a unique intertwiner: the so-called Barrett-Crane intertwiner. Explicitly, setting $j_f=0$ in (\ref{eprl-so4}) we get 
\ba\label{BC-so4}
 Z_{BC}(\Delta)&=&\sum \limits_{ j_{f}^{+}}  \ \prod\limits_{f \in \Delta^{\star}} {\rm d}_{j^{+}}^2
\begin{array}{c}
\includegraphics[width=5cm]{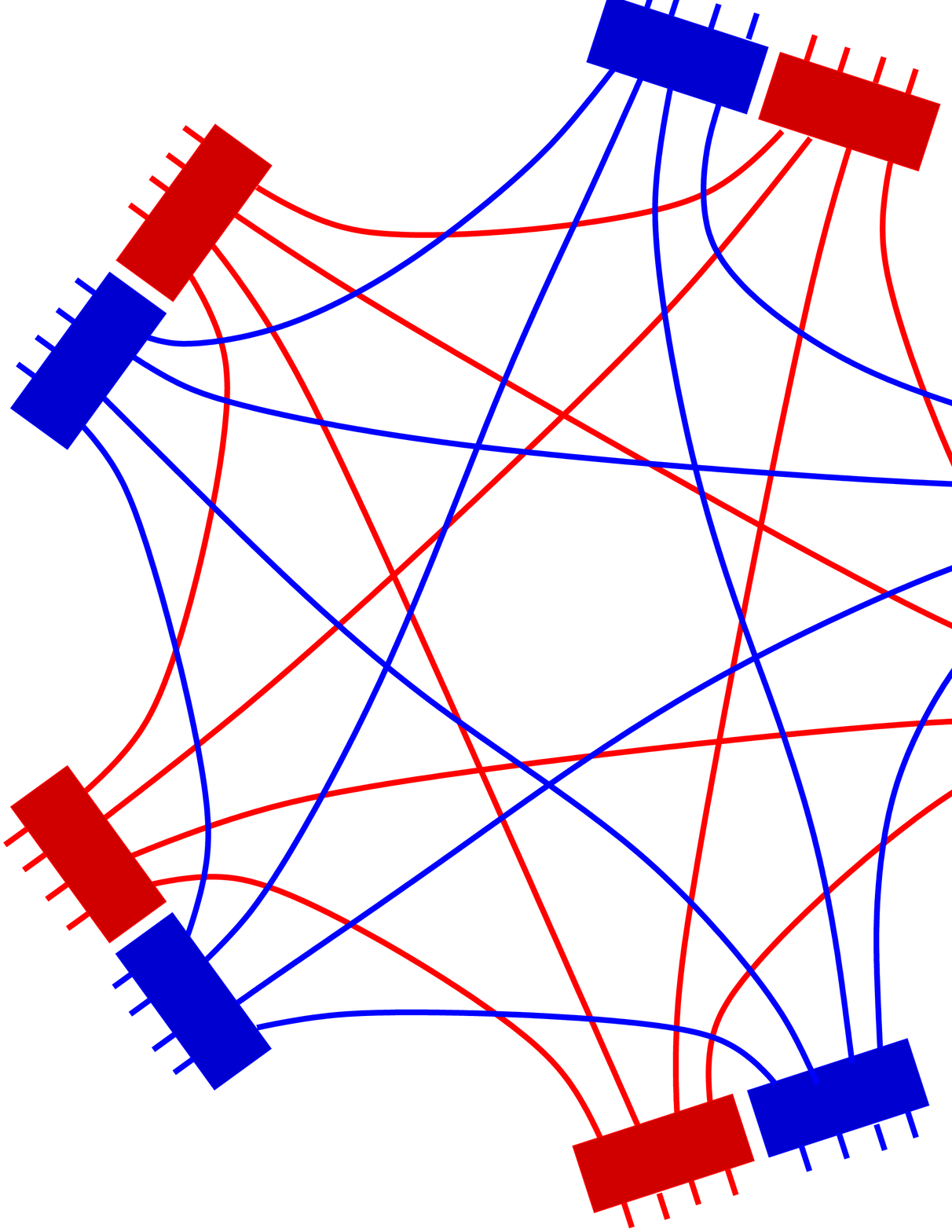}
\end{array} ,\ea
where since $j_f^{+}=j_f^{-}$ we are arbitrarily choosing $j_f^{+}$ as the spins summed over in the state sum. Eliminating redundant  group integrations  we get the previous amplitude in the spin foam 
representation, namely 
\be\label{BC-sf}
Z_{BC}(\Delta)=\sum \limits_{ j^{+}_f}  \\ \prod\limits_{f \in \Delta^{\star}} {\rm d}_{j_f^{+}}^2
\prod_{v\in \Delta^{\star}}\begin{array}{c}
\includegraphics[width=4cm]{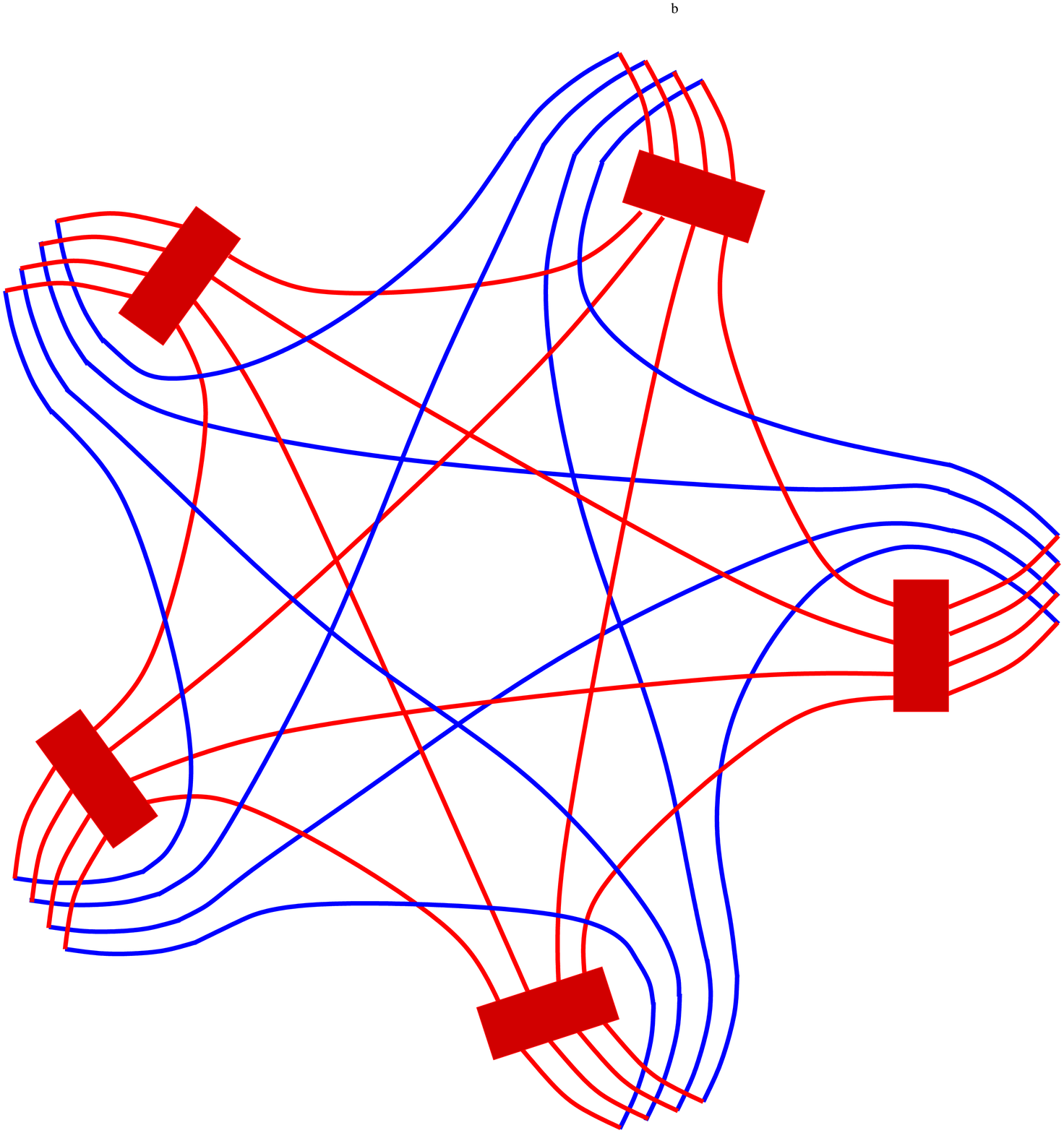}
\end{array}.
\ee 
Observe that the vertex amplitude depend only on the 10 corresponding $j_f^{+}$, this is the Riemannian Barrett-Crane vertex. It is interesting to write
the analog of equation (\ref{eprl-projection}). The Barrett-Crane model is obtained by the modification of the BF projectors in (\ref{bf4}) so that 
\be
 Z_{BC}(\Delta)=\sum \limits_{ j_{f}^{+}}  \ \prod\limits_{f \in \Delta^{\star}} {\rm d}_{j^{+}}^2 \prod_{e} P_{BC}^{e}(j^+_1,\cdots,j^+_4)
\ee
with
\be\label{BC-projection}
P_{BC}^{e}(j_1\cdots j_4)=|BC\rangle\langle BC|=
\begin{array}{c}
\includegraphics[width=1.4cm]{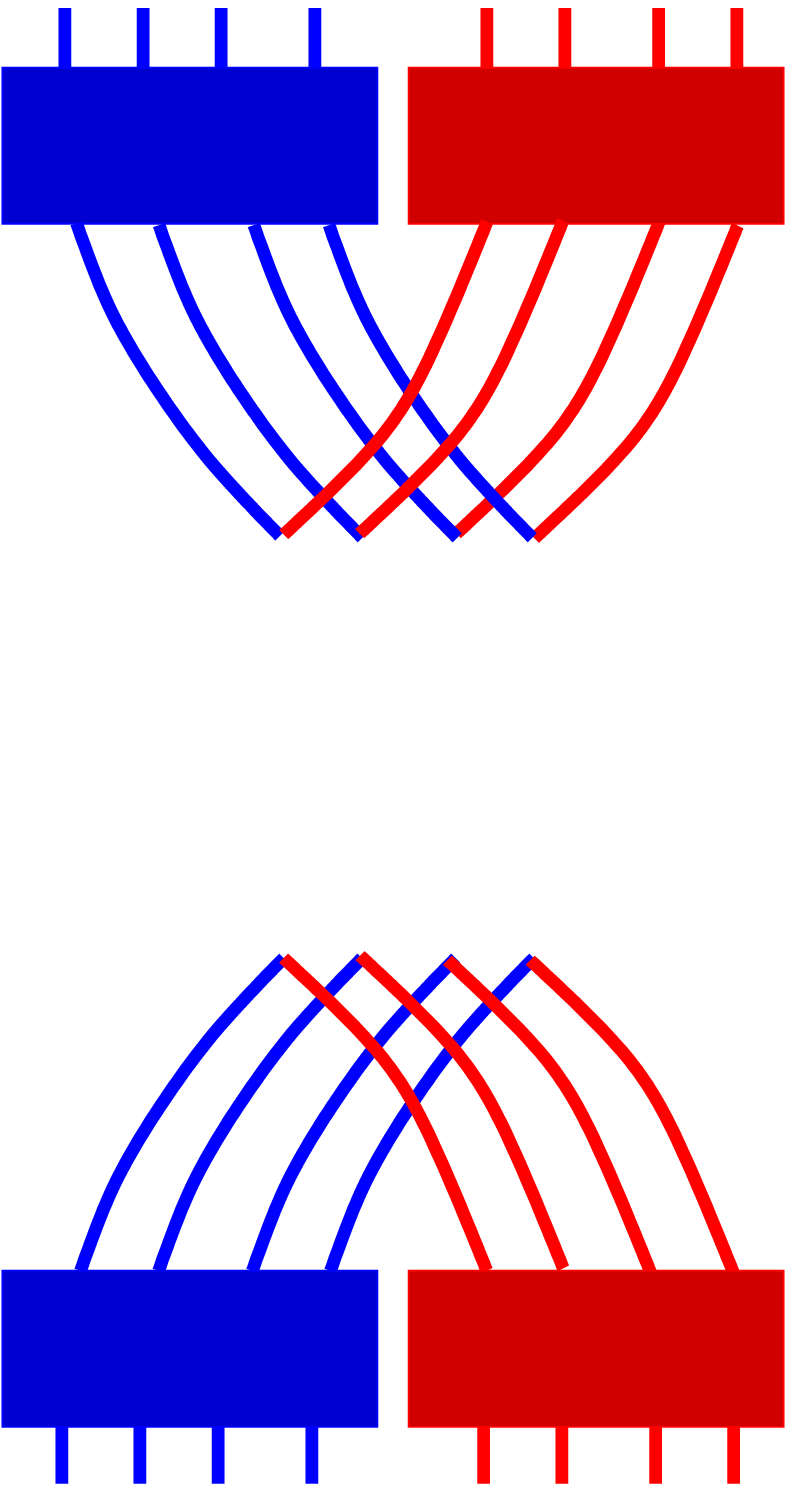}
\end{array}=\sum_{\iota}\sum_{\iota^{\prime}}\begin{array}{c}\psfrag{a}{$\iota$}\psfrag{b}{$\iota^{\prime}$}
\includegraphics[width=1.4cm]{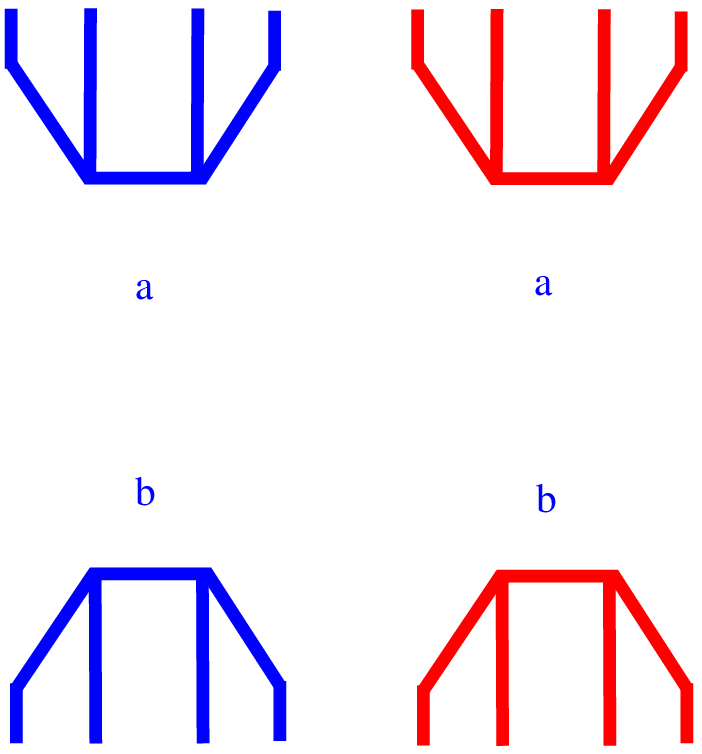},
\end{array}
\ee 
where $|BC\rangle$ is the Barrett-Crane intertwiner.
The tetrahedra state space is one-dimensional once the four quantum numbers $j_{f}^{+}$ has been fixed. 
Indeed, it has been shown that the state space of strong solutions of the 
the quadratic quantum Plebanski constraints  is one dimensional and it is given by the
BC intertwiner $|BC\rangle$ \cite{reis3}.

\subsection{The coherent states representation of the BC amplitude}

One can insert coherent states resolution of the identity (\ref{patacu})  in the 10 wires obtaining the following  coherent state representation of the Barrett--Crane amplitude
\ba
\label{BC-cohe}Z_{BC}=\sum \limits_{ j^+_f }  \ \prod\limits_{f \in \Delta^{\star}} {\rm d}_{j_f^+}^2 \int \prod_{e\in  \Delta^{\star}} {\rm d}_{j_{ef}} dn_{ef} dg^{+}_{ev}
\ \exp{(S^{\va BC}_{j^{+},\bn}[g^{+}])}, \ea
where the discrete action \be\label{BC-discrete-action}
S^{\va BC}_{j^{+},\bn}[g^{+}]=\sum_{v\in \Delta^{\star}}\ \sum\limits^{5}_{a < b=1} 2j^{+}_{ab} \ln \, \la n_{ab}| (g^{+}_a)^{-1} g^{+}_b| \, n_{ab} \ra.
\ee
Notice that there are only 10 normals $n_{ab}$ as in this case $n_{ab}=n_{ba}$. The large spin asymptotics of the Barrett--Crane vertex has been investigated in~\cite{Barrett:1998gs}.

\section{Boundary Data for the New Models and Relationship with the Canonical Theory}

So far we have considered cellular complexes with no boundary. Transition amplitudes
are expected to be related to the definition of the physical scalar product as discussed in Section~\ref{valin}.
In order to define them one needs to consider complexes with boundaries. Boundary states are defined on the boundary of the dual two-complex $\Delta^{\star}$ that we denote
$\partial \Delta^{\star}$.  The object $\partial \Delta^{\star}$ is a one-complex (a graph). According to the construction of the model (Section \ref{eprl-r} and \ref{eprl-l}) boundary states 
are in one-to-one correspondence with $SU(2)$ spin networks.
This comes simply from the fact that links (one-cells) $\ell\in \partial\Delta^{\star}$ inherit the
 spins labels (unitary irreducible representations of the subgroup $SU(2)$) of the boundary faces while 
 nodes (zero-cells) $n\in \partial \Delta^{\star}$ inherit the intertwiner levels of boundary edges.
 
 At this stage one can associate the boundary data with elements of a Hilbert space. Being in one-to-one correspondence with $SU(2)$ spin networks, a 
 natural possibility is to associate to them an element of the kinematical Hilbert space of LQG. More precisely, with a given coloured boundary graph $\gamma$ with  links labelled by spins $j_{\ell}$ and nodes labelled by interwiners $\iota_n$  
 we associate a cylindrical function $\Psi_{\gamma,\{j_\ell\},\{\iota_n\}}\in {\sL}^2(SU(2)^{N_{\ell}})$, where here $N_{\ell}$ denotes number of links in the graph $\gamma$ (see Equation \ref{cyly}). 
 In this way, the boundary Hilbert space associated with $\partial \Delta^{\star}$ is isomorphic (if one used the natural AL measure)
 with the Hilbert space of LQG truncated to that fixed graph. Moreover, geometric operators such as volume and area
 defined in the covariant context are shown to coincide with the corresponding operators defined in the canonical formulation \cite{Ding:2010ye, Ding:2009jq}.
 Now, if cellular complexes are  dual to triangulations then  the boundary spin networks can have at most four valent nodes. 
 This limitation can be easily overcome: as in BF theory the EPRL amplitudes can be generalized to arbitrary complexes with boundaries given by graphs with
 nodes of arbitrary valence. The extension of the model to arbitrary complexes has been first studied in \cite{Kaminski:2009qb, Kaminski:2009fm}, it has also been revisited in 
 \cite{Ding:2010fw}.
 
 Alternatively, one can associate the boundary states with elements of
 $\sL^2(Spin(4)^{N_{\ell}})$ (in the Riemannian models) -- or 
 carefully define the analog of spin network states as distributions in the Lorentzian case\epubtkFootnote{The definition of a polymer-like gauge invariant 
 states as elements of a Hilbert space of a gauge theory with non compact group is an open issue. The naive attempts fail, the basic problem is that gauge invariant states are not square integrable functions with respect to the obvious generalization of the AL measure  for fixed graph: this is due to the infinite volume of the gauge group (the invariant measure is not normalizable). For  that reason the discussion 
 of the Lorentzian sector with $G=SL(2,\C)$ is usually formal. See \cite{Freidel:2002xb} for some insights on the problem of defining a gauge invariant Hilbert space of graphs for non compact gauge groups. }.
 In this case one gets special kind of spin network states that are a subclass of the so-called projected spin networks introduced in
 \cite{Alexandrov:2002br, Livine:2002ak} in order to define an heuristic quantization of the (non-commutative and very complicated) Dirac algebra of a Lorentz connection formulation of the phase space of gravity 
 \cite{Alexandrov:2006wt, Alexandrov:2005ng, Alexandrov:2002br, Alexandrov:2002xc,Alexandrov:2001wt, Alexandrov:2000jw, Alexandrov:1997yk}.
 The fact that these special subclass of projected spin networks  appear naturally as boundary states of the new spin foams is shown in \cite{Dupuis:2010jn}. 
  
Due to their similarity for $\gamma<1$ the same relationship between boundary data and elements of the kinematical Hilbert space hold for
the FK model. However, the such simple relationship does not hold for the model in the case $\gamma>1$.

The role of knotting in the Hilbert space of the EPRL model is studied in \cite{Bahr:2010my}.

\section{Further developments and related models}

\subsection{The measure}
\label{medida}

The spin foam amplitudes discussed in the previous sections have been introduced by 
constraining the BF histories through the simplicity constraints. However, in the 
path integral formulation, the presence of constraints has the additional effect of 
modifying the weights with which  those histories are to be summed: second class constraints
modify the path integral measure (in the spin foam context this issue was raised in~\cite{myo}).
As pointed out before, this question has not been completely settled yet in the spin foam community.
The explicit modification of the formal measure in terms of continuous variables 
for the Plebansky action was presented in~\cite{karim3}.  
A systematic investigation of the measure in the spin foam context was attempted in ~\cite{Engle:2009ba} and~\cite{Han:2009bb}.
As pointed out in~\cite{myo}, there are restrictions in the manifold of possibilities coming from the requirement of
background independence. The simple BF measure chosen in the presentation of the amplitudes in the previous sections
satisfy these requirements.   There are other consistent possibilities; see for instance~\cite{Bianchi:2010fj}  for
a modified measure which remains extremely simple and is suggested from the structure of LQG.

\subsection{Relation with the canonical formulation: the scalar constraint}

An important question is the relationship between the spin foam amplitudes and the canonical operator formulation. The question of wether one can reconstruct the Hamiltonian constraints out of spin foam amplitudes 
has been analysed in detail in three dimensions. For the study of quantum three dimensional gravity from the BF perspective see~\cite{Noui:2004iy}, we will in fact present this perspective in detail in the three dimensional part of this article.
For the relationship with the canonical theory using variables that are natural from the Regge gravity perspective see~\cite{Bonzom:2011hm, Bonzom:2011tf}.
In four dimensions the question has been investigated in~\cite{Alesci:2008yf} in the context of the new spin foam models.
In the context of group field theories  this issue is explored in~\cite{Livine:2011yb}. 
Finally, spin foams can in principle be obtained directly from the implementation of the Dirac program 
using path integral methods this avenue has been explored in~\cite{Han:2009aw, Han:2009ay} 
from which a discrete path integral formulation followed~\cite{Han:2009az}.
The question of the relationship between covariant and canonical formulations in the discrete setting has been analyzed
also in~\cite{Dittrich:2009fb}.

\subsection{Spin foam models with timelike faces}

By construction all tetrahedra in the FK and EPRL models are embedded in a spacelike
hypersurface and hence have only spacelike triangles. It seem natural to ask the question of whether a more 
general construction allowing for timelike faces is possible. 
The models described in previous sections have been generalized in
order to include timelike faces in the work of
Conrady~\cite{Conrady:2010sx, Conrady:2010vx, Conrady:2010kc}. An
earlier attempt to define such models  in the context of the
Barrett--Crane model can be found in~\cite{a8}.

\subsection{Coupling to matter}

The issue of the coupling of the new spin foam models to matter  remains to a large extend un-explored territory.
Nevertheless some results can be found in the literature. The coupling of the Barrett--Crane model (the $\gamma\to \infty$ limit of the EPRL model) to Yang--Mills fields 
was studied in~\cite{ori4}. More recently the coupling of the EPRL model to fermions has been investigated in~\cite{Han:2011as, Bianchi:2010bn}. 
A novel possibility of unification of the gravitational and gauge fields was recently proposed in~\cite{Alexander:2011jf}.

\subsection{Cosmological constant}

The introduction of a cosmological constant in the construction of four dimensional  spin foam models has a long history. 
Barrett and Crane introduced a vertex amplitude~\cite{BC1} in terms of the Crane and Yetter model~\cite{crane0} 
for BF theory with cosmological constant. The Lorentzian quantum deformed version of the previous model was 
studied in~\cite{Noui:2002ag}.   For the new models the coupling  with a cosmological constant is explored in terms of the quantum deformation 
of the internal gauge symmetry in~\cite{Ding:2011hp,Han:2010pz,Han:2011aa} as well as (independently) in~\cite{Fairbairn:2010cp}.

\subsection{Spin foam cosmology} 

The spin foam approach applied to quantum cosmology has been explored
in~\cite{Bianchi:2011ym, Vidotto:2010kw, Henderson:2010qd,
  Bianchi:2010zs, Rovelli:2009tp, Rovelli:2008dx}. The spin foam
formulation can also be obtained from the canonical picture provided
by loop quantum cosmology (see~\cite{Bojowald:2006da} and references
therein). This has been explored systematically
in~\cite{Ashtekar:2010gz, Ashtekar:2010ve, Ashtekar:2009dn,
  Campiglia:2010jw}. Very recent results are \cite{Livine:2011up}

\subsection{Constraints are imposed semiclassically}

As we have discussed in the introduction of the new models, Heisenberg uncertainty principle precludes the strong imposition of the Plebanski constraints that reduce BF theory 
to general relativity. The results on the semiclassical limit of these models seem to indicate that metric gravity should be recovered in the low energy limit. However, its seems likely that the semiclassical limit could be related to
certain  modifications of Plebanski's formulation of gravity~\cite{Krasnov:2006du, Krasnov:2007cq, Krasnov:2008fm, Krasnov:2009iy, Krasnov:2009ip}. A simple interpretation of the new models in the
context of the bi-gravity  paradigm proposed in~\cite{Speziale:2010cf} could be of interest.

\subsection{Group field theories associated to the new spin foams}

As already pointed out in~\cite{baez7} spin foams can be
interpreted in close analogy to Feynman diagrams. Standard Feynman
graphs are generalized to $2$-complexes and the labeling of
propagators by momenta to the assignment of spins to faces.
Finally, momentum conservation at vertices in standard
feynmanology is now represented by spin-conservation at edges,
ensured by the assignment of the corresponding intertwiners. In
spin foam models the non-trivial content of amplitudes is
contained in the vertex amplitude  which in the language of Feynman
diagrams can be interpreted as an interaction. This analogy is indeed realized in the formulation of spin foam
models in terms of a group field theory (GFT)~\cite{reis1,reis2}.

The GFT formulation resolves by definition the two fundamental conceptual
problems of the spin foam approach: diffeomorphism gauge symmetry and
discretization dependence. The difficulties are shifted to the question of the
physical role of $\lambda$ and the convergence of the corresponding perturbative series.

In three dimensions this idea has been studied in more detail. 
In~\cite{Magnen:2009at} scaling properties of the modification of the Boulatov group field theory 
introduced in~\cite{fre10} was studied in detail. In a further modification of the previous model (known as coloured tensor models~\cite{Gurau:2009tw}) 
new techniques based on a suitable $1/N$ expansion imply that
amplitudes are dominated by spherical topology~\cite{Gurau:2010ba};
moreover, it seem possible that the continuum limit  might be critical
as in certain matrix models~\cite{Gurau:2011tj, Bonzom:2011zz,
  Gurau:2011xq, Gurau:2011aq, Ryan:2011qm}. However, it is not yet
clear if there is a sense in which these models correspond to a
physical theory. The naive interpretation of the models is that they
correspond to a formulation of 3d quantum gravity including a
dynamical topology.

\subsection{Results on the semiclassical limit of  EPRL-FK models}
\section{The semiclassical limit}
\label{semiclas}

Having introduced the relevant spin foam models in the previous sections we now 
present the results on the large spin asymptotics of the spin foam amplitudes suggesting
that on a fixed discretization  the semiclassical limit of the EPRL-FK models is 
given by Regge's discrete formulation of general relativity \cite{Barrett:2009gg, Barrett:2009mw}.

The semiclassical limit of spin foams is based on the  study of the the large spin limit asymptotic behaviour of coherent state spin foam amplitudes. The notion of large spin can be 
defined by the rescaling of quantum numbers and Planck constant according to $j\to \lambda j$ and  $\hbar \to \hbar/\lambda$ and taking $\lambda >>1$.
In this limit the quantum geometry approximates the classical one when tested with suitable states (e.g. coherent states).
However, the geometry remains discrete during this limiting process as the limit is taken on a fixed regulating cellular structure. That is why one usually makes a clear distinction between  
semiclassical limit and the continuum limit. In the semiclassical analysis presented here one can only hope to make contact with 
discrete formulations of classical gravity; hence the importance of Regge calculus in the discussion of this section. 

The key technical ingredient in this analysis is the representation of spin foam amplitudes 
in terms of the coherent state basis introduced in Section \ref{cohecohe}.  Here we follow \cite{Barrett:2009gg, Barrett:2009mw, Barrett:2009cj, Barrett:2009as, Barrett:2010ex}.
The idea of using coherent states and discrete effective actions for the study of the large spin asymptotics of spin foam amplitudes was
put forward in \cite{Conrady:2008mk, Conrady:2008ea}.
The study of the large spin asymptotics has a long tradition in the context of quantum gravity dating back  
to the studied of Ponzano--Regge \cite{ponza}.  More directly related to our discussion here are the early works \cite{Barrett:2002ur, Barrett:1998gs}.
The key idea is to use asymptotic stationary phase methods for the amplitudes written in terms of the discrete actions presented in the previous section.

\subsection{Vertex amplitudes asymptotics}

In this section we review the results of the analysis of the large spin asymptotics of
the EPRL vertex amplitude for both the Riemannian and Lorentztian models. We follow the notation and terminology 
of \cite{Barrett:2009gg} and related papers. 

\subsubsection{$\SU(2)$ $\ftj$-symbol asymptotics}

 As $SU(2)$ BF theory is quite relevant for the construction of the EPRL-FK models, the study of the
 large spin asymptotics of the $SU(2)$ vertex amplitude is a key ingredient in the analysis of \cite{Barrett:2009gg}. 
 The coherent state vertex amplitude is
 \ba
 \ftj(j,\nb)=\int \prod_{a=1}^5 dg_{a} \prod_{1\le a\le b\le 5} \langle n_{ab} |g^{-1}_{a} g_{b}| n_{ba}\rangle^{2j_{ab}},\ea
 which depends on $10$ spins $j_{ab}$ and $20$ normals $n_{ab}\not=n_{ba}$. The previous amplitude can be expressed 
 as
  \ba\label{qq}
 \ftj(j,\nb)=\int \prod_{a=1}^5 dg_{a} \prod_{1\le a\le b\le 5} \exp S_{j,\bn}[g],\ea
\be \label{v-a}
S_{j,\bn}[g] = \sum\limits^{5}_{a < b=1} 2j_{ab} \ln \, \la n_{ab}| g^{-1}_a g_b| \, n_{ba} \ra,
\ee
and the indices $a,b$ label the five edges of a given vertex. The previous expression is exactly equal to the form (\ref{coloring4}) of the BF amplitude. In the case of the EPRL model studied in Sections \ref{eprl-r} and \ref{eprl-l},
the coherent state representation -- see equations \ref{eprl-cohe}, \ref{eprl-cohe-g}, and \ref{eprl-cohe-l} -- is  the basic tool for the study of the semiclassical limit of the models and the relationship with Regge discrete formulation of general relativity. 

In order to study the asymptotics of  (\ref{qq}) one needs to use extended stationary phase methods due to the fact the the action (\ref{v-a}) is complex (see \cite{Conrady:2008mk, Conrady:2008ea}).
The basic idea is that in addition to stationarity one requires real part of the action to be maximal. Points satisfying these two conditions are called \emph{critical points}. 
As the real part of the action is negative definite, the action at critical points is purely imaginary.  

Notice that the action (\ref{v-a}) depends parametrically on 10 spins $j$ and $20$ normals $\bn$. These parameters define the so-called boundary data for the
four simplex $v\in \Delta^{\star}$. Thus, there is an action principle for every given boundary data. The number of critical points and their properties depend on these 
boundary data, hence the asymptotics of the vertex amplitude is a function of the boundary data.  Different cases are studied in detail in \cite{Barrett:2009gg}, here we present their results in the
special case where the boundary data describe a non-degenerate Regge geometry for the boundary of a four simplex, these data are referred to as Regge-like, and satisfy the 
gluing constraints. For such boundary data the action (\ref{v-a}) has exactly two critical points leading to the asymptotic formula
\be 
15j(\lambda j,\bn) \sim  \frac{1}{\lambda^{12}} \left[ N_+ \exp ( i \sum\limits_{a < b}\lambda  j_{ab} \Theta_{ab}^E )+ N_- \exp ( -i  \sum\limits_{a < b}\lambda  j_{ab} \Theta_{ab}^E )\right],
\ee
where 
 $\Theta_{ab}$ the appropriate diahedral angles
defined by the four simplex geometry; finally the 
$N_{\pm}$ are constants that do not scale with $\lambda$.

\subsubsection{The Riemannian EPRL vertex asymptotics} 

The previous result together with the fact that the  EPRL amplitude for $\gamma<1$ is a product of
$SU(2)$ amplitudes with the same $\bn$ in the coherent state representation (\ref{eprl-cohe}) implies the asymptotic formula for the vertex amplitude to be given by the unbalanced square of the above formula \cite{Barrett:2009cj}, namely
\ba \n && 
A^{\va eprl}_v\sim \frac{1}{\lambda^{12}} \left[ N_+  e^{ i \frac{(1-\gamma)}{2} \sum\limits_{a < b}\lambda  j_{ab} \Theta_{ab}^E  } + N_- \ e^{- i  \frac{(1-\gamma)}{2} \sum\limits_{a < b}\lambda  j_{ab} \Theta_{ab}^E } \right]\times \\ && \n \ \ \ \ \ \ \ \ \ \ \ \ \ \ \ \  \ \ \ \ \ \ \ \ \ \ \ \ \ \ \ \  \left[ N_+ \ e^{ i \frac{(1+\gamma)}{2}\sum\limits_{a < b}\lambda  j_{ab} \Theta_{ab}^E  } + N_- \ e^{- i \frac{(1+\gamma)}{2} \sum\limits_{a < b}\lambda  j_{ab} \Theta_{ab}^E  } \right].
\ea
One can write the previous expression as
\ba
A^{\va eprl}_v\sim \frac{1}{\lambda^{12}} \left[2 N_{+}N_{-} \cos \left(   S^{\va E}_{\va Regge} \right) + N_{+}^2  \ e^{i \frac{1}{\gamma}S^{\van E}_{\va Regge} } + N_{-}^2\  e^{ -i \frac{1}{\gamma} S^{\va E}_{\va Regge} } \right].
\ea
where
\be
S^{\van E}_{\va Regge}=\sum_{a < b}\lambda \gamma  j_{ab} \Theta_{ab}^E 
\ee
is the Regge like action for  $\lambda\gamma j_{ab}=A_{ab}$ the ten triangle areas (according to the LQG area spectrum (\ref{aarreeaa})).
Remarkably, the above asymptotic formula is also valid for the case $\gamma>1$ \cite{Barrett:2009gg}. The first term in the vertex asymptotics is 
in essence the expected one: it is the analog of the $6j$ symbol asymptotics in three dimensional spin foams. Due to their explicit dependence on the Immirzi parameter, the last two 
terms are somewhat strange from the point of view of the continuum  field theoretical view point. However, this seems to be a 
peculiarity of the Riemannian theory alone as the results of \cite{Barrett:2009mw} for the Lorentzian models show.
{ Non geometric configurations are exponentially surpressed}. Finally, recently is has been shown that it is possible to restrict the state 
sum in order to obtain complex amplitudes \cite{Engle:2012yg} and \cite{Mikovic:2011zx, Mikovic:2011zv}.

\subsubsection{Lorentzian EPRL model}

To each solution one can associate a second solution corresponding to a parity related $4$-simplex and, consequently, the asymptotic formula has two terms. It is given, up to a global sign, by the expression
\be 
A^{\va eprl}_{v} \sim  \frac{1}{\lambda^{12}} \left[ N_+ \exp \left( i \lambda \gamma \sum\limits_{a < b} j_{ab} \Theta_{ab}^L \right) + N_- \exp \left(- i \lambda \gamma \sum\limits_{a < b} j_{ab} \Theta_{ab}^L \right) \right],
\ee
where $N_{\pm}$ are constants that do not scale. 
{Non geometric configurations are exponentially surpressed}. By taking into account discrete symmetries it might be possible to further improve the asymtotics of the model \cite{Rovelli:2012yy}.

\subsection{Full spin foam amplitudes asymptotics}

In \cite{Conrady:2008ea} Freidel and Conrady gave a detailed description of the coherent state representation of the various spin foam models described so far.
In particular they provided the definition of the effective discrete actions associated to each case which we presented in (\ref{fk-cohe}). This provides the basic elements for setting up the asymptotic 
analysis presented in \cite{Conrady:2008mk} (the first results on the semiclassical limit of the new spin foam models) which is similar to the studies of the asymptotic of the vertex amplitude reviewed above but more general
in the sense that the semiclassical limit of a full spin foam configuration (involving many vertices) is studied.
The result is technically more complex as one studies now critical points of the action associated to a coloured complex which in addition of depending on group variables  
$g$ it depends on the coherent state parameters $\bn$. The authors of \cite{Conrady:2008mk} write Equation (\ref{fk-cohe}) in the following way
\ba
\label{fk-semi}Z^{\va \gamma}_{fk}=\sum \limits_{ j_f }  \ \prod\limits_{f \in \Delta^{\star}} {\rm d}_{(1-\gamma)\frac{j_f}{2}}{\rm d}_{(1+\gamma)\frac{j_f}{2}} W_{\Delta^{\star}}^{\gamma} (j_f), \ea
where
\be
W_{\Delta^{\star}}^{\gamma}(j_f)=\int \prod_{e\in  \Delta^{\star}} {\rm d}_{|1-\gamma|\frac{j_{ef}}{2}} {\rm d}_{(1+\gamma)\frac{j_{ef}}{2}}  dn_{ef} dg^{-}_{ef}dg^{+}_{ef}
\ \exp{(S^{\va fk\ \gamma}_{j^{\pm},\bn}[g^{\pm}])}.
\ee
They show that those solutions of the equations of motion of the effective discrete action that are non geometric (i.e. the contrary of Regge like) are not critical and hence exponentially suppressed in the scaling $j_{f}\to \lambda j_f$ with $\lambda >>1$. If configurations are geometric (i.e. Regge like) one has two kind of contributions to the amplitude assymptotics: those coming from degenerate and non-degenerate configurations. If one (by hand) restricts to the non-degenerate configurations then one has
\be
W_{\Delta^{\star}}^{\gamma}(j_f)\sim \frac{c}{\lambda^{(33 n_e-6n_v-4n_f)}} \exp(i\lambda S^{\va E}_{\va Regge}(\Delta^{\star},j_f)),
\ee
where $n_e$, $n_v$, and $n_f$ denote the number of edges, vertices, and faces in the two complex $\Delta^{\star}$ respectively.
For more recent results on asymptotics of spin foam amplitudes of arbitrary simplicial complexes see \cite{Han:2011rf}.

\subsection{Fluctuations: two point functions.}\label{physics}

The problem of computing the two point function and higher correlation functions in the context of spin foam 
has received lots of attention recently. The framework for the definition of the correlation functions in the background independent 
setting has been generally discussed by Rovelli in \cite{Rovelli:2005yj} and correspods to a special application of a more general 
proposal investigated by Oeckl \cite{Oeckl:2011yi, Oeckl:2011qd, Oeckl:2010ra, Colosi:2009cp, Colosi:2008jf, Colosi:2008fv, Colosi:2007bj, Oeckl:2006rs}. It was then applied to the Barrett-Crane model in \cite{Alesci:2007tg, Alesci:2007tx, Bianchi:2006uf}, where it was discovered that certain components of the two point function could not yield the expected result 
compatible with Regge gravity in the semiclassical limit. This was used as the main motivation of the weakening of the
imposition of the Plebanski constraints leading to the new models. Soon thereafter it was argued that the difficulties of the Barrett-Crane model
where indeed absent in the EPRL model \cite{Alesci:2008ff}. The two point function for the EPRL model was calculated in \cite{Bianchi:2009ri} and it was shown to produce
a result in agreement with that of Regge calculus\cite{Bianchi:2008ae,Magliaro:2008zz}
in the limit $\gamma\to 0$. 

The fact that, for the new model,  the double scaling limit  $\gamma\to 0$ and $j\to \infty$ with $\gamma j$=constant  defines the appropriate regime where the fluctuation  behave as in Regge gravity (in the leading order) has been further clarified in \cite{Magliaro:2011dz}. This indicates that the quantum fluctuations in the new models are more general than simply metric fluctuations. The fact the the new models are not metric at all scales should not be surprising as we know that the Plebanski constraints that produce metric general relativity out of BF theory has been implemented only semiclassically (in the large spin limit). At the deep Planckian regime fluctuations are more general than metric. {However, it not clear at this stage why this is controlled by the Immirzi parameter.}

All the previous calculations involve a complex with a single four-simplex. The first computation involving more than one simplex was performed in 
\cite{Mamone:2009pw,Bianchi:2006uf} for the case of the Barrett--Crane model. 
Certain peculiar properties were found and it is not clear at this stage whether these issues remain in the EPRL model. 
Higher order correlation functions have been computed in \cite{Rovelli:2011kf}, the results are in agreement with Regge gravity in the $\gamma\to 0$ limit.

\clearpage

\part{Three dimensional gravity}
\label{part3d}

In this part we shall review the quantization of three dimensional
gravity introducing in this simpler context what in this case can be
called the \emph{spin foam representation} of the quantum dynamics. We
present this case in great detail as it is completely solvable. More
precisely, it allows for an explicit realization of the path integral
quantization for a generally covariant system as generally discussed
in Section~\ref{valin}. Moreover, the path integral takes the form of
a sum of spin foam amplitudes which share all the \emph{kinematical}
quantum geometric properties with their four dimensional relatives.

The simplicity of this theory allows for the clear-cut illustration of
some central conceptual difficulties associated with the quantization
of a generally covariant system. We will use the insights provided by
3d gravity to support the discussion of certain difficult conceptual
issues for the spin foam approach in Section~\ref{sci}.

\clearpage

\section{Spin Foams for 3-Dimensional Gravity}
\label{sfm3d}

Three dimensional gravity is an example of BF theory for which the
spin foam approach can be implemented in a rather simple way.
Despite its simplicity the theory allows for the study of many of
the conceptual issues to be addressed in four dimensions. In
addition spin foams for BF
theory are the basic building block of 4-dimensional gravity
models. For a beautiful presentation of BF theory and its relation
to spin foams see~\cite{baez5}. For simplicity we study the
Riemannian theory; the Lorentzian generalization of the
results of this section have been studied in~\cite{fre1}.

\subsection{\emph{Spin foams} in 3d quantum gravity}
\label{pipo}

Here we derive the \emph{spin foam representation} of LQG in a simple
solvable example: 2+1 gravity. For the definition other approaches to
3d quantum gravity see Carlip's book~\cite{carlip}.

\subsection{The classical theory}

Riemannian gravity in 3 dimensions is a theory with no local degrees
of freedom, i.e., a topological theory. Its action (in the first order
formalism) is given by
\begin{equation}
\label{bfaction}
 S_{}[e,\omega]=\int
\limits_{ M}{\rm Tr}(e\wedge F(\omega)),
\end{equation}
where $M=\Sigma\times \R$ (for $\Sigma$ an arbitrary Riemann
surface), $\omega$ is an $SU(2)$-connection and the triad $e$ is
an $su(2)$-valued 1-form. The gauge symmetries of the action are
the local $SU(2)$ gauge transformations
\begin{equation}
\label{gauge1}
\delta e = \left[e,\alpha \right], \ \ \ \ \ \ \ \ \ \delta \omega
= d_{\omega} \alpha,
\end{equation}
where $\alpha$ is a ${{su(2)}}$-valued 0-form, and the
`topological' gauge transformation
\begin{equation}\label{gauge2}
\delta e = d_{\va \omega} \eta, \ \ \ \ \ \ \ \ \ \delta \omega =
0,
\end{equation}
where $d_{\va \omega}$ denotes the covariant exterior derivative
and $\eta$ is a ${\tt su(2)}$-valued 0-form. The first
invariance is manifest from the form of the action, while the
second is a consequence of the Bianchi identity, $d_{\va
\omega}F(\omega)=0$. The gauge symmetries are so large that all
the solutions to the equations of motion are locally pure gauge.
The theory has only global or topological degrees of freedom.

Upon the standard 2+1 decomposition, the phase space in these
variables is parametrized by the pull back to $\Sigma$ of $\omega$
and $e$. In local coordinates one can express them in terms of the
2-dimensional connection $A_a^{i}$ and the triad field
$E^b_j=\epsilon^{bc} e^k_c \delta_{jk}$ where $a=1,2$ are space
coordinate indices and $i,j=1,2,3$ are $su(2)$ indices. The Poisson
bracket is given by
\be
\{A_a^{i}(x),
E^b_j(y)\}=\delta_a^{\, b} \; \delta^{i}_{\, j} \;
\delta^{(2)}(x,y).
\end{equation}
Local symmetries of the theory are generated by the first class
constraints
\be 
D_b E^b_j = 0, \qquad F_{ab}^i(A) = 0,
\end{equation}
which are referred to as the Gauss law and the curvature constraint
respectively. This simple theory has been quantized in various ways in
the literature~\cite{carlip}, here we will use it to introduce the
\emph{spin foam representation}.

\subsection{\emph{Spin foams} from the Hamiltonian formulation}
\label{SFH}

The physical Hilbert space, $\Hhp$, is defined by those `states in
$\Hk$' that are annihilated by the constraints. As discussed
in~\cite{Rovelli:2004tv,bookt}, spin network states solve the Gauss
constraint -- $\widehat{ D_a E^a_i}|s\rangle=0$ -- as they are
manifestly $SU(2)$ gauge invariant. To complete the quantization one
needs to characterize the space of solutions of the quantum curvature
constraints $\widehat F^i_{ab}$, and to provide it with the physical
inner product. As discussed in Section~\ref{valin}, we can achieve
this if we can make sense of the following formal expression for the
generalized projection operator $P$:
\be
\PP=\int D[N] \ {\rm exp}(i\int
 \limits_{\Sigma} {\rm Tr}[ N \widehat{F}(A)])=\prod \limits_{x\subset \Sigma}
 \delta[\widehat {F(A)}],
\label{pis}
\end{equation}
where $N(x)\in {\rm su(2)}$. Notice that this is just the field
theoretical analog of equation (\ref{pipi}). $P$ will be defined below
by its action on a dense subset of test-states called the cylindrical
functions ${\rm Cyl}\subset \Hk$. If $P$ exists then we have
\be
\label{exists}
\langle s PU[N],s^{\prime}\rangle=\langle s P,s^{\prime}\rangle\  \forall\ \  s,s^{\prime}\in {\rm Cyl}, \ 
N(x)\in su(2)
\ee
where $U[N]=\exp(i\int {\rm Tr}[i N\hat F(A)])$. $P$ can be viewed as
a map $P:{\rm Cyl}\rightarrow K_F\subset {\rm Cyl}^{\star}$ (the space
of linear functionals of $\rm Cyl$) where $K_F$ denotes the kernel of
the curvature constraint.  The physical inner product is defined as
\be
\label{meo} \langle s^{\prime},s\rangle_p:=\langle s^{\prime} P,s
\rangle,
\ee
where $\langle,\rangle$ is the inner product in $\Hk$, and the
physical Hilbert space as
\be
\Hhp:={\overline{{\rm Cyl}/J}} \ \ \ {\rm for}\ \ \ J:=\{s \in {\rm
Cyl}\ \ {\rm s.t.} \ \ \langle s,s\rangle_p=0\}
\label{null},
\ee
where the bar denotes the standard Cauchy completion of the quotient
space in the physical norm.

One can make (\ref{pis}) a rigorous definition if one introduces a
regularization.  A regularization is necessary to avoid the naive UV
divergences that appear in QFT when one quantizes non-linear
expressions of the canonical fields such as $F(A)$ in this case (or
those representing interactions in standard particle physics). A
rigorous quantization is achieved if the regulator can be removed
without the appearance of infinities, and if the number of ambiguities
appearing in this process is under control.  We shall see that all
this can be done in the simple toy example of this section.

\epubtkImage{}{%
\begin{figure}[htbp]
\centerline{\includegraphics[width=7cm]{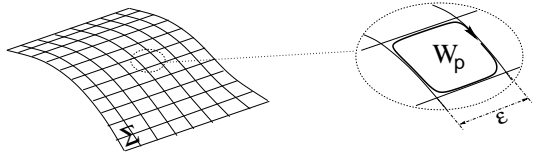}}
\caption{Cellular decomposition of the space manifold $\Sigma$ (a
  square lattice of size $\epsilon$ in this example), and the
  infinitesimal plaquette holonomy $W_p[A]$.}
\label{regu}
\end{figure}}

We now introduce the regularization. Given a partition of $\Sigma$ in
terms of 2-dimensional plaquettes of coordinate area $\epsilon^2$
(Figure~\ref{regu}) one can write the integral
\be
 \label{***} F[N]:=\int\limits_{\Sigma} {\rm Tr}[ N {F}(A)]=
\lim_{\epsilon\rightarrow 0}\ \sum_{p} \epsilon^2 {\rm
Tr}[N_{p} F_{p}]
\end{equation}
as a limit of a Riemann sum, where $N_{p}$ and $F_{p}$ are values of
the smearing field $N$ and the curvature $\epsilon^{ab}F_{ab}^i[A]$ at
some interior point of the plaquette $p$ and $\epsilon^{ab}$ is the
Levi-Civita tensor. Similarly the holonomy $W_{p}[A]$ around the
boundary of the plaquette $p$ (see Figure~\ref{regu}) is given by 
\be
W_{p}[A]=\mathbbm{1}+ \epsilon^2 F_{p}(A)+{\cal O}(\epsilon^2).
\end{equation}
The previous two equations imply that $F[N]=\lim_{\epsilon\rightarrow
  0}\sum_{p} {\rm Tr}[N_pW_p]$, and lead to the following definition:
given $s, s^{\prime} \in {\rm Cyl}$ (think of \emph{spin network}
states) the physical inner product (\ref{meo}) is given by
\be
\label{new}
\langle s^{\prime}P,s\rangle := \lim_{\epsilon\rightarrow 0} \ \ \langle s \prod_{p} \ \int \ dN_{p}
\ {\rm exp}(i {\rm Tr}[ N_{p} {W}_{p}]), s\rangle.
\end{equation}
The partition is chosen so that the links of
the underlying spin network graphs border the plaquettes. One can
easily perform the integration over the $N_{p}$ using the
identity (Peter--Weyl theorem)
\be
\label{pw}
\int \ dN \
 {\rm exp}(i {\rm Tr}[ N {W}])=\sum_{j} \ (2j + 1) \ {\rm
   Tr}[\stackrel{j}{\Pi}\!(W)],
\end{equation}
where $\stackrel{j}{\Pi}\!(W)$ is the spin $j$ unitary irreducible
representation of $SU(2)$. Using the previous equation 
\be
\label{final}
 \langle s^{\prime}P,s\rangle := \lim_{\epsilon\rightarrow 0} \ \ \prod^{n_p(\epsilon)}_{p} \sum_{j_p}
 (2j_p+1)\ \langle s^{\prime}\
{\rm Tr}[\stackrel{ j_p}{\Pi}\!({W}_{p})]), s\rangle,
\end{equation}
where the spin $j_p$ is associated to the p-th plaquette, and
$n_p(\epsilon)$ is the number of plaquettes. Since the elements of the
set of Wilson loop operators $\{W_p\}$ commute, the ordering of
plaquette-operators in the previous product does not matter. The limit
$\epsilon \rightarrow 0$ exists and one can give a closed expression
for the physical inner product. That the regulator can be removed
follows from the orthonormality of $SU(2)$ irreducible representations
which implies that the two spin sums associated to the action of two
neighboring plaquettes collapses into a single sum over the action of
the \emph{fusion} of the corresponding plaquettes (see
Figure~\ref{figurin}). One can also show that it is
finite\epubtkFootnote{The physical inner product between spin network
  states satisfies the following inequality
\[
\left| \langle s,s^{\prime}\rangle_p\right|\le C \sum_{j}
(2j+1)^{2-2g},
\]
for some positive constant $C$. The convergence of the sum for genus
$g\ge 2$ follows directly. The case of the sphere $g=0$ and the torus
$g=1$ can be treated individually\cite{karim}.}, and satisfies all the
properties of an inner product~\cite{karim}.

\epubtkImage{}{%
\begin{figure}[htbp]
 \centerline{\hspace{0.0cm} \( \sum \limits_{j k} (2j+1)(2k+1)\!\!\!\!
\begin{array}{c}
\includegraphics[height=2cm]{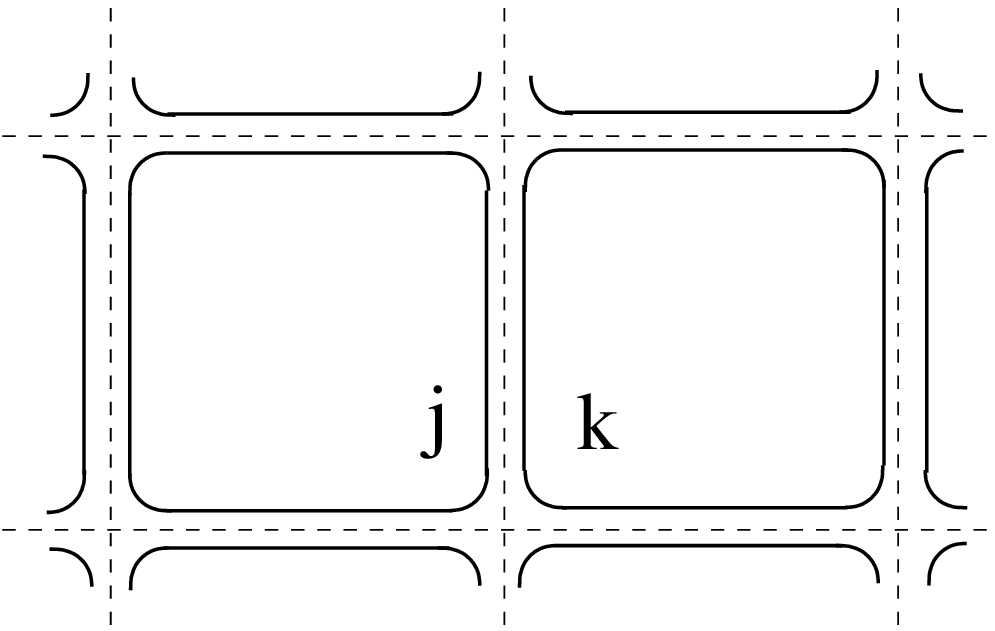}
\end{array} \!\!\!\! =  \sum \limits_k (2k+1)\!\!\!\!
\begin{array}{c}
\includegraphics[height=2cm]{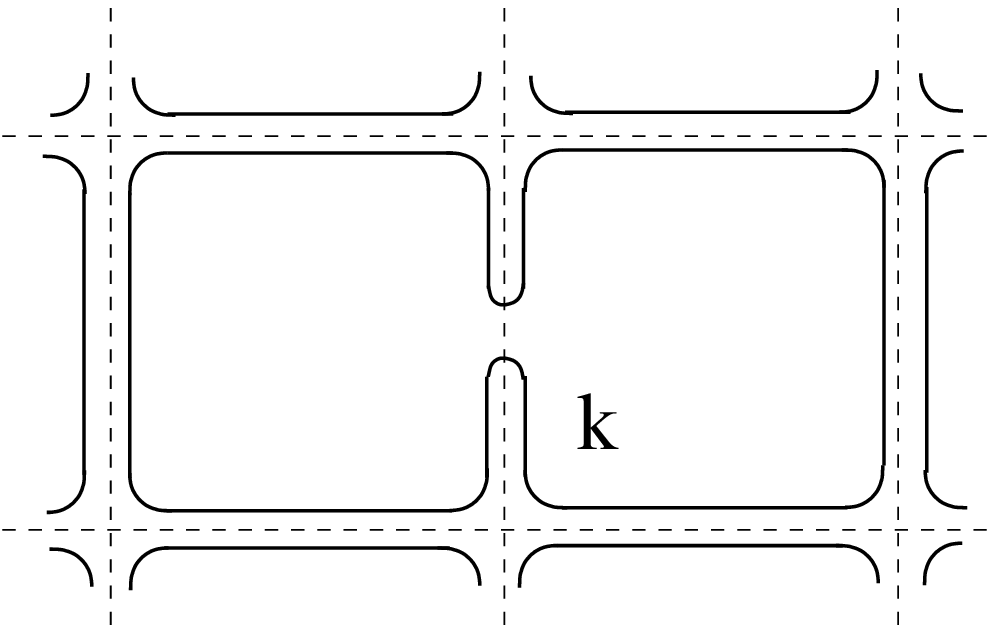}
\end{array}
\) }
\caption{In two dimensions the action of two neighboring
 plaquette-sums on the vacuum is equivalent to the action of a single
 larger plaquette action obtained from the fusion of the original
 ones. This implies the trivial \emph{scaling} of the physical inner
 product under refinement of the regulator and the existence of a well
 defined limit $\epsilon\rightarrow 0$.}
\label{figurin}
\end{figure}}

\subsection{The spin foam representation}

Each ${\rm Tr}[\stackrel{j_p}{\Pi}\!(W_{p})]$ in (\ref{final})
acts in $\Hk$ by creating a closed loop in the $j_{p}$ representation
at the boundary of the corresponding plaquette (Figure~\ref{pito} and
\ref{pitolon}).

\epubtkImage{}{%
\begin{figure}[htbp]
\centerline{\hspace{0.5cm} \( {\rm Tr}[\stackrel{k}{\Pi}\!(W_{p})]
\rhd \!\!\!\!\!\!\!\!\!\!\!\!\!\!\!\!\begin{array}{c}
\includegraphics[width=3.6cm]{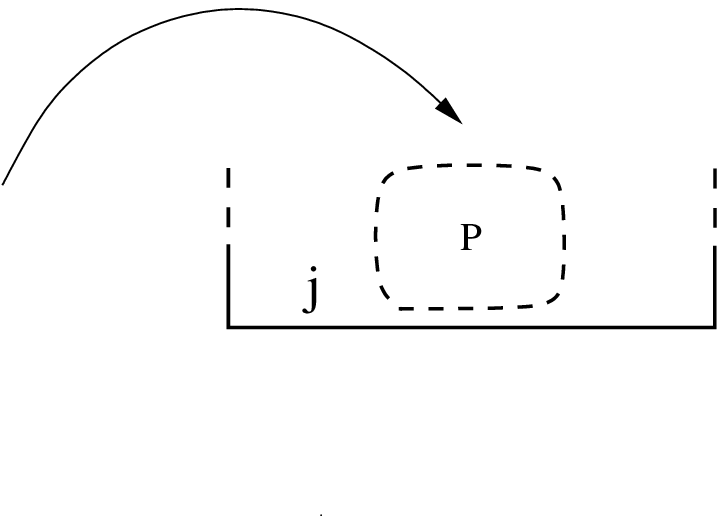}
\end{array}
=
\sum \limits_{m} N_{j,m,k}
\begin{array}{c}\includegraphics[width=2.5cm]{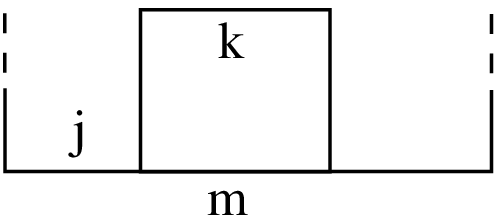}
\end{array}
\) }
\caption{Graphical notation representing the action of one
plaquette holonomy on a \emph{spin network} state. On the right is
the result written in terms of the \emph{spin network} basis. The
amplitude $N_{j,m,k}$ can be expressed in terms of Clebsch-Gordan
coefficients.}
\label{pito}
\end{figure}}

Now, in order to obtain the \emph{spin foam representation} we
introduce a non-physical (\emph{coordinate time}) as follows: Instead
of working with one copy of the space manifold $\Sigma$ we consider $n_p(\epsilon)$ copies as  a
discrete folliation $\{\Sigma_p\}_{p=1}^{n_p(\epsilon)}$. Next we represent 
each of the ${\rm Tr}[\stackrel{j_p}{\Pi}\!(W_{p})]$ in (\ref{final})
on the corresponding $\Sigma_p$.
If one inserts the partition of unity between in
$\Hk$ between the slices, graphycally \be \!\!\!\mathbbm{1}=\!\!\!\!\!\!\! \sum
\limits_{\gamma\subset \Sigma, \{j\}_{\gamma}} \!\!\!\!
|\gamma,\{j\}\rangle\langle\gamma,\{j\}|\begin{array}{c}
\includegraphics[width=5cm]{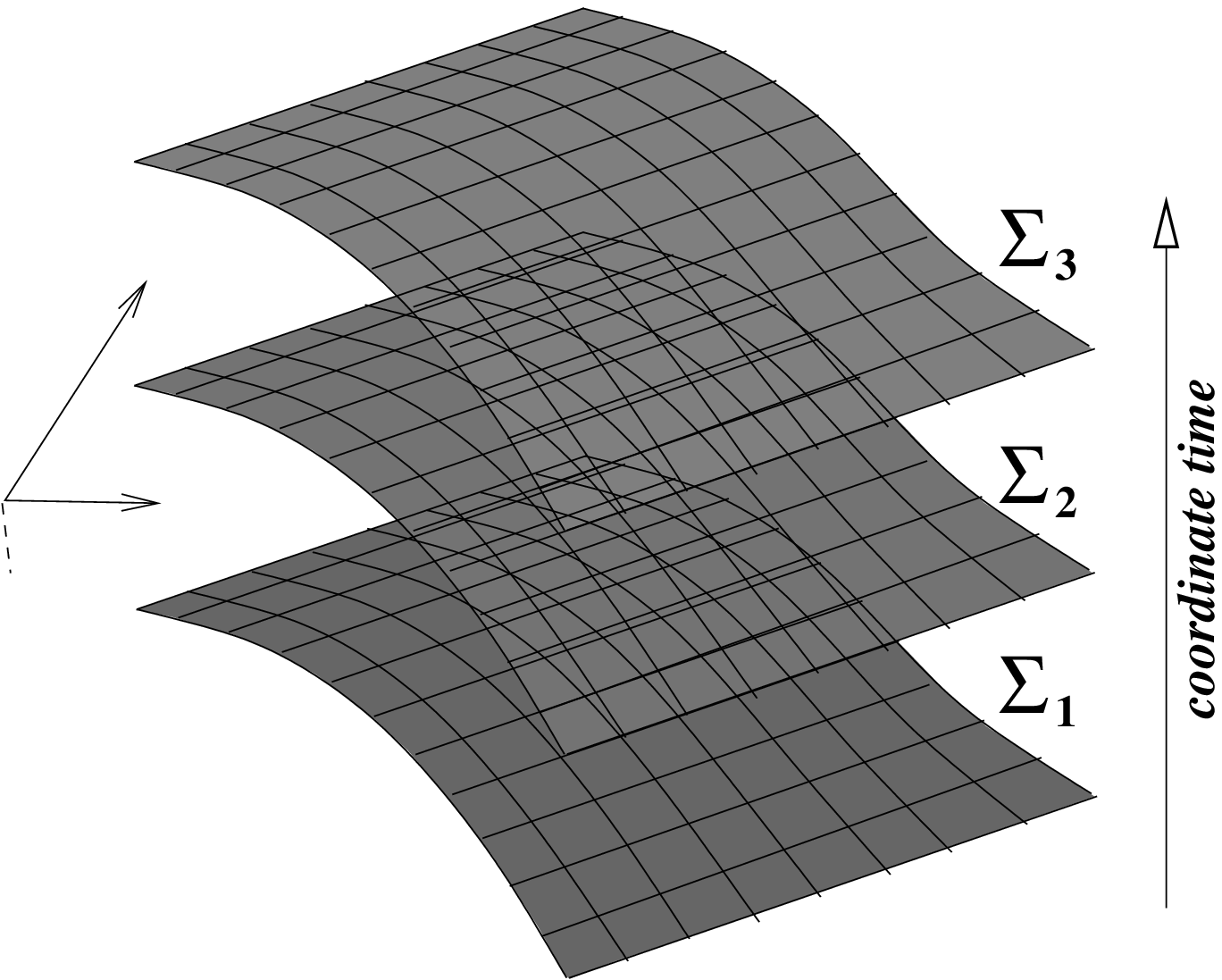}
                \end{array}
\label{e} \end{equation} where the sum is over the complete basis
of \emph{spin network} states $\{|\gamma,\{j\}\rangle\}$ -- based on all
graphs $\gamma \subset \Sigma$ and with all possible spin
labelling -- one arrives at a sum over
spin-network histories representation of $\langle s,s^{\prime}\rangle_p$. More
precisely, $\langle s^{\prime},s\rangle_p$ can be expressed as a sum over
amplitudes corresponding to a series of transitions that can be viewed
as the `time evolution' between the `initial' \emph{spin network}
$s^{\prime}$ and the `final' \emph{spin network} $s$. This is illustrated in
the two simple examples of Figure~\ref{lupy} and \ref{vani}); on the
r.h.s.\ we illustrate the continuum spin foam picture obtained when
the regulator is removed in the limit $\epsilon\rightarrow 0$.

\epubtkImage{}{%
\begin{figure}[htbp]
 \centerline{\hspace{0.0cm}\(
\begin{array}{ccc}
\includegraphics[height=1cm]{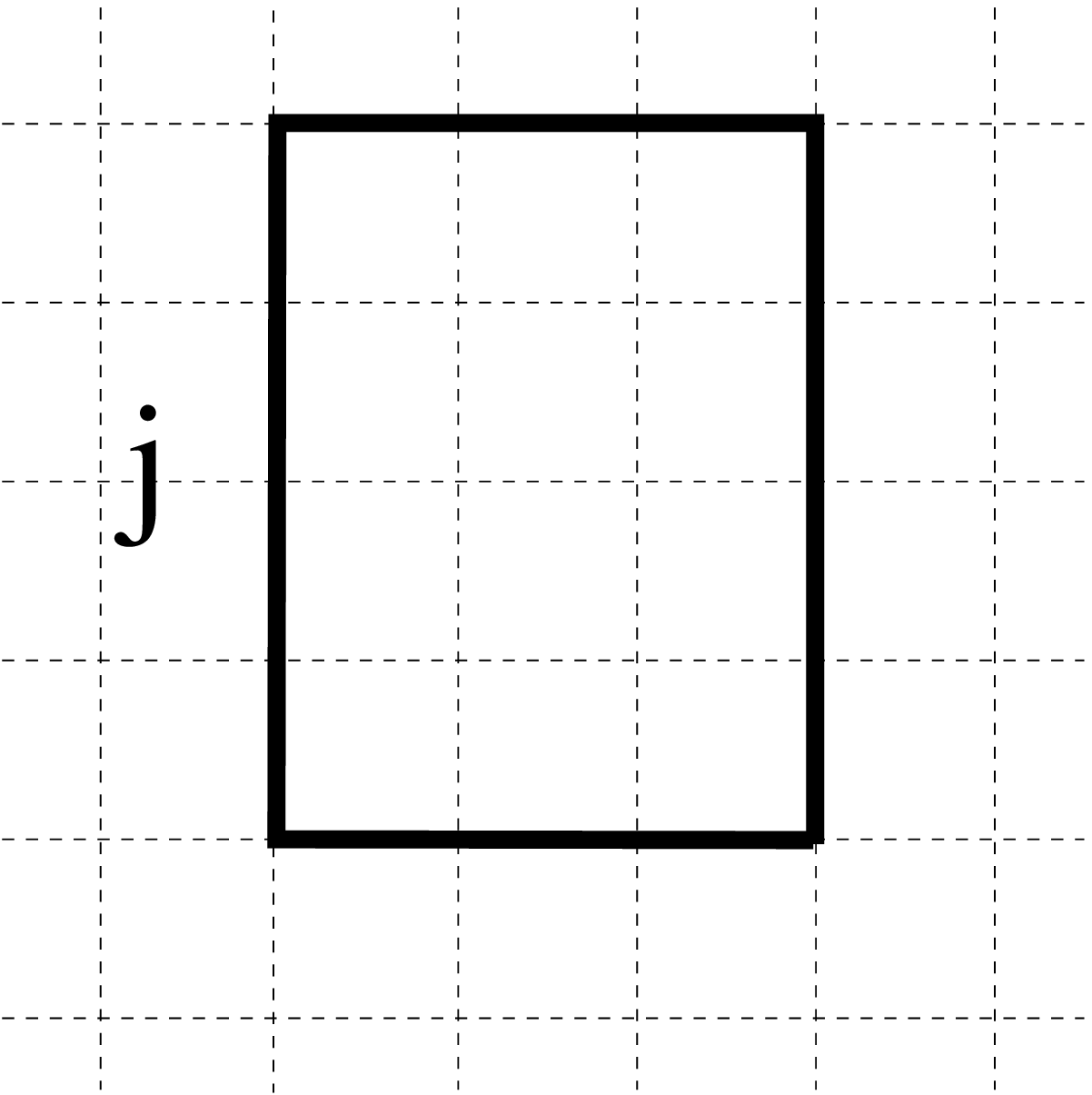} \\
\includegraphics[height=1cm]{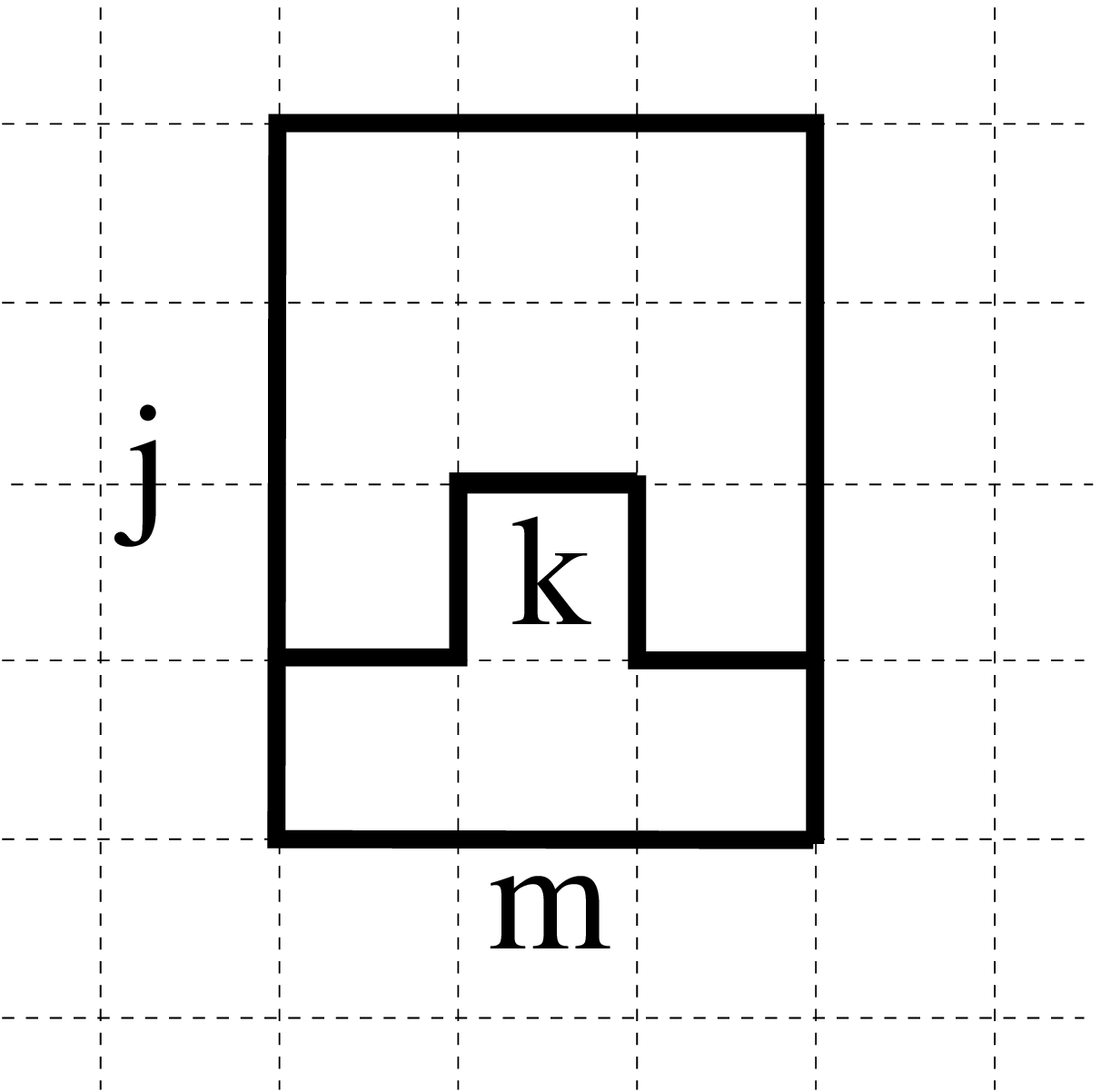}\\
\includegraphics[height=1cm]{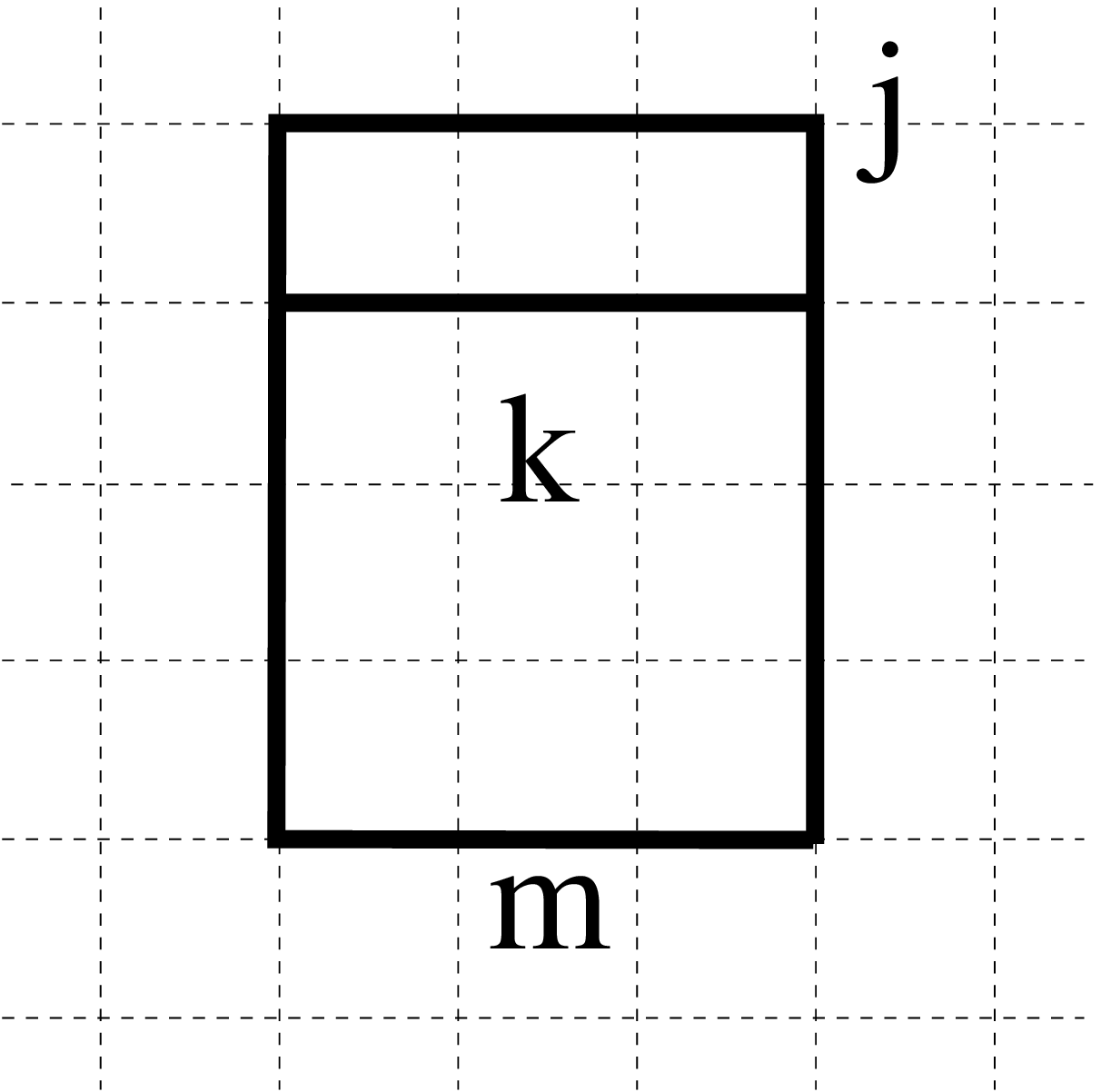}
\end{array}
\begin{array}{ccc}
\includegraphics[height=1cm]{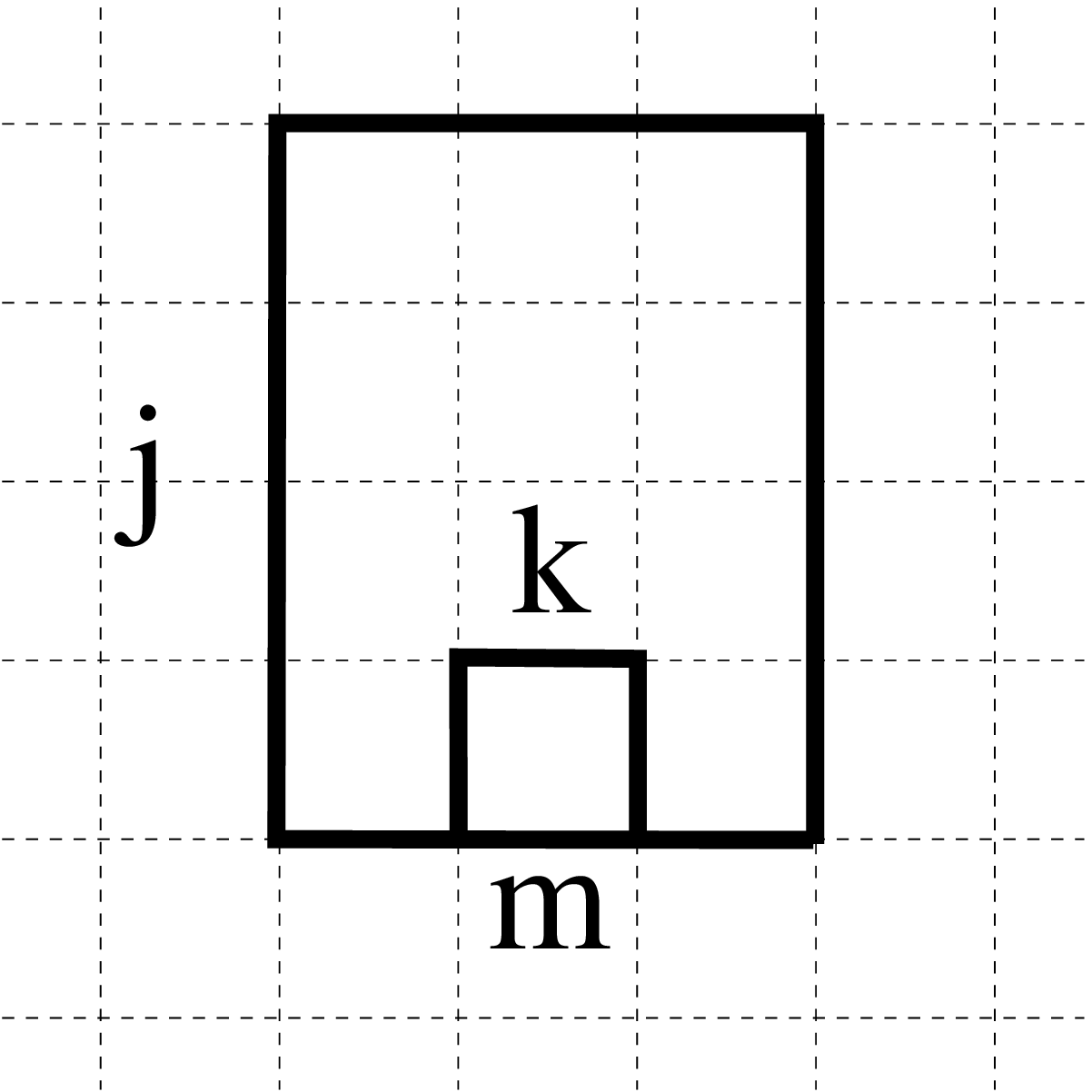} \\
\includegraphics[height=1cm]{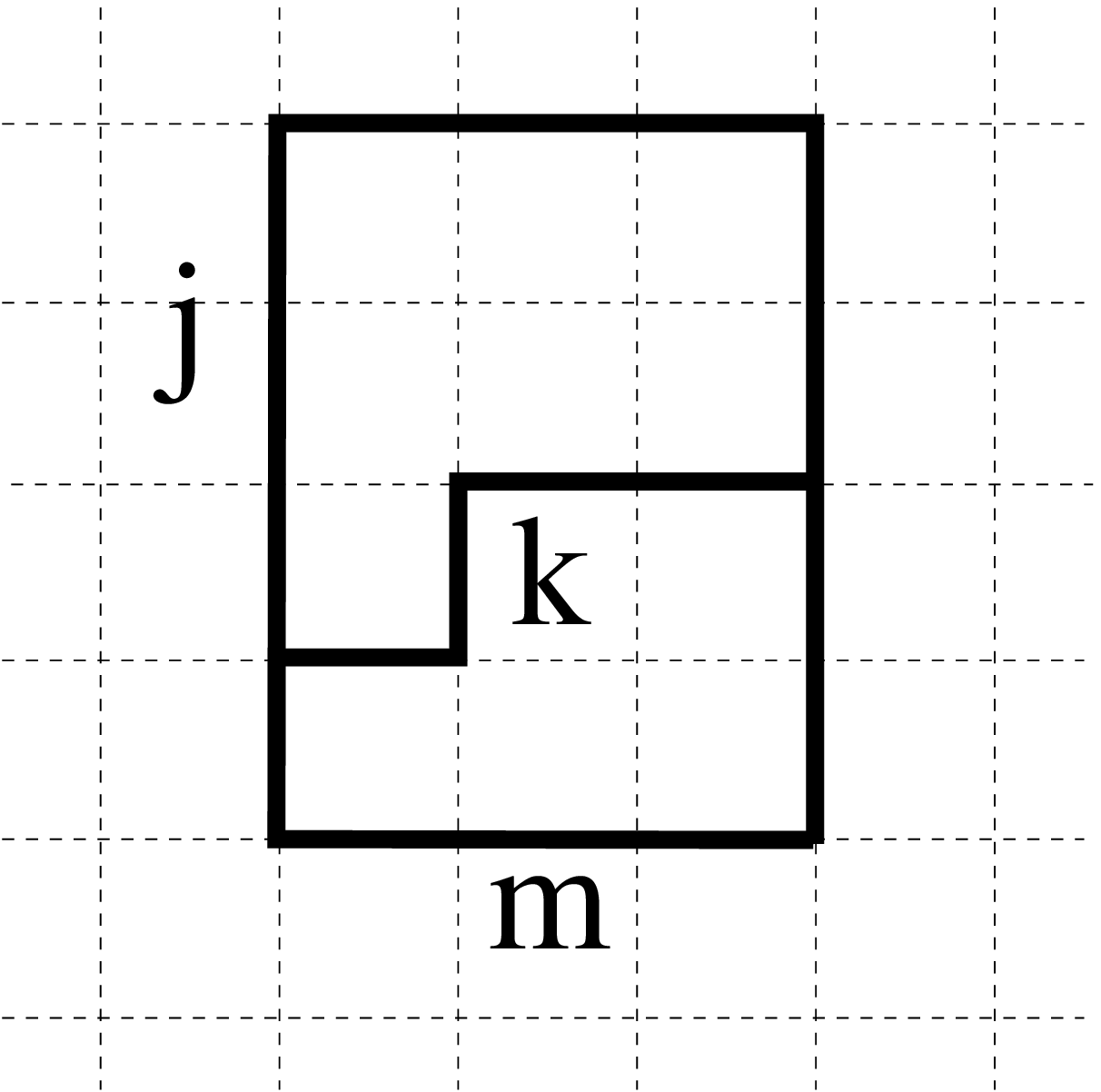}\\
\includegraphics[height=1cm]{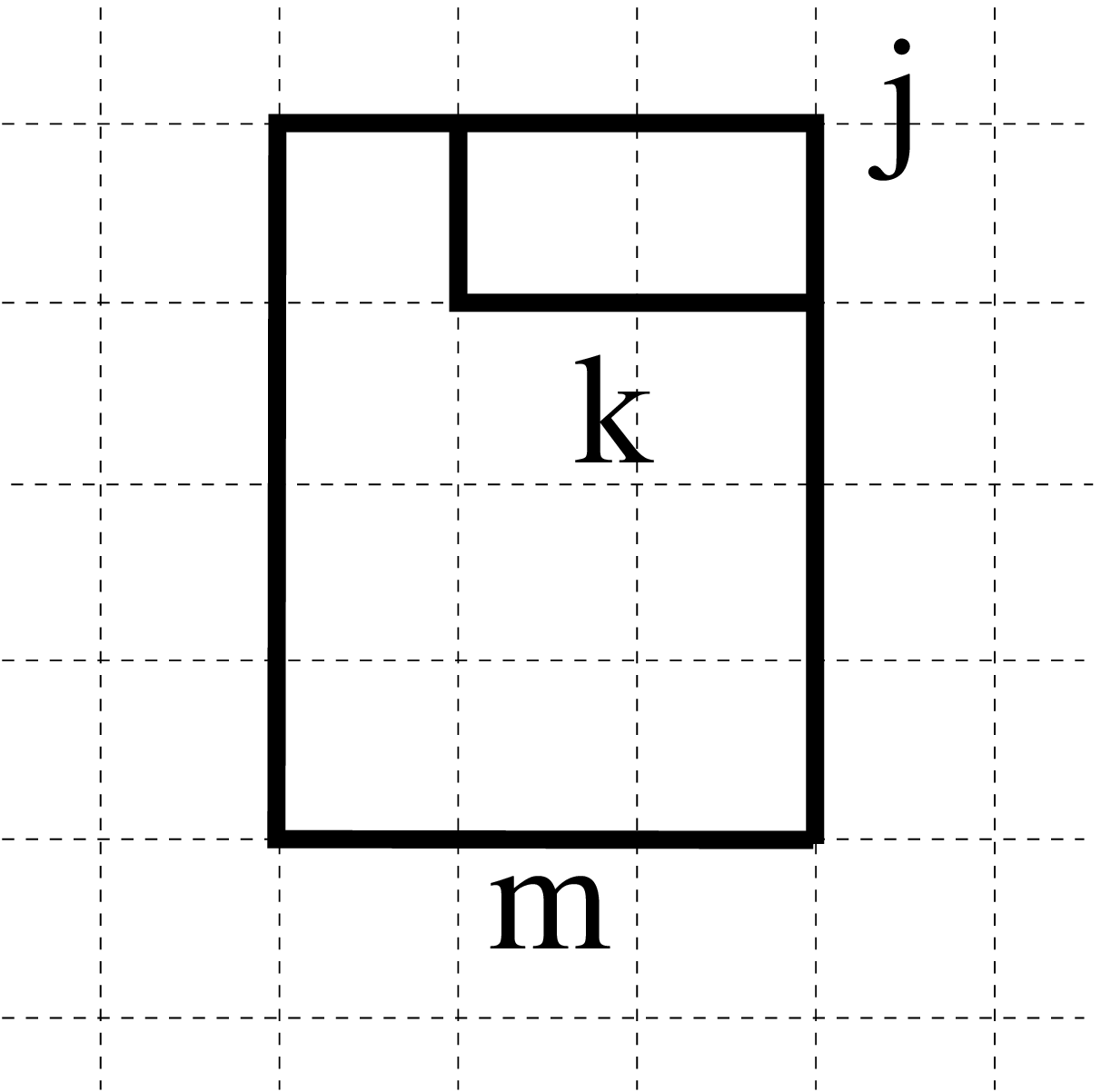}
\end{array}
\begin{array}{ccc}
\includegraphics[height=1cm]{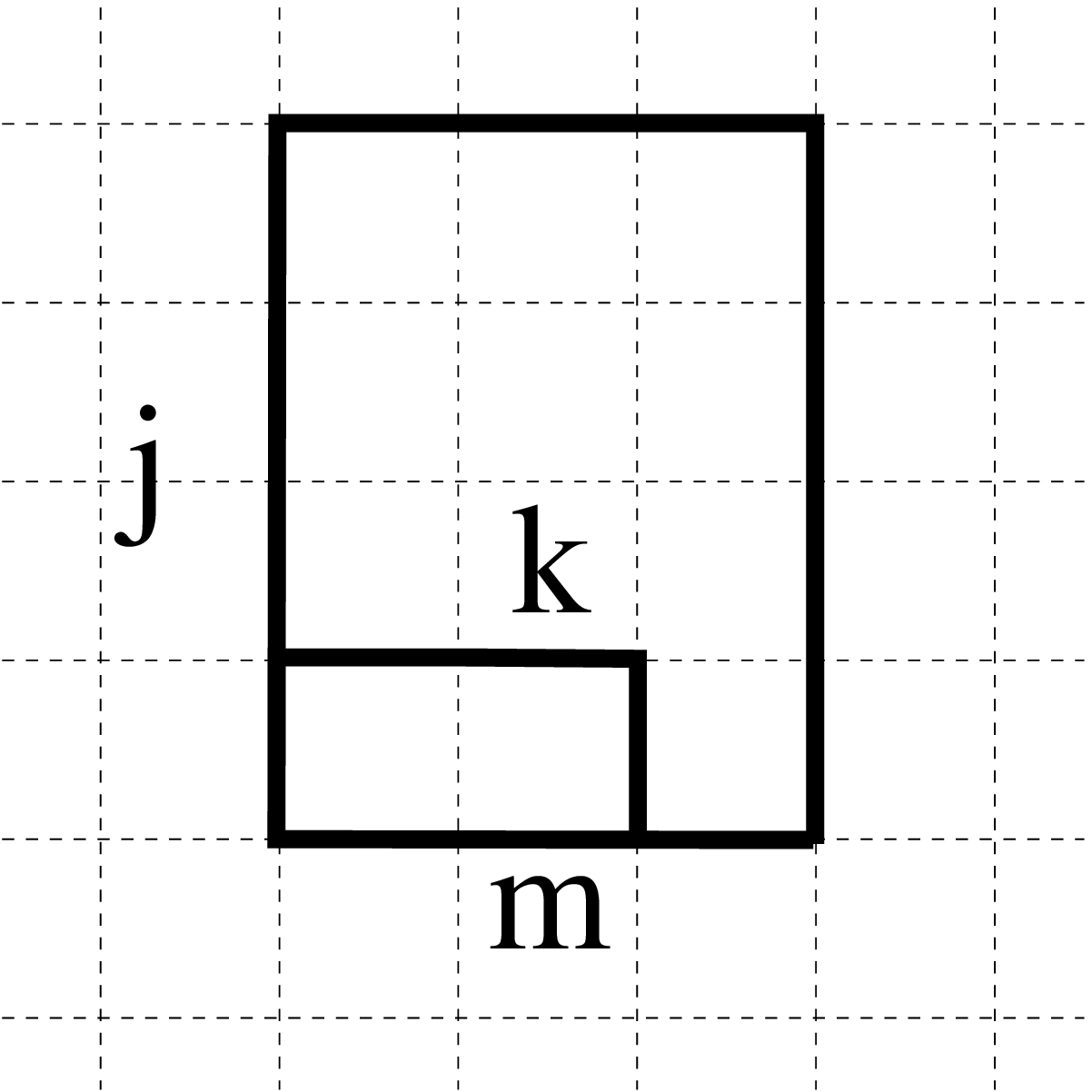} \\
\includegraphics[height=1cm]{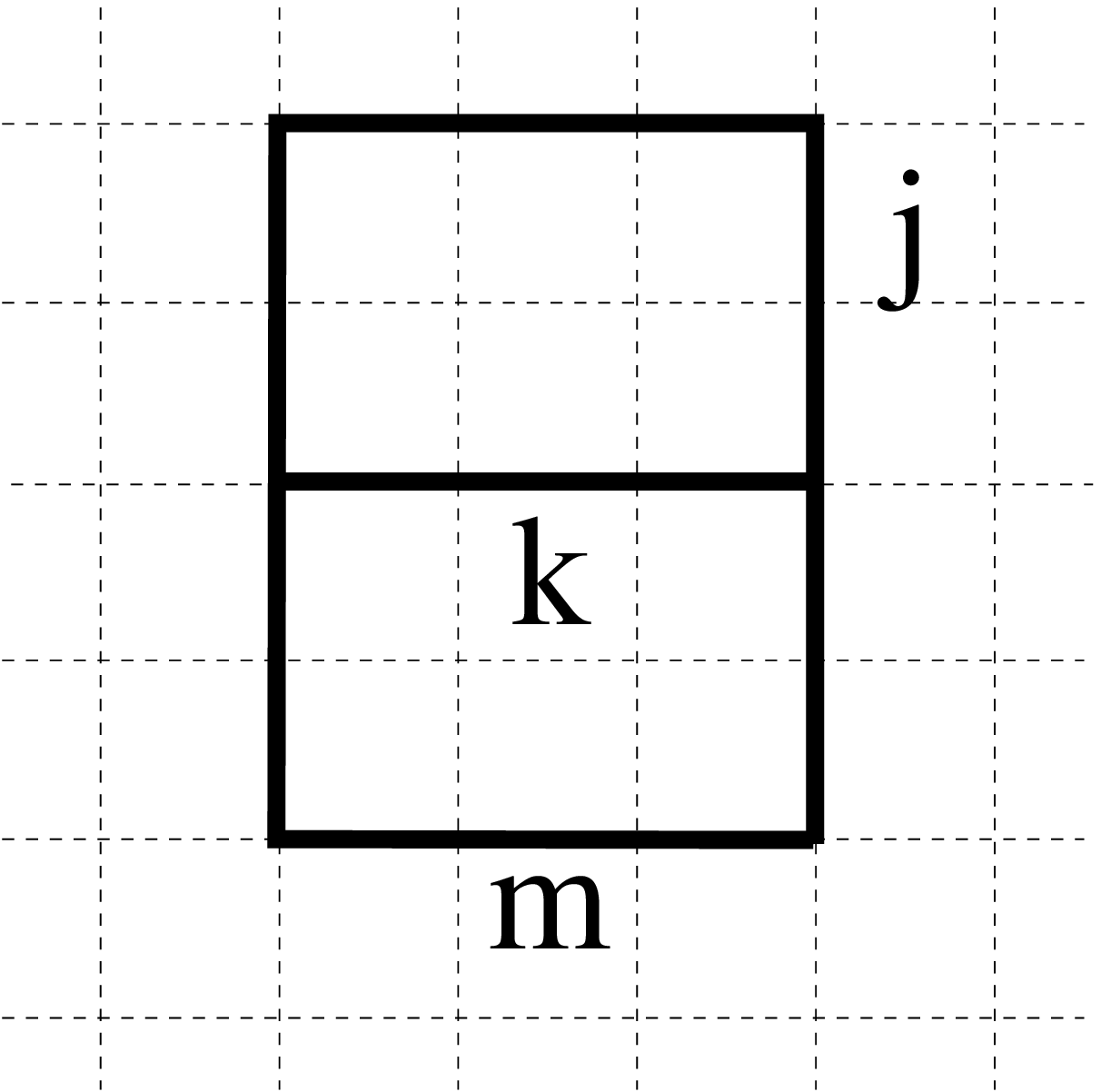}\\
\includegraphics[height=1cm]{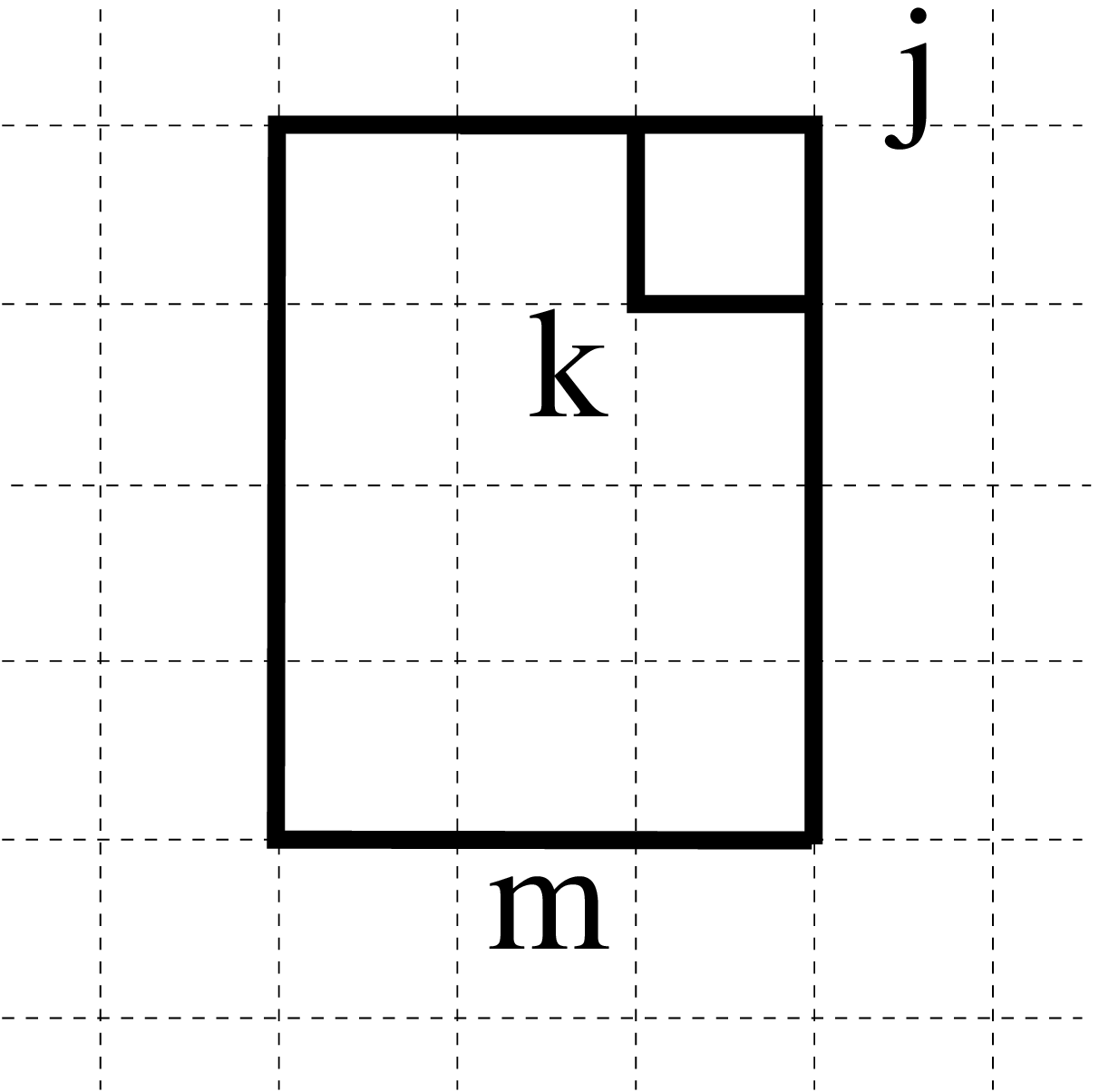}
\end{array}
\begin{array}{ccc}
\includegraphics[height=1cm]{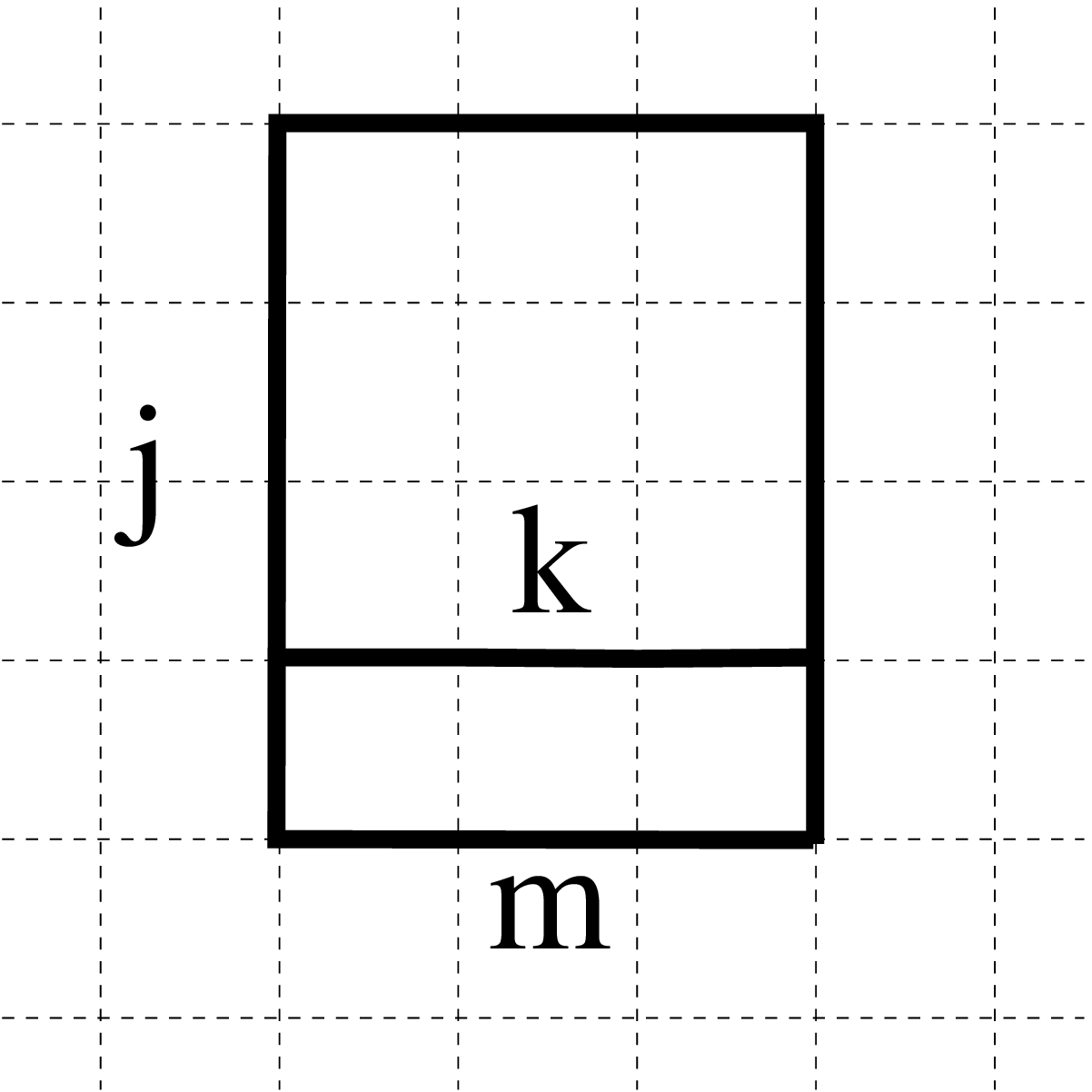} \\
\includegraphics[height=1cm]{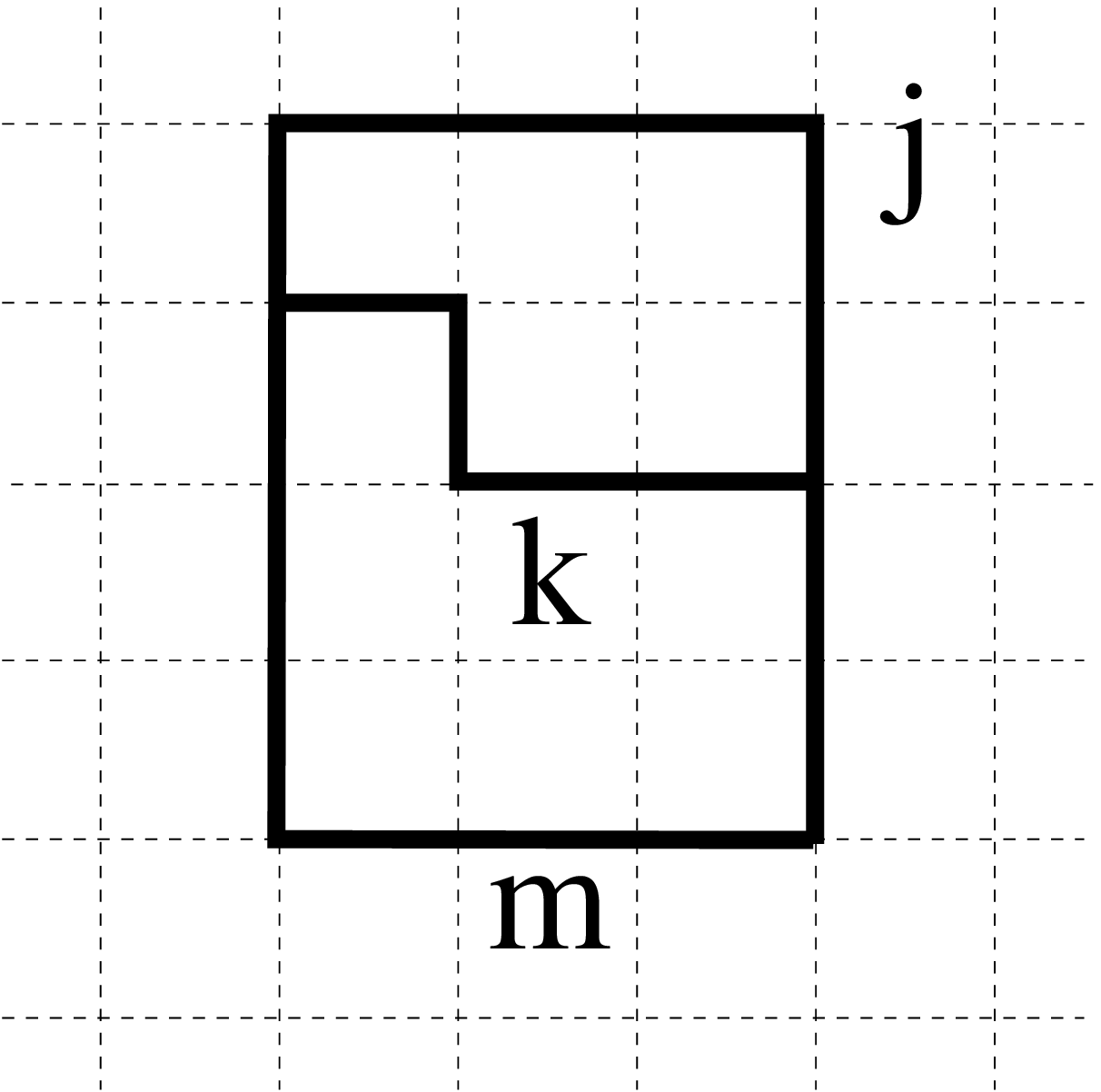}\\
\includegraphics[height=1cm]{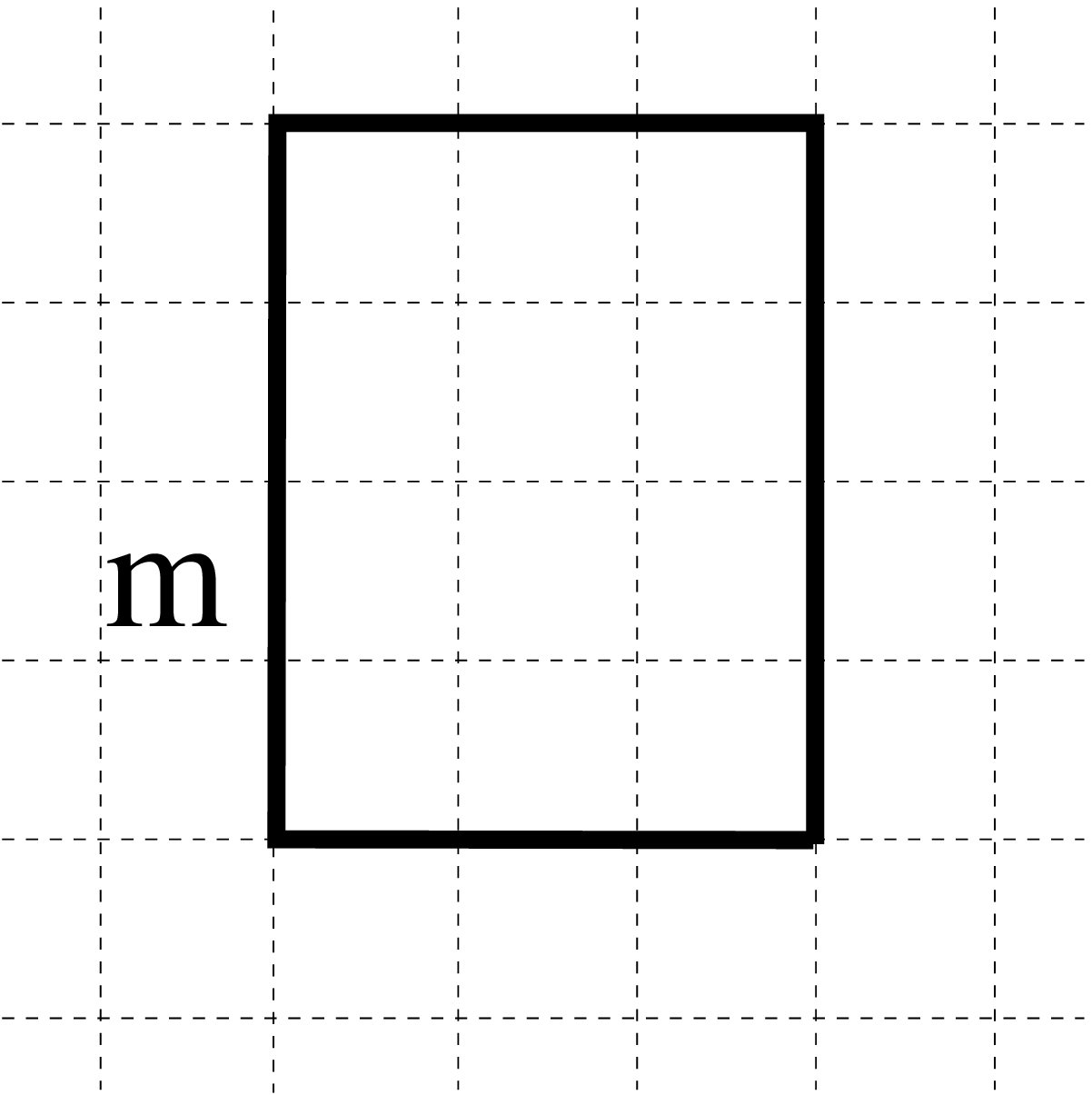}
\end{array}\begin{array}{c}
\includegraphics[height=.3cm]{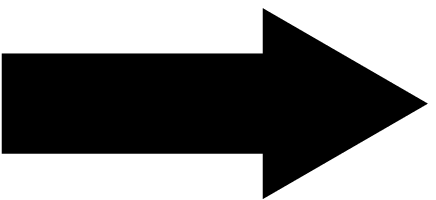}
\end{array}\!\!\!
\begin{array}{c}
\includegraphics[height=3cm]{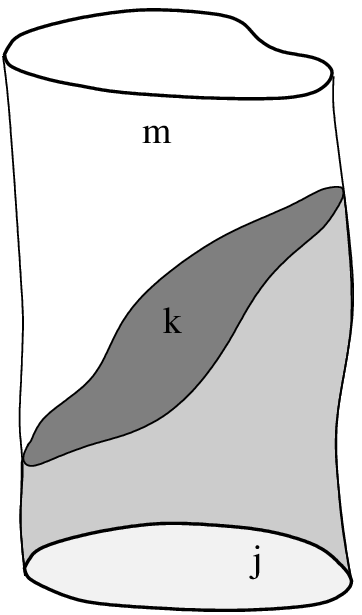}
\end{array}
\)}
\caption{A set of discrete transitions in the
loop-to-loop physical inner product obtained by a series of
transitions as in Figure~\ref{pito}. On the right, the continuous
\emph{spin foam representation} in the limit $\epsilon\rightarrow
0$.}
\label{lupy}
\end{figure}}

Spin network nodes evolve into edges while \emph{spin network}
links evolve into 2-dimensional faces. Edges inherit the
intertwiners associated to the nodes and faces inherit the spins
associated to links. Therefore, the series of transitions can be
represented by a 2-complex whose 1-cells are labelled by
intertwiners and whose 2-cells are labelled by spins. The places
where the action of the plaquette loop operators create new links
(Figure~\ref{pitolon} and \ref{vani}) define  0-cells or
vertices. These foam-like structures are the so-called \emph{spin
foams}. The \emph{spin foam} amplitudes are purely combinatorial
and can be explicitly computed from the simple action of the loop
operator in $\Hk$.

\epubtkImage{}{%
\begin{figure}
\centerline{\hspace{0.5cm} \( {\rm
Tr}[\stackrel{n}{\Pi}\!(W_{p})]\rhd
\!\!\!\!\!\!\!\!\!\!\!\!\!\!\!\!\begin{array}{c}
\includegraphics[width=2.5cm]{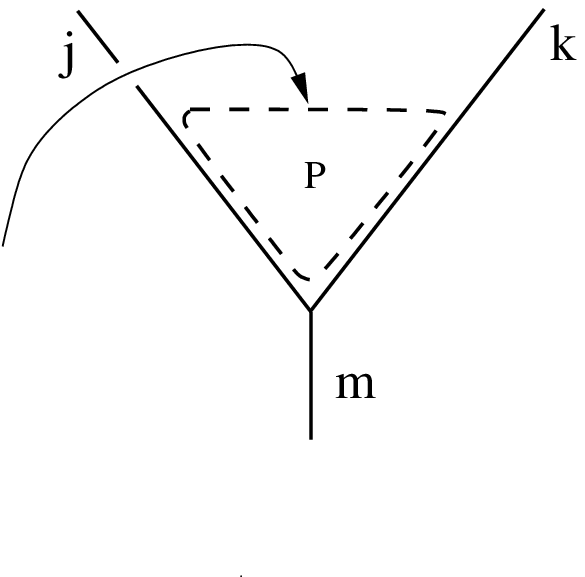}
\end{array}
=
\sum\limits_{o,p} \frac{1}{\Delta_n \Delta_j \Delta_k \Delta_m}
\left\{\begin{array}{ccc}j\ \ k \ \ m\\ n\ \  o\ \  p
\end{array}\right\}
\begin{array}{c}
\includegraphics[width=2.5cm]{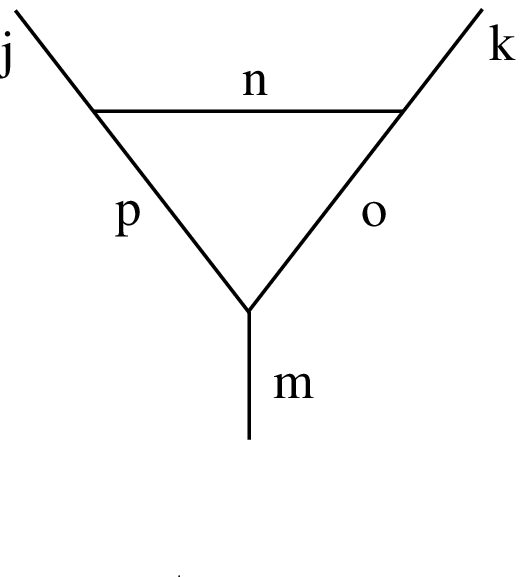}
\end{array}
\) }
\caption{Graphical notation representing the action of one
plaquette holonomy on a \emph{spin network} vertex. The object in
brackets ($\{\}$) is a $6j$-symbol and $\Delta_j:=2j+1$.}
\label{pitolon}
\end{figure}}

The physical inner product takes the standard Ponzano--Regge form when
the \emph{spin network}
states $s$ and $s^{\prime}$ have only 3-valent nodes. Explicitly,
\be \label{3dc} \langle s,s^{\prime}\rangle_p = \sum
\limits_{ F_{s\rightarrow s^{\prime}}} \ \prod_{f \subset F_{s\rightarrow s^{\prime}}} (2
j_f+1)^{\frac{\nu_f}{2}}
                \prod_{v\subset F_{s\rightarrow s^{\prime}}}
                \!\!\!\!\begin{array}{c}
                \includegraphics[width=2.3cm]{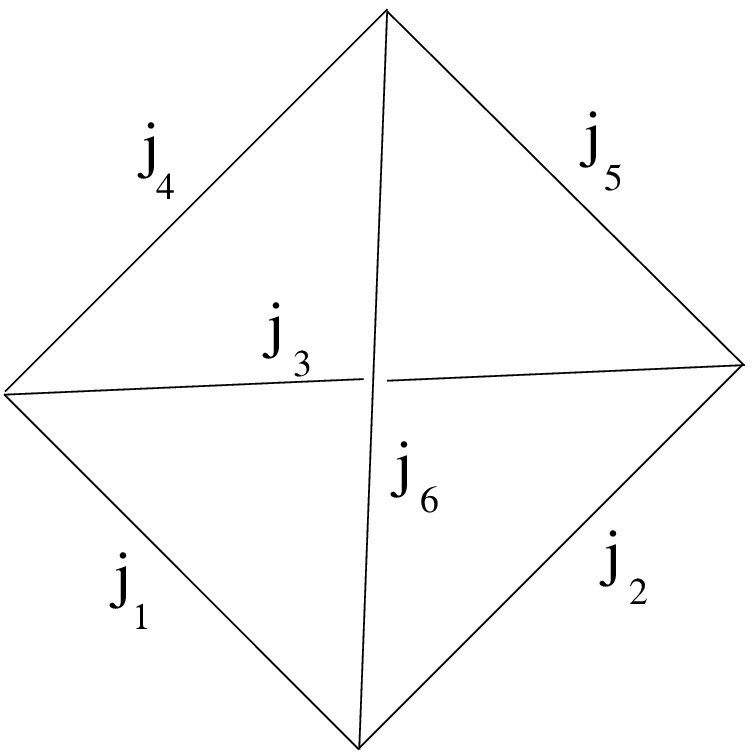}
                \end{array},
\end{equation} where the sum is over all the 
spin foams interpolating between $s$ and $s^{\prime}$ (denoted
$F_{s\rightarrow s^{\prime}}$, see Figure~\ref{spino}), $f\subset F_{s\rightarrow s^{\prime}}$ denotes
the faces of the spin foam (labeled by the spins $j_f$), $v\subset F_{s\rightarrow s^{\prime}}$
denotes vertices, and $\nu_f=0$ if $f
\cap s \not= 0 \wedge f \cap s^{\prime}\not= 0$, $\nu_f=1$ if $f
\cap s \not= 0 \vee f \cap s^{\prime}  \not= 0$,  and $\nu_f=2$ if
$f \cap s = 0 \wedge f \cap s^{\prime}= 0$. The tetrahedral
diagram denotes a $6j$-symbol: the amplitude obtained by means of
the natural contraction of the four intertwiners corresponding to
the 1-cells converging at a vertex.  More generally, for arbitrary
\emph{spin networks}, the vertex amplitude corresponds to
$3nj$-symbols, and $\langle s,s^{\prime}\rangle_p$ takes the same
general form.

\epubtkImage{}{%
\begin{figure}[htbp]
 \centerline{\hspace{0.5cm}\(
\begin{array}{ccc}
\includegraphics[height=1.4cm]{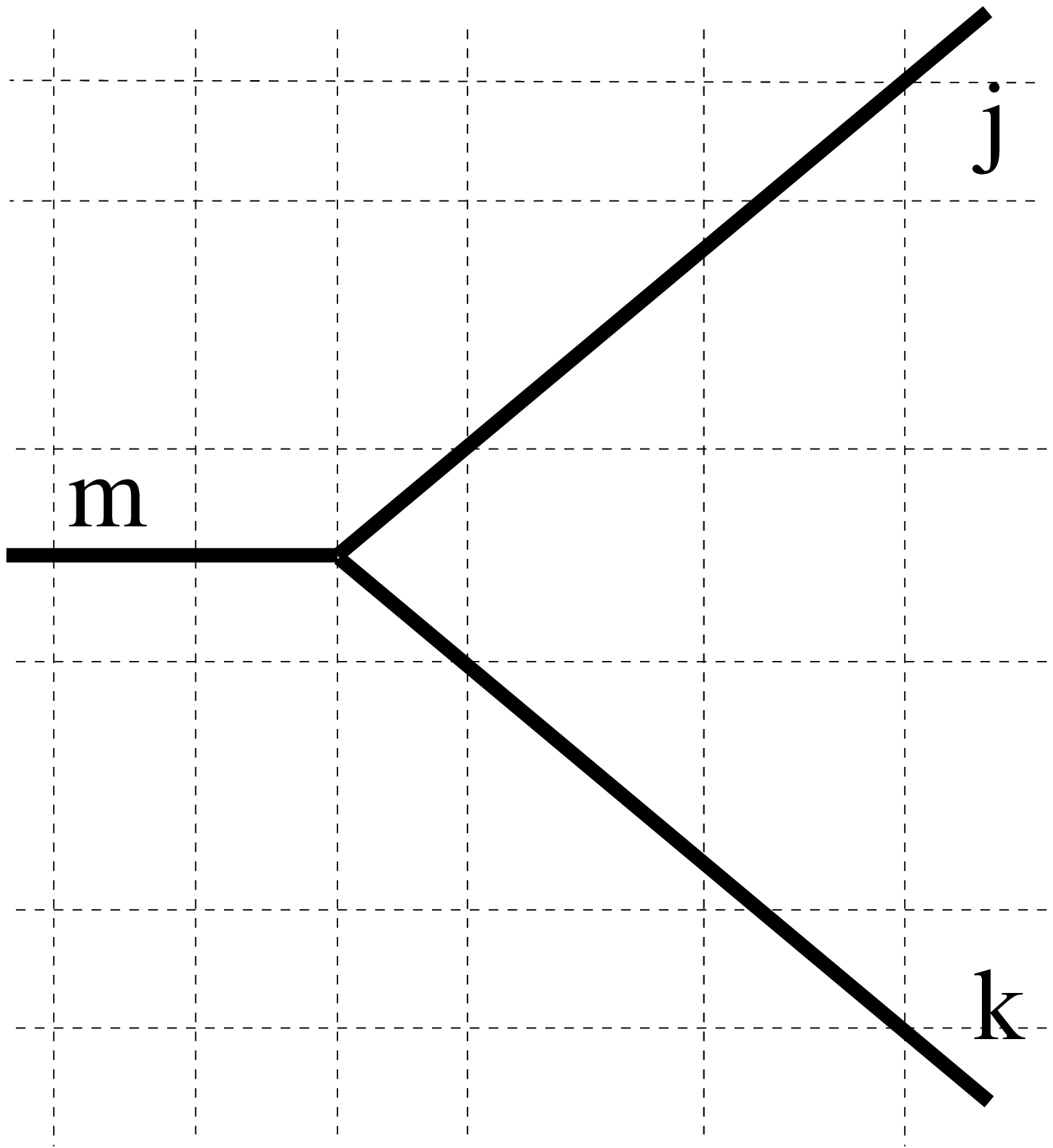} \\
\includegraphics[height=1.4cm]{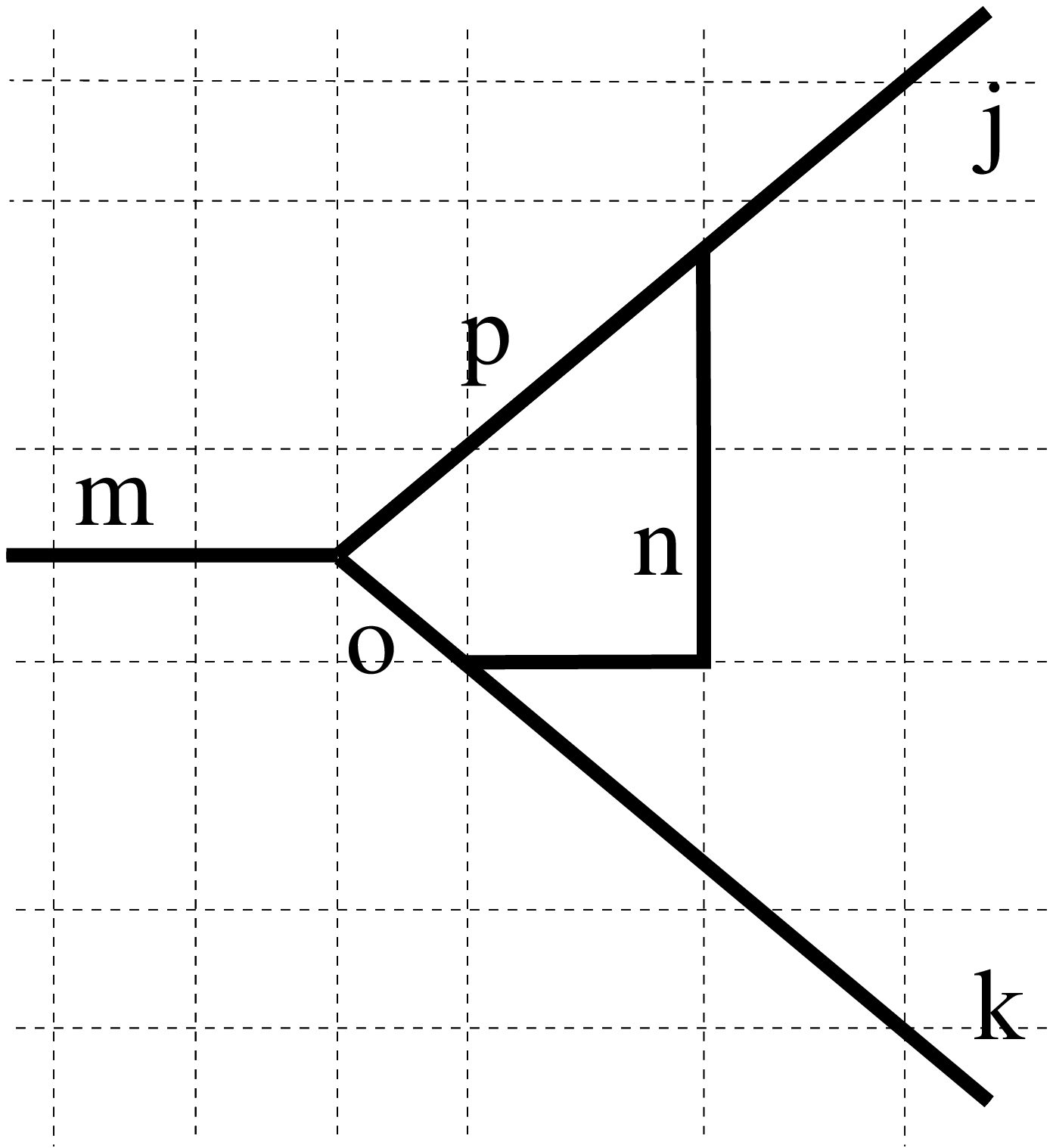} \\
\includegraphics[height=1.4cm]{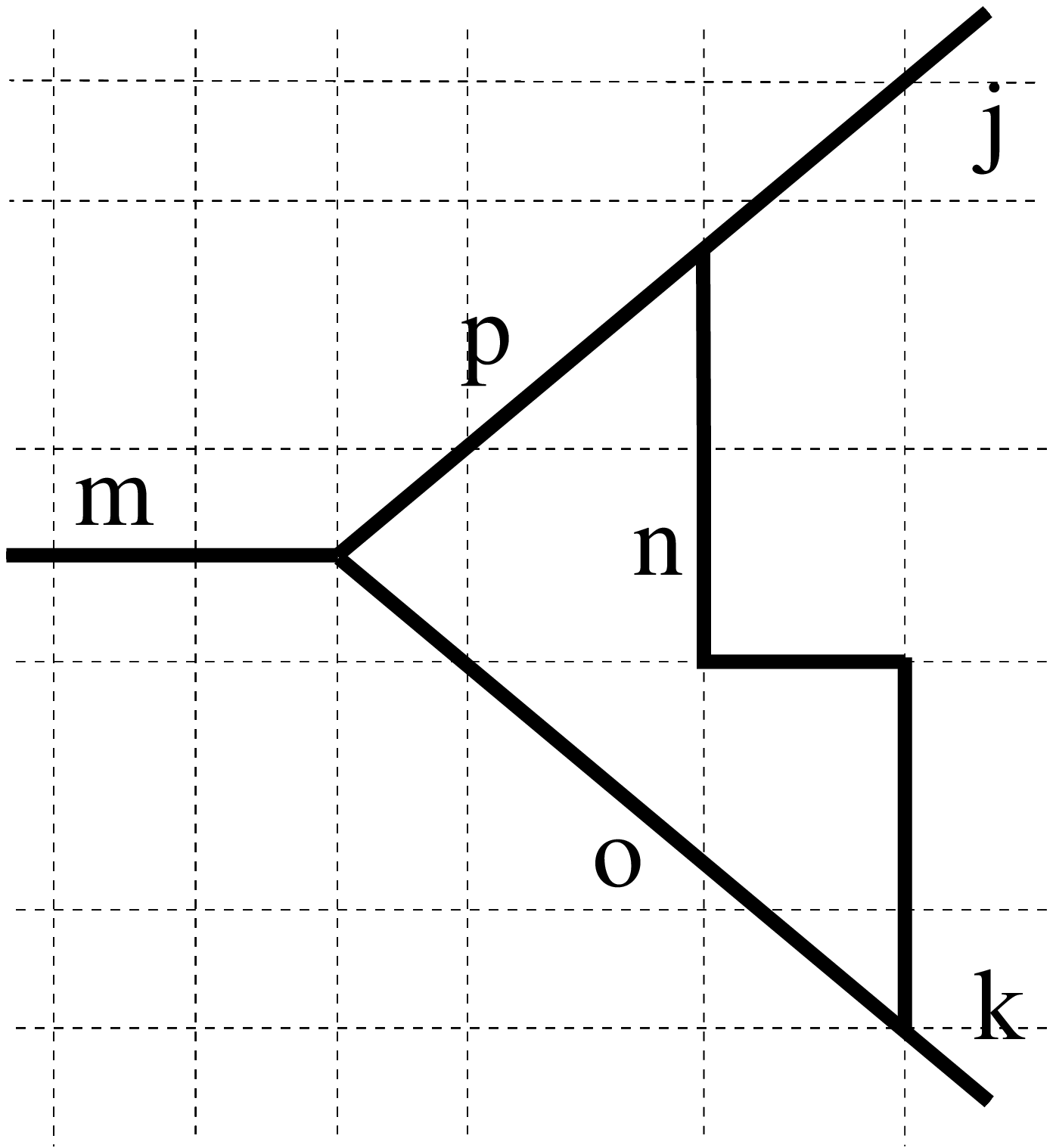}
\end{array}
\begin{array}{c}
\includegraphics[height=1.4cm]{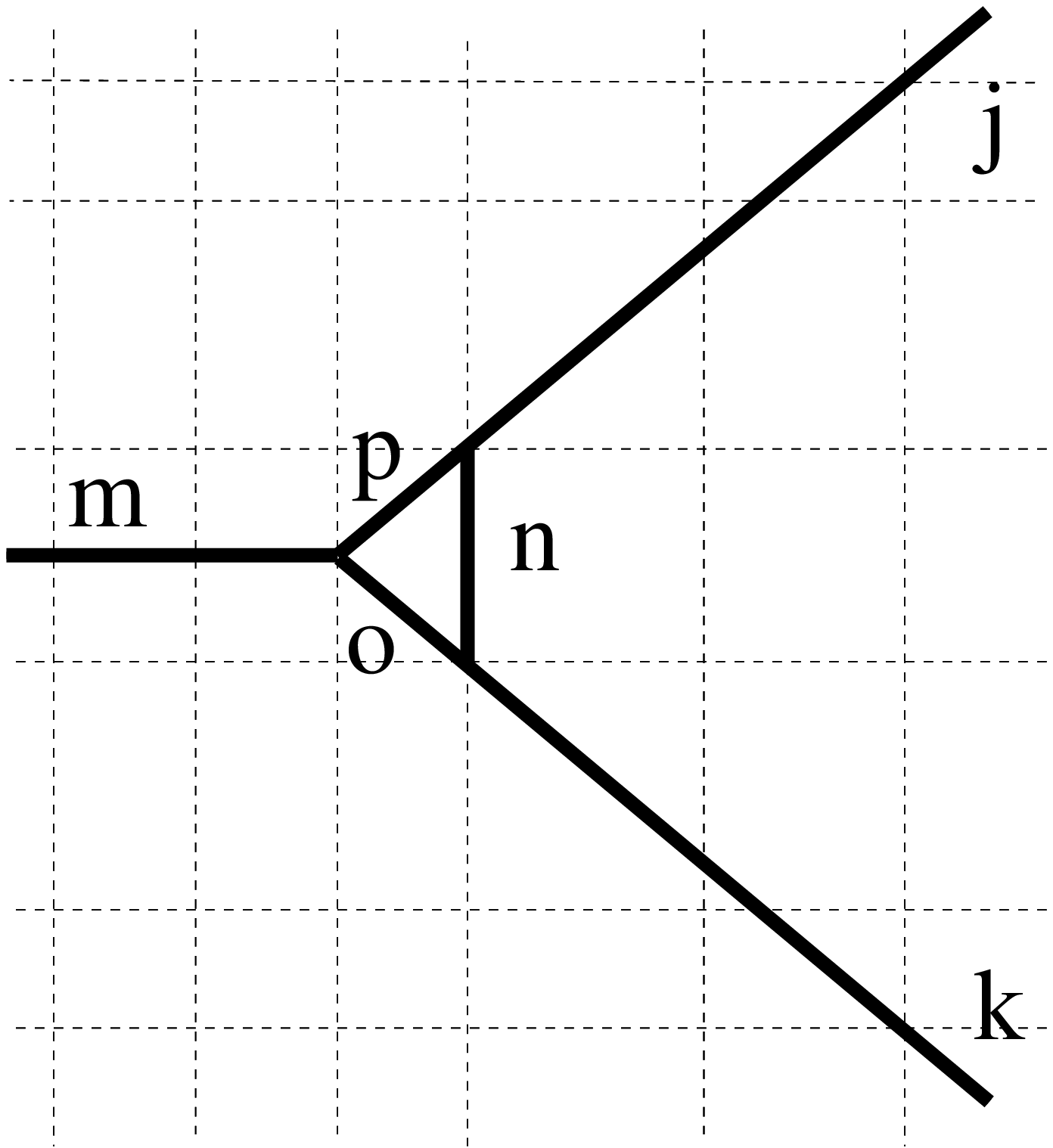}\\
\includegraphics[height=1.4cm]{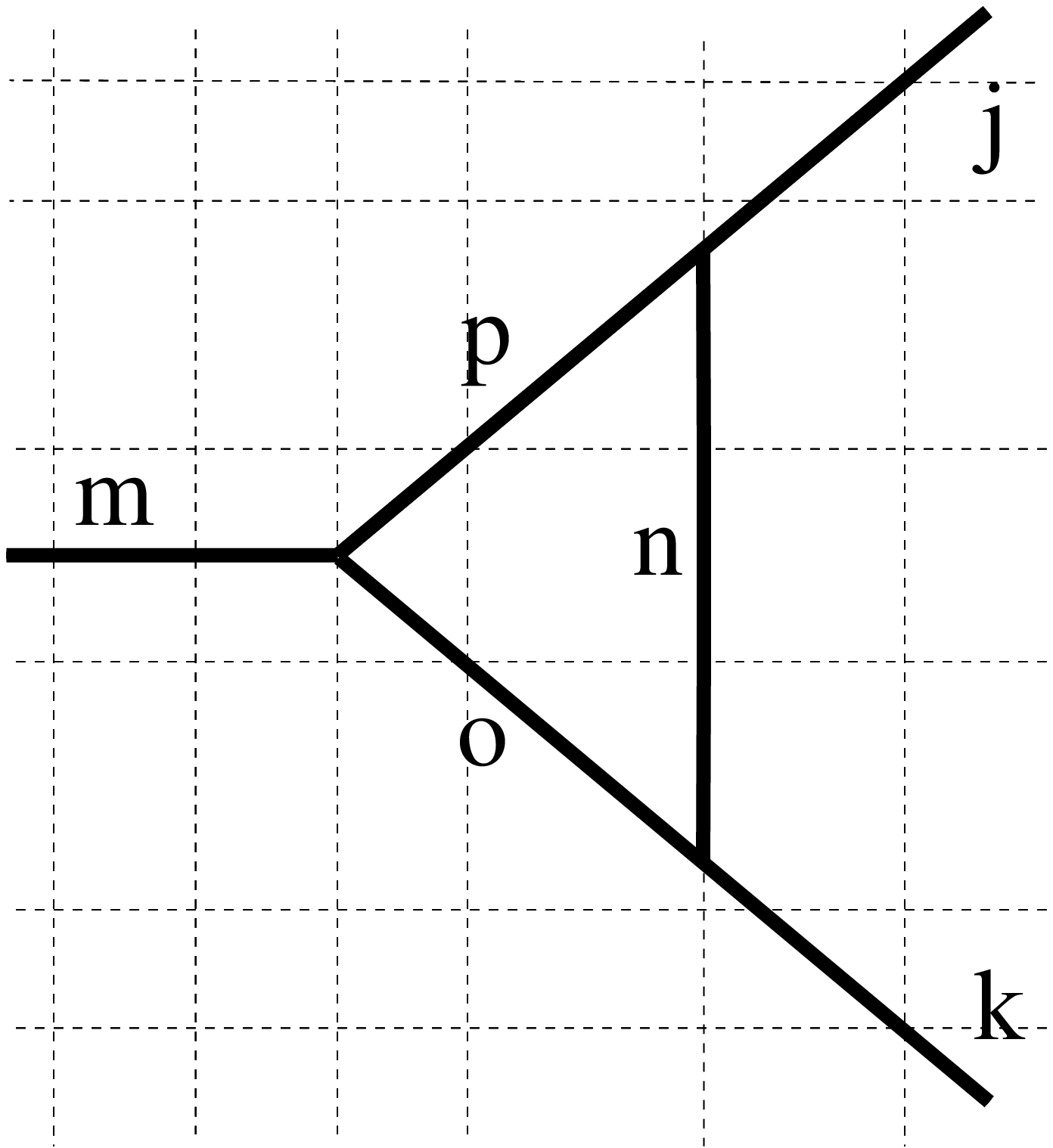}\\
\includegraphics[height=1.4cm]{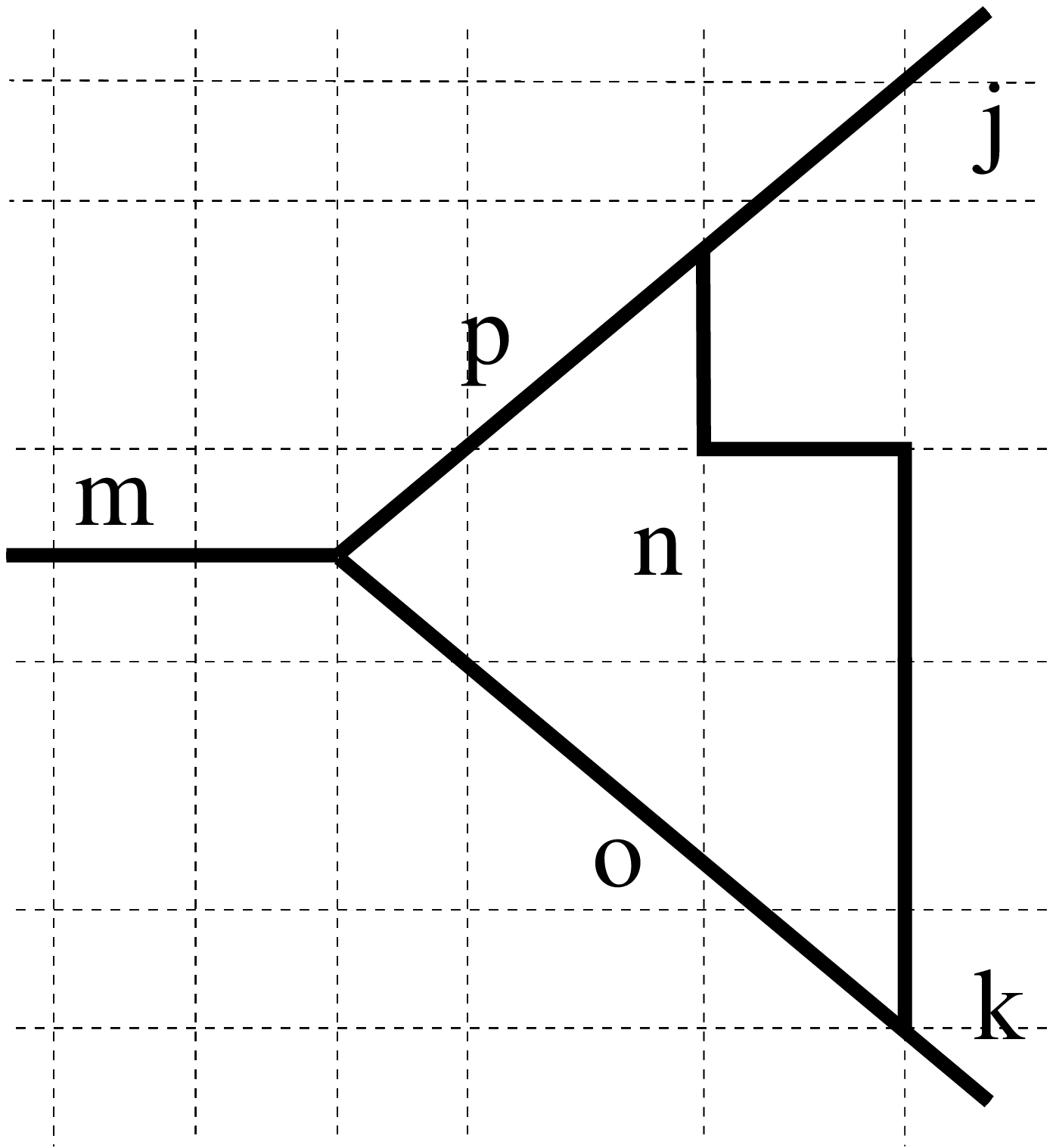}
\end{array}
\begin{array}{c}
\includegraphics[height=1.4cm]{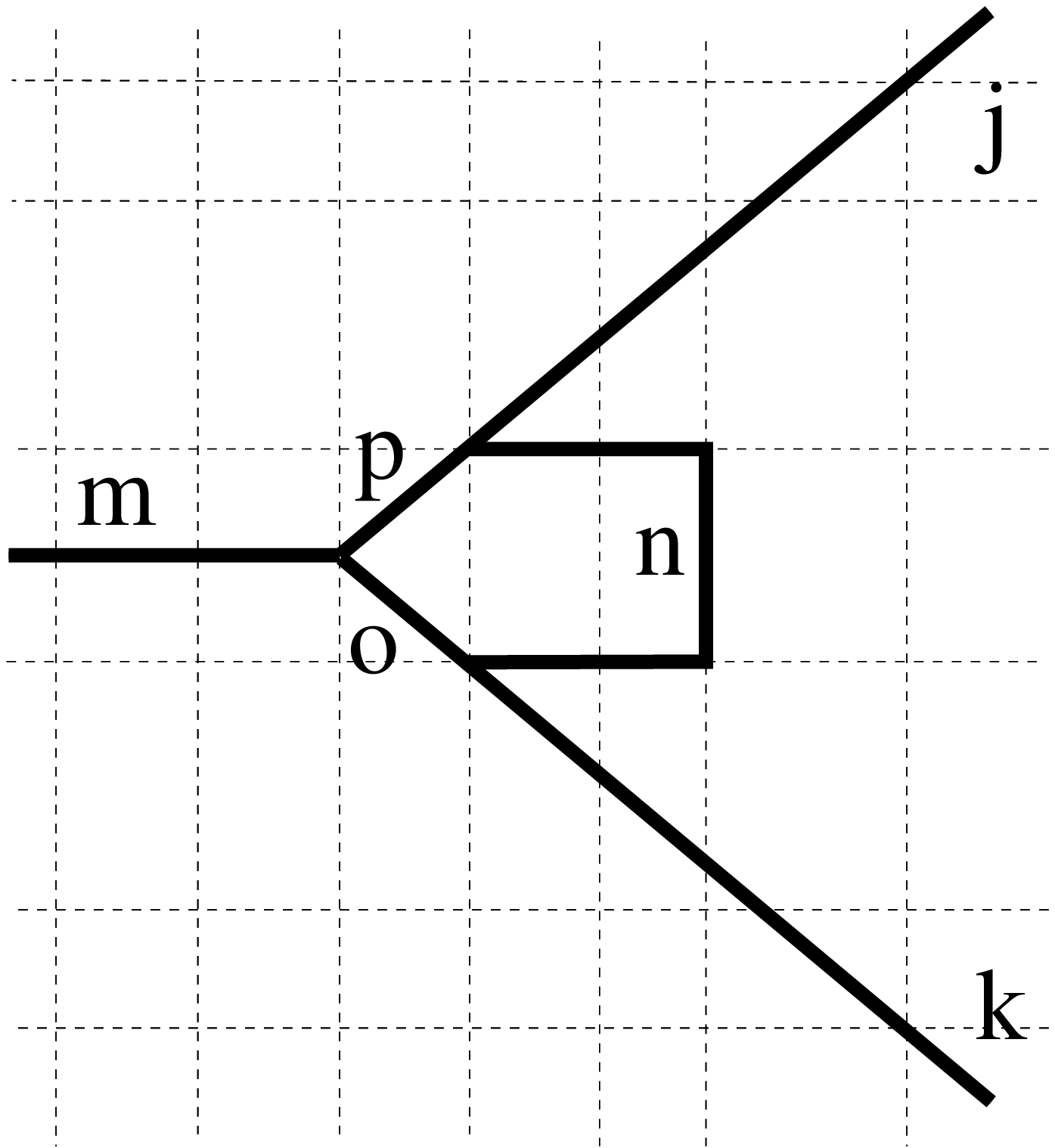}\\
\includegraphics[height=1.4cm]{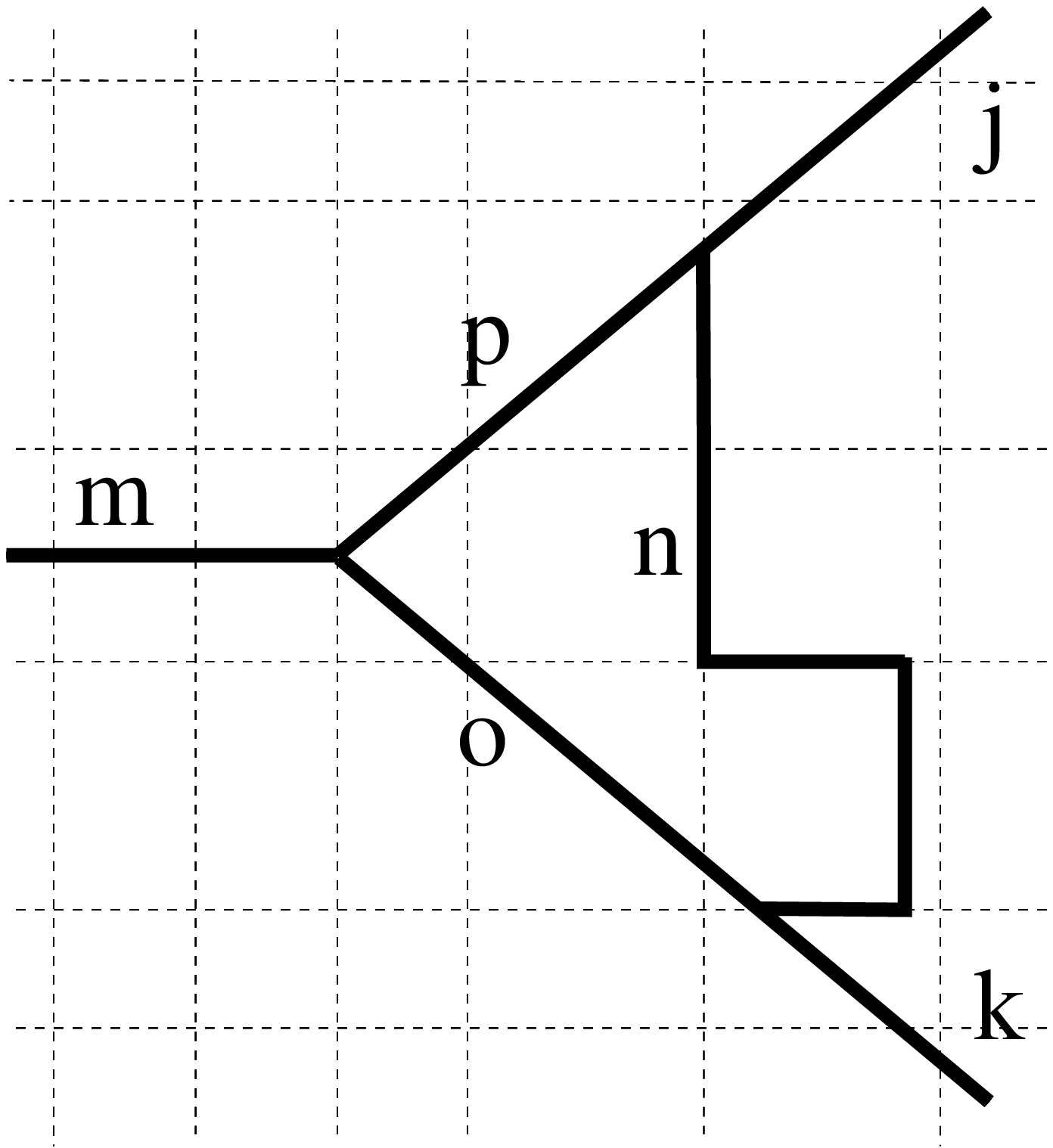}\\
\includegraphics[height=1.4cm]{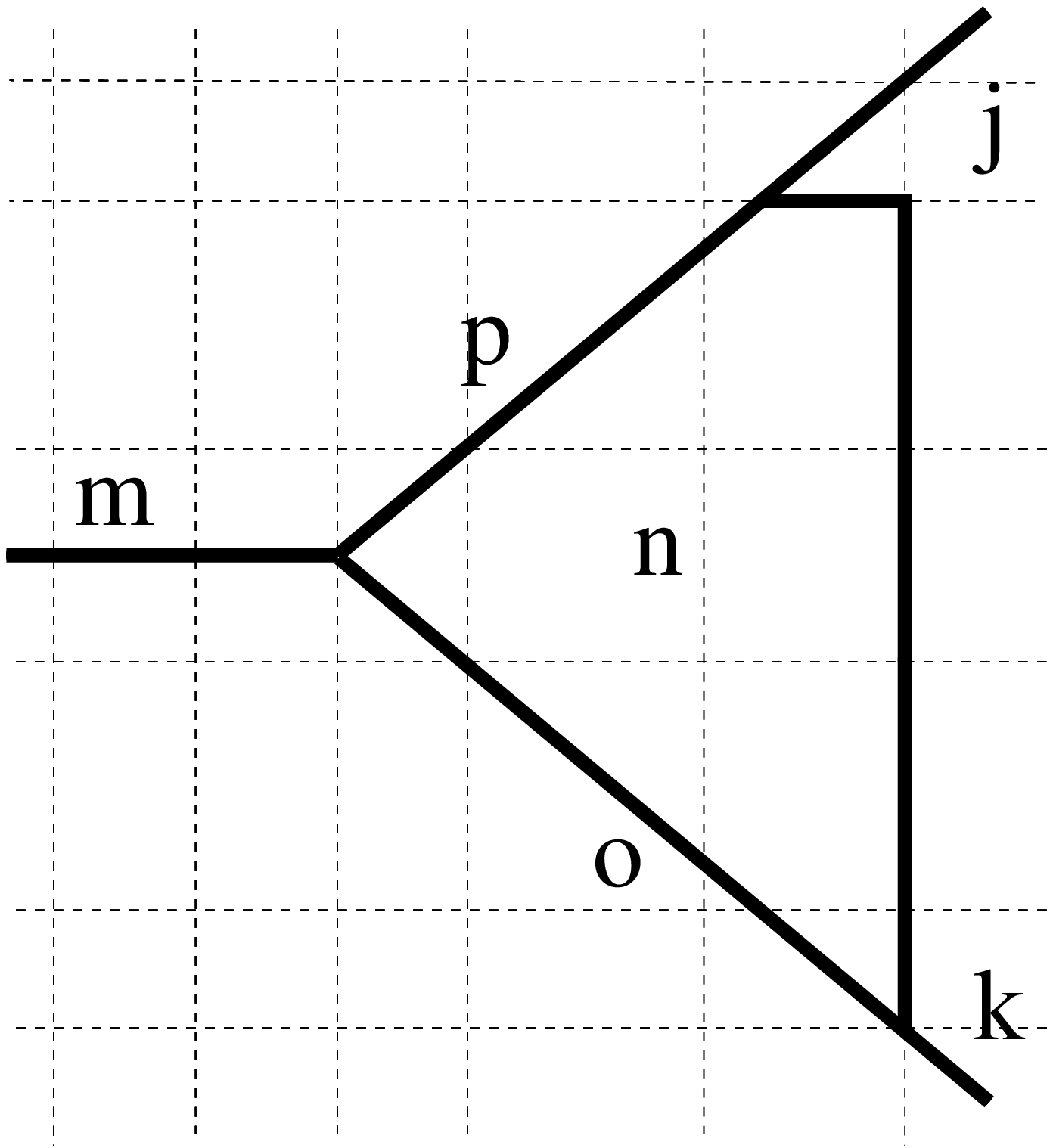}
\end{array}\begin{array}{c}
\includegraphics[height=.3cm]{flecha}
\end{array}
\begin{array}{c}
\includegraphics[height=4cm]{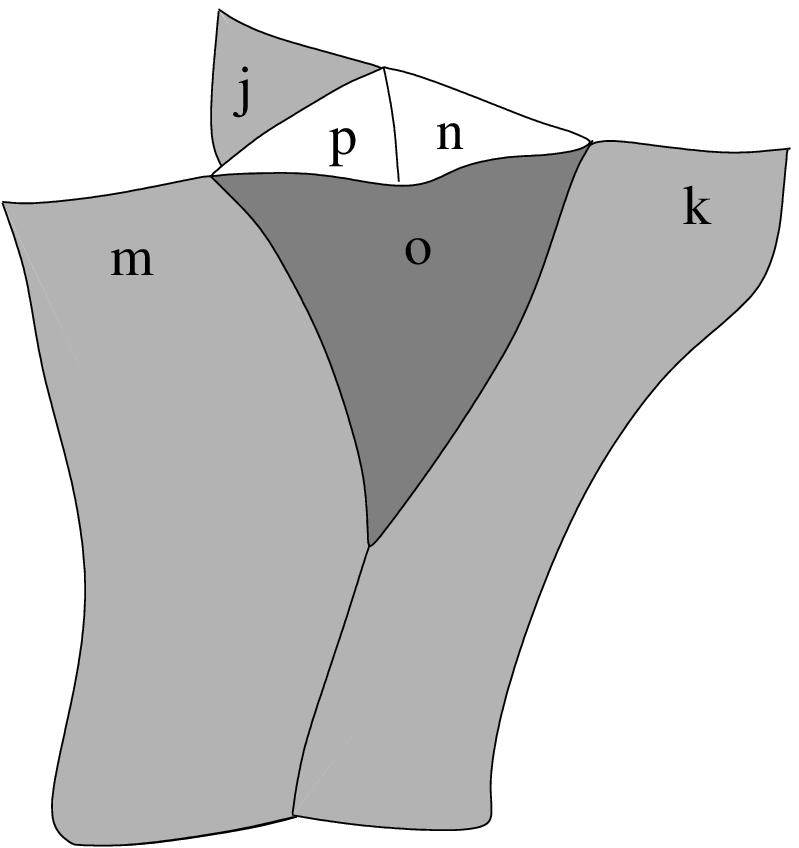}
\end{array}
\)}
\caption{A set of discrete transitions representing one of the
contributing histories at a fixed value of the regulator. On the
right, the continuous \emph{spin foam representation} when the
regulator is removed.}
\label{vani}
\end{figure}}

Even though the ordering of the plaquette actions does not affect
the amplitudes, the \emph{spin foam representation} of the terms in the
sum (\ref{3dc}) is highly dependent on that ordering. This is
represented in Figure\ref{defy} where a spin foam equivalent to that
of Figure\ref{lupy} is obtained by choosing an ordering of
plaquettes where those of the central region act first. One can
see this freedom of representation as an analogy of the gauge
freedom in the spacetime representation in the classical theory.

\epubtkImage{}{%
\begin{figure}[h!!!!!!!!!!!!!]
 \centerline{\hspace{0.0cm}\(
\begin{array}{ccc}
\includegraphics[height=1cm]{lupo} \\
\includegraphics[height=1cm]{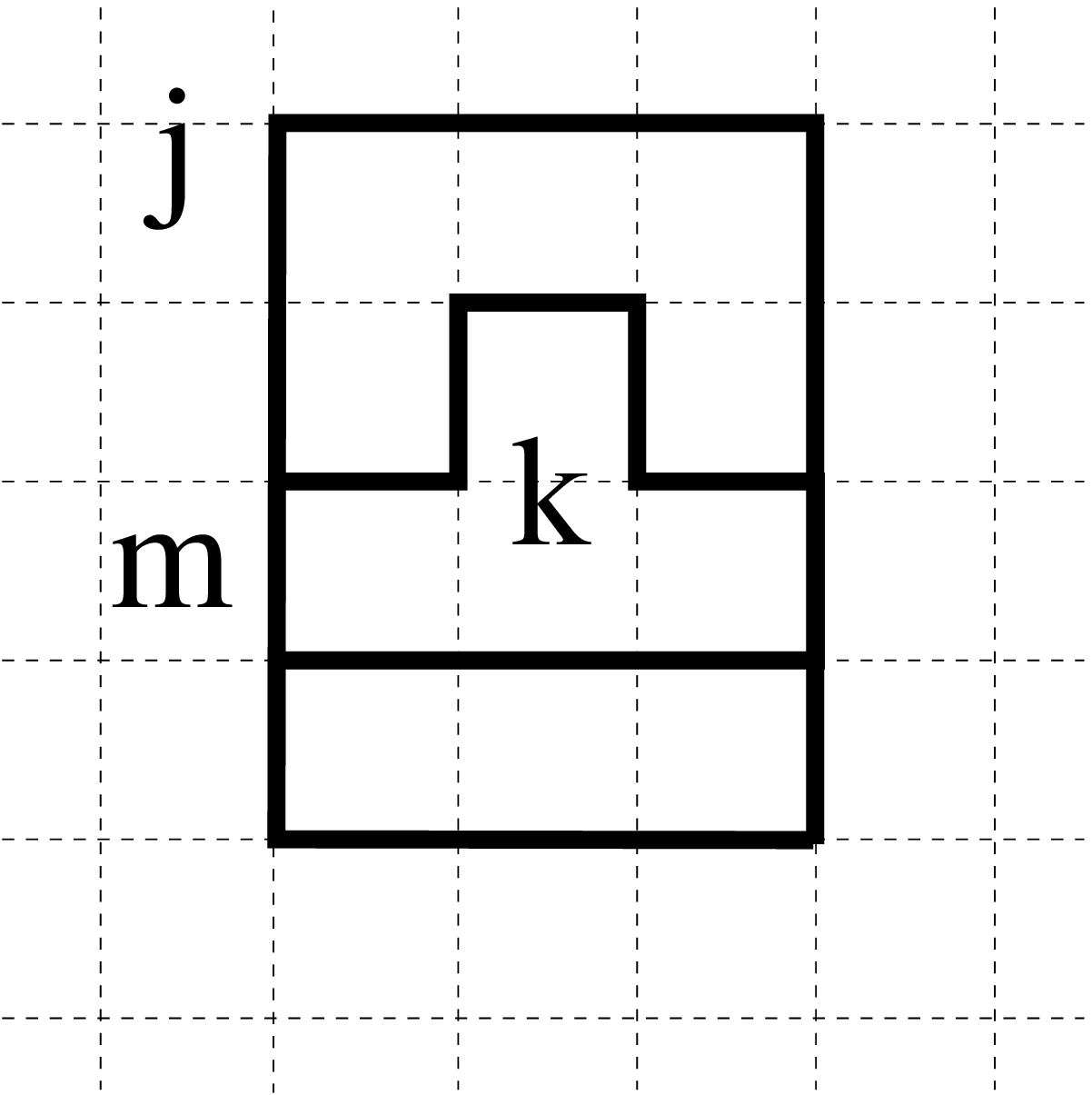}\\
\includegraphics[height=1cm]{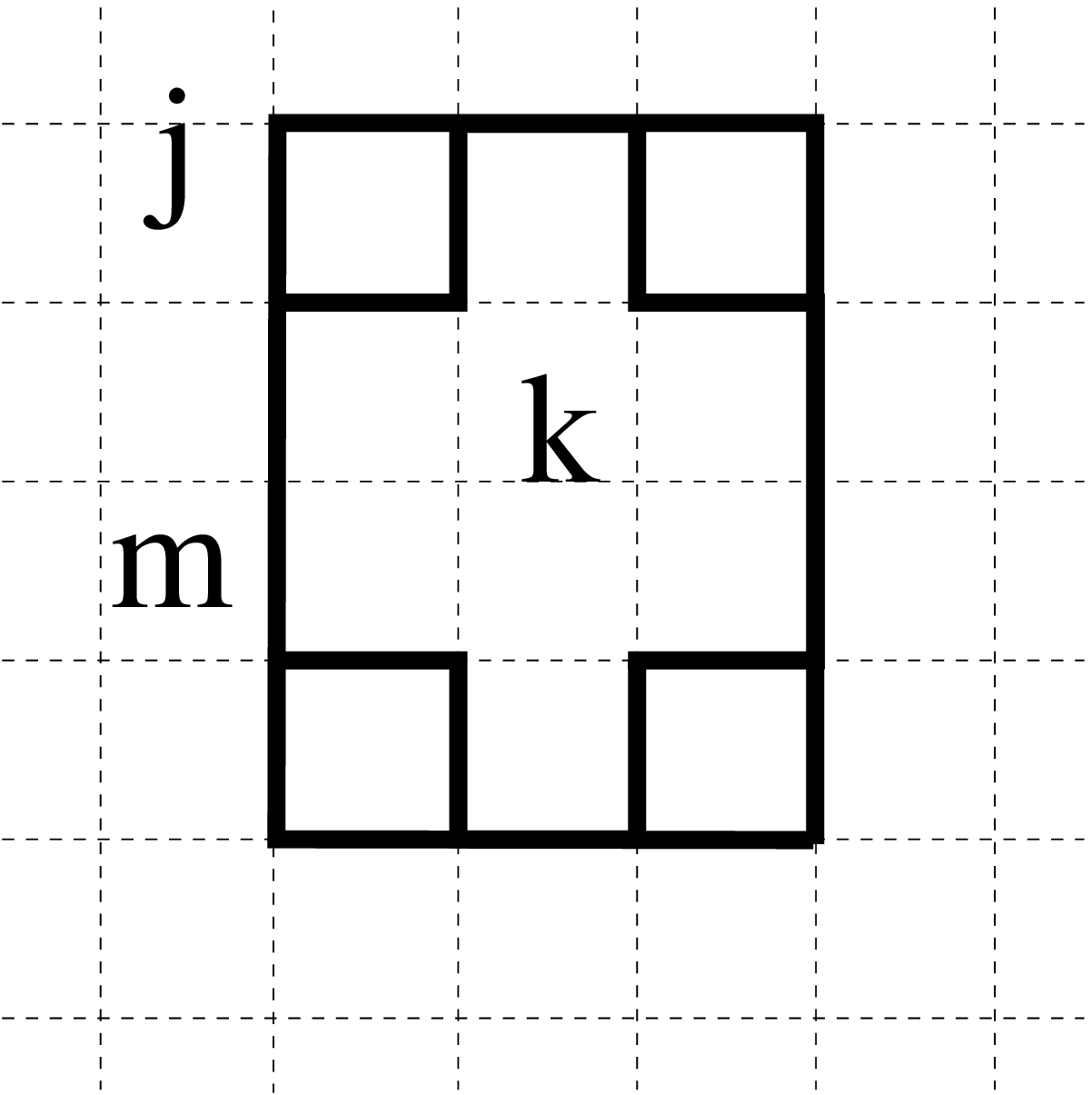}
\end{array}
\begin{array}{ccc}
\includegraphics[height=1cm]{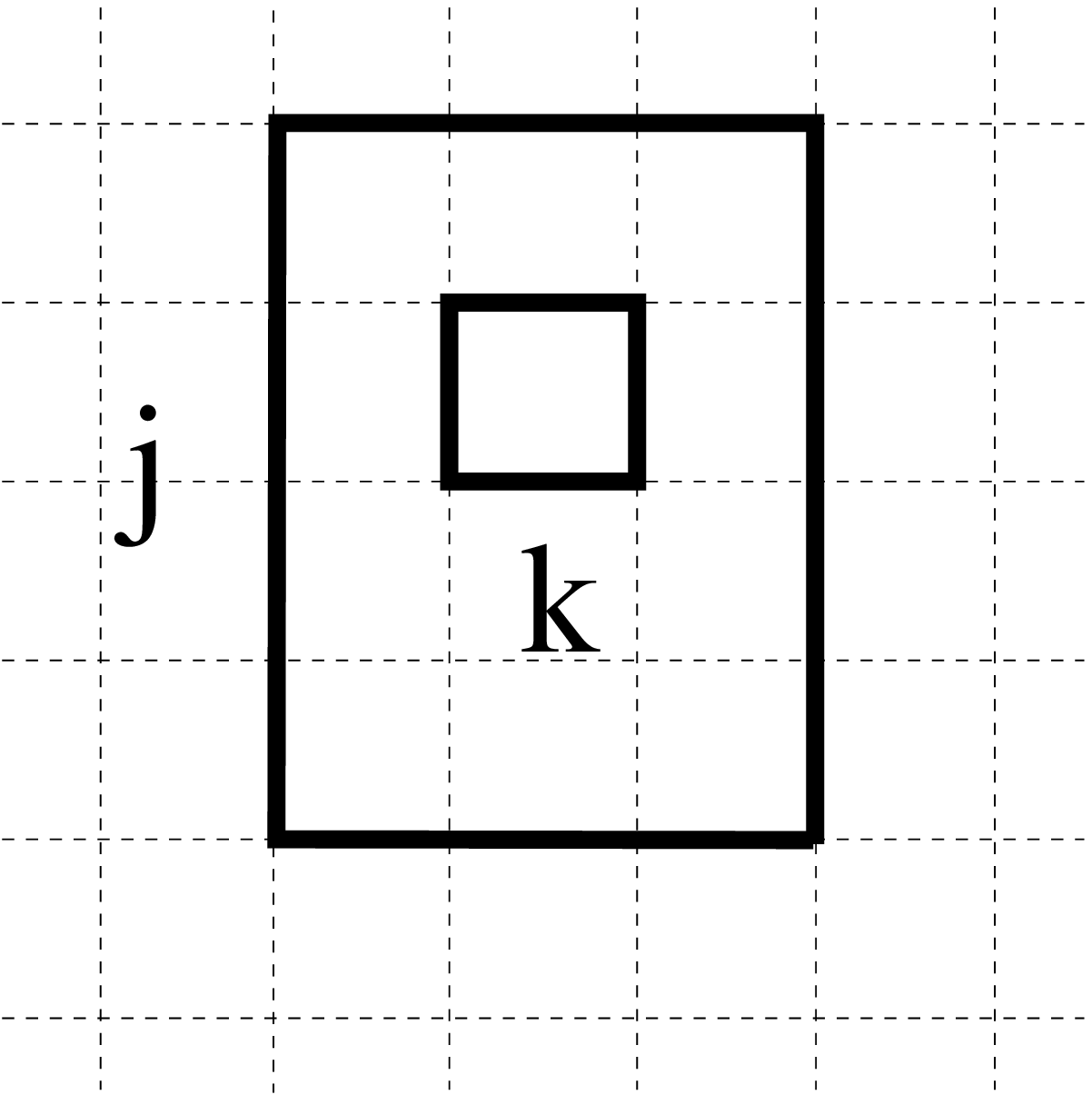} \\
\includegraphics[height=1cm]{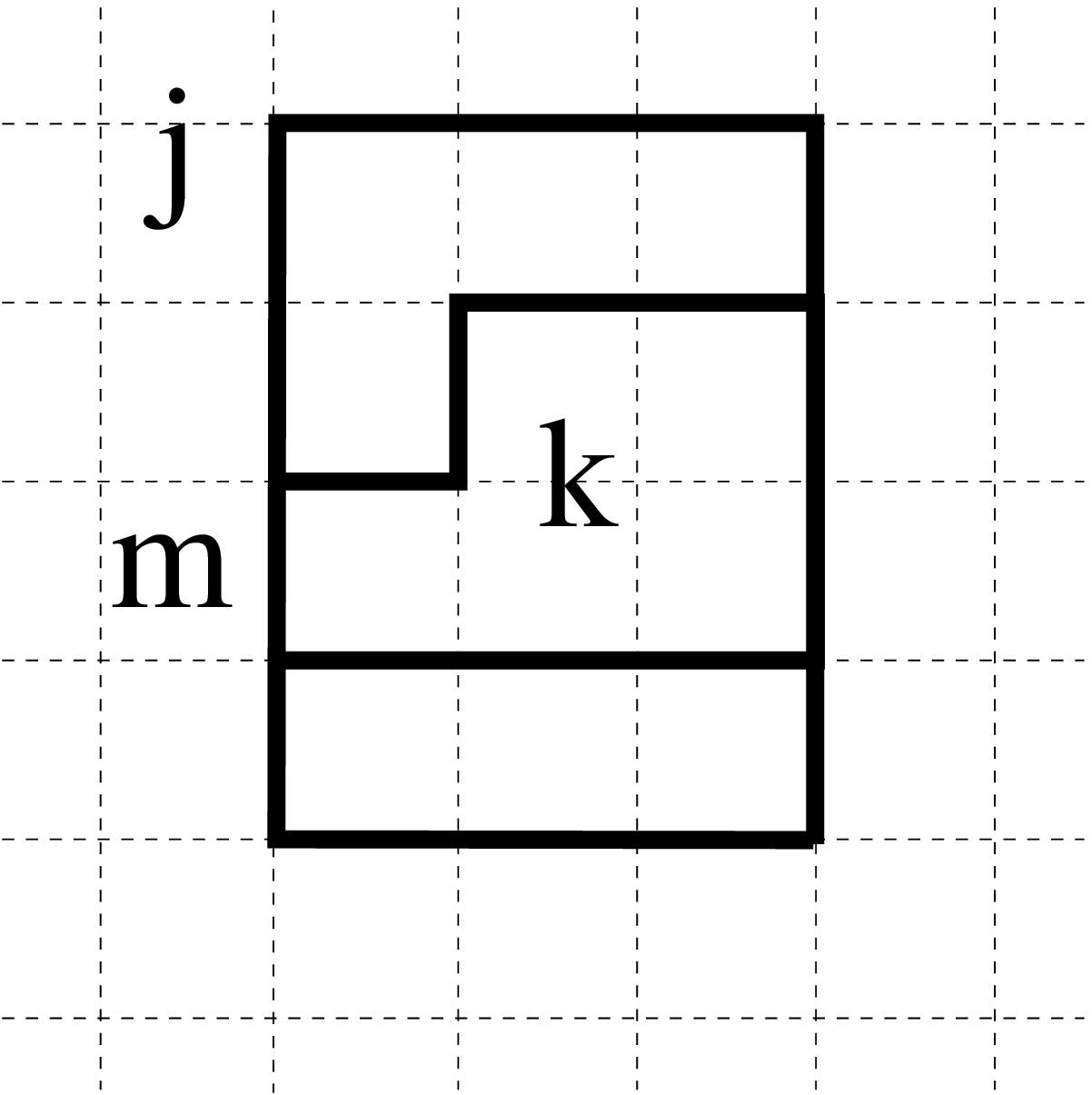}\\
\includegraphics[height=1cm]{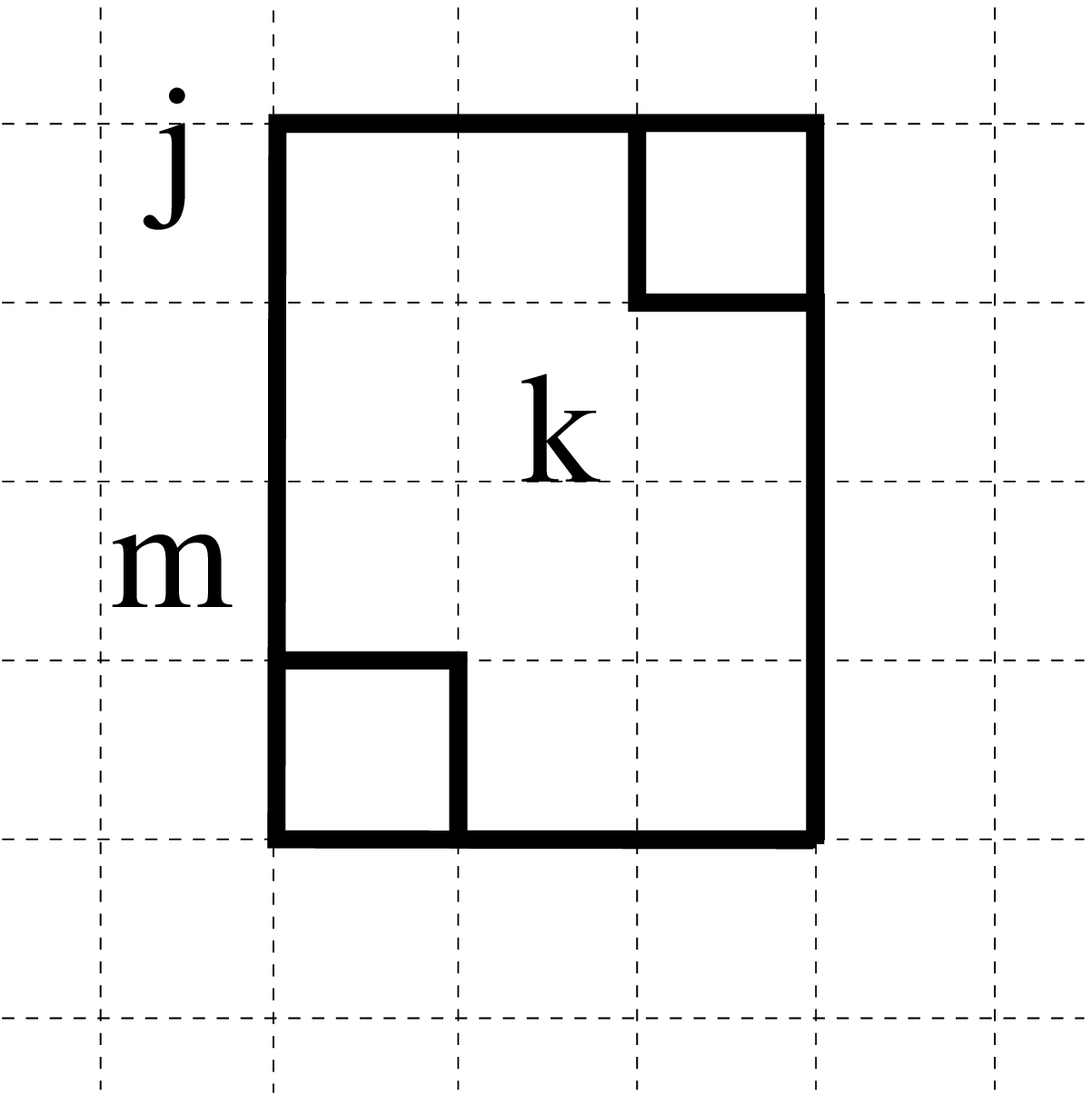}
\end{array}
\begin{array}{ccc}
\includegraphics[height=1cm]{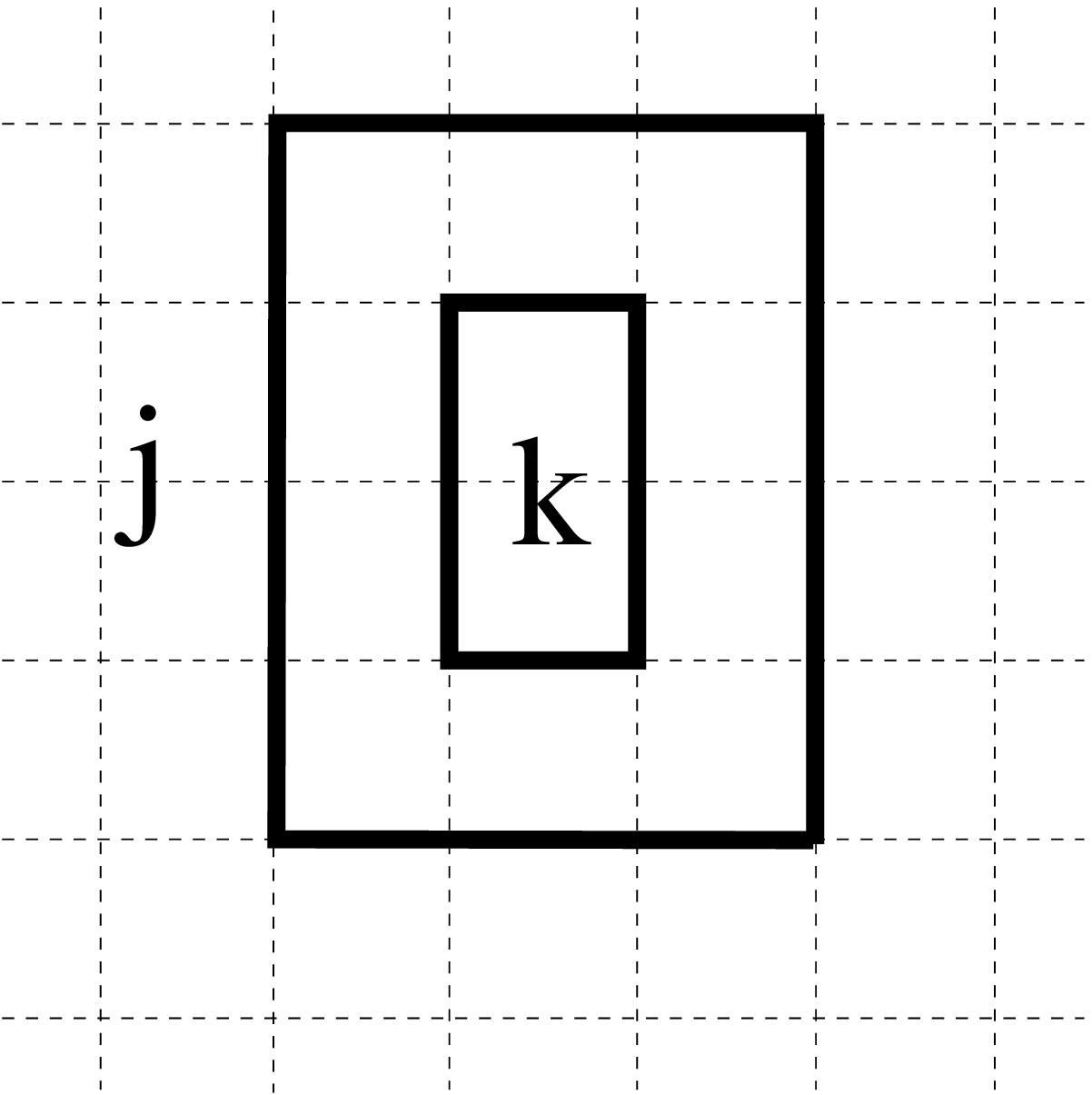} \\
\includegraphics[height=1cm]{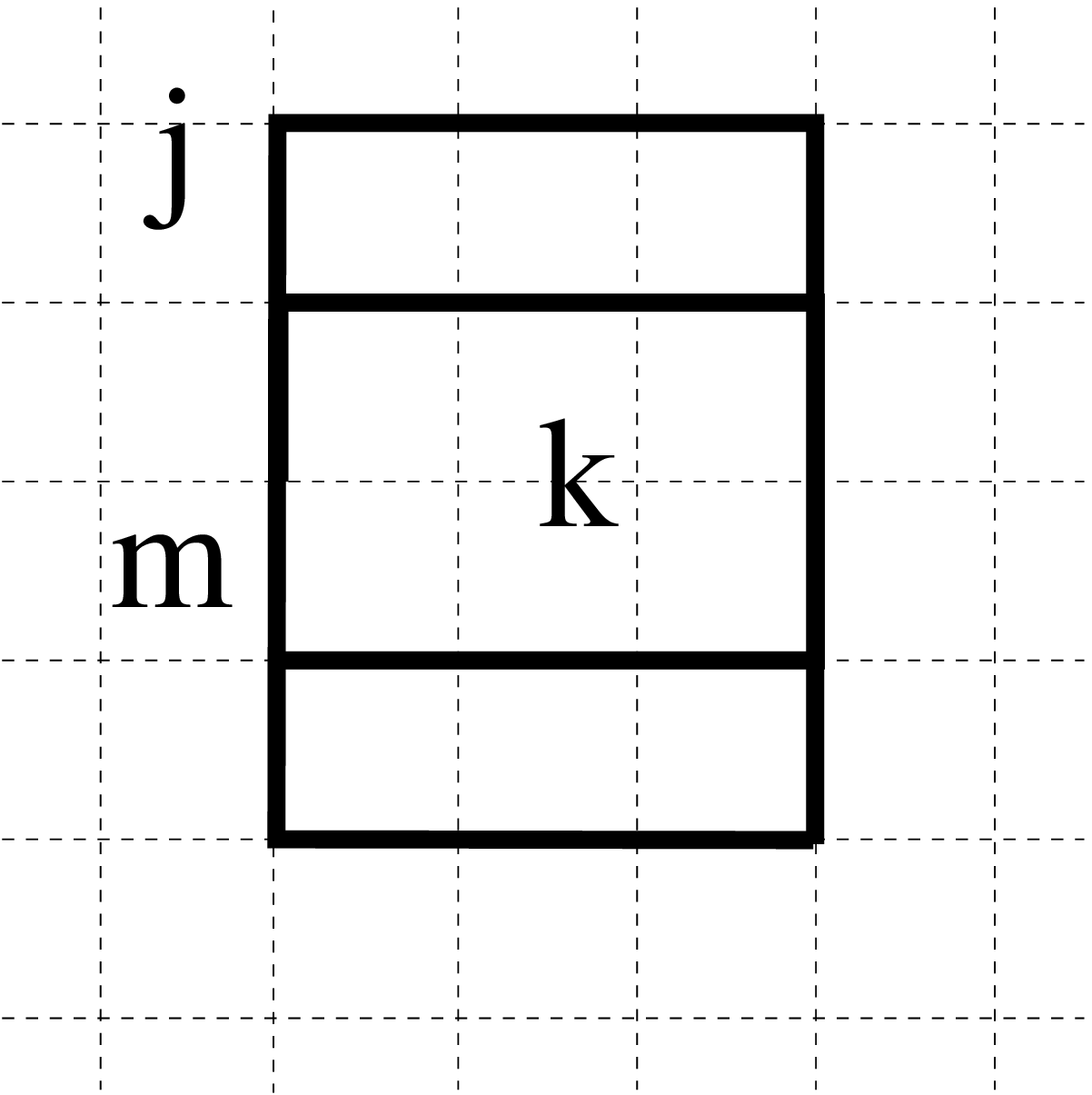}\\
\includegraphics[height=1cm]{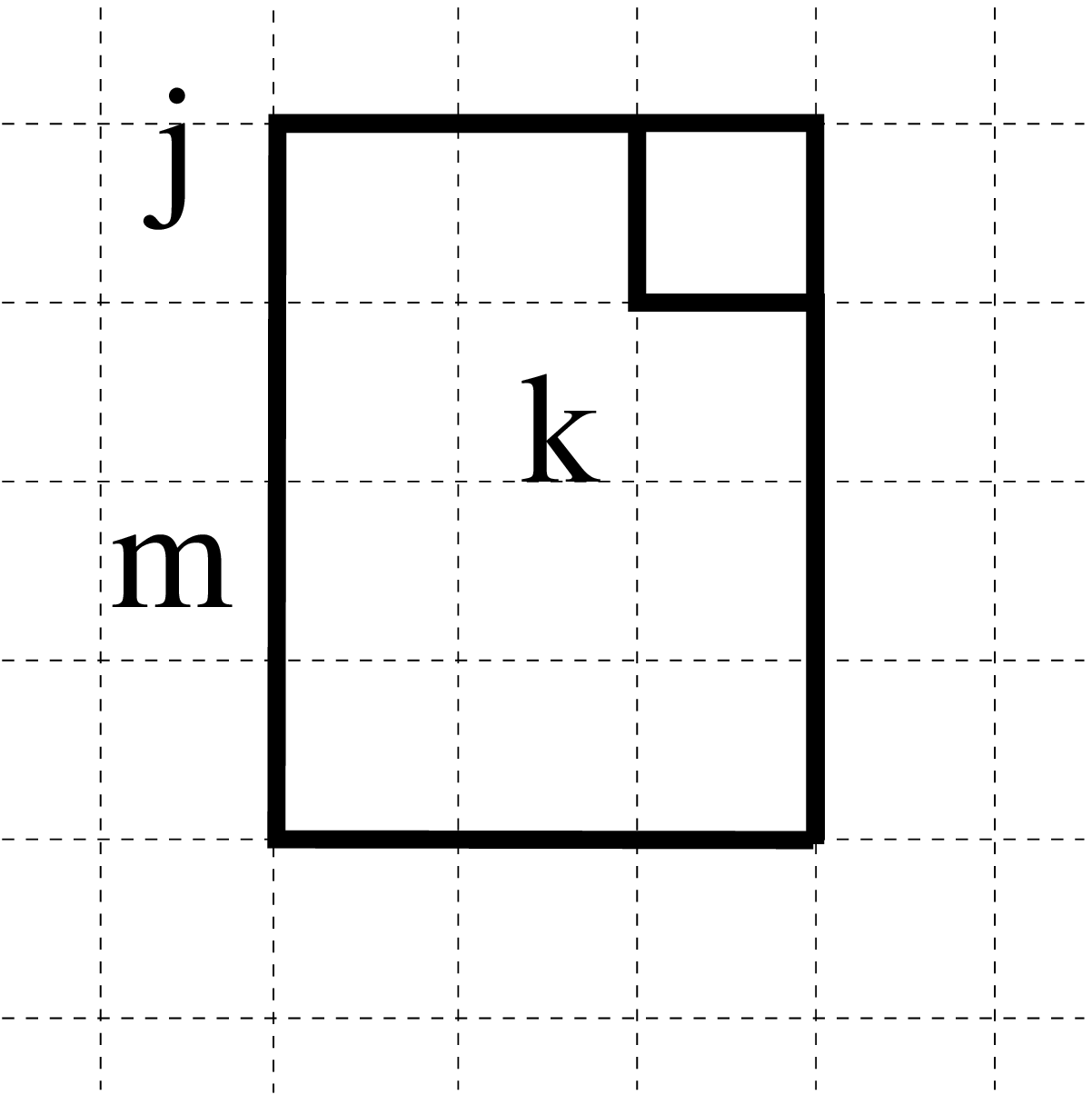}
\end{array}
\begin{array}{ccc}
\includegraphics[height=1cm]{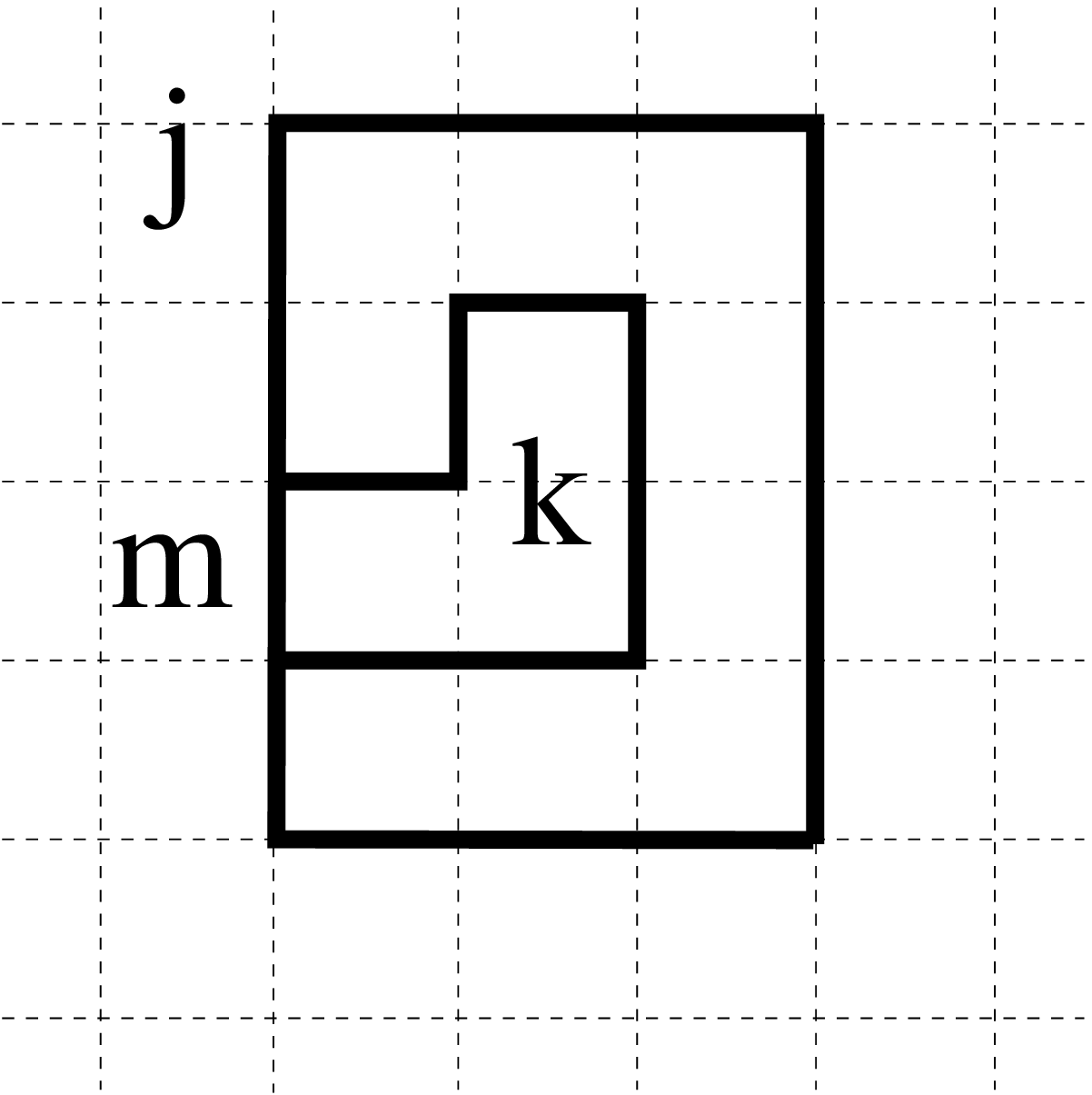} \\
\includegraphics[height=1cm]{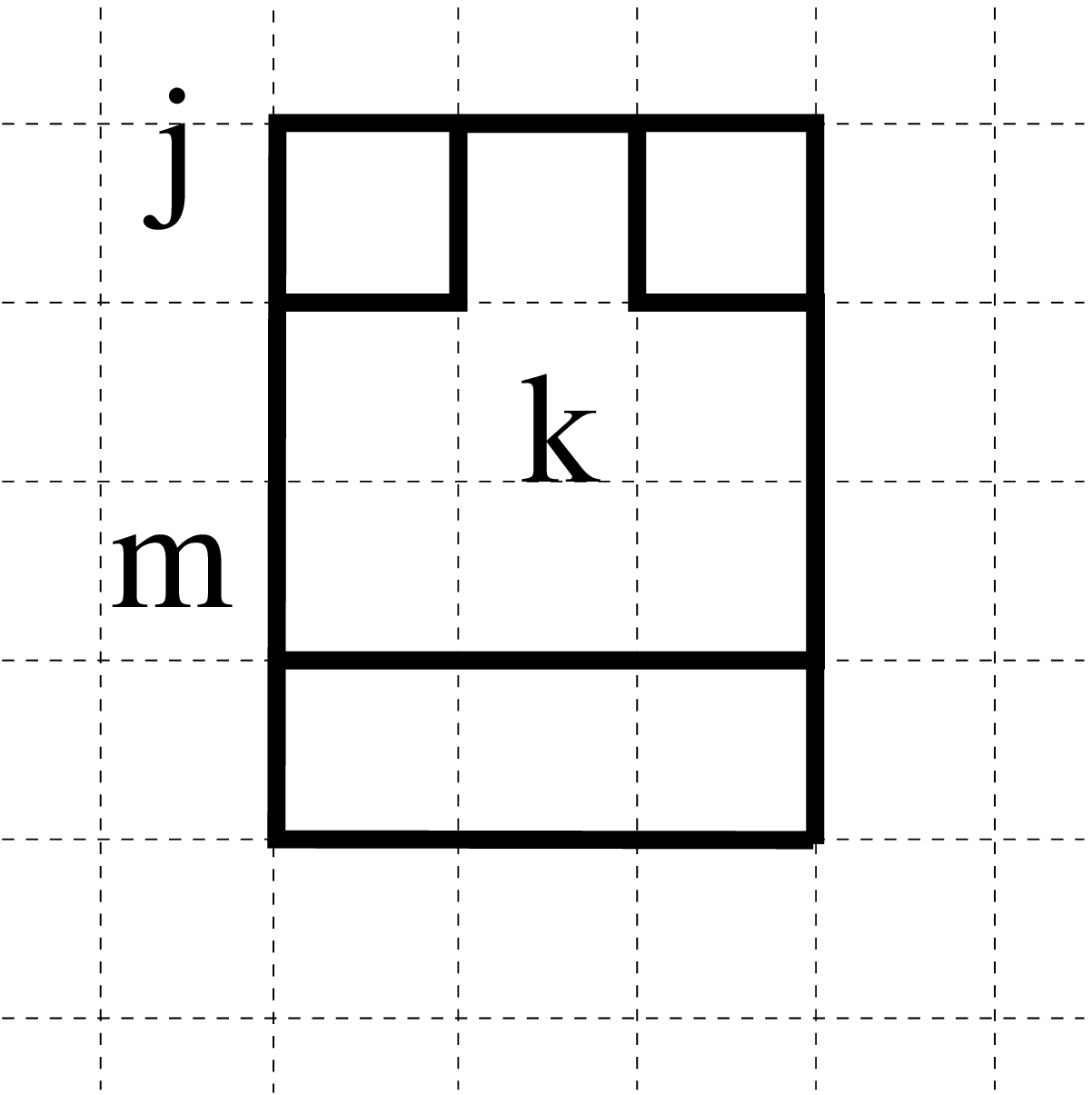}\\
\includegraphics[height=1cm]{lupo12}
\end{array}\begin{array}{c}
\includegraphics[height=.3cm]{flecha}
\end{array}\!\!\!
\begin{array}{c}
\includegraphics[height=3cm]{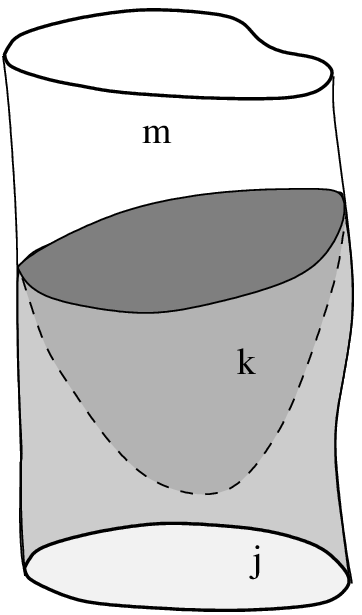}
\end{array}
\)}
\caption{A different representation of the transition of Figure~\ref{lupy}. This spin-foam
is obtained by a different ordering choice in (\ref{final}).} \label{defy}
\end{figure}}

One can in fact explicitly construct a basis of $\Hhp$ by choosing
an linearly independent set of representatives of the equivalence
classes defined in (\ref{null}). One of such basis is illustrated in
Figure~\ref{toron}. The number of quantum numbers necessary to label the basis
element is $6g-6$ corresponding to the dimension of the moduli space of $SU(2)$ flat
connections on a Riemann surface of genus $g$. This is the number of degrees
of freedom of the classical theory. In this way we
arrive at a fully combinatorial definition of the standard $\Hhp$
by reducing the infinite degrees of freedom of the kinematical
phase space to finitely many by the action of the generalized
projection operator $P$.

\epubtkImage{}{%
\begin{figure}[htbp]
\centerline{\includegraphics[width=0.8\textwidth]{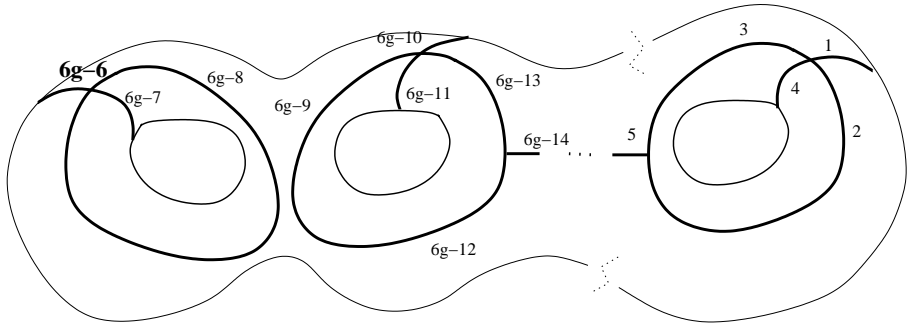}}
\caption{A spin-network basis of physical states for an arbitrary
  genus $g$ Riemann surface. There are $6g-6$ spins labels (recall
  that 4-valent nodes carry an intertwiner quantum number).}
\label{toron}
\end{figure}}

\subsection{Spin foam quantization of 3d gravity}
\label{sec:qbf}

Here we apply the general framework of Sections~\ref{BF} and
\ref{BF3}. This has been first studied by Iwasaki (in the spin foam
framework) in~\cite{iwa3,iwa4}. The partition function, ${\cal Z}$, is
formally given by\epubtkFootnote{We are dealing with Riemannian
  3-dimensional gravity. This should not be confused with the approach
  of Euclidean quantum gravity formally obtained by a Wick rotation of
  Lorentzian gravity. Notice the imaginary unit in front of the
  action. The theory of Riemannian quantum gravity should be regarded
  as a toy model with no obvious connection to the Lorentzian sector.}
\begin{equation}
\label{zbf}
{\cal Z}=\int  {\cal D}[e] {\cal D}[A]\ \ e^{i \int_{\va \cal M}
{\rm Tr}[e\wedge F(A)]},
\end{equation}
where for the moment we assume ${\cal M}$ to be a compact and
orientable. Integrating over the $e$ field in (\ref{zbf}) we
obtain
\begin{equation}
\label{VA}
{\cal Z}=\int {\cal D}[A] \ \ \delta \left(F(A)\right).
\end{equation}
The partition function ${\cal Z}$ corresponds to the `volume' of
the space of flat connections on $\cal M$.

In order to give a meaning to the formal expressions above, we replace
the 3-dimensional manifold ${\cal M}$ with an arbitrary cellular
decomposition $\Delta$. We also need the notion of the associated dual
2-complex of $\Delta$ denoted by $\Delta^{\star}$. The dual 2-complex
$\Delta^{\star}$ is a combinatorial object defined by a set of
vertices $v\in \Delta^{\star}$ (dual to 3-cells in $\Delta$) edges
$e\in \Delta^{\star}$ (dual to 2-cells in $\Delta$) and faces $f\in
\Delta^{\star}$ (dual to $1$-cells in $\Delta$). The fields $e$ and
$A$ have support on these discrete structures. The $su(2)$-valued
1-form field $e$ is represented by the assignment of an $e \in
{su(2)}$ to each 1-cell in $\Delta$. The connection field $A$ is
represented by the assignment of group elements $g_e \in SU(2)$ to
each edge in $\Delta^{\star}$.


The partition function is defined by
\begin{equation}
\label{Zdiscrete}
{\cal Z}(\Delta)=\int \prod_{f \in \Delta^{\star}} de_f \
\prod_{e \in \Delta^{\star}} dg_e  \ e^{i {\rm Tr}
\left[e_f U_f\right]},
\end{equation}
where $de_f$ is the regular Lebesgue measure on $\R^3$,
$dg_e$ is the Haar measure on $SU(2)$, and $U_f$
denotes the holonomy around faces, i.e., $U_f=g^1_e\dots g^{\va
N}_e$ for $N$ being the number of edges bounding the corresponding
face. Since $U_f \in SU(2)$ we can write it as $U_f=u^0_f\ {\mathbbm{1}} + F_f$ 
where $u^0_f\in \C$ and $F_f \in su(2)$. $F_f$ is interpreted as 
the discrete curvature around the face $f$. Clearly ${\rm Tr}[e_f U_f]={\rm Tr}[e_f F_f]$.
An arbitrary orientation is assigned to faces when computing
$U_f$. We use the fact that faces in $\Delta^{\star}$ are
in one-to-one correspondence with $1$-cells in $\Delta$ and label
$e_f$ with a face subindex.

Integrating over $e_f$, we obtain
\begin{equation}
\label{Zdiscrete0}
{\cal Z}(\Delta)=\int \ \prod_{e \in \Delta^{\star}} dg_e \
\prod_{f \in \Delta^{\star}}{\huge \delta}(g^1_e\dots
g^{\va N}_e),
\end{equation}
where $\delta$ corresponds to the delta distribution defined on
${\cal L}^2(SU(2))$. Notice that the previous equation corresponds
to the discrete version of equation (\ref{VA}).

The integration over the discrete connection ($\prod_e dg_e$) can
be performed expanding first the delta function in the previous
equation using the Peter--Weyl decomposition
\begin{equation}
\label{deltarep}
\delta(g)=\sum \limits_{j \in {\rm irrep}(SU(2))} \Delta_{j} \
{\rm Tr} \left[ j(g)\right],
\end{equation}
where $\Delta_{j}=2j+1$ denotes the dimension of the unitary
representation $j$, and $j(g)$ is the corresponding representation
matrix. Using equation (\ref{deltarep}), the partition function
(\ref{Zdiscrete0}) becomes
\begin{equation}
\label{coloring}
{\cal Z}(\Delta)=\sum \limits_{{\cal C}:\{j\} \rightarrow \{ f\}}
\int \ \prod_{e \in \Delta^{\star}} dg_e \ \prod_{f \in
\Delta^{\star}} \Delta_{j_f} \ {\rm Tr}\left[j_f(g^1_e\dots
g^{\va N}_e)\right],
\end{equation}
where the sum is over coloring of faces in the notation of
(\ref{sixteen}).

Going from equation (\ref{Zdiscrete}) to (\ref{coloring}) we have
replaced the continuous integration over the $e$'s by the sum over
representations of $SU(2)$. Roughly speaking, the degrees of
freedom of $e$ are now encoded in the representation being summed
over in (\ref{coloring}).

Now it remains to integrate over the lattice connection $\{g_e\}$.
If an edge $e\in \Delta^{\star}$ bounds $n$ faces there are
$n$ traces of the form ${\rm Tr}[j_f(\cdots g_e\cdots)]$ in
(\ref{coloring}) containing $g_e$ in the argument. The relevant
formula is
\begin{equation}\label{3dp}
P^{n}_{inv}:= \int dg\ {j_1(g)}\otimes j_2(g) \otimes \cdots \otimes j_n(g)=
\sum_{\iota} {C^{\va \iota}_{\va j_1 j_2 \cdots j_n} \ C^{*{\va
\iota}}_{\va j_1 j_2 \cdots j_n}},
\end{equation}
where $P^{n}_{inv}$ is the projector onto ${\rm Inv}[j_1\otimes j_2
  \otimes \cdots \otimes j_n]$. On the RHS we have chosen an
orthonormal basis of invariant vectors (intertwiners) to express the
projector. Notice that the assignment of intertwiners to edges is a
consequence of the integration over the connection. This is not a
particularity of this example but rather a general property of local
spin foams. Finally, (\ref{Zdiscrete0}) can be written as a sum over
spin foam amplitudes
\begin{equation}\label{statesum}
{\cal Z}(\Delta)=\sum \limits_{ {\cal C}:\{j\} \rightarrow \{ f\}
} \ \sum \limits_{ {\cal C}:\{\iota\} \rightarrow \{ e\} }\
\prod_{f \in \Delta^{\star}} \Delta_{j_f} \prod_{v\in {\cal
J}_{\va \Delta}} A_v(\iota_v,j_v),
\end{equation}
where $A_v(\iota_v,j_v)$ is given by the appropriate trace of the
intertwiners $\iota_v$ corresponding to the edges bounded by the
vertex and $j_v$ are the corresponding representations. This
amplitude is given in general by an $SU(2)$ $3Nj$-symbol
corresponding to the flat evaluation of the spin network defined
by the intersection of the corresponding vertex with a $2$-sphere.
When $\Delta$ is a simplicial complex all the edges in ${\cal
J}_{\Delta}$ are 3-valent and vertices are 4-valent (one such
vertex is emphasized in Figure~\ref{3g}, the intersection with the
surrounding $S^2$ is shown in dotted lines). Consequently, the
vertex amplitude is given by the contraction of the corresponding
four 3-valent intertwiners, i.e., a $6j$-symbol. In that case
the partition function takes the familiar
Ponzano--Regge~\cite{ponza} form
\begin{equation}\label{statesum}
{\cal Z}(\Delta)=\sum \limits_{ {\cal C}:\{j\} \rightarrow \{ f\}
} \ \prod_{f \in \Delta^{\star}} \Delta_{j_f} \prod_{v\in
\Delta^{\star}} \begin{array}{c}
\includegraphics[width=3cm]{tetras}\end{array},
\end{equation}
were the sum over intertwiners disappears since ${\rm dim}({\rm
Inv}[j_1\otimes j_2 \otimes j_3])=1$ for $SU(2)$ and there is only
one term in (\ref{3dp}). Ponzano and Regge originally defined the
amplitude (\ref{statesum}) from the study of the asymptotic
properties of the $6j$-symbol.

\subsubsection{Discretization independence}
\label{disind}

A crucial property of the partition function (and transition
amplitudes in general) is that it does not depend on the
discretization $\Delta$. Given two different cellular
decompositions $\Delta$ and $\Delta^{\prime}$ (not necessarily
simplicial) \begin{equation}\label{rucu} \tau^{-n_0}
Z(\Delta)=\tau^{-n^{\prime}_0} Z(\Delta^{\prime}),
\end{equation}
where $n_0$ is the number of 0-simplexes in $\Delta$ (hence the
number of bubbles in ${\cal J}_{\Delta}$), and $\tau=\sum_j (2j+1)$
is clearly divergent which makes discretization independence a
formal statement without a suitable regularization.

The sum over spins in (\ref{statesum}) is typically divergent, as
indicated by the previous equation. Divergences occur due to
infinite volume factors corresponding to the topological gauge
freedom (\ref{gauge2})(see~\cite{frei8})\epubtkFootnote{\label{u1} For
  simplicity we concentrate on the Abelian case $G=U(1)$. The analysis
  can be extended to the non-Abelian case.  Writing $g\in U(1)$ as
  $g=e^{i\theta}$ the analog of the gravity simplicial action is
\begin{equation}
S(\Delta, \{e_f\},\{\theta_e\})=\sum_{f \in \Delta^{\star}}
e_f F_f(\{\theta_e\}),
\end{equation}
where $F_f(\{\theta_e\}) = \sum_{e\in f} \theta_{e}$. Gauge
transformations corresponding to (\ref{gauge1}) act at the end points
of edges $e\in {\cal J}_{\Delta}$ by the action of group elements
$\{\beta\}$ in the following way
\begin{eqnarray}
\nonumber && B_f \rightarrow B_f,\\
&& \theta_{e} \rightarrow \theta_{e}+\beta_{s}-\beta_{t},
\end{eqnarray}
where the sub-index $s$ (respectively $t$) labels the source
vertex (respectively target vertex) according to the orientation
of the edge. The gauge invariance of the simplicial action is
manifest. The gauge transformation corresponding to (\ref{gauge2})
acts on vertices of the triangulation $\Delta$ and is given by
\begin{eqnarray}
\nonumber && B_f \rightarrow B_f + \eta_s -\eta_t,\\
&& \theta_{e} \rightarrow \theta_{e}.
\end{eqnarray}
According to the discrete analog of Stokes theorem 
\[\sum_{f \in \mathrm{Bubble}} F_f(\{\theta_e\})=0,\] 
which implies the invariance of the
action under the transformation above. The divergence of the
corresponding spin foam amplitudes is due to this last freedom.
Alternatively, one can understand it from the fact that Stokes
theorem implies a redundant delta function in (\ref{Zdiscrete0})
per bubble in ${\cal J}_{\Delta}$.}. The factor
$\tau$ in (\ref{rucu}) represents such volume factor. It can also
be interpreted as a $\delta(0)$ coming from the existence of a
redundant delta function in (\ref{Zdiscrete0}). One can partially
gauge fix this freedom at the level of the discretization. This
has the effect of eliminating bubbles from the 2-complex.

In the case of simply connected $\Sigma$ the gauge fixing is
complete. One can eliminate bubbles and compute finite transition
amplitudes. The result is equivalent to the physical scalar product
defined in the canonical picture in terms of the delta
measure\epubtkFootnote{If ${\cal M}=S^2\times [0,1]$ one can construct
  a cellular decomposition interpolating any two graphs on the
  boundaries without having internal bubbles and hence no
  divergences.}.

In the case of gravity with cosmological constant the state-sum
generalizes to the Turaev--Viro model~\cite{TV} defined in terms of
$SU_q(2)$ with $q^n=1$ where the representations are finitely
many. Heuristically, the presence of the cosmological constant
introduces a physical infrared cutoff. Equation~(\ref{rucu}) has been
proved in this case for the case of simplicial decompositions
in~\cite{TV}, see also~\cite{tur,kau}. The generalization for
arbitrary cellular decomposition was obtained in~\cite{a1}.

\subsubsection{Transition amplitudes}

Transition amplitudes can be defined along similar lines using a
manifold with boundaries. Given $\Delta$, ${\cal J}_{\Delta}$ then
defines graphs on the boundaries. Consequently, spin foams induce
spin networks on the boundaries. The amplitudes have to be
modified concerning the boundaries to have the correct composition
property (\ref{cobordism}). This is achieved by changing the face
amplitude from $(\Delta_{j_{f}})$ to $(\Delta_{j_{\ell}})^{1/2}$
on external faces.

The crucial property of this spin foam model is that the amplitudes
are independent of the chosen cellular
decomposition~\cite{tur,a1}. This allows for computing transition
amplitudes between any spin network states $s=(\gamma,
\{j\},\{\iota\})$ and $s^{\prime}=(\gamma,
\{j^{\prime}\},\{\iota^{\prime}\})$ according to the following
rules\epubtkFootnote{Here we are ignoring various technical issues in
  order to emphasize the relevant ideas. The most delicate is that of
  the divergences due to gauge factors mentioned above. For a more
  rigorous treatment see~\cite{za1, Zapata:2002eu}.}:

\begin{itemize}
\item Given ${\cal M}=\Sigma \times [0,1]$ (piecewise linear) and
spin network states $s=(\gamma, \{j\},\{\iota\})$ and
$s^{\prime}=(\gamma, \{j^{\prime}\},\{\iota^{\prime}\})$ on the
boundaries -- for $\gamma$ and $\gamma^{\prime}$ piecewise linear
graphs in $\Sigma$ -- choose any cellular decomposition $\Delta$
such that the dual 2-complex ${\cal J}_{\Delta}$ is bordered by
the corresponding graphs $\gamma$ and $\gamma^{\prime}$
respectively (existence can be shown easily).
%
%
\item Compute the transition amplitude between $s$ and $s^{\prime}$
by summing over all spin foam amplitudes (rescaled as in (\ref{rucu}))
for the spin foams $F:s\rightarrow s^{\prime}$ defined on the $2$-complex ${\cal
J}_{\Delta}$.
\end{itemize}

\subsubsection{The generalized projector}\label{gp}

We can compute the transition amplitudes between any element of the
kinematical Hilbert space ${\cal H}$\epubtkFootnote{The sense in which
  this is achieved should be apparent from our previous definition of
  transition amplitudes. For a rigorous statement see~\cite{za1,
    Zapata:2002eu}.}. Transition amplitudes define the physical scalar
product by reproducing the skein relations of the canonical
analysis. We can construct the physical Hilbert space by considering
equivalence classes under states with zero transition amplitude with
all the elements of ${\cal H}$, i.e., null states.

\epubtkImage{}{%
\begin{figure}[htbp]
  \centerline{
    \includegraphics[width=2cm]{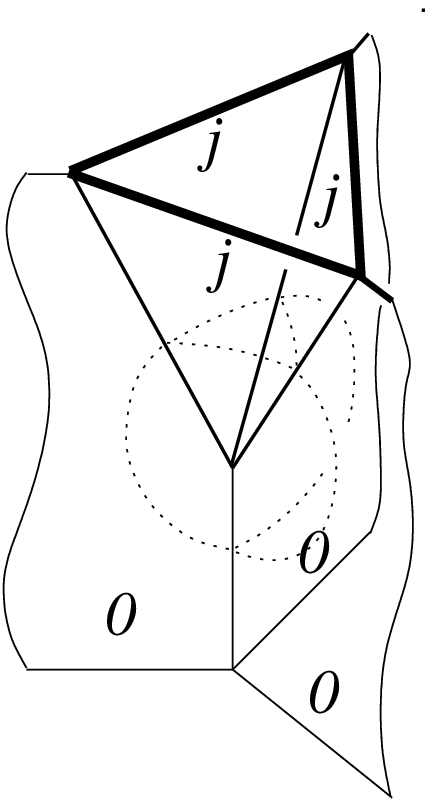}\qquad
    \includegraphics[width=5cm]{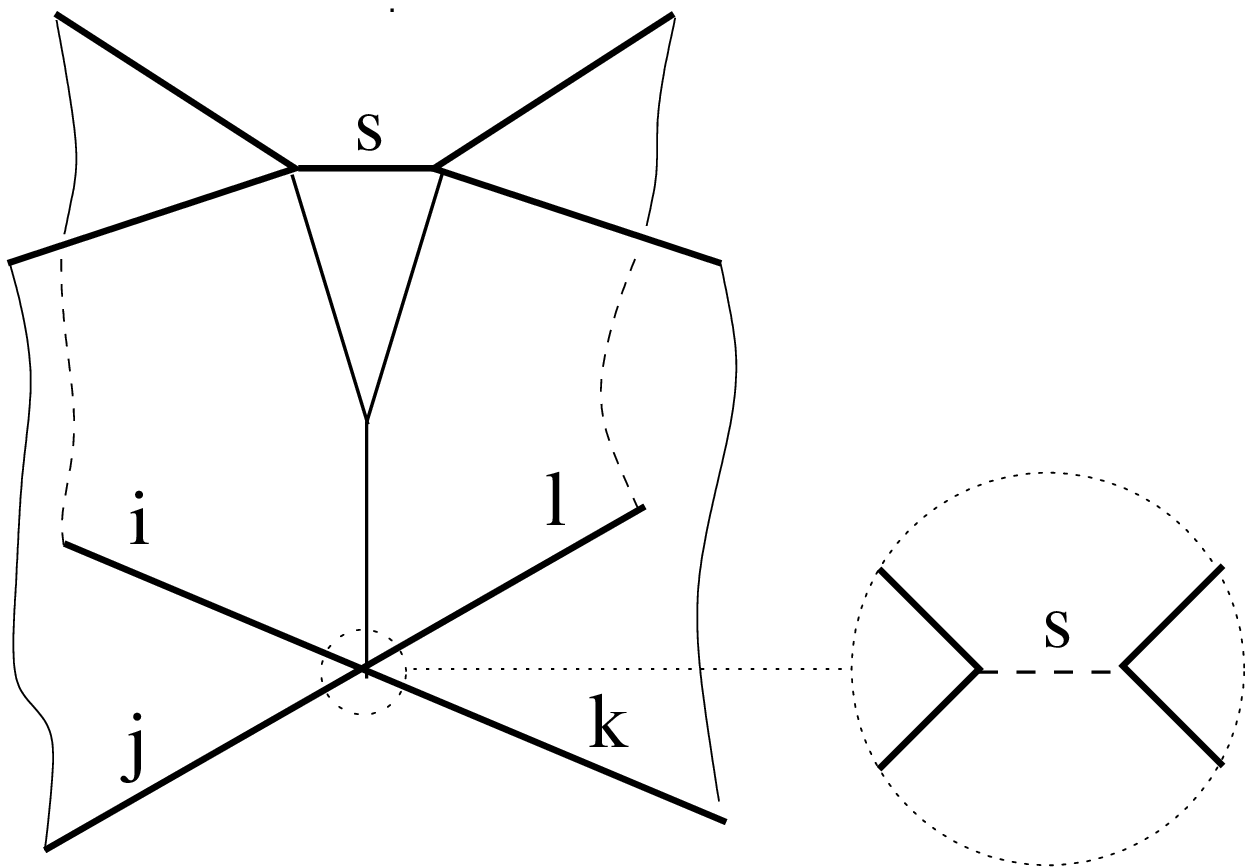}\qquad
    \includegraphics[width=3cm]{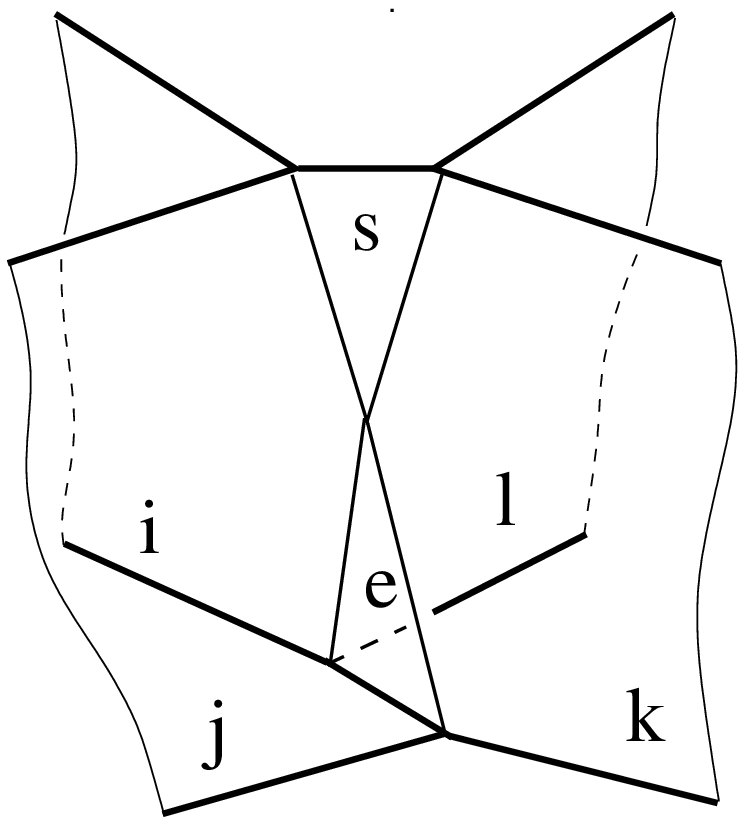}}
  \caption{Elementary spin foams used to prove skein relations.}
  \label{deltasf}
\end{figure}}

Here we explicitly construct a few examples of null states. For
any contractible Wilson loop in the $j$ representation the state
\begin{equation} \psi=(2j+1)\ s - \begin{array}{c}
\includegraphics[width=1cm]{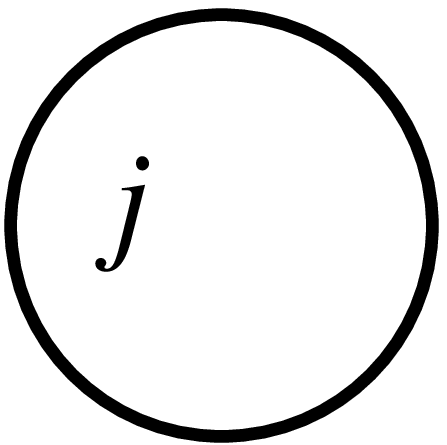}\end{array} \otimes s \phys 0,
\end{equation}
for any spin network state $s$, has vanishing transition amplitude
with any element of ${\cal H}$. This can be easily checked by
using the rules stated above and the portion of spin foam
illustrated in Figure~\ref{deltasf} to show that the two terms in
the previous equation have the same transition amplitude (with
opposite sign) for any spin-network state in ${\cal H}$. Using the
second elementary spin foam in Figure~\ref{deltasf} one can
similarly show that
\begin{eqnarray}
 \begin{array}{ccc}
\includegraphics[width=2cm]{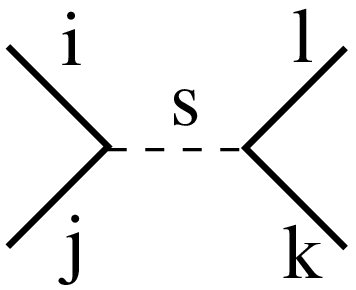}\end{array}
 - \begin{array}{c}
\includegraphics[width=2cm]{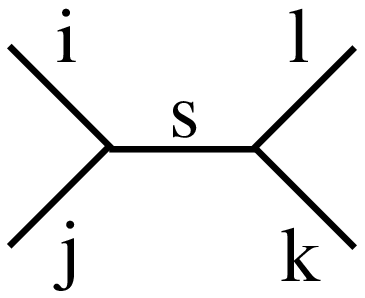}\end{array} \phys 0,
\end{eqnarray}
or the re-coupling identity using the elementary spin foam on the right of Figure~\ref{deltasf}
\begin{eqnarray}\label{42} \begin{array}{ccc}
\includegraphics[width=2cm]{intu1}\end{array}
 - \sum \limits_e \sqrt{2s+1}\sqrt{2e+1} \left\{\begin{array}{ccc}i\ \ j \ \ s\\ k\ \  l\ \  e  \end{array}\right\}
\begin{array}{c}
\includegraphics[height=2cm]{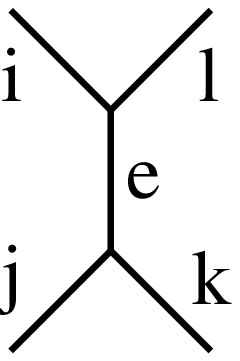}\end{array} \phys 0,
\end{eqnarray}
where the quantity in brackets represents an $SU(2)$ $6j$-symbol.
All skein relations can be found in this way. The transition
amplitudes imply the skein relations that define the physical
Hilbert space! The spin foam quantization is equivalent to the
canonical one.

\subsubsection{The continuum limit}

Recently, Zapata~\cite{za1, Zapata:2002eu} formalized the idea of a
continuum spin foam description of 3-dimensional gravity using
projective techniques inspired by those utilized in the canonical
picture~\cite{ash3}. The heuristic idea is that due to the
discretization invariance one can define the model in an `infinitely'
refined cellular decomposition that contains any possible spin network
state on the boundary (this intuition is implicit in our rules for
computing transition amplitudes above). Zapata concentrates on the
case with non-vanishing cosmological constant and constructs the
continuum extension of the Turaev--Viro model.

\subsection{Conclusion}

We have illustrated the general notion of the spin foam
quantization in the simple case of 3 dimensional Riemannian
gravity (for the generalization to the Lorentzian case
see~\cite{fre1}). The main goal of the approach is to provide a
definition of the physical Hilbert space. The example of this section
sets the guiding principles of what one would like to realize in four
dimensions. However, as should be expected, there are various new
issues that make the task by far more involved.

\subsection{Further results in 3d quantum gravity}

In this part of the article we have reviewed $SU(2)$ BF theory as from
the pespective of classical and quantum gravity in three dimensions
(for a classic reference see~\cite{carlip}). The state sum as
presented above matches the quantum amplitudes first proposed by
Ponzano and Regge in the 1960s based on their discovery of the
asymptotic expressions of the $6j$-symbols~\cite{ponza} and is often
referred to as the Ponzano--Regge model. Divergences in the above
formal expression require regularization. We have seen in
Section~\ref{SFH} that transition amplitudes are indeed finite in the
canonical framework where $M=\Sigma\times R$. Natural regularizations
are available in more general cases~\cite{Barrett:2008wh, Noui:2004iy,
  frei8}. For a detailed study of the divergence structure of the
model see~\cite{Bonzom:2010ar, Bonzom:2010zh, Bonzom:2011br}. The
quantum deformed version of the above amplitudes  lead to the so
called Turaev--Viro model~\cite{TV}  which is expected to correspond
to the quantization of three dimensional Riemannian gravity in the
presence of a non vanishing positive cosmological constant.

The topological character of BF theory can be preserved by the
coupling of the theory with topological defects playing the role of
point particles. In the spin foam literature this has been considered
form the canonical perspective in~\cite{Noui:2004jb, a21} and from the
covariant perspective extensively by Freidel and
Louapre~\cite{Freidel:2004vi}. These theories have been shown by
Freidel and Livine to be dual, in a suitable sense, to certain
non-commutative fields theories in three
dimensions~\cite{Freidel:2005bb, Freidel:2005me}.

Concerning coupling BF theory with non topological matter
see~\cite{Fairbairn:2006dn, Dowdall:2010ej} for the case of fermionic
matter, and~\cite{Speziale:2007mt} for gauge fields. A more radical
perspective for the definition of matter in 3d gravity is taken
in~\cite{Fairbairn:2007sv}. For three dimensional supersymmetric BF
theory models see~\cite{Livine:2003hn, Baccetti:2010xd}.

Recursion relations for the $6j$ vertex amplitudes have been
investigated in~\cite{Bonzom:2011jh, Dupuis:2009qw}. They provide a
tool for studying dynamics in spin foams of 3d gravity and might be
useful in higher dimensions~\cite{Bonzom:2009zd}.

\clearpage

\clearpage
\part{Conceptual issues and open problems}
\label{sci}

In this last part of this review we discuss some important conceptual
issues that remain open questions to a large degree in the
formulation. The description of the completely solvable model of the
previous section  will serve as the main example to illustrate some of
these issues.

\clearpage

\subsection{Quantum spacetime and gauge-histories}
\label{gaugy}

What is the geometric meaning of the spin foam configurations? Can we
identify the spin foams with \emph{``quantum spacetime
  configurations''}?  The answer to the above questions is, strictly
speaking, in the negative in agreement with our discussion at the end
of Section~\ref{valin}. Physical degrees of freedom are extracted from
the huge set of kinematical ones by the sum over gauge-histories that
is realized in the spin foam representation.

This conclusion is illustrated  in an exact way the simple example in
2+1 gravity where $M=S^2\times \R$ ($g=0$) described in detail in
previous sections.  In this case the spin foam configurations
appearing in the transition amplitudes look locally the same to those
appearing in the representation of $P$ for any other topology.
However, a close look at the physical inner product defined by $P$
permits to conclude that the physical Hilbert space is one dimensional
-- the classical theory has zero degree of freedom and so there is no
non-trivial Dirac observable in the quantum theory. This means that
the sum over spin foams in Eq.~(\ref{3dc}) is nothing else but a sum over
\emph{pure gauge degrees of freedom} and hence no physical
interpretation can be associated to it. The spins labelling the
intermediate spin foams do not correspond to any measurable
quantity. For any other topology this still holds true, the true
degrees of freedom being of a global topological character. This means
that in general (even when local excitations are present as in 4d) the
spacetime geometric interpretation of the spin foam configurations is
subtle.  This is an important point that is often overlooked in the
literature: one cannot interpret the spin foam sum of Eq.~(\ref{3dc})
as a sum over geometries in any obvious way. Its true meaning instead
comes from the averaging over the gauge orbits generated by the
quantum constraints that defines $P$ -- recall the classical picture
Figure~\ref{phase}, the discussion around Eq.~(\ref{pipi}), and the
concrete implementation in 2+1 where $U(N)$ in (\ref{exists}) is the
unitary transformation representing the orbits generated by $F$. Spin
foams represent a \emph{gauge history} of a kinematical state. A sum
over gauge histories is what defines $P$ as a means for extracting the
true degrees of freedom from those which are encoded in the
kinematical boundary states.

\epubtkImage{}{%
\begin{figure}[htbp]
\centerline{\includegraphics[width=0.8\textwidth]{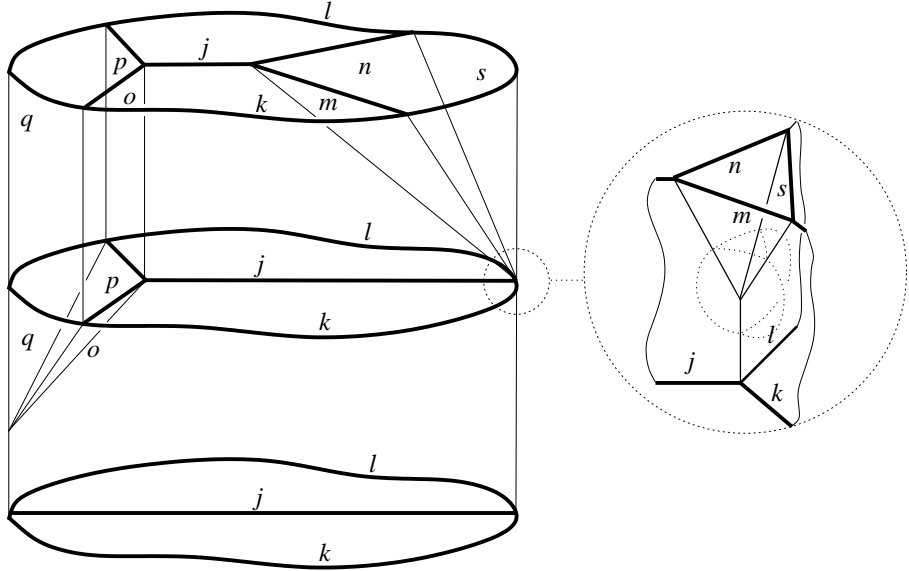}}
\caption{A \emph{spin foam} as the `colored' 2-complex representing
  the transition between three different \emph{spin network} states:
  it represents a pure gauge-history interpolating between kinematical
  boundary quantum geometry states. The sum over such histories in the
  spin foam representation of the path integral is meant to project
  out gauge degrees of freedom in order to extract the true physical
  information out of the kinematical Hilbert space.}
\label{spino}
\end{figure}}

Here we are discussing the interpretation of the \emph{spin foam
  representation} in the precise context of our solvable 3d example;
however, the validity of the conclusion is of general character and
holds true in the case of physical interest: four dimensional LQG.
Although, the quantum numbers labelling the spin foam configurations
correspond to eigenvalues of \emph{kinematical} geometric quantities
such as length (in 2+1) or area (in 3+1) LQG, their physical meaning
and {\em measurability} depend on dynamical considerations (for
instance the naive interpretation of the spins in 2+1 gravity as
quanta of physical length is shown here to be of no physical
relevance).  Quantitative notions such as time, or distance as well as
qualitative statements about causal structure or time ordering are
misleading (at best) if they are naively constructed in terms of
notions arising from an interpretation of \emph{spin foams} as quantum
spacetime configurations.

Now this does not mean that kinematical features (that usually admit
simpler geometric interpretation) is completely meaningless. It may be
that such kinematical quantum geometric considerations might reflect
true physical features when Dirac observables are considered. For
instance, the simplicial geometry interpretation of spin foam
histories (e.g., the total area of a polyhedron associated to a spin
network vertex) is the analog of the computation of some  classical
geometric quantity in general relativity (e.g. the area of some closed
surface embedded in spacetime). Both the former and the latter have no
intrinsic physical meaning: in the first case as argued above in the
second case because of the absent diffeomorphism invariant
characterization of the surface. However, in classical general
relativity, the ability of computing area of surfaces is useful to
construct a physical meaningful quantity such as the area of the event
horizon in Schwarzschild  spacetime which is a fully diffeomorphism
invariant property of the BH spacetime. Similarly, all the machinery
of quantum geometry (and its simplicial geometric interpretation in
suitable cases) might play an important role in the construction of
the physically meaningful objects (Dirac observables) of quantum
gravity \epubtkFootnote{An interesting example where this is strictly
  realized is the midi-superspace formulation of black holes in the
  context of loop quantum gravity known as the \emph{isolated horizon}
  formulation (see~\cite{Ashtekar:2004cn} for a review). In this
  framework the area of what represents the black hole horizon is a
  Dirac observable and has a discrete spectrum inherited from the area
  spectrum of the kinematical area in the full
  theory~\cite{Ashtekar:2000eq}. Such discreteness id crucial in
  recovering the Bekenstein--Hawking area law for the black hole
  entropy in LQG~\cite{Engle:2009vc, Ashtekar:1997yu}.}.

It is well known (and can be illustrated by many simple examples) that
gauge invariant properties can often be obtained by the suitable
combination of gauge covariant quantities. For example, in Yang--Mills
theory the holonomy $h(\gamma_{AB})\in G$ from a point $A$ to a point
$B$ along a path $\gamma_{AB}$ connecting the two points is not
physically meaningful. Nor is the value of some field mutlplets
$\Psi(A)$ and $\Psi(B)$ at points $A$ and $B$ respectively. A
physically meaningful quantity arises from the gauge invariant
combination of the above quantities, namely
$\langle\Psi(A)h(\gamma_{AB})\Psi(B)\rangle$. The pure gauge character
of the quantum geometric interpretation pointed out above in the
simple 3d context corresponds to the holonomy $h(\gamma_{AB})\in G$
in the above analogy. As  soon as we couple 2+1 gravity with matter
fields (corresponding to the $\Psi$ in the above analogy) non-trivial
Dirac observables become available (in fact such theory would have
local degrees of freedom). These will have to encode the
\emph{relationship} between the gauge dependent geometry and the gauge
dependent field content in a gauge invariant fashion.

In four dimensions we do not need to add anything as gravity has
already its local degrees of freedom yet physically meaningful
questions are still \emph{relational} (for more discussion see \cite{} and ). Unfortunately, no background
independent quantum theory with local degrees of freedom is
sufficiently  simple to illustrate this with an example. There is no
lucky example as what are \emph{free quantum field theories} for
standard QFT (non linearity and background independence may go
hand-in-hand). In four dimensions, and in particular, in the case of
the spin foam models described in this review, the difficult question
of extracting physics from the models is very important and presents a
major challenge for the future. The references discussed in
Section~\ref{physics} are important encouraging steps into this
direction.

\subsection{Anomalies and gauge fixing}
\label{anom}

As we mentioned before and illustrated with the example of three
dimensional gravity (see discussion in Section~\ref{gaugy}) the spin
foam path integral is meant to provide a definition of the physical
Hilbert space. Spin foam transition amplitudes are not interpreted as
defining propagation in time but rather as defining the physical
scalar roduct. This interpretation of spin foam models is the one
consistent with general covariance. However, in the path integral
formulation, this property relies on the gauge invariance of the path
integral measure. If the measure meets this property we say it is
\emph{anomaly free}. It is well known that in addition to the
invariance of the measure, one must provide appropriate gauge fixing
conditions for the amplitudes to be well defined. In this section we
analyze these issues in the context of the spin foam approach.

Since we are interested in gravity in the first order formalism, in
addition to diffeomorphism invariance one has to deal with the gauge
transformations in the internal space. Let us first describe the
situation for the latter. If this gauge group is compact then anomaly
free measures are defined using appropriate variables and invariant
measures. In this case gauge fixing is not necessary for the
amplitudes to be well defined. Examples where this happens are: the
models of Riemannian gravity considered in this paper (the internal
gauge group being $SO(4)$ or $SU(2)$), and standard lattice gauge
theory. In these cases, one represents the connection in terms of
group elements (holonomies) and uses the (normalized) Haar measure in
the integration. In the Lorentzian sector (internal gauge group
$SL(2,\C)$) the internal gauge orbits have infinite volume and the
lattice path integral would diverge without an appropriate gauge
fixing condition. These conditions generally exist in spin foam models
and are used to regularize the vertex amplitudes of the Lorentzian
models (we have illustrated this at the end of Section~\ref{alala};
for a general treatment see~\cite{frei8}).

The remaining gauge freedom is diffeomorphism invariance. 
To illustrate this issue we concentrate on the case of a model defined
on a fixed discretization $\Delta$ as it is the usual setting for
calculations dealing with  the models in four dimensions.

Let us start by considering the spin network states, at $\partial \Delta^{\star}$: boundary of $\Delta^{\star}$,
for which we want to define the transition amplitudes. According to 
what we have learned from the canonical approach, 3-diffeomorphism 
invariance is implemented by considering (diffeomorphism) equivalence classes of 
spin-network states. In the context of spin foams, the underlying discretization $\Delta$
restricts the graphs on the boundary to be contained on the dual 1-skeleton of the boundary 
complex $\partial\Delta^{\star}$. These states are regarded as representative elements of 
the corresponding 3-diffeomorphism equivalence class. The discretization can be interpreted, 
in this way, as a gauge fixing of 3-diffeomorphisms on the boundary.
This gauge fixing is partial in the sense that, generically, there will remain
a discrete symmetry remnant given by the discrete symmetries of
the spin network. This remaining symmetry has to be factored out when computing
transition amplitudes (in fact this also plays a role in the definition of the
kinematical Hilbert space of LQG).

A natural view point (consistent with LQG and quantum geometry) is
that this should naturally generalize to 4-diffeomorphisms for spin
foams. The underlying 2-complex ${\cal J}_{\Delta}$ on which spin
foams are defined represents a partial gauge fixing for the
configurations (spin foams) entering in the path integral. The
remaining symmetry, to be factored out in the computation of
transition amplitudes, corresponds simply to the finite group of
discrete symmetries of the corresponding spin
foams\epubtkFootnote{Baez~\cite{baez7} points out this equivalence
  relation between spin foams as a necessary condition for the
  definition of the \emph{category of spin foams}.}. This
factorization is well defined since the number of equivalent spin
foams can be characterized in a fully combinatorial manner, and is
finite for any spin foam defined on a finite discretization. In
addition, a spin foam model is anomaly free if the amplitudes are
invariant under this discrete symmetry. We have seen that this
requirement (advocated in~\cite{myo}) can  be met by suitable
definitions of the transition amplitudes~\cite{Bahr:2010bs,
  Kaminski:2009cc} (see Section~\ref{anofree}).

However, it is expected -- from the experience in lower dimensional
background independent models (recall the discussion of the previous
section) -- that there will be remnants of the gauge symmetries acting
non trivially on spin foams histories by doing more than simply
changing the embedding in the discrete regulating structure $\Delta$
as described above. After all, in gravity the Hamiltonian constraint
generates gauge transformations that hide in themselves the non
trivial dynamics of the theory. This is illustrated precisely in the
simplest scenario of 3d gravity where it is well known that bubble
divergencies are directly  linked to the infinite volume of the gauge
orbits generated by the curvature constraint~\cite{frei8}. The action
of such gauge symmetry related spin foams that do not differ only by
their embedding (see Section~\ref{sfm3d}).

It is also important to point out that the view that spin foams represent diffeomorphism equivalence classes of geometries,
which sometimes can be loosely stated in the community, is incompatible with the idea that spin foams should act as a projector onto the 
solutions of constraints. A path integral would only act as a projector if it includes gauge symmetries in the sense of group averaging. 
This related to the fact that physical states (those left invariant by the projector) are outside the kinematical Hilbert space, i.e. non-normalizable.



The discretization of the manifold $\Delta$ is seen as a regulator
introduced to define the spin foam model. Even when the regulator (or
the discretization dependence) eventually has to be removed (see next
subsection), the theory is presumed to remain discrete at the
fundamental level.  The smooth manifold diffeomorphism invariant
description is expected to emerge in the (low energy) continuum
limit. Fundamental excitations are intrinsically discrete. From this
viewpoint, the precise meaning of the gauge symmetries of such a
theory would have to be formulated directly at the discrete level. We
have seen that this can be achieved in the case of 3-dimensional
gravity (recall Section~\ref{disind}).

The previous paragraph raises the question of whether the
discretization procedures used in the derivation of the spin foams are
compatible with the expectation that one is approximating a
diffeomorphism invariant fundamental theory. More precisely, can one
tell whether the diffeomorphism invariance is broken or not by our
regularization procedure. This question is a quite important one and
has been one of the central concerns of the work led by B.~Dittrich
and collaborators for the last few years~\cite{Dittrich:2008va, Dittrich:2007wm,
  Dittrich:2008ar, Dittrich:2008pw, Bahr:2009mc}. There exist indeed
possible discretizations of a field theory which maintain the full
symmetry content of their continuum relatives. These regularizations
are called \emph{perfect actions} and even when they are difficult to
construct explicitly in general cases there are in principle methods
(based on renormalization group ideas) to approach them
\cite{Dittrich:2008pw, Bahr:2009ku} (see also~\cite{Bahr:2011uj} and
\cite{Bahr:2010cq} for more recent discussion and explicit
calculations in the perturbative context).

There are however indications that the discretization procedures used
generically do break the general covariance of gravity. In the simple
case of 3d gravity with cosmological constant this can be shown
explicitly at the classical level~\cite{Bahr:2009qc} as well as at the
quantum level~\cite{Perez:2010pm}. In the first reference it is shown
nonetheless that  there exist a well defined perfect action in this
case. At the quantum level the results of~\cite{Noui:2011im} indicate
a possible resolution of the anomaly issue. In four dimensions, this
was first validated by the results obtained by Gambini and Pullin et
al. in the so-called \emph{consistent discretization}
approach~\cite{Gambini:2005pg, Gambini:2005jm, Gambini:2005vn,
  Gambini:2004vz, Gambini:2004bm, DiBartolo:2004cg, DiBartolo:2004dn,
  Gambini:2004ew, Gambini:2003tn, Gambini:2001bd}. They study the
canonical formulation of theories defined on a lattice from the
onset. The consistent discretization approach provides a way to
analyzing the meaning of gauge symmetries directly \`a la Dirac. Their
results indicate that diff-invariance is indeed broken by the
discretization in the sense that there is no infinitesimal generator
of diffeomorphism. This is consistent with the covariant picture of
discrete symmetries above. In their formulation the canonical
equations of motion fix the value of what were Lagrange multipliers in
the continuum (e.g., lapse and shift). This is interpreted as a
breaking of diffeomorphism invariance; however, the solutions of the
multiplier equations are highly non unique. The ambiguity in selecting
a particular solution corresponds to the remnant diffeomorphism
invariance of the discrete theory.  More recently, the issue of the
breaking of the diffeomorphism symmetry by the regularizations used in
spin foams has been studied by the Dittrich group (see references
above). A possibility of using very simple models to study these
questions has been open in~\cite{Bahr:2011yc}.

An important question is whether the possible breaking of the
diffeomorphims gauge symmetry by the regularization can be made to
disappear in the continuum limit. The breaking of a gauge symmetry
implies the existence of a new host of degrees of freedom that might
be in conflict with the low energy interaction one is trying to
recover in that limit\epubtkFootnote{Breaking gauge symmetries comes
  hand-in-hand with introducing (unwanted) new degrees of freedom. For
  example, longitudinal photon modes would appear if $U(1)$ gauge
  symmetry is violated in electromagnetism; similar spurious modes
  come to life if diffeormorphism invariance is broken in gravity
  theories.}. In addition the breaking of non-compact symmetries by a
regulating structure poses serious difficulties for the low energy
regime even in the case of global symmetries~\cite{Collins:2004bp}
(see~\cite{Gambini:2011nx, } and \cite{Polchinski:2011za} for a recent
discussion). We mention this point here because it is in our opinion
of a major importance. Even though the question seems quite difficult
to address in the context of the present models at this stage it must
be kept in mind in search of opportunities of further insights.

However, the notion of anomaly freeness evoked at the beginning of
this section should be strengthened. In fact according to our
tentative definition, an anomaly free measure can be multiplied by any
gauge invariant function and yield a new anomaly free measure. This
kind of ambiguity is not wanted; however, it is in fact present in
most of the spin foam models defined so far: In standard QFT theory,
the formal (phase space) path integral measure in the continuum has a
unique meaning (up to a constant normalization) emerging from the
canonical formulation. Provided an appropriate gauge fixing, the
corresponding Dirac bracket determines the formal measure on the gauge
fixed constraint surface. In the discussion in Section~\ref{medida} we
have provided  references where this issue is analyzed.

\subsection{Discretization dependence}
\label{dd}

The spin foam models we have introduced so far are defined on a
fixed cellular decomposition of $\cal M$. This is to be
interpreted as an intermediate step toward the definition of the
theory. The discretization reduces the infinite dimensional
functional integral to a multiple integration over a finite number
of variables. This cutoff is reflected by the fact that only a
restrictive set of spin foams (spin network histories) is allowed
in the path integral: those that can be obtained by all possible
coloring of the underlying 2-complex. In addition it restricts
the number of possible 3-geometry states (spin network states) on
the boundary by fixing a finite underlying boundary graph. This
represents a truncation in the allowed fluctuations and the set of
states of the theory playing the role of a regulator.
However, the nature of this regulator is fundamentally different
from the standard concept in the background independent framework:
since geometry is encoded in the coloring (that can take any spin
values) the configurations involve fluctuations all the way to
Plank scale\epubtkFootnote{Changing the label of a face from $j$ to
$j+1$ amounts to changing an area eigenvalue by an amount of the
order of Planck length squared according to (\ref{aarreeaa}).}.
This scenario is different in lattice gauge theories where the
lattice introduces an effective UV cutoff given by the lattice
spacing. Transition amplitudes are however discretization
dependent now. A consistent definition of the path integral using
spin foams should include a prescription to eliminate this
discretization dependence.

A special case is that of topological theories such as gravity in
3 dimensions. In this case, one can define the sum over spin foams
with the aid of a fixed cellular decomposition $\Delta$  of the
manifold. Since the theory has no local excitations (no
gravitons), the result is independent of the chosen cellular
decomposition. A single discretization suffices to capture the
degrees of freedom of the topological theory.

In lattice gauge theory the solution to the problem is implemented
through the so-called continuum limit. In this case the existence
of a background geometry is crucial, since it allows one to define the
limit when the lattice constant (length of links) goes to zero. In
addition the possibility of working in the Euclidean regime allows
the implementation of statistical mechanical methods.

None of these structures are available in the background
independent context. The lattice (triangulation) contains only
topological information and there is no geometrical meaning
associated to its components. As we mentioned above this has the
novel consequence that the truncation cannot be regarded as an UV
cutoff as in the background dependent context. This in turn
represents a conceptual obstacle to the implementation of standard
techniques. Moreover, no Euclidean formulation seems meaningful in
a background independent scenario. New means to eliminate the
truncation introduced by the lattice need to be developed.
For a recent analysis where the difference between lattice regularization of background dependent versus
background independent  theories is carefully considered see \cite{Rovelli:2011eq}.
For a more general discussion see \cite{Rovelli:2011fk}.

This is a major issue where concrete results have not been
obtained so far beyond the topological case. Here we explain the
two main approaches to recover general covariance corresponding to
the realization of the notion of `summing over discretizations' of
 \cite{c2}.

\begin{itemize}
\item \emph{Refinement of the discretization:}

According to this idea topology is fixed by the simplicial
decomposition. The truncation in the number of degrees of freedom
should be removed by considering triangulations of increasing
number of simplexes for that fixed topology. The flow in the space
of possible triangulations is controlled by the Pachner moves. The
 idea is to take a limit in which the number of four
simplexes goes to infinity together with the number of tetrahedra
on the boundary. Given a $2$-complex ${\cal J}_2$ which is a
refinement of a  $2$-complex ${\cal J}_1$ then the set of all
possible spin foams defined on ${\cal J}_1$ is naturally contained
in those defined on ${\cal J}_2$ (taking into account the
equivalence relations for spin foams mentioned in the previous
section). The refinement process should also enlarge the space of
possible 3-geometry states (spin networks) on the boundary
recovering the full kinematical sector in the limit of infinite
refinements. An example where this procedure is well defined is 
Zapata's treatment of the Turaev--Viro model
 \cite{za1}. The key point in this case is that amplitudes are
independent of the discretization (due to the topological
character of the theory) so that the refinement limit is trivial.
In the general case the definition of the refinement limit
has been recently been studied and formalized in \cite{Rovelli:2010qx}.
There is no evidence as to whether the amplitudes of any of the present models
for 4d converges in such limit.

It has been often emphasized that such refinement limit may be studied from the Wilsonian renormalization view point. 
In the past a renormalization approach for
spin foams have been proposed by Markopoulou \cite{fot4,fot5}.
Also Oeckl \cite{Oeckl:2002ia} has studied the issue of 
renormalization in the context of spin foam models containing 
a coupling parameter. These models include generalized covariant gauge theories \cite{Oeckl:2001wm, Pfeiffer:2001ig, Oeckl:2000hs}, 
the Reisenberger model, and the so called \emph{interpolating model} 
(defined by Oeckl). The latter is given by a one-parameter family of  models that interpolate 
between the trivial BF topological model and the Barrett-Crane model according to the 
value of a `coupling constant'. Qualitative aspects of the renormalization 
groupoid flow of the couplings are studied in the various models.

\item\emph{Spin foams as Feynman diagrams:}

This idea has been motivated by the generalized matrix models of
Boulatov and Ooguri  \cite{bu,oo}. The fundamental observation is
that spin foams admit a dual formulation in terms of a field
theory over a group manifold  \cite{fre2,reis1,reis2}. The duality
holds in the sense that spin foam amplitudes correspond to Feynman
diagram amplitudes of the GFT. The perturbative Feynman expansion
of the GFT (expansion in a fiducial coupling constant $\lambda$)
provides a definition of \emph{sum over} discretizations which is
fully combinatorial and hence independent of any manifold
structure\footnote{This is more than a `sum over topologies' as
many of the 2-complex appearing in the perturbative expansion
cannot be associated to any manifold  \cite{pietri1}.}. The latter
is most appealing feature of this approach. For a recent reference proposing a GFT for the EPRL type of models see \cite{Baratin:2011hp}.

However, the convergence issues clearly become  more involved. The perturbative series
are generically divergent. This is not necessarily a definite obstruction as
divergent series can often be given an asymptotic meaning and provide physical 
information. Moreover, there are standard techniques that can allow to `re-sum' a
divergent series in order to obtain non perturbative information \cite{Rivasseau:1991ub}. 
 Freidel and Louapre \cite{fre10} have shown that this is indeed
possible for certain GFT's in three dimensions.
Other possibilities have been proposed in \cite{reis1}. 

 Diffeomorphism equivalent configurations (in the
discrete sense described above) appear at all orders in the
perturbation series\epubtkFootnote{The GFT formulation is clearly non
trivial already in the case of topological theories. There has
been attempts to make sense of the GFT formulation dual to BF
theories in lower dimensions \cite{a5}.}. From this perspective
(and leaving aside the issue of convergence) the sum of different
order amplitudes corresponding to equivalent spin foams should be interpreted 
as the definition of the physical amplitude of that particular spin foam. The discussion of the
previous section does not apply in the GFT formulation, i.e.,
there is no need for gauge fixing.

The GFT formulation could resolve by definition the two fundamental conceptual
problems of the spin foam approach: diffeomorphism gauge symmetry and
discretization dependence. The difficulties are shifted to the question of the
physical role of $\lambda$ and the convergence of the corresponding perturbative series.
A prescription that avoids this last issue is the radical proposal of Freidel \cite{Freidel:2005qe}
where the tree-level GFT amplitudes are use to define the physical inner product 
of quantum gravity.

In three dimensions this idea has been studied in more detail. 
In \cite{Magnen:2009at} scaling properties of the modification of the Boulatov group field theory 
introduced in \cite{fre10} was studied in detail. In a further modification of the previous model (known as coloured tensor models \cite{Gurau:2009tw}) 
new techniques based on a suitable $1/N$ expansion imply that amplitudes are dominated by spherical topology  
\cite{Gurau:2010ba}; moreover, it seem possible that the continuum limit  might be critical as in certain matrix models
\cite{Gurau:2011tj, Bonzom:2011zz, Gurau:2011xq, Gurau:2011aq, Ryan:2011qm}. However, it is not yet clear if there is a sense 
in which these models correspond to a physical theory. The appealing possible interpretation of the models is that they correspond to a
formulation of 3d quantum gravity including a dynamical topology. No much is known about gravity models in four dimensions at this stage.

\end{itemize}

Finally, discretization independence is not only a problem of spin foams but also of Regge gravity. Even linearized classical Regge calculus is not discretization independent in 4d as shown in \cite{Dittrich:2011vz} where a general discussion of the issues raised in this section are considered.
It is also natural to expect the issue of discretization independence to be connected to the issue of diffeomorphism symmetry discussed in the previous section. There is evidence \cite{Bahr:2010cq, Dittrich:2011zh, Dittrich:2012qb} that implementation of diffeomorphism symmetry into the models will lead to discretization independence. 

\subsection{Lorentz invariance in the effective low energy regime}\label{clew}

Finally a very important question for loop quantum gravity and spin foams (closely related to the consistency of their low energy limit)
is that of the fate of Lorentz symmetry in the low energy regime. This question  has received lots of attention in recent years as it seems to provide
an open window of opportunity for observations of quantum gravity effects (see \cite{Mattingly:2005re, Jacobson:2005bg} and references therein).
In the context of loop quantum gravity a systematic first-principles analysis of such properties of the low energy regime 
remains open. This is mainly  due to the presently unsolved  difficulties associated with a quantitative dynamical description of
physical states. The spin foam approach discussed in this review is meant to address the dynamical question 
and the definition of the new models may provide the necessary insights to tackle the relevant aspects for the question
of Lorentz invariance at low energies. This question is perhaps within reach at the present stage of development of the theory, so it deserves 
all the attention in future investigations.

In the absence of detailed formulations, certain early pioneering model calculations \cite{Gambini:1998it} (see also general treatment \cite{Sahlmann:2002qj, Sahlmann:2002qk}) indicated the possibility for certain Lorentz invariance violating (LIV) effects 
associated to a quantum gravity granularity of spacetime measured at rest by a preferred frame of observers.
The effects take the form of corrections to the effective Lagrangian with LIV terms of dimension 5 or higher
that are suppressed by negative powers of the Planck mass. This makes them in principle negligible at low energies so that the potential quantum gravity effects 
would be observed as corrections to the standard QFT well tested physics. However, it was soon realised \cite{Collins:2004bp} that
the framework of effective field theories severely restricts the possibility that LIV remains small when interactions in QFT are taken into account. In fact
dimension 5 or higher LIV terms will generically (in the absence of some fundamental protecting mechanism) 
generate dimension 4 an lower LIV  terms (with factors in front that are at best quadratic in standard model coupling constants). Such unsuppressed contributions are in flagrant conflict with the observed Lorentz invariance of particle physics at low energies.

Therefore, the early calculations in LQG cited above need to be revised under the light of the strong constraints on the way LIV can arise in quantum gravity.
A logical possibility is that these LIV terms are there thanks to the  protecting effect of some custodial mechanism restricting the LIV  to dimension 5 and higher operators \footnote{For an example of protecting symmetry see \cite{Cohen:2006ky}; however, notice that 
the non-compact nature of the protecting symmetry in this case makes it seemingly unsuitable for a sensible possibility in LQG.}. For example such possibility  has been considered in \cite{Gambini:2011nx} where it is shown that
the large LIV effects can be avoided in the context of Euclidean field theory (the protecting mechanism being here the trading of the non-compact Poincare group by the Euclidean group). This example is interesting but unfortunately does not provide an answer to  the question for the physical Lorentzian theory (see discussion in \cite{Collins:2006bw} and more recently
\cite{Polchinski:2011za}).  

There is thus a challenge for LQG and spin foams consisting of showing that the low energy limit of the theory is compatible with the observed low energy
LI. This can be put in the form of two conservative possibilities. The description of the low energy physics in terms of
\begin{enumerate} 
\item an effective action with no LIV terms, or 
\item an effective action where LIV terms appear starting from dimension 5 operators on due to some fundamental protecting mechanism in LQG 
for such structure to be stable under radiative corrections.
\end{enumerate} 
My view, based on the ignorance of a suitable protecting mechanism in LQG,  is that only the first 
possibility is sensible. This would mean that the way the fundamental Planckian discreteness of LQG
must enter physics is through reference frame independent phenomena (as mass enters a Klein--Gordon theory).
In my opinion the main ingredient that must be taken into account in considering this issue is the
construction of the low energy limit of quantum gravity in terms of physical or gauge invariant notions (see discussion in Section \ref{gaugy}). The LIV effects obtained in LQG so far promote by assumption the kinematical discreteness of quantum geometry 
to discreteness of the physical background in a literal fashion (on the one hand discreteness of geometric can coexist with Lorentz invariance \cite{Rovelli:2002vp}, on the other hand, at the level of gauge invariant observables some discrete aspect could disappear \cite{Dittrich:2007th}). The limitations of such naive view point on kinematical discreteness versus physical properties in the theory is, in my view, best manifested in the the setting of pure 3d gravity as discussed above. 
 
For completeness we add that in addition to the above (conservative) possibilities, some less standard scenario could 
be realized. This has been the view point of some part of the quantum gravity community. For a recent 
proposal of a completely new theoretical framework see \cite{AmelinoCamelia:2011bm} and references therein.

%


\section{Acknowledgements}
\label{section:acknowledgements}

I would like to thank the help of many people in the field that have
helped me in various ways. I am grateful to  Benjamin Bahr, Eugenio Bianchi, Bianca Dittrich, Carlo
Rovelli and Simone Speziale for the many discussions on aspects and
details of the recent literature. Many detailed calculations that
contributed to the presentation of the new models in this review where
done in collaboration with Mercedes Vel\'azquez to whom I would like to
express my gratitude. I would also like to thank You Ding, Florian
Conrady, Laurent Freidel, Muxin Han, Merced Montesinos, Jose Antonio Zapata for help and
valuable interaction.



\newpage
\bibliography{refs}

\end{document}